%% file: EXOT-2016-08-PAPER.tex
\pdfoutput=1

\newcommand*{\ATLASLATEXPATH}{latex/}

\documentclass[atlasdraft=false,cernpreprint, texlive=2016, UKenglish]{\ATLASLATEXPATH atlasdoc}

\usepackage{\ATLASLATEXPATH atlaspackage}
\usepackage{\ATLASLATEXPATH atlasbiblatex}
\usepackage{\ATLASLATEXPATH atlasphysics}

\usepackage{multirow}
\usepackage{verbatim}
\usepackage{pdflscape}
\usepackage{array,ragged2e}
\usepackage{makecell}
\usepackage{siunitx}
\usepackage{lineno}

\graphicspath{{logos/}{figures/}}

\usepackage{EXOT-2016-08-PAPER-defs}

\input{EXOT-2016-08-PAPER-metadata}

\hypersetup{pdftitle={ATLAS document},pdfauthor={The ATLAS Collaboration}}

\begin{document}

\maketitle

\input{LQIntro.tex}

\input{Detector.tex}

\input{FirstSecondGen.tex}

\input{Conclusions.tex}

\clearpage
\appendix
\part*{Appendix}
\addcontentsline{toc}{part}{Appendix}
\input{Appendix}

\section*{Acknowledgements}

\input{acknowledgements/Acknowledgements}

\printbibliography

\clearpage 
\input{atlas_authlist}

\end{document}

%% file: EXOT-2016-08-PAPER-metadata.tex
\AtlasTitle{Searches for scalar leptoquarks and differential cross-section measurements in
dilepton--dijet events in proton--proton collisions at a centre-of-mass energy of  $\boldmath{\sqrt{s}}$ = 13\,\TeV\  with the ATLAS experiment}

\AtlasAbstract{Searches for scalar leptoquarks pair-produced in proton--proton collisions at
 $\sqrt{s}=13$~\TeV\ at the Large Hadron Collider are performed by the ATLAS experiment. A data set corresponding to an 
integrated luminosity of
36.1~fb$^{-1}$ is used. Final states containing two electrons or two muons and two or more jets are studied, as are states with one electron or muon, missing transverse momentum and two or more jets. No statistically significant
excess above the Standard Model expectation is observed.
The observed and expected lower limits on the leptoquark
mass at 95\% confidence level extend up to 1.29~\TeV\ and 1.23~\TeV\ for first- and second-generation leptoquarks, respectively, as postulated in the  minimal 
Buchm\"uller--R\"uckl--Wyler model, assuming a
branching ratio into a charged lepton and a quark of 50\%. 
In addition, measurements of particle-level fiducial and differential
cross sections are presented for the \Zee, \Zmm and \ttbar\ processes in several regions related to the search control 
regions.
Predictions from a range of generators are compared with the measurements, and good agreement is seen for many of the observables.  
However, the predictions for the $Z\rightarrow\ell\ell$ measurements in observables sensitive to jet energies disagree with the data. 
}

 \author{The ATLAS Collaboration}

\AtlasRefCode{EXOT-2016-08}

\PreprintIdNumber{CERN-EP-2018-262}

\arXivId{1902.00377}

\HepDataRecord{https://www.hepdata.net/record/83968}

\AtlasJournal{EPJC}
\AtlasJournalRef{Eur. Phys. J. C 79 (2019) 733}
\AtlasDOI{10.1140/epjc/s10052-019-7181-x}

%% file: LQIntro.tex
\section{Introduction}

Leptoquarks (LQs) are features of a number of extensions of the Standard 
Model 
(SM)~\cite{BSM,techni2,techni3,string,comp,pati_salam_colour,georgi_glashow_unification,Diaz:2017lit} 
and 
may 
provide an explanation for the similarities between the quark and lepton 
sectors in the SM. They also appear in models addressing 
some of the  
recent $b$-flavour 
anomalies~\cite{Gripaios:2014tna,Cline:2017aed,DiLuzio:2017chi}. 
They are 
colour-triplet bosons with fractional electric charge and possess non-zero baryon and lepton 
number \cite{Schrempp:1984nj}. Scalar and vector LQs have been proposed and are expected to decay directly into lepton--quark pairs. The lepton can be either electrically charged or neutral.
 
A single Yukawa coupling, $\lambda_{\mathrm{LQ} \rightarrow \ell q}$,  
determines the coupling strength between scalar LQs and the lepton--quark pair \cite{LQlq_coupling}. Two 
additional coupling constants due to magnetic moment and 
electric quadrupole moment interactions are needed for 
vector LQs~\cite{Hewett:1997ce}.

\begin{figure}[h!]
\centering
\subfloat[]{
    \includegraphics[width=0.39\textwidth]{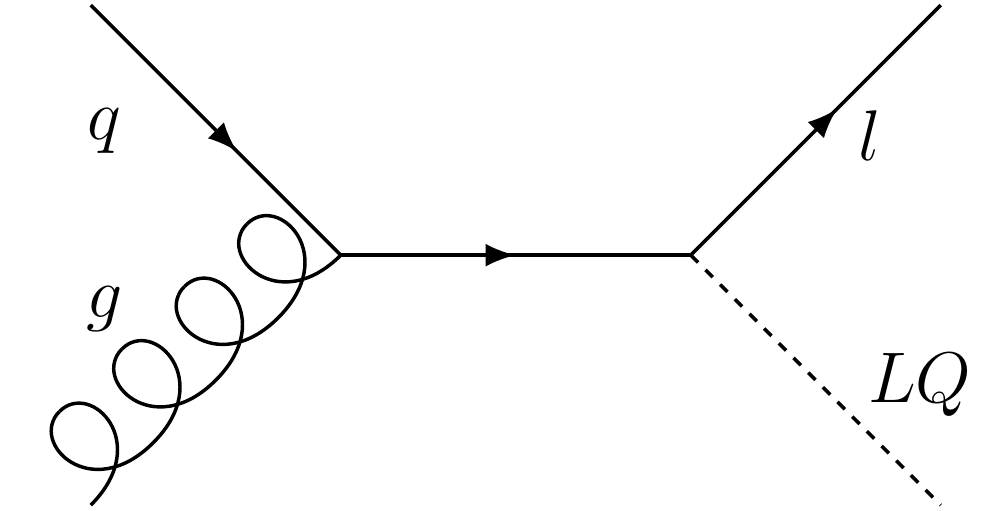}}
\hspace{2.0cm}
\subfloat[]{
    \includegraphics[width=0.38\textwidth]{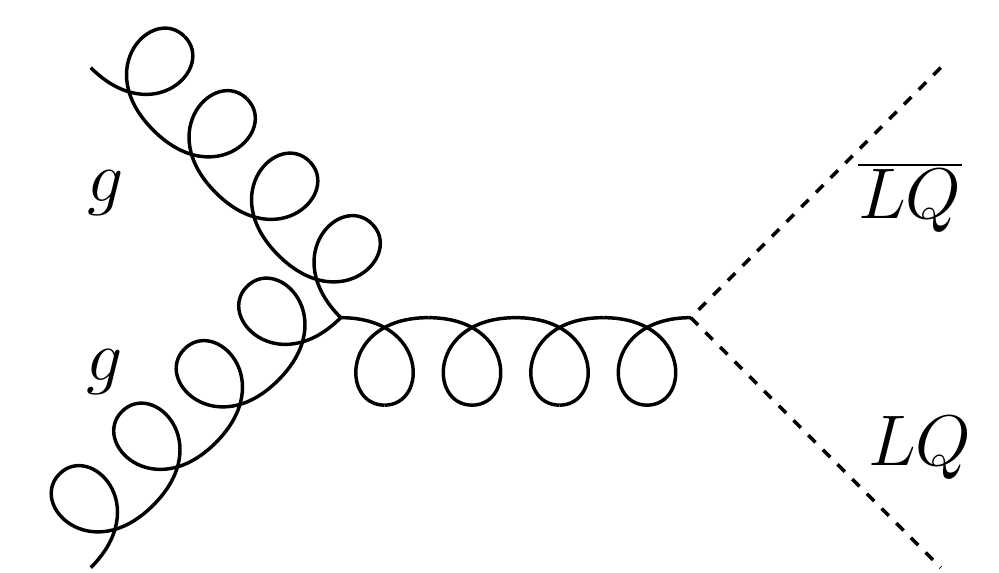}}
\\
\caption{Diagrams for representative (a) single  and  (b) pair production of LQs.}
  \label{fig:FeynmanDiagrams}
\end{figure}

LQs can be produced both singly and 
in pairs in proton--proton ($pp$)  
interactions; diagrams showing representative single- and pair-production processes are shown in Figure~\ref{fig:FeynmanDiagrams}. 
The single-LQ production cross-section 
depends on $\lambda_{\mathrm{LQ} \rightarrow \ell q}$ whereas the 
pair-production cross-section is largely insensitive to this coupling.  For \textit{pp} interactions 
with a centre-of-mass energy $\sqrt{s}$ = 13\,\TeV, gluon fusion represents the 
dominant pair-production mechanism for LQ masses below around 1~\TeV. The 
contribution of the $q\bar{q}$-annihilation process rises with LQ mass.
Only scalar LQ production is considered in this paper because this is less model dependent than vector LQ production. The production of 
vector LQs depends on additional parameters and a full interpretation in such models is beyond the scope of this analysis. The results 
obtained here can, however, be regarded as conservative estimates of limits on vector LQ production, since the production cross-section 
for vector LQs is typically much larger than for scalar LQs, while the kinematic properties used to search for their signature are very 
similar for both spin hypotheses~\cite{Diaz:2017lit}.

The benchmark signal model used in this paper is the minimal 
Buchm\"uller--R\"uckl--Wyler (BRW) model \cite{Buchmuller:1986zs}. 
This model imposes a number of constraints on the properties of the LQs. Couplings are purely chiral and LQs belong to three families,  
corresponding to the three SM generations, such that only leptons and quarks within a given generation can interact. 
The requirement of same-generation interactions excludes flavour-changing neutral currents (FCNC) \cite{Mitsou:2004hm}. 
The branching ratio (BR) of a LQ into different states is taken as a free parameter. 
In this paper, $\beta$ denotes the BR for a LQ decay into a charged lepton and quark, $\LQ \rightarrow \ell^{\pm}q$. 
The BR for a LQ decay into a neutrino and quark is $(1 - \beta)$. 

This paper presents a search for LQs pair-produced in $pp$ collisions at $\sqrt{s}$ = 13~\TeV, performed by the ATLAS experiment at the Large 
Hadron Collider (LHC). 
In previous searches for pair-produced LQs made by 
ATLAS~\cite{EXOT-2010-08,EXOT-2011-17,EXOT-2011-26,EXOT-2011-26,EXOT-2014-03,EXOT-2015-19}, most recently with 3.2~fb$^{-1}$  of
data collected at $\sqrt{s}$ = 13~\TeV, 
the existence of 
scalar LQs with
 masses below 1100 (900)~\GeV\ for first-generation LQs at  $\beta$  = 1 (0.5) and, for second-generation LQs, with masses below 1050 (830)~\GeV\
 at  $\beta$  = 1 (0.5) was excluded~\cite{EXOT-2015-19}.   
Searches have also been made by the CMS Collaboration~\cite{ 
CMS-EXO-10-005,CMS-EXO-10-006,CMS-EXO-10-007,CMS-EXO-11-028,CMS-EXO-12-041,CMS-EXO-17-009,CMS-EXO-17-003} using the same model assumptions as in this work.
That experiment found that scalar
LQs with masses below 1435 (1270)~\GeV\ for first-generation LQs at  $\beta$  = 1 (0.5) and below 1530 (1285)~\GeV\
for second-generation LQs at  $\beta$  = 1 (0.5) are excluded at
95\% confidence level (CL) using a data sample of 35.9~fb$^{-1}$ collected at
 $\sqrt{s}$ = 13~\TeV~\cite{CMS-EXO-17-009,CMS-EXO-17-003}. An overview of limits on
LQ production and masses can be found in Ref.~\cite{Patrignani:2016xqp}.

The decay of pair-produced LQs can lead to final states containing 
a pair of charged leptons and jets, or  
one charged lepton, a neutrino and jets. 
For down-type leptoquarks with a charge of $\frac{1}{3}e$, assumed here as a benchmark, this can be written more explicitly as
$\LQ\overline{\LQ}\longrightarrow\ell^{-}q\ell^{+}\bar{q}$ for the dilepton channel, and 
$\LQ\overline{\LQ}\longrightarrow\ell^{-}q\bar{\nu}\bar{q'}$ or $\LQ\overline{\LQ}\longrightarrow\nu q\ell^{+}\bar{q'}$ for the lepton-neutrino channel, respectively. 
The four search 
channels in this paper correspond to the above decays. The \eejj (\mumujj) 
channel comprises exactly two electrons (muons) and at least two jets, and 
the \enujj (\munujj) channel requires exactly one electron (muon), missing momentum in the transverse plane 
and at least two jets.   The electron and muon channels are treated 
separately as they correspond to LQs of different generations. 
The dilepton and lepton--neutrino channels are combined in order to increase the 
sensitivity to the parameter of the BRW model,  $\beta$. 
While the main target of this paper are first- and second-generation LQs, there are also 
dedicated searches looking for 3rd generation LQ 
production~\cite{EXOT-2014-03,EXOT-2017-30,CMS-EXO-17-016,CMS-EXO-17-029,CMS-B2G-16-027,CMS-B2G-16-028}.

The main background processes in this search are $Z/\gamma^*$+jets, 
$W$+jets and 
\ttbar{} production. The shapes of these background distributions are 
estimated from Monte Carlo (MC) simulations while their 
normalisation is obtained from a simultaneous background-only profile 
likelihood fit to three data control regions (CRs). The CRs are defined 
with specific selections such that they are orthogonal to the signal 
regions (SRs), free from any LQ signal of interest, and enriched in
their relevant background process.  
For the particular benchmark model chosen, further signal and background discrimination could be achieved by exploiting 
flavour tagging in order to reject events with $b$-hadrons in the  search region. This is not used, in order to maintain 
sensitivity to a $\LQ$ coupling to a $b$-quark.   The shapes and
normalisations of other (subdominant) backgrounds are estimated from MC
simulation. Background contributions from misidentified leptons are also sub-dominant, and are estimated from data when their 
contribution is significant.

Several variables that provide separation between signal and background processes are combined into
 a single discriminant using a boosted decision tree (BDT) method~\cite{Speckmayer:2010zz}. 
For each LQ-mass hypothesis and for each lepton flavour, two BDTs are used: one for the dilepton channel and one for the 
lepton--neutrino channel.  
The BDT output score is binned and the signal is accumulated at high score values.  
A single bin at a high score defines the SR for the respective mass hypothesis. The bin range is chosen such that a 
sensitivity estimator based on the expected number of 
signal and background events is maximised.  

A  profile likelihood fit is performed simultaneously to both 
the dilepton and lepton-neutrino channel for each lepton flavour to extract a limit on the signal yield. 
This is done for different assumptions of LQ mass and branching ratio  $\beta$, resulting in 
exclusion bounds in terms of these parameters.

In addition to the search results, this paper presents a set of particle-level fiducial and differential 
cross-section measurements in six dilepton--dijet regions which are related to the analysis CRs.
The measurements are made for the 
dominant process in each region ($Z$+jets or \ttbar) exclusively. 
The motivation for making these measurements is as follows.
Search CRs are typically in kinematic regions which are not directly covered by dedicated measurements of SM processes. 
For instance, the ATLAS measurement of $Z+\textrm{jets}$ production described in Ref.~\cite{STDM-2016-01} has much looser selections on the transverse momentum 
(\pT) of leptons and jets than the region that includes $Z$ bosons for this search. 
Event generators are tuned and validated to ensure that they agree with the available SM data. However, in some cases, although good agreement is seen for the regions 
covered by dedicated SM measurements, significant disagreements remain possible in more extreme regions where no measurements are available to validate the 
generator predictions. Indeed, although no disagreement was seen when validating generated $Z$+jets samples against various 
SM measurements~\cite{ATL-PHYS-PUB-2017-006}, significant disagreement was observed in the $Z$+jets CR
for this search. Extracting particle-level cross-sections in these CRs and related more extreme regions of phase space can therefore provide 
complementary information to validate the performance of event generators. Such measurements may lead to improved background modelling in future searches with final states using similar CRs (see for example Ref.~\cite{EXOT-2016-29}).

The paper is organised as follows. 
Section~\ref{sec:atlas} gives an overview of the ATLAS detector, while Section~\ref{sec:DataMC} summarises information on the data set and simulations used. 
The physics objects used in the analysis are defined in Section~\ref{sec:ObjSel}. 
The background estimation and CRs used in the process are described in Section~\ref{sec::CtrlRegion}, and 
Section~\ref{sec:unfoldingmethod} provides details on the cross-section extraction. 
Section~\ref{sec:BDT&SR} gives an overview over the multivariate analysis employed for the search and the SRs defined with it. 
The systematic uncertainties are summarised in Section~\ref{sec:Systematics}, and the statistical procedure employed for the search is described in Section~\ref{sec:method}. 
Results from both the measurement and the search are given in Section~\ref{secl:FResults:StatAnalysis}. 
Section~\ref{sec:conclusion} concludes the paper.

%% file: Detector.tex
\section{The ATLAS detector}
\label{sec:atlas}

The ATLAS experiment \cite{PERF-2007-01} is a multipurpose  detector with a forward--backward symmetric cylindrical geometry and nearly 4$\pi$ coverage in solid 
angle.\footnote{ATLAS uses a right-handed coordinate system with its origin at the nominal interaction point (IP)
in the centre of the detector and the $z$-axis along the beam pipe. The $x$-axis points from the IP to the centre of the LHC
ring, and the $y$-axis points upward. Cylindrical coordinates $(r,\phi)$ are used in the transverse plane, $\phi$ being the
azimuthal angle around the beam pipe. The pseudorapidity is defined in terms of the polar angle $\theta$ as $\eta=-\ln\tan(\theta/2)$.
The quantity $\Delta R = \sqrt{(\Delta \phi)^2 + (\Delta \eta)^2}$ is used to define a cone size. }
The three major subcomponents are the tracking detector, the calorimeter and the muon spectrometer.  Charged-particle tracks and vertices are reconstructed by the inner detector (ID) tracking system, comprising silicon pixel and microstrip detectors covering the pseudorapidity
range $|\eta|$~$<$~2.5, and a transition radiation straw-tube tracker that covers $|\eta|$~$<$~2.0 and provides 
electron identification.
The ID is immersed in a homogeneous 2\,T magnetic field 
provided by a solenoid.  Electron, photon, and jet  
energies are measured with sampling calorimeters.  The 
calorimeter system covers a pseudorapidity range of $|\eta|$~$<$~4.9. 
Within the region $|\eta|$~$<$~3.2, barrel and endcap high-granularity lead/liquid
argon (LAr) electromagnetic calorimeters are deployed, with an additional 
thin LAr presampler covering $|\eta|$~$<$~1.8,
 to correct for energy loss in material upstream of the calorimeters. Hadronic calorimetry is provided by a 
steel/scintillator-tile calorimeter, segmented into three 
barrel structures within $|\eta|$~$<$~1.7, and two 
copper/LAr hadronic endcap calorimeters. The forward region 
(3.1~$<$~$|\eta|$~$<$~4.9) is instrumented with a LAr 
calorimeter with copper (electromagnetic) and tungsten 
(hadronic) absorbers.  Surrounding the calorimeters is a 
muon spectrometer (MS) with air-core toroid magnets to provide 
precise muon identification
and momentum measurements, consisting of a 
system of precision tracking chambers providing coverage over $|\eta|$~$<$~2.7, and detectors with 
triggering capabilities over $|\eta|$~$<$~2.4.  A two-level
trigger system \cite{TRIG-2016-01}, the first level using custom hardware and followed by a software-based level, is used to
reduce the event rate to a maximum of around 1\,kHz for offline storage.

%% file: FirstSecondGen.tex
\input{DataBG}

\section{Object definition and event pre-selection}
\label{sec:ObjSel}

Electrons are reconstructed from ID tracks which are matched to energy clusters found in the electromagnetic calorimeter. 
The reconstruction efficiency is higher than 97\% for candidates with \pt{} greater than 30\,\GeV. 
An object is identified as an electron following 
requirements made on the quality of the associated track, shower shapes exploiting the longitudinal segmentation of the electromagnetic calorimeter, leakage into the hadronic calorimeter, the quality of the 
track-to-cluster matching and measurements of transition radiation made with the TRT~\cite{ATLAS-CONF-2016-024}. The transverse energy of the electron candidates in the $eejj$ 
($e\nu jj$) 
channel must exceed $40${} ($65$)~\GeV. Only electron candidates in the pseudorapidity region  
$|\eta|<2.47$ and excluding the transition region between the barrel  and endcap EM  
calorimeters ($1.37<|\eta|<1.52$) are used.  
The electron identification uses a multivariate 
technique to define different 
working points of selection efficiency. In the dielectron channel, the `medium' identification working point with an efficiency above 90\% for the
candidates considered in this analysis is used. To achieve better rejection against jets misidentified as electrons, a tighter selection is used 
in the electron--neutrino   channel.
For the electron--neutrino  channel, the `tight' identification working point is used with an  efficiency of 85\% or more for candidates with \pt{} above 65\,\GeV. 
The electrons are required to be isolated, using both calorimeter- and track-based criteria such that the isolation efficiency is 98\%. 
Finally, the electron candidates have to be compatible with originating from the primary vertex.

Muon tracks are reconstructed independently in the ID and the MS~\cite{PERF-2015-10}. Tracks are required to have a minimum number of hits in each 
system, and must be compatible in terms of geometrical and momentum matching. 
Information from both the ID and MS is used in a combined fit to refine the measurement of the momentum of each muon over $|\eta| <2.5$~\cite{PERF-2014-05}.  
The efficiency for reconstructing muons is 98\%. As for the electron channels, muon candidates  are required to 
have \pt~$>$~40~\GeV\ 
and \pt~$>$~65~\GeV\ for the \mumujj and \munujj searches respectively. 
They are further required to be compatible with originating from the primary vertex.
A track-based isolation requirement is applied to the muons, yielding a selection efficiency of 99\%.

Jets are reconstructed using the anti-$k_{t}$ algorithm
\cite{AntiKt} with a radius parameter $R=0.4$ from topological clusters of 
calorimeter cells which are noise-suppressed and 
calibrated to the electromagnetic scale. They 
are calibrated using energy- and $\eta$-dependent
correction factors derived from simulation and with residual corrections 
from {in situ} measurements~\cite{PERF-2016-04}. 
The jets must satisfy $\pt$~>~$60$~\GeV\ and $|\eta|$~<~$2.5$. 
Additional jet quality criteria are also applied to remove 
fake jets caused by detector effects~\cite{ATLAS-CONF-2015-029}. 
Furthermore, to eliminate jets containing a large energy contribution from pile-up, jets are tested 
for compatibility with
the hard scatter vertex with a jet vertex tagger discriminant, utilising information from the ID
tracks associated with the jet~\cite{PERF-2014-03}.

Jets containing $b$-hadrons ($b$-jets) are identified with an algorithm which is based on multivariate
techniques. The algorithm combines information from the impact parameters of displaced tracks and from topological
properties of secondary and tertiary decay vertices reconstructed within the jet. 
A working point at which the $b$-tagging efficiency is around 77\% for jets 
originating from a $b$-quark is chosen, 
as determined with a MC simulation of 
\ttbar{} processes~\cite{PERF-2012-04,ATL-PHYS-PUB-2016-012}. Scale factors needed to match the MC performance to 
data are compatible with unity~\cite{Aaboud:2018xwy}.

The missing transverse momentum is
defined as the negative vector transverse momentum sum of the
reconstructed and calibrated physics objects, plus an additional soft 
term~\cite{PERF-2014-04,ATL-PHYS-PUB-2015-027}. The soft term is built
from tracks that are not associated with any reconstructed electron, muon
or jet, but which are associated
with the primary vertex. The absolute value of the missing 
transverse momentum is denoted by \met{}. In the \lnujj{} channel, an  
\met{}-related quantity, 
calculated 
as
$S=\met/\sqrt{\jetonept+\jettwopt+\ptl}$ is used to further reduce
backgrounds from objects wrongly reconstructed as leptons. 
Here, \jetonept{} (\jettwopt) is the \pt{} of the leading (subleading) jet and \ptl{} is the \pt{} of the charged lepton.

Ambiguities in the object identification which arise during reconstruction, i.e.\ when a reconstructed object
can match multiple object hypotheses (electron, muon, jet), are resolved in several steps. 
First, electrons are removed if they share a track with a muon. 
An algorithm is also used to remove jet--lepton ambiguities based on the proximity of identified leptons and jets.

For the dilepton channel, events are selected such that they contain at least two jets and two same-flavour leptons fulfilling the requirements above for the respective flavour. 
Selected lepton-neutrino events have to contain at least two jets, one lepton, and  a missing transverse energy of more than 40\,GeV.

\section{Control regions and background estimation}
\label{sec::CtrlRegion}

The major SM background processes to the LQ signal correspond to the production of (off-shell)
$Z/\gamma^*$+jets, $W$+jets and \ttbar{} 
events in which at least one top quark decays 
leptonically. 
These backgrounds are determined using data in selected CRs, as described below. 
Subdominant contributions arising from diboson, single-top, $W\rightarrow\tau\nu$ and $Z\rightarrow\tau\tau$ production  are estimated entirely from simulation. 
There are also background contributions due to objects being misidentified as leptons or non-prompt 
leptons that are   produced in the decay of hadrons. 
These backgrounds are collectively referred to as the fake (lepton) background and are estimated in a 
data-driven way where relevant.

Three CRs, independent of the SRs used for the leptoquark search (see Section ~\ref{sec:BDT&SR}),  in 
which any high-mass LQ signal contributions 
are expected to 
be negligible  
are defined in order to estimate the main backgrounds. 
First, the $Z/\gamma^*$+jets control region ($Z$ CR) in the dilepton channels is defined by restricting the dilepton invariant mass, \mll{}, to values between 70\,\GeV\ and 110\,\GeV.
These CRs contain samples of 
$Z/\gamma^*$+jets events of purity 92\% (94\%) for the electron (muon) channel.  
Next, the $W$ control region ($W$ CR) in the  lepton--neutrino channels requires 
the transverse mass\footnote{The transverse mass is defined as $\mt=\sqrt{2\cdot
p_{\text{T}}^{\ell}\cdot\met\cdot(1-\cos(\Delta\phi(\ell,\met)))}$ where $\Delta\phi(\ell,\met)$ is the azimuthal separation between the charged lepton and the \met{}.}, \mt{}, to be between  
40\,\GeV\ and 
130\,\GeV. In addition, \met{} has to be greater than 40\,\GeV\ and  
the observable $S$ greater than 4. Events that contain one 
or more $b$-jets are vetoed to reduce the \ttbar{} contribution. The purity of the $W$ CR is 74\% (80\%) for the electron (muon) channel.

\begin{figure}
  \centering
\subfloat[]{
    \includegraphics[width=0.42\textwidth]{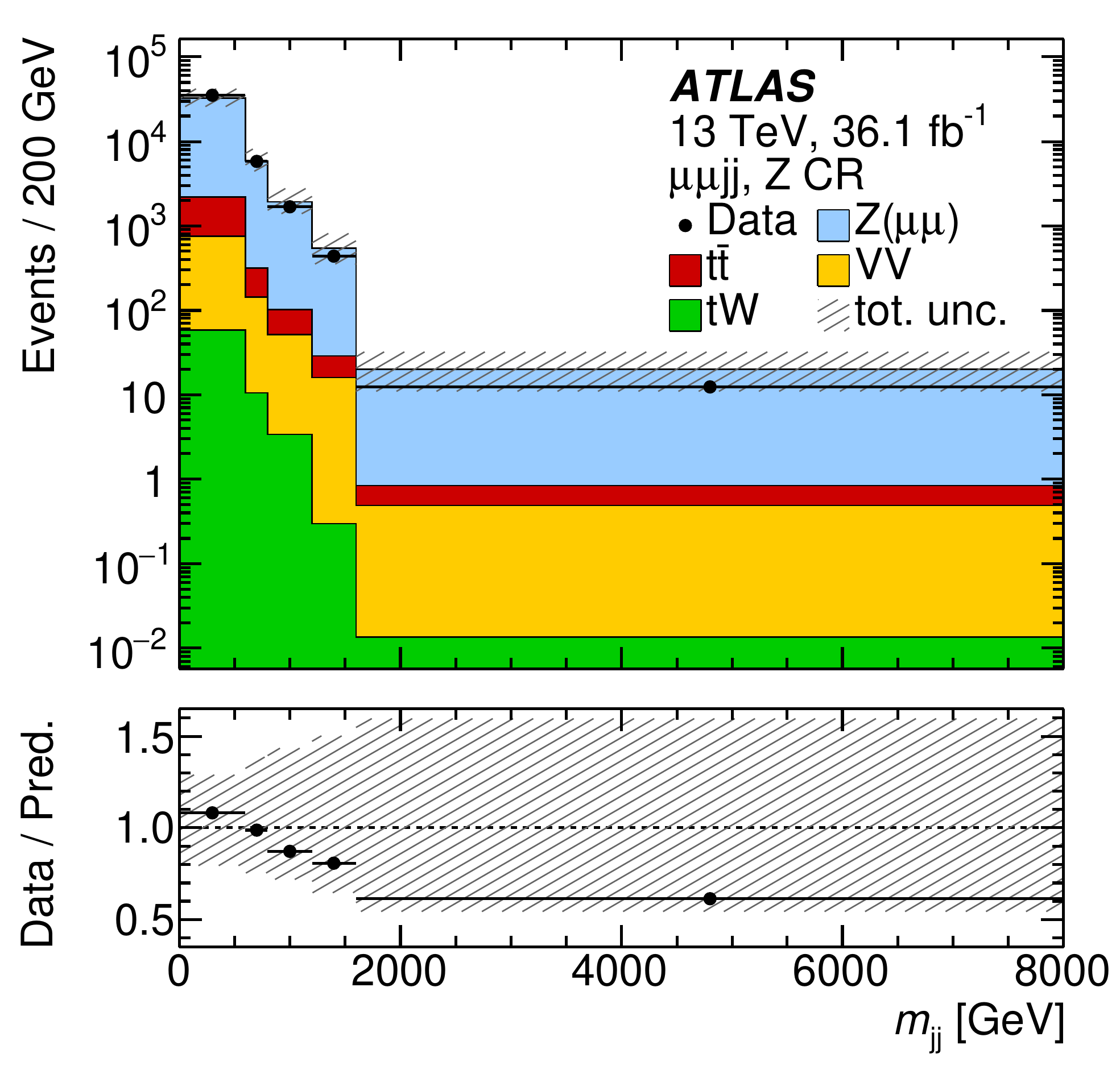} }
\subfloat[]{
    \includegraphics[width=0.42\textwidth]{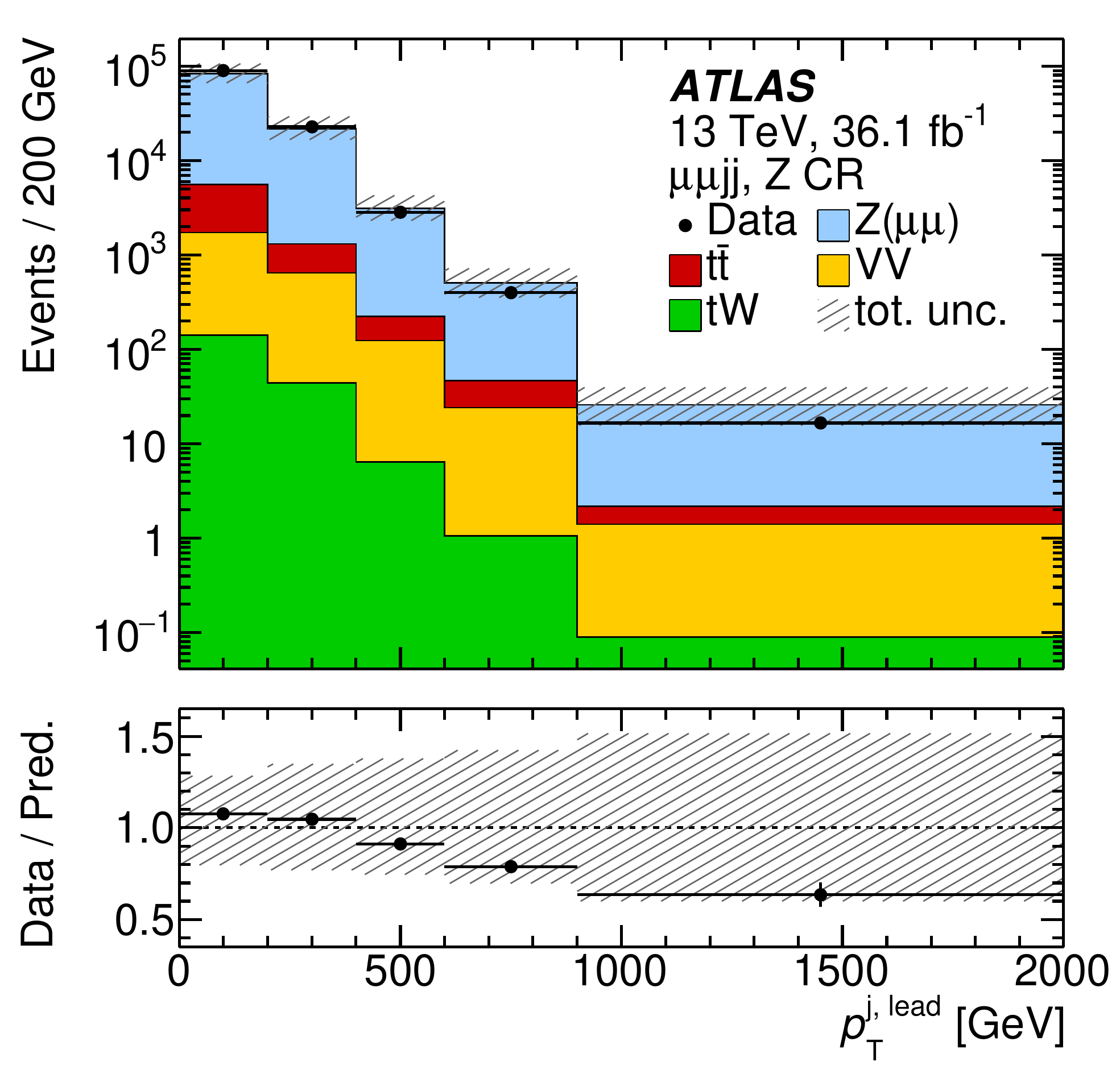}} \\
\subfloat[]{
    \includegraphics[width=0.42\textwidth]{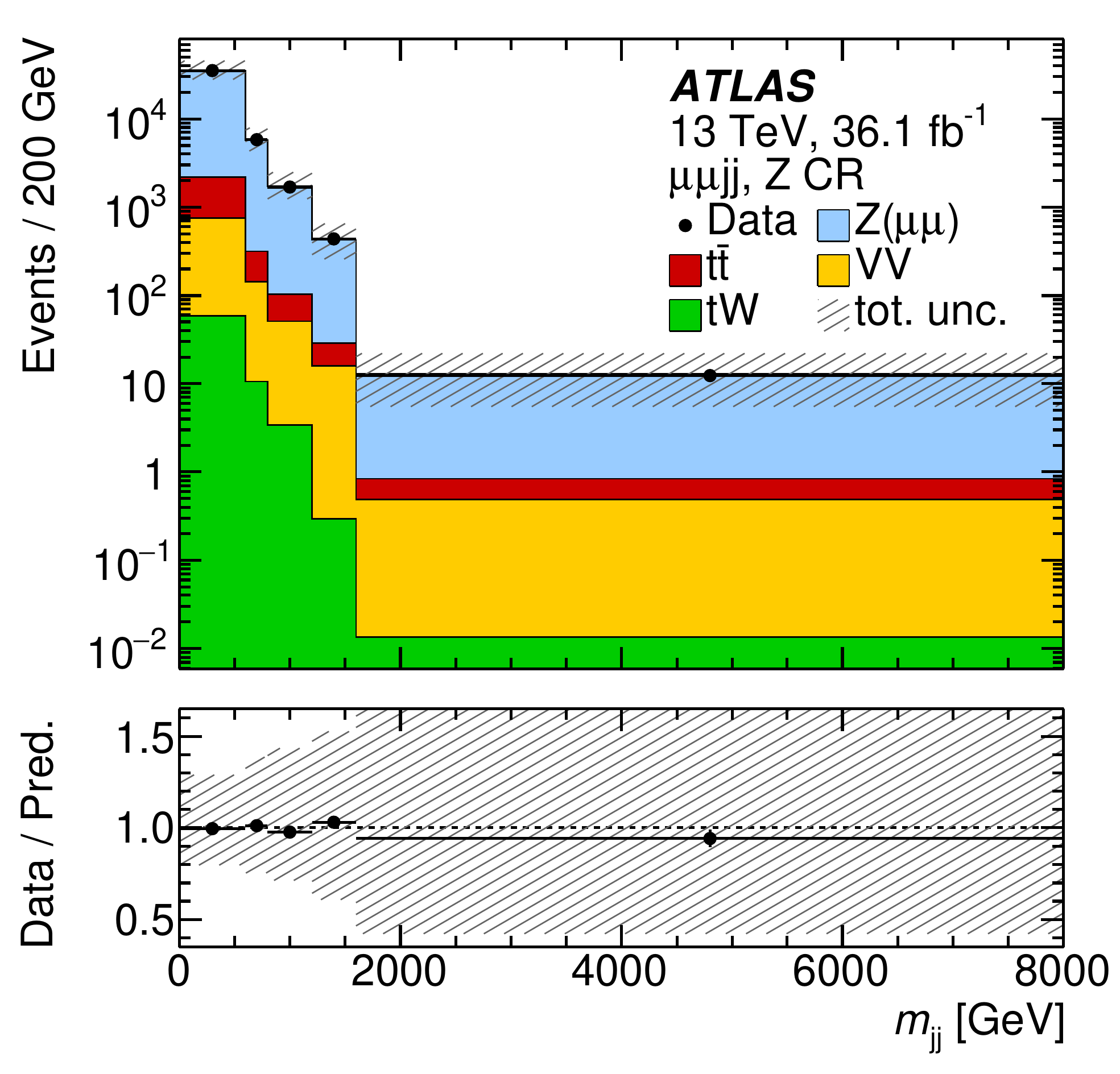} }
\subfloat[]{
    \includegraphics[width=0.42\textwidth]{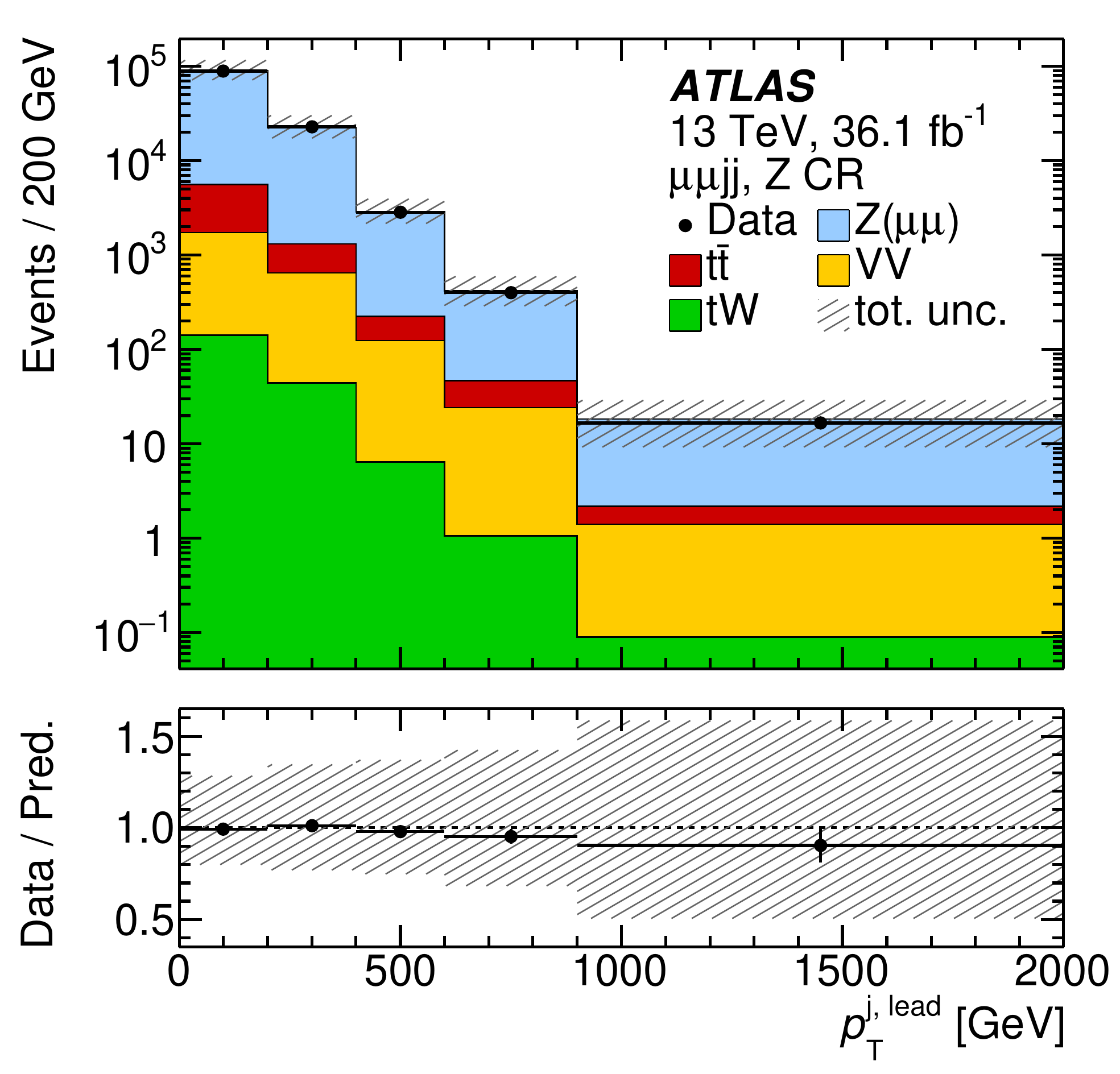}}
  \caption{Distributions of \mjj{} (left) and leading jet \pt{} (right) in the $Z$ CR for the  \mumujj channel. 
Data are shown together with MC contributions corresponding to  $Z \rightarrow \mu\mu$, \ttbar{}, diboson ($VV$) and  single-top
($tW$) processes. The MC distributions are cumulatively stacked. The bottom panels show the ratio of data to expected background. The grey hatched band represents
the total uncertainty. The \textsc{Sherpa
2.2.1} predictions for $Z \rightarrow \mu\mu$  
are shown before (top) and after (bottom) the reweighting procedure (see text). }
  \label{fig:reweighting}
\end{figure}

To improve the description of data, a reweighting of the $Z/\gamma^*$+jets and $W$+jets predictions by \textsc{Sherpa 2.2.1} is 
performed. The weights in this procedure are parameterised 
as a function of the  dijet mass, $m_{jj}$, in the respective CR, and applied to both CRs and SRs (see 
Section~\ref{sec:BDT&SR}). 
The parameterisation is derived by fitting the ratio between data and prediction with second-degree polynomials.  
Figure~\ref{fig:reweighting} shows distributions of \mjj{} and leading jet \pt{} in the $Z$ CR for the  \mumujj 
channel before and after reweighting. The data are better described by 
the Monte Carlo simulations following this reweighting. The total uncertainty (statistical, systematic and theory) 
is shown. The total uncertainty for CR distributions 
is dominated by the uncertainty in 
the theoretical predictions which is correlated between bins. Experimental uncertainties from jet energy scale and jet energy resolution are 
typically 
around 1\%. Similarly, the  \textsc{Sherpa 2.2.1} predictions of  $W$+jets require reweighting to match the data. The functional form 
of the reweighting algorithm is similar to that used for the $Z/\gamma^*$+jets  sample.   
The signal regions differ from the control regions mainly by the cut on \mll{} or \mt{}.
It is verified that applying weights derived in one mass region to the MC simulation in a different mass region provides a good description of the data in that region, indicating that the same correction is also valid in the signal region. 
All mass regions used in this test do not overlap with the signal regions. 
A systematic uncertainty is included to account for the reweighting procedure (Section~\ref{sec:Systematics}). 

In the 
statistical method used to 
evaluate the exclusion limits  
 which is described in Section~\ref{sec:method}, the normalisations of background 
estimates are adjusted in the fitting procedure. These normalisations are not applied to any of the distributions shown prior 
to 
that section.  Finally, the \ttbar{} control region (\ttbar\ CR) for 
the $\munujj$ and $\enujj$ channels is defined with  the same requirements on \mt{} and \met{} as the $W$ CR, but here the presence of at least two 
$b$-jets is required. The purity of the \ttbar{} CR is 86\% (89\%) for the electron (muon) channel.
The CRs are summarised in Table~\ref{tab:AnaStrat_Regions}.
The $Z$ CRs and other measurement regions are used to extract particle-level differential cross-sections (see Section~\ref{sec:unfoldingmethod}).

\begin{table}
  \caption{Definition of the CRs used in the dilepton and lepton--neutrino channel, respectively. Selections common to both 
channels are shown at the top.}
  \centering
  \begin{tabular}{|l|c|c|c|}
    \hline
    \multirow{2}{1.8cm}{Common selections} & \multicolumn{3}{c|}{$\geq2$ jets, $\pt > 60$~\GeV, $|\eta|<2.5$} \\
                                                           & \multicolumn{3}{c|}{$|\eta_{muon}|<2.5$, $|\eta_{elec}|<2.47$} \\ \hline
    & $\ell\ell jj$ & \multicolumn{2}{c|}{$\ell\nu jj$} \\ \hline
      & $\etl > 40$\,\GeV &  \multicolumn{2}{c|}{$\etl > 65$\,\GeV} \\     
    \hline
    \zcr & $70 < \mll < 110$\,\GeV &  \multicolumn{2}{c|}{-} \\
    \hline
    \multirow{3}{*}{\wcr}  & \multirow{3}{*}{-}  &  \multirow{6}{3.4cm}{\makecell{$40 < \mt < 130$\, \GeV \\ \met{} > 40\, \GeV \\  $S>4$}}  & \multirow{3}{*}{0 $b$-jets} \\ 
       &  &  & \\
       &  &  & \\
    \cline{1-2} \cline{4-4}
    \multirow{3}{*}{\ttbarcr}  & \multirow{3}{*}{-}  &   & \multirow{3}{*}{$\geq2$  $b$-jets} \\ 
       &  &   &\\
       &  &  & \\
    \hline
  \end{tabular}
  \label{tab:AnaStrat_Regions}
\end{table}

This analysis makes use of the matrix method~\cite{EXOT-2016-06} to estimate the background from misidentified electrons for the first-generation LQ 
search. This background is negligible for the second generation search. 
The matrix method is based on the estimation of the probabilities for prompt electrons and fake electron candidates that pass identification and 
isolation selections that are looser than the nominal selection described in Section~\ref{sec:ObjSel}, to also pass the 
nominal selection. 
These probabilities are referred to as the prompt rate and fake rate, respectively. 
In the loose selection, several criteria that are used to suppress fake contributions in the nominal selection are relaxed. 
In particular, no isolation is required in the loose selection. 
The prompt rate is estimated to a good approximation from MC simulations, while the fake rate is estimated from a data sample enriched in fake 
backgrounds. 
To suppress contributions from prompt electrons in this sample, events are rejected if they contain at least two electron candidates that pass the 
`medium' identification criteria. 
In both channels, events are also rejected if there are two electron candidates that fulfil the `loose' object definition specific to the respective channel and have an invariant mass in an interval of $\pm 20$~\GeV\ around the $Z$ boson mass.
The contribution from the fake electron background is at most 24\%, but generally much less. 

Backgrounds with non-prompt muons arise from decays of heavy-flavour hadrons inside jets. Their contribution
to the dimuon final states is negligible as in the search
documented in Ref.~\cite{EXOT-2016-05}. In lepton--neutrino final states, a contamination of about 5\% 
was found~\cite{EXOT-2016-06}, corresponding to about half the size of the fake electron background.  
Tests showed that a contamination of 5\% would not alter the results of this analysis, and therefore this background is neglected. 
The uncertainty in the total background estimation due to neglecting this background is taken to be half the size of the background from fake electrons.  

\begin{figure}
  \centering
\subfloat[]{
    \includegraphics[width=0.42\textwidth]{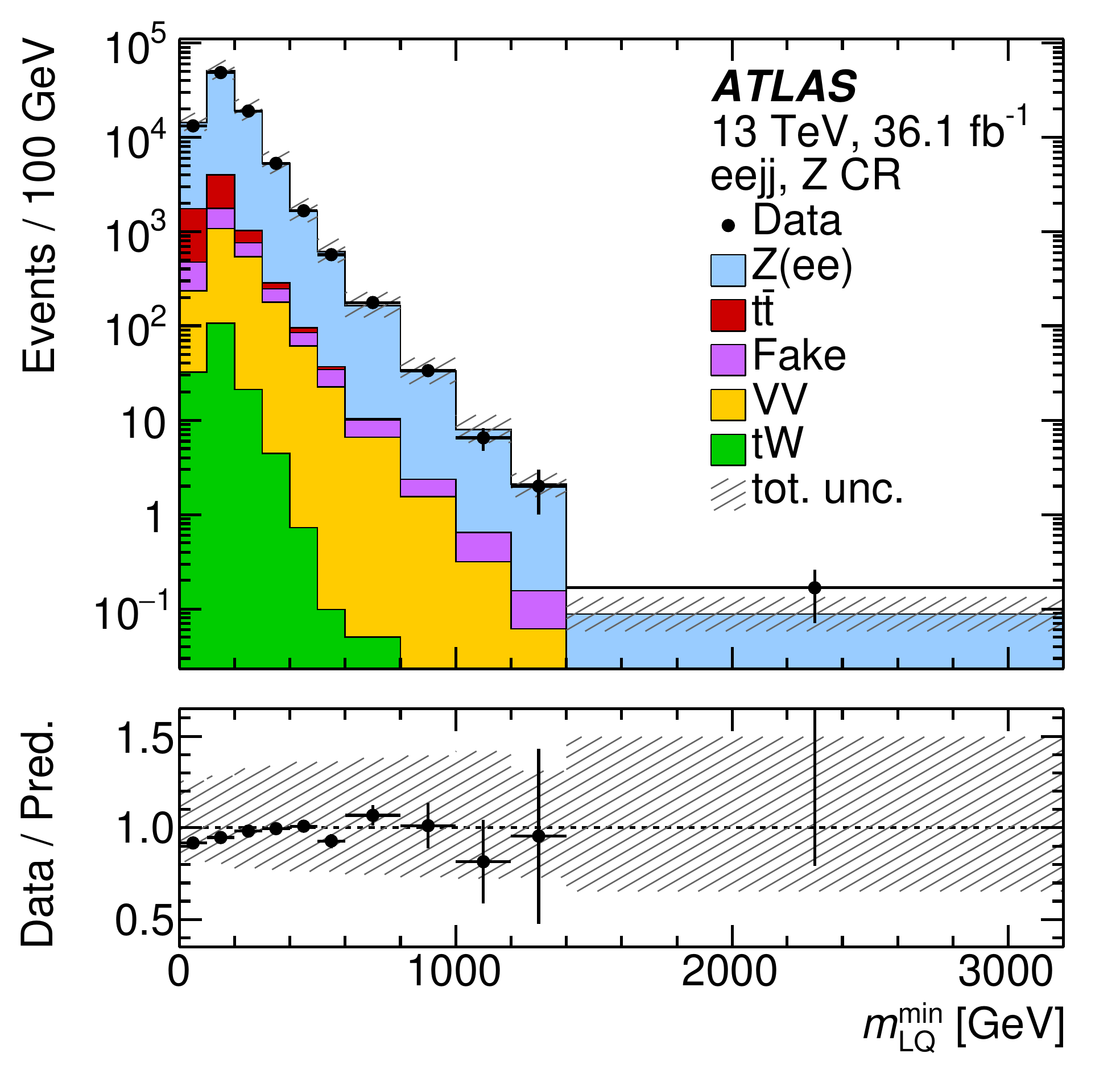}}
\subfloat[]{
    \includegraphics[width=0.42\textwidth]{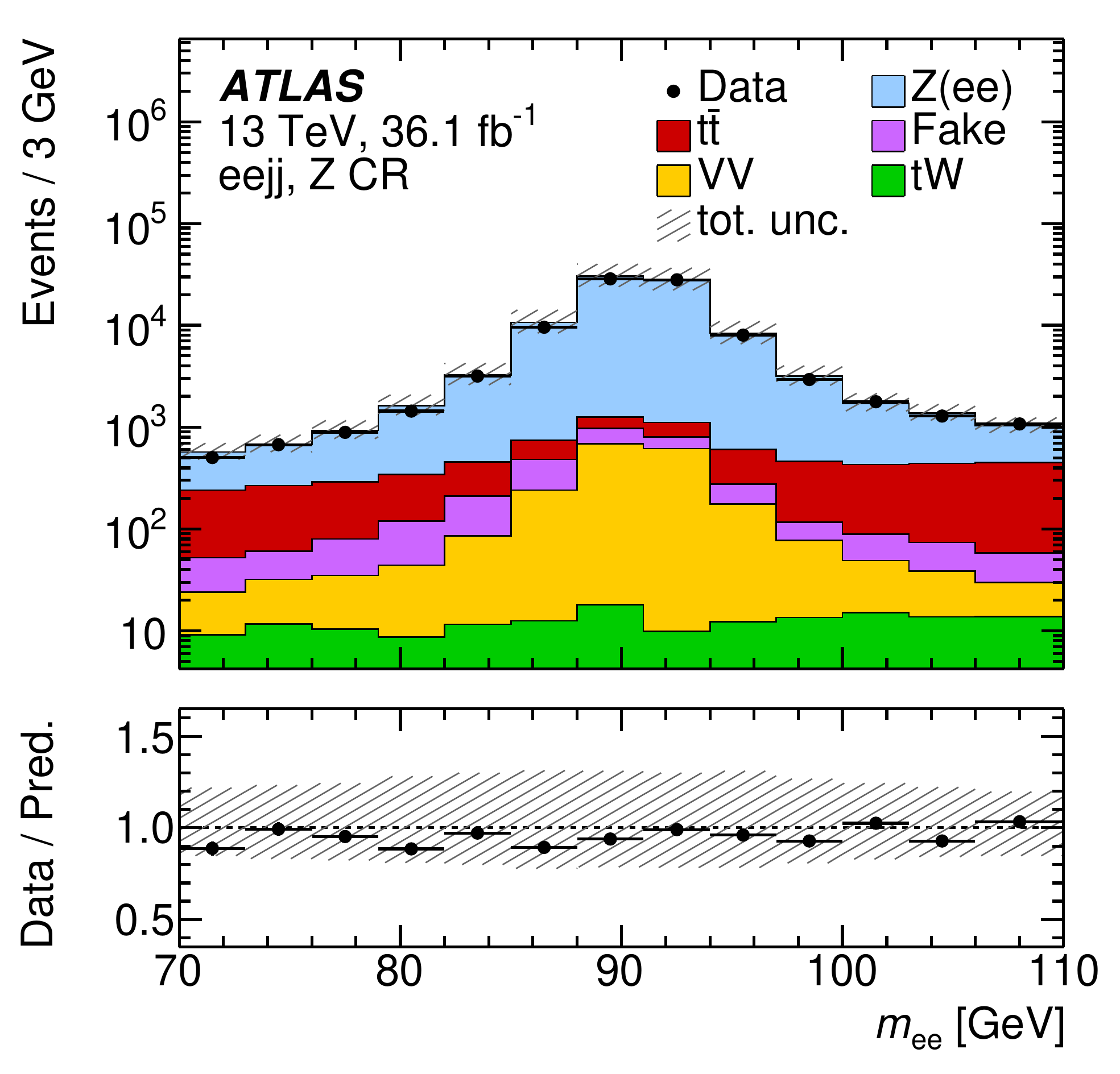}} 
\\
\subfloat[]{
    \includegraphics[width=0.42\textwidth]{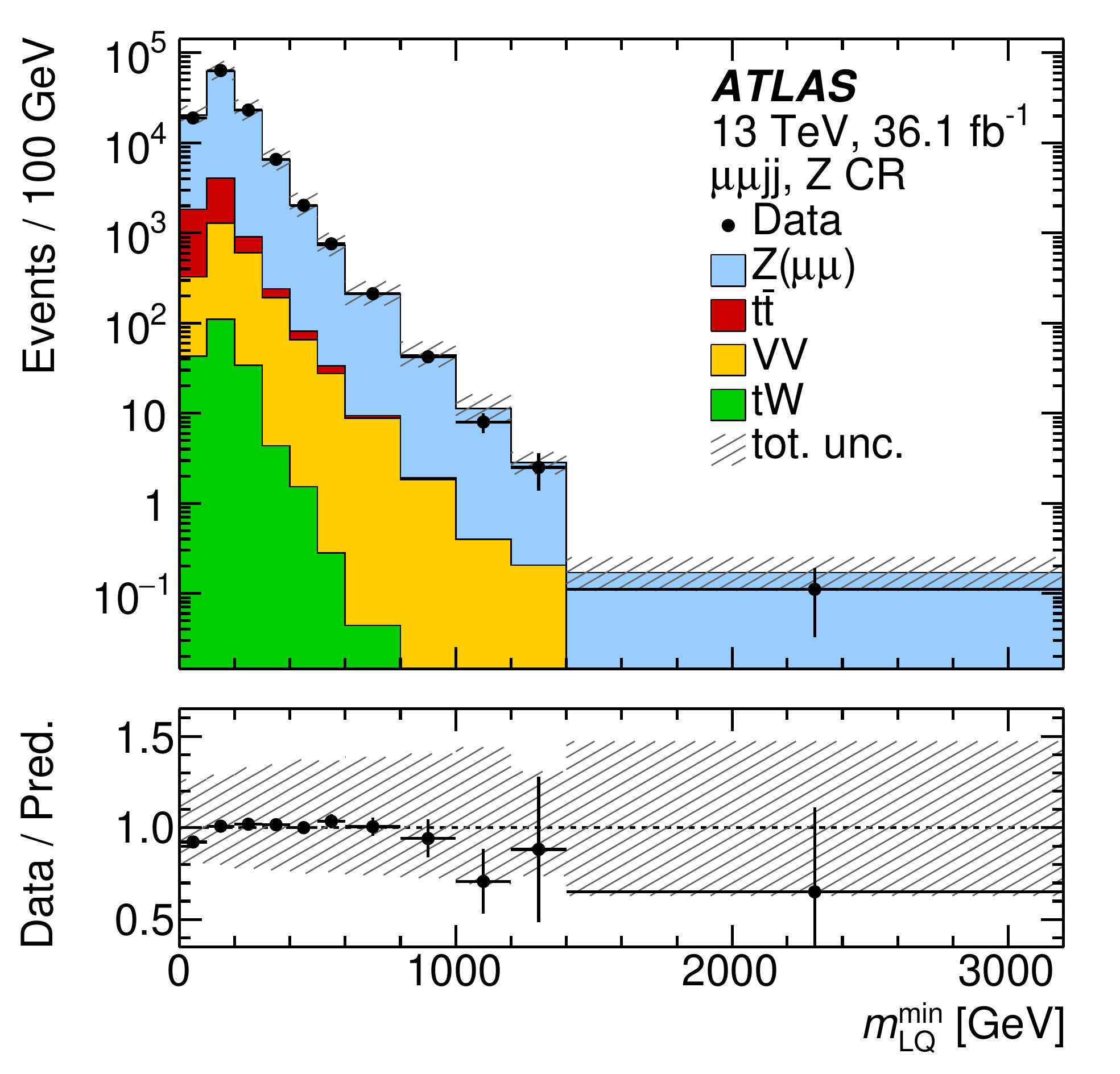} }
\subfloat[]{
    \includegraphics[width=0.42\textwidth]{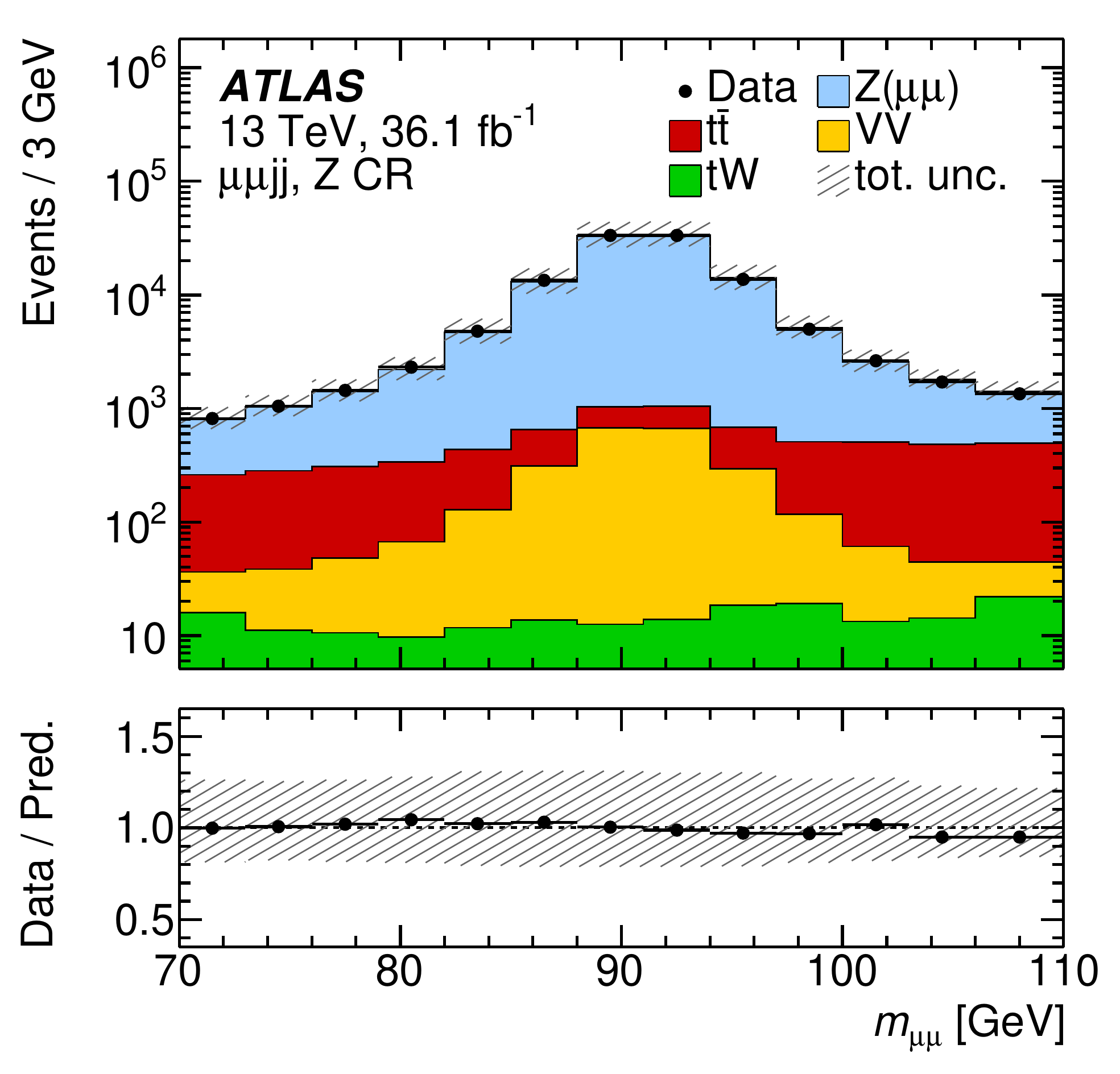}}

\caption{Distributions of \mlqmin{} and \mll{} in the $Z$ CR for the \eejj (top) and \mumujj (bottom) channels. 
Data are shown together with background contributions corresponding to \ttbar{}, diboson ($VV$), $Z \rightarrow ee$, $Z \rightarrow \mu\mu$, single-top 
($tW$) processes, and fake electrons. The background distributions are cumulatively stacked. The bottom panels show the ratio of data to expected background. The grey hatched band represents 
the total uncertainty. A reweighting as a function of \mjj{} is applied to the $Z$+jets simulation. The other MC predictions are not reweighted. 
}
  \label{fig:ZCR_mumujj_2}
\end{figure}

In Figure~\ref{fig:ZCR_mumujj_2}, the distributions of \mlqmin{} and the dilepton invariant mass in the $Z$ CR are shown for the \mumujj{} and \eejj{} 
channels.  The quantities \mlqmin{} ane \mlqmax{} are defined as the lower and higher of the two
invariant masses which can be reconstructed using the two lepton--jet pairs
in the dilepton channels.  
In the analysis, the lepton--jet pairing is chosen such that the absolute mass difference between the two
$\LQ$ candidates is minimised.  The full set of experimental and theoretical uncertainties is considered for the uncertainty band. The spectra correspond to 
the predictions prior to the fit described in Section~\ref{sec:method}.

\begin{figure}
  \centering
\subfloat[]{
    \includegraphics[width=0.42\textwidth]{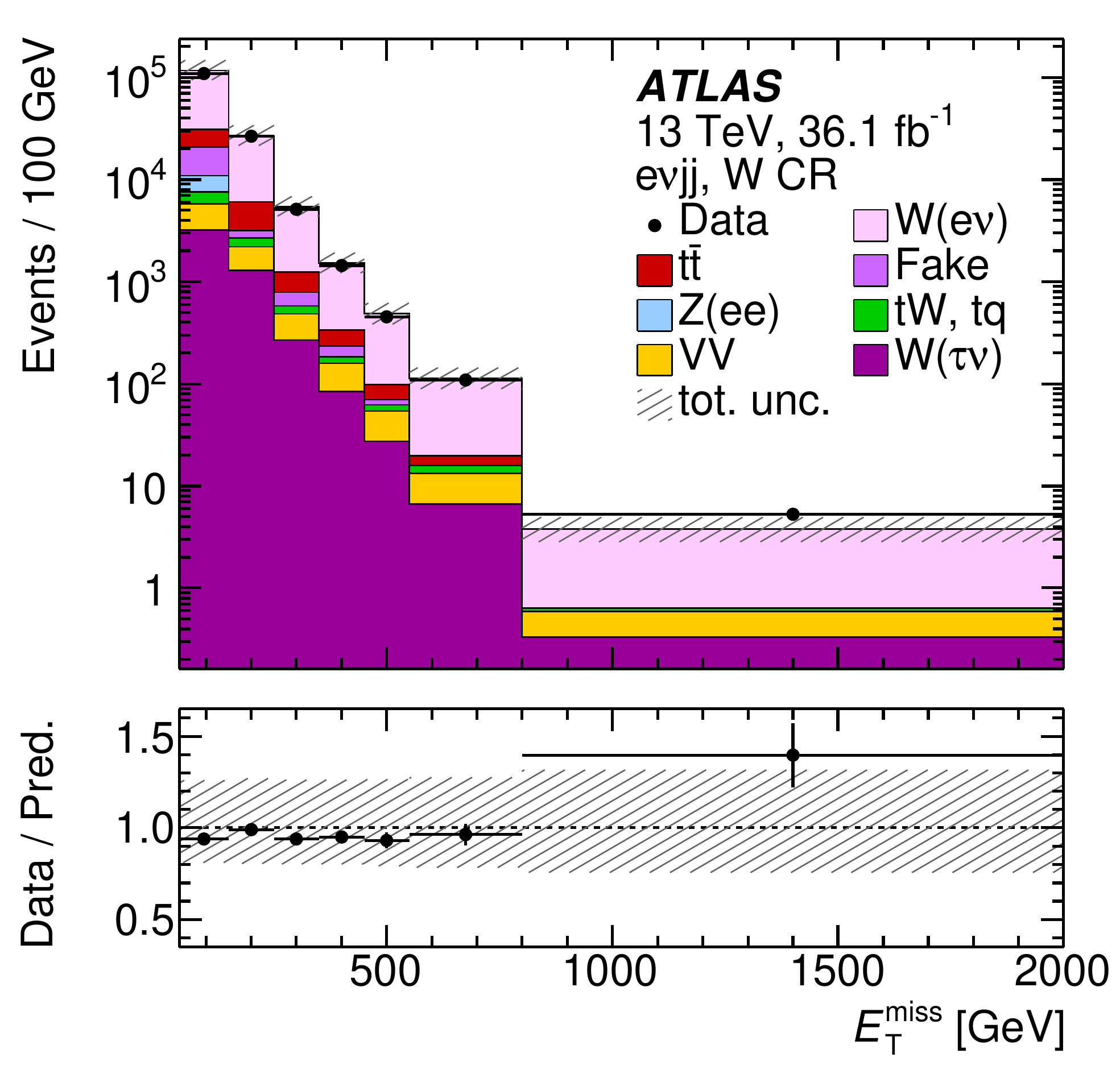} }
\subfloat[]{
    \includegraphics[width=0.42\textwidth]{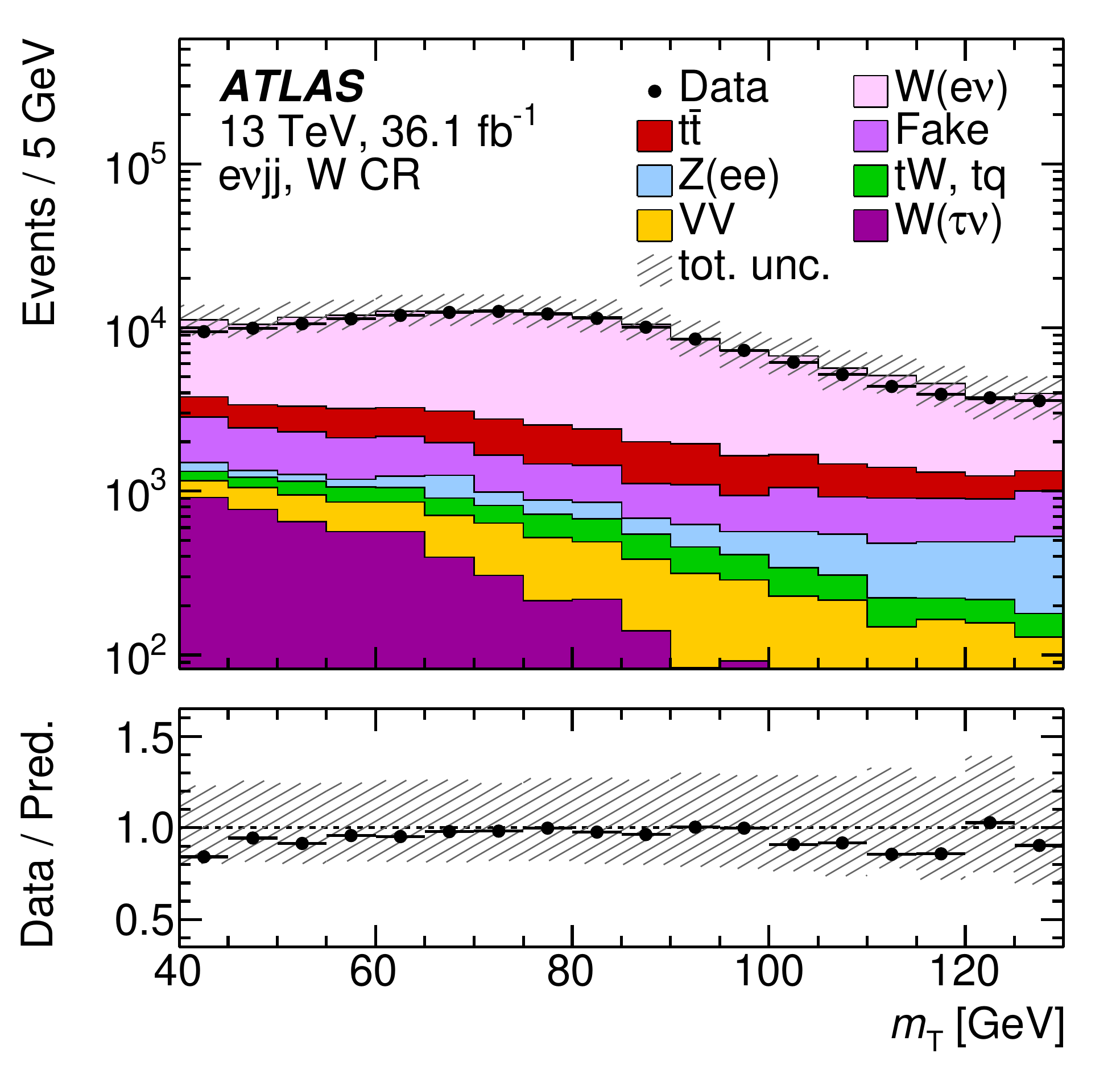}}
\\
\subfloat[]{
    \includegraphics[width=0.42\textwidth]{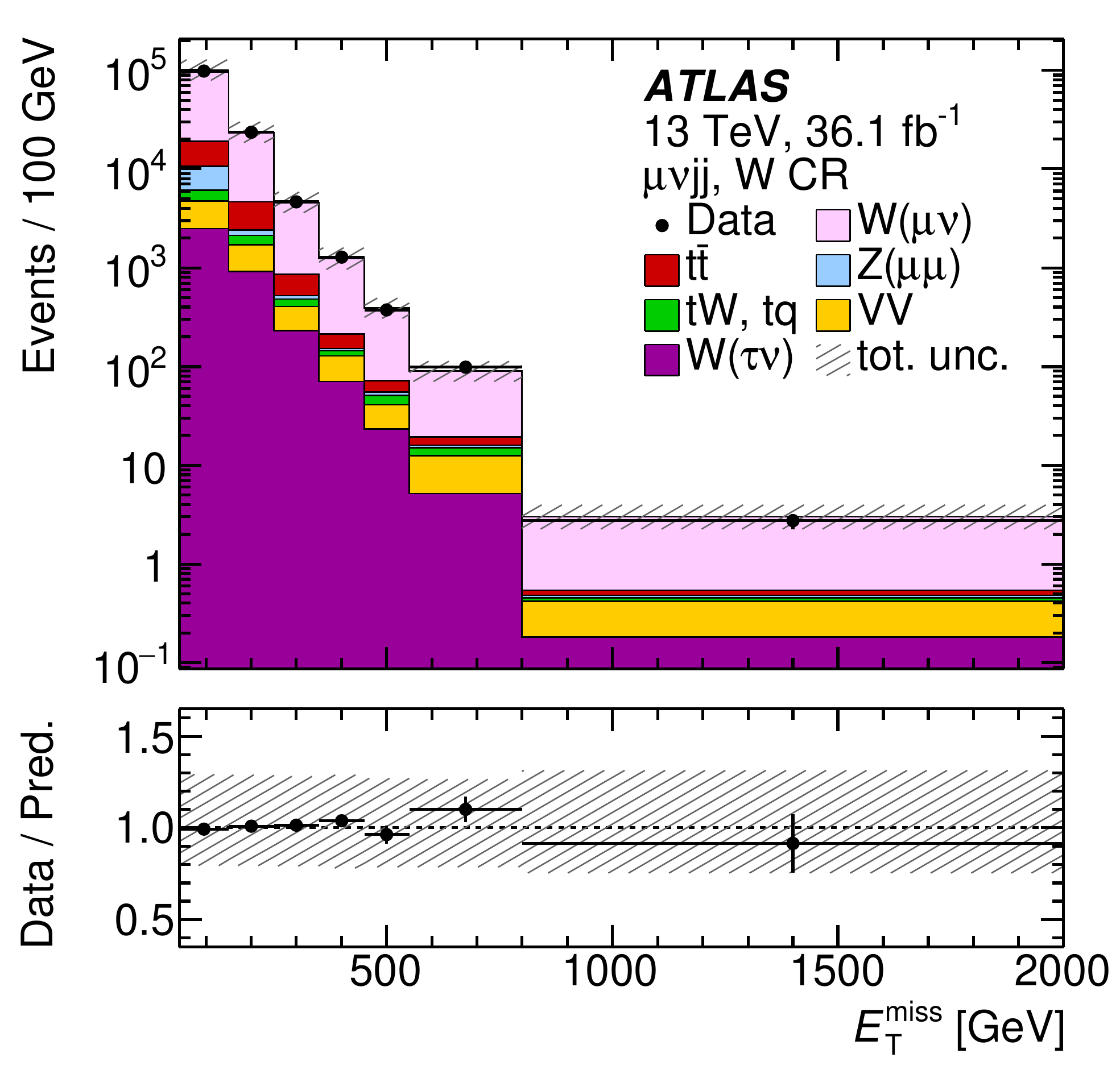}}
\subfloat[]{
    \includegraphics[width=0.42\textwidth]{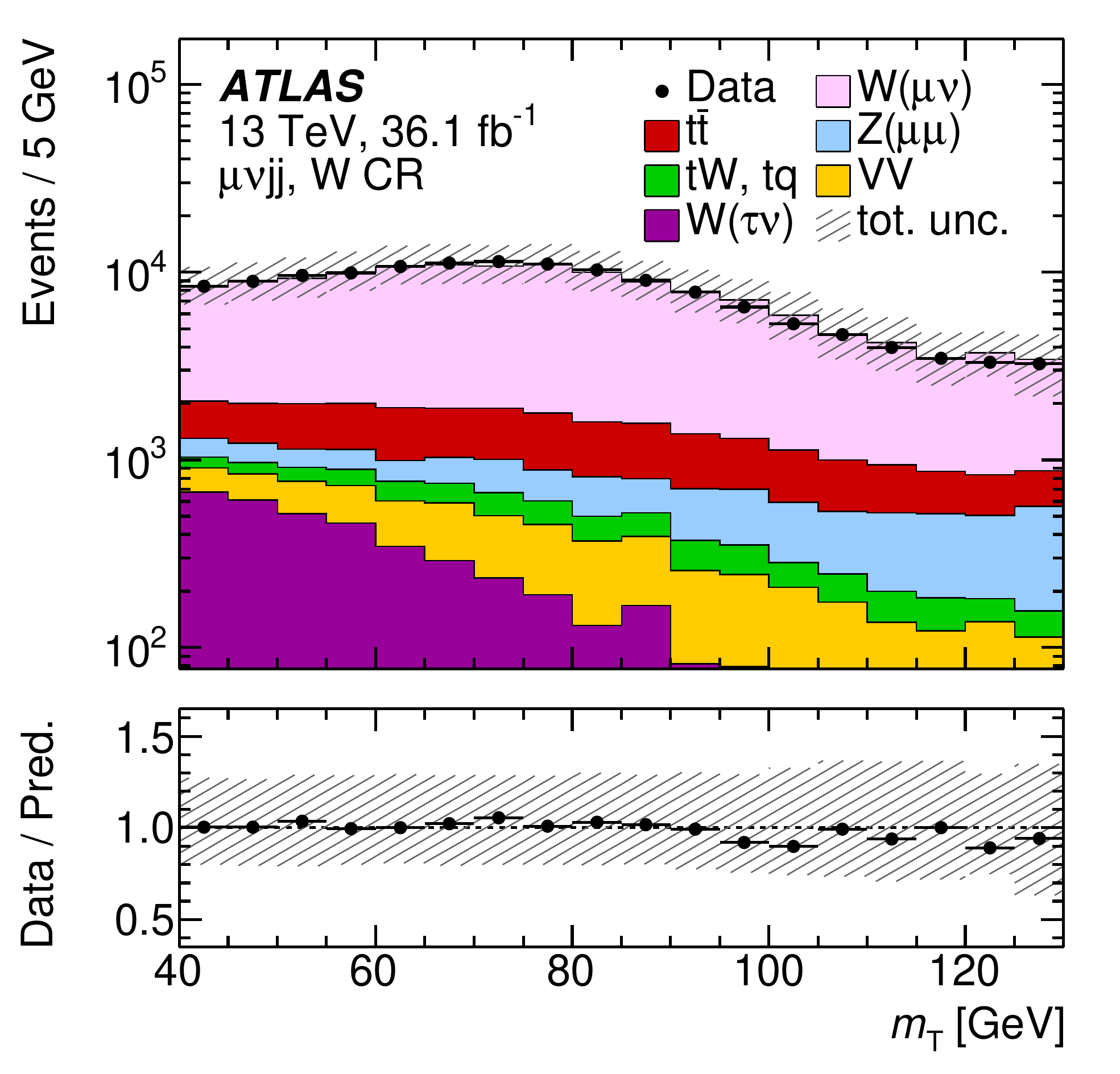}}
  \caption{Distributions of \met{} and \mt{} in the $W$ CR for the \enujj (top) and \munujj (bottom) channels.
Data are shown together with background contributions corresponding to \ttbar{}, diboson ($VV$), $Z\rightarrow ee$,  $Z\rightarrow \mu\mu$, $W\rightarrow 
\mu \nu$,  $W\rightarrow e\nu$, $W\rightarrow \tau\nu$, single-top production ($tW$, $tq$) processes, and fake electrons. The background distributions are cumulatively stacked. The bottom panels show the 
ratio of 
data to expected background.  The grey hatched band represents 
the total uncertainty.  A reweighting as a function of \mjj{} is applied to the $W$+jets and $Z$+jets simulation.  The other MC predictions are 
not 
reweighted. }
  \label{fig:WCR}
\end{figure}

\begin{figure}
  \centering
\subfloat[]{
    \includegraphics[width=0.42\textwidth]{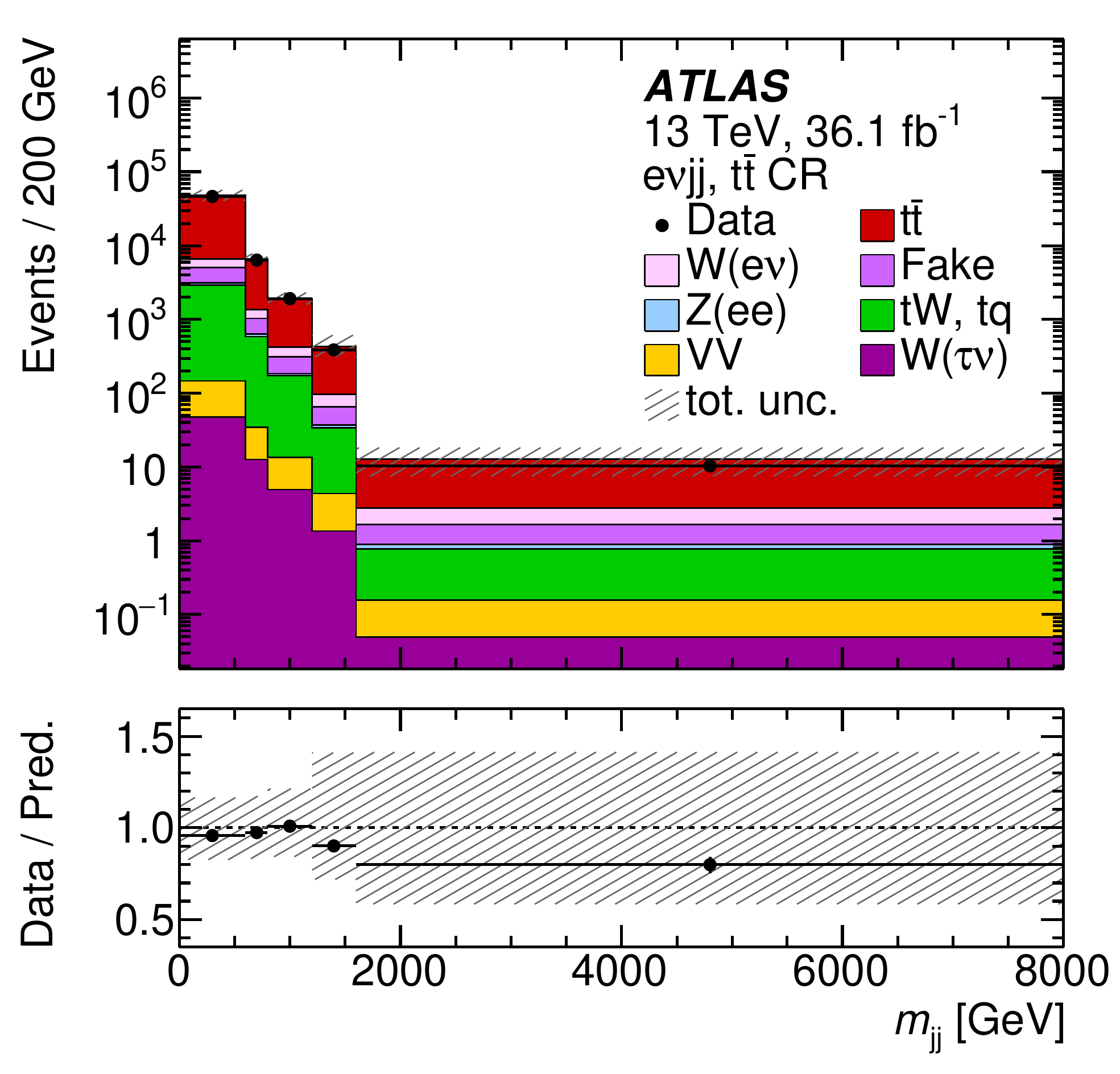} }
\subfloat[]{
    \includegraphics[width=0.42\textwidth]{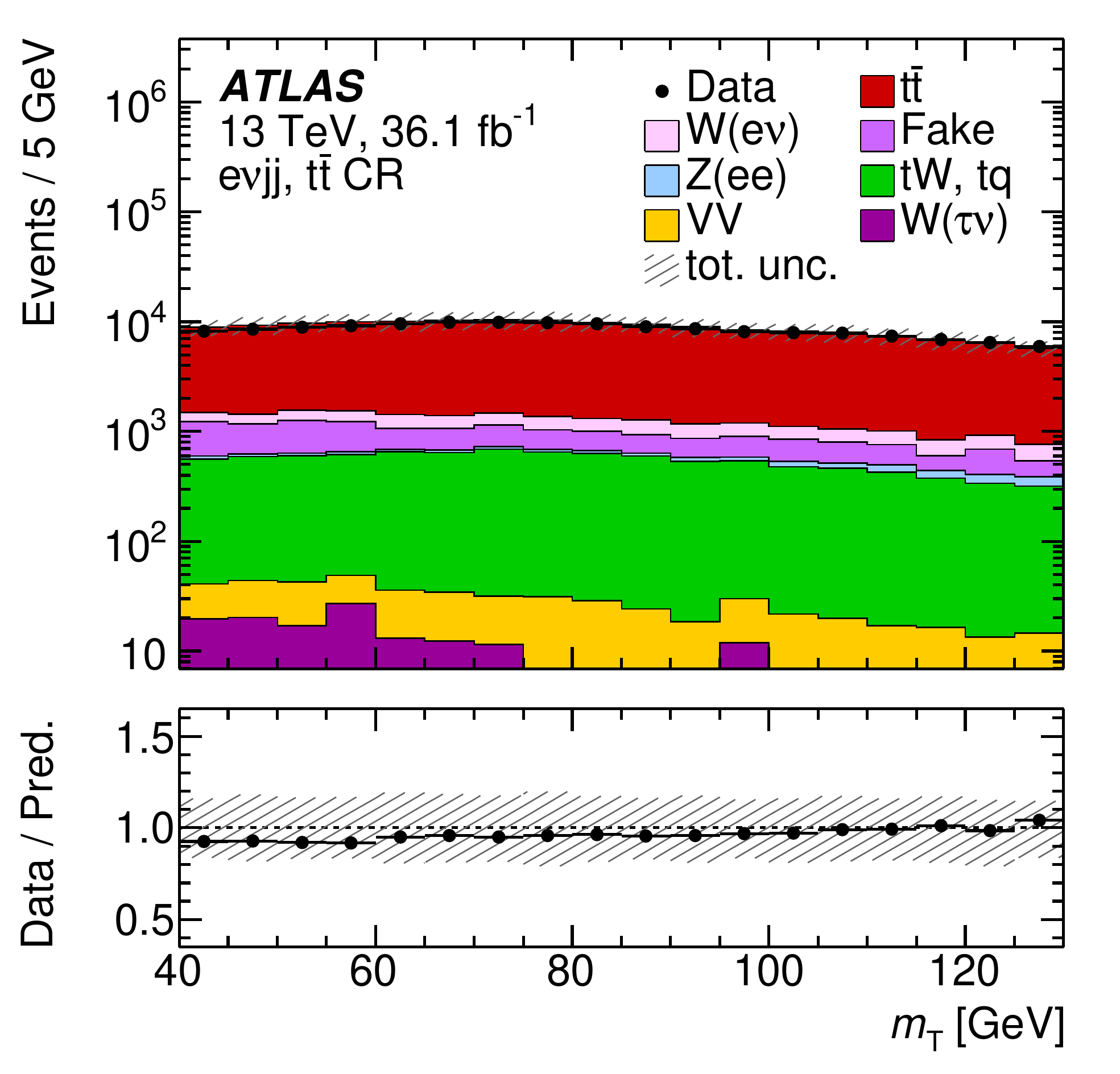}}
\\
\subfloat[]{
    \includegraphics[width=0.42\textwidth]{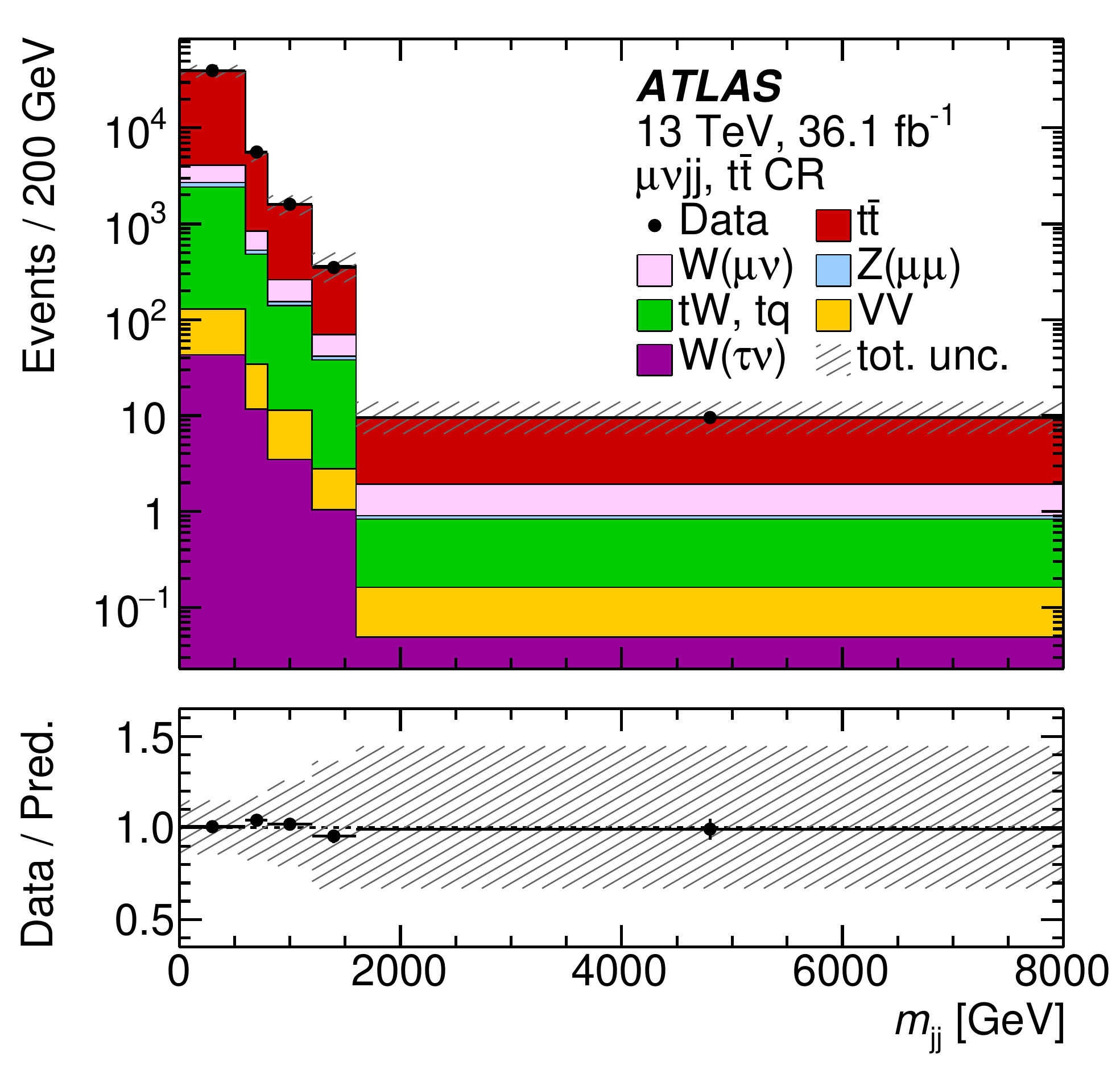}}
\subfloat[]{
    \includegraphics[width=0.42\textwidth]{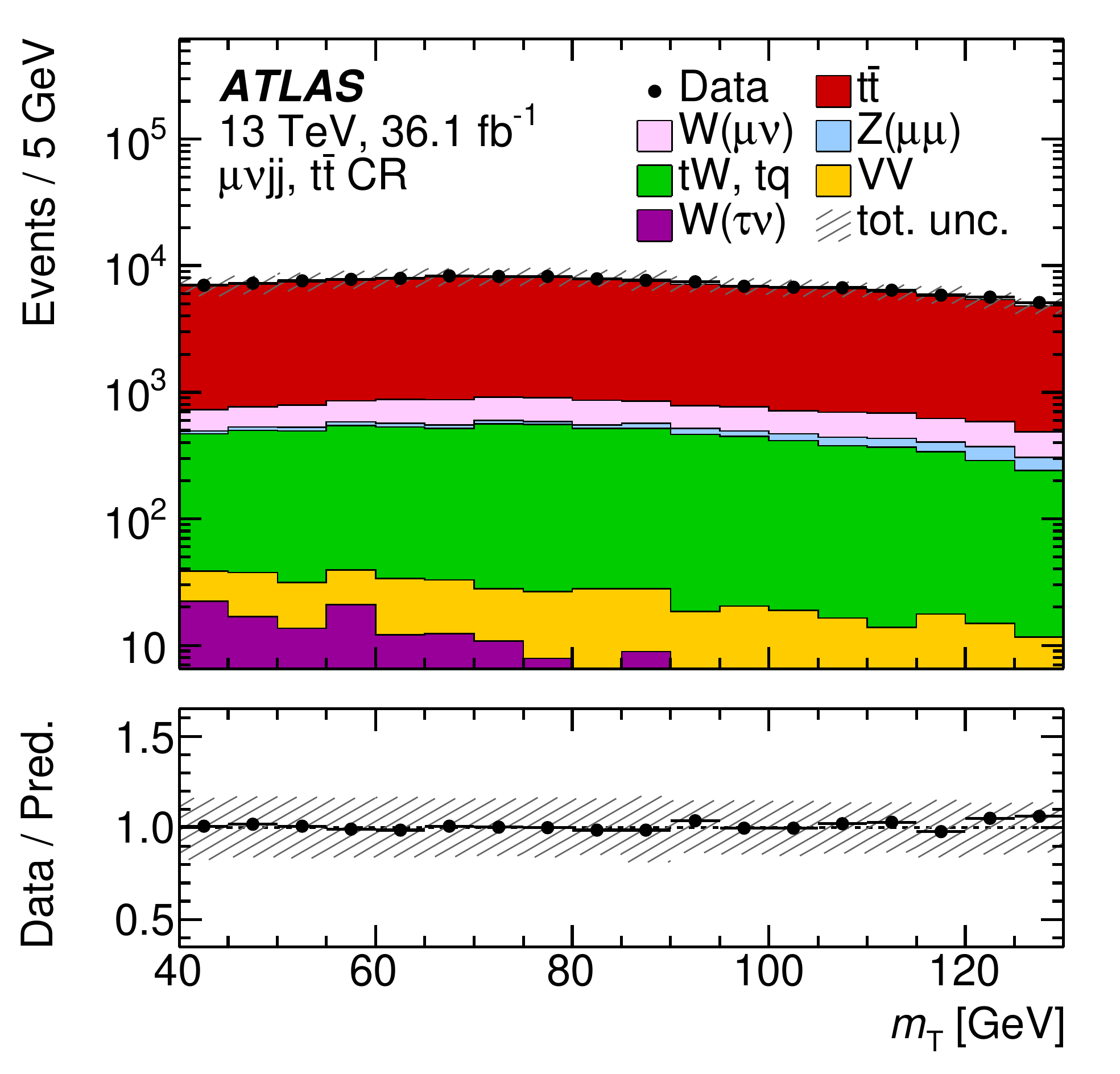}}
  \caption{  Distributions of \mjj{} and \mt{} in the \ttbar\ CR for the \enujj (top) and \munujj (bottom) channels.
Data are shown together with background contributions corresponding to \ttbar\, diboson ($VV$), $Z\rightarrow ee$,  $Z\rightarrow \mu\mu$, $W\rightarrow
\mu \nu$,  $W\rightarrow e\nu$ $W\rightarrow \tau\nu$, single-top production ($tW$, $tq$) processes, and fake electrons. The background distributions are cumulatively stacked. The bottom panels show 
the ratio of data to expected background.  The grey hatched band represents 
the total uncertainty. A reweighting as a function of \mjj{} is applied to the $Z$+jets and $W$+jets simulations. The other MC predictions are not 
reweighted. }
  \label{fig:ttbarCR}
\end{figure}

Distributions for the \enujj and \munujj channels are similarly shown for the $W$ and \ttbar{} CRs in Figures 
~\ref{fig:WCR} and ~\ref{fig:ttbarCR}, respectively. The \met{} and \mt{} spectra are shown in 
Figure~\ref{fig:WCR}, while the \mjj{} and \mt{} spectra are shown in Figure~\ref{fig:ttbarCR}. 
The data and the predictions in the various CRs are in agreement within 
uncertainties (described in Section~\ref{sec:Systematics}) following the reweighting of the 
\textsc{Sherpa 2.2.1} predictions for  $Z/\gamma^*$+jets and $W$+jets processes. 
After the reweighting,  the overall normalisation of the simulation is in very good agreement with the data in the $V$+jets CRs.

\section{Extraction of particle-level cross-sections from the measurement regions} 
\label{sec:unfoldingmethod}

Measurements of the particle-level fiducial and differential cross-sections are made in several regions related to the search CRs described in the previous section.
These measurement regions (MRs) correspond to kinematic regions where new physics signals have already been excluded. They are chosen 
because several searches with related final states 
use similar selections for their CRs (see for example Ref.~\cite{EXOT-2016-29}). 

\subsection{Particle-level objects} 

In order to extract fiducial and differential cross-sections in the MRs, particle-level object selections are defined, and chosen to 
be as close as possible to their detector-level counterparts described in Section~\ref{sec:ObjSel}.

Prompt dressed\footnote{Dressed leptons are obtained by adding the four-vectors of all photons found within a cone of size $\Delta R$ = 0.1 around the lepton. Photons from hadron decays are excluded.} particle-level electrons and muons are used to define the fiducial region. They are required to  
have $\pt>40\gev$ and $|\eta|<2.5$.

Particle-level jets are formed by clustering stable\footnote{Particles 
with $c\tau>10$~mm are designated as stable.} particles using the 
anti-$k_t$ ($R=0.4$) algorithm~\cite{AntiKt}, excluding muons and neutrinos. Particle-level jets are required to have $\pT>60\GeV$ and $|\eta|<2.5$. 
The electrons used in the jet clustering are not dressed. 
To match the requirements of the detector-level jet-lepton ambiguity resolution procedure, any particle-level leptons within $\Delta R <0.4$ from a selected jet are removed. 

\begin{table}[pt!]
\caption{Requirements at particle-level for all measurement regions (MRs) where particle-level differential cross-sections are extracted. The purity refers 
to the proportion of the particle-level yield of the MR which is accounted for by the dominant process. 
The variable $S_{T}$ refers to the scalar sum of the \pt\ of the jets and leptons.}
\resizebox{\textwidth}{!}{
  \centering
  \begin{tabular}{|l|l|l|l|l|l|l|l|}
    \hline
    
		MR         & Dominant process (purity) & Required leptons and jets     & \mll selection       & \st\ selection & Remark \\
		\hline
		\eejj      & \Zee (93\%)             & $= 2e$ ; $\geq 2 \textrm{jets} $       & $70 <\mll <110\GeV$ & -             & Identical to $Z$~CR \\
		\mumujj    & \Zmm (94\%)             & $= 2\mu$  ; $\geq 2 \textrm{jets} $                         & $70 <\mll <110\GeV$ & -             & Identical to $Z$~CR \\
		\emujj     & $\ttbar\rightarrow e\mu$ (93\%)          & $= 1\mu,1e$  ; $\geq 2 \textrm{jets} $                        & -                     & -            & - \\
		Extreme \eejj    & \Zee (94\%)             & $= 2e$     ; $\geq 2 \textrm{jets} $    & $70 <\mll <110\GeV$  & $\st>600\GeV$ & - \\
		Extreme \mumujj     & \Zmm (94\%)             & $= 2\mu$     ; $\geq 2 \textrm{jets} $   & $70 <\mll <110\GeV$  & $\st>600\GeV$ & - \\
		Extreme \emujj      & $\ttbar\rightarrow e\mu$  (86\%)          & $= 1\mu,1e$   ; $\geq 2 \textrm{jets} $  & -                     & $\st>600\GeV$ & - \\
		\hline
  \end{tabular}
	}
    \label{tab:MRs}
\end{table}
\subsection{Measurement region definitions} 
\label{sec:unfoldingregions}

The first two MRs are identical to the \eejj and \mumujj$Z$~CRs described in Table~\ref{tab:AnaStrat_Regions}. 
The third is an \emujj region where the leptons are required to have different flavours, and no requirement on the dilepton mass is applied. 
Finally, three additional regions are defined as for the \eejj, \mumujj and \emujj MRs, and labelled as \emph{extreme} with the additional requirement that the scalar sum of the \pt\ of the leptons and the two leading jets (\st) be above 600~\gev. The requirements for events entering all MRs are summarised in Table~\ref{tab:MRs}.

The MRs are defined both at detector level and particle level, where the requirements are identical but use the objects passing the detector-level and particle-level selections respectively.
The $m_{jj}$ reweighting is also applied to events from the simulated Drell--Yan sample at particle level, where the particle-level value of $m_{jj}$ is used to calculate the weight. 
In each MR, the cross-section measurements are made for dominant processes. The contributions from all other processes are subtracted from the data yield, as 
explained in Section~\ref{sec:unfoldingmethodn}.

\subsection{Particle-level measurement method} 
\label{sec:unfoldingmethodn}

Each MR is designed such that it is dominated by a single SM process (accounting for more than 85\% of the yield, see Table~\ref{tab:MRs}), 
for which the cross-section can be measured exclusively. 
A simple procedure to extract particle-level cross-sections, based on bin-by-bin correction factors, is used.
In this method, the particle-level differential cross-section for the dominant process $p$ in bin $i$ of the distribution of variable 
$X$ can be expressed as:
\begin{equation} 
 \frac{{\mathrm{d}} \sigma_i^p}{{\mathrm{d}}X} = \frac{(N_i - \sum_{q\neq p} R_i^q) \cdot \frac{T_i^p}{R_i^p}}{w_i \cdot L} ,
 \label{eq:diffXsec}
\end{equation} 
where the sum runs over all sub-dominant SM processes which contribute to the yield, $N_i$ is the number of events observed in the bin, $R_i^p$ ($T_i^p$) is the yield at
 detector level (particle level) for process $p$, $w_i$ is the width of the bin and $L$ is the integrated luminosity of the data sample.  
The bin-by-bin correction factors $T_i^p/R_i^p$ may be thought of as the inverse of the reconstruction efficiency in bin $i$, assuming no bin-by-bin migrations.
The bin-by-bin correction procedure doesn't account for migration between bins due to resolution effects when going from particle-level to detector-level distributions.
For this reason, the binning of the distributions is chosen to be much broader than the experimental resolution to minimise the migration between bins. This binning is optimised for the measurements to ensure that at least 90\% of events passing the particle-level selection remain in the same bin at detector level (for events passing both selections). This binning is not used for the search.

Differential cross-section measurements are made for the following observables: 
$\pt$ of the dilepton system; $\Delta \phi$ between the leptons; minimum $\Delta \phi$ between lepton and leading jet; minimum $\Delta \phi$ between lepton and subleading jet; \st; leading jet \pt; subleading jet \pt; $\Delta \phi$ and $\Delta \eta$ between the leading and subleading jets; scalar sum of \pt\ of leading and subleading jets ($H_\mathrm{T}$); and invariant mass of the dijet system.

\section{Multivariate analysis and signal regions}
\label{sec:BDT&SR}

In order to discriminate between signal and background in the LQ search, the TMVA~\cite{Speckmayer:2010zz} implementation of a BDT is used.
A number of variables are expected to provide discrimination between signal and background.
Generally, the \pt{} of the leptons and jets as well as related variables such as \st{} will
possess higher values for the signal than for the background, in particular for high LQ mass.
Since the signature arises from the decay of parent LQs with well-defined masses, mass-sensitive 
discriminating variables are further candidates for BDT input variables.
In the dilepton channels, the lepton-jet masses \mlqmin{} and \mlqmax{} are reconstructed, as described in 
Section~\ref{sec::CtrlRegion}.
In the lepton--neutrino channels, the mass of one $\LQ$ (\mlq{}) and the transverse
mass of the second $\LQ$ (\mlqt{}) can be reconstructed. The transverse
mass is defined as $\mlqt=\sqrt{2\cdot p_T^j \cdot \met \cdot (1-\cos(\Delta
\phi(j,\met)))}$, where $\Delta \phi(j,\met)$ is the
azimuthal angle
between the jet and the direction of missing transverse momentum vector.
The procedure is analogous to that used in the dilepton channel; both
possible pairings of jet and lepton (or \met{}) are tested and the pairing that results in the
smaller absolute difference between \mlq{} and \mlqt{} is chosen.

All background processes are used in the BDT 
training. 
For the electron and muon channels separately, the training is done for each $\LQ$ mass hypothesis for which signal simulations 
were produced.  
To ensure independence from the CRs, in the dilepton channel, only events with a dilepton invariant mass above 
130\,\GeV\ are 
considered in the training. 
Similarly, only events with a transverse mass greater than 130\,\GeV\ are considered in the lepton--neutrino 
channel. Here, to further suppress the fake-electron contribution, 
it is required that \met{}$>150$\,\GeV\ and that $S>3$.
These requirements define the training regions (TRs), i.e. event samples used to train the BDTs. 

To ensure that no bias is introduced, two BDTs are used for each mass point; one is trained on one 
half of the events and evaluated on the other half, and vice versa for the second BDT. 
Different methods of dividing the simulation samples into training and 
test samples are tested and compared, giving consistent results such that no overtraining or other 
biases are introduced in the process.

Table~\ref{tab:BDTvar} lists the variables that were found to give optimal sensitivity, ordered by their discriminating power in the BDT. In the dilepton channel, 
these are both reconstructed 
$\LQ$ candidate masses, \mlqmin{} and \mlqmax{}, the dilepton invariant mass, \mll{}, the subleading jet \pt{} (\jettwopt{}) and the subleading lepton
 \pt{} (\ltwopt{}). 
The lepton--neutrino channel also makes use of both of the LQ mass variables, \mlq{} and \mlqt{}, and the other variables are the transverse mass, \mt{}, the missing transverse 
momentum, \met{}, the subleading jet \pt{}, and the charged-lepton \pt{}, \ptl{}. 
Distributions of a selection of the variables used for training are shown in Appendix~\ref{app:trplots}. 
It was checked and confirmed that not only the individual input variables but also their pairwise correlations are modelled well in the simulation.
\begin{table}[tb]
\caption{BDT input variables in the dilepton and lepton--neutrino channels.}
  \setlength\extrarowheight{4pt}
  \centering
  \begin{tabular}{|l|c|}
    \hline
    Channel & Input variables \\ \hline
    \lljj{} & \mlqmin{},  \hspace{2.0mm} \mll{},  \hspace{2.0mm} \jettwopt{},  \hspace{2.0mm} \ltwopt{},  \hspace{2.0mm} \mlqmax{} \\ 
    \lnujj{} &  \mlq{},  \hspace{2.0mm} \mlqt{},  \hspace{2.0mm} \mt{},  \hspace{2.0mm} \met{},  \hspace{2.0mm} \jettwopt{},  
\hspace{2.0mm} 
\ptl{} \\ 
    \hline
  \end{tabular}
 \label{tab:BDTvar}
\end{table}

\begin{figure}
  \centering
\subfloat[]{
    \includegraphics[width=0.42\textwidth]{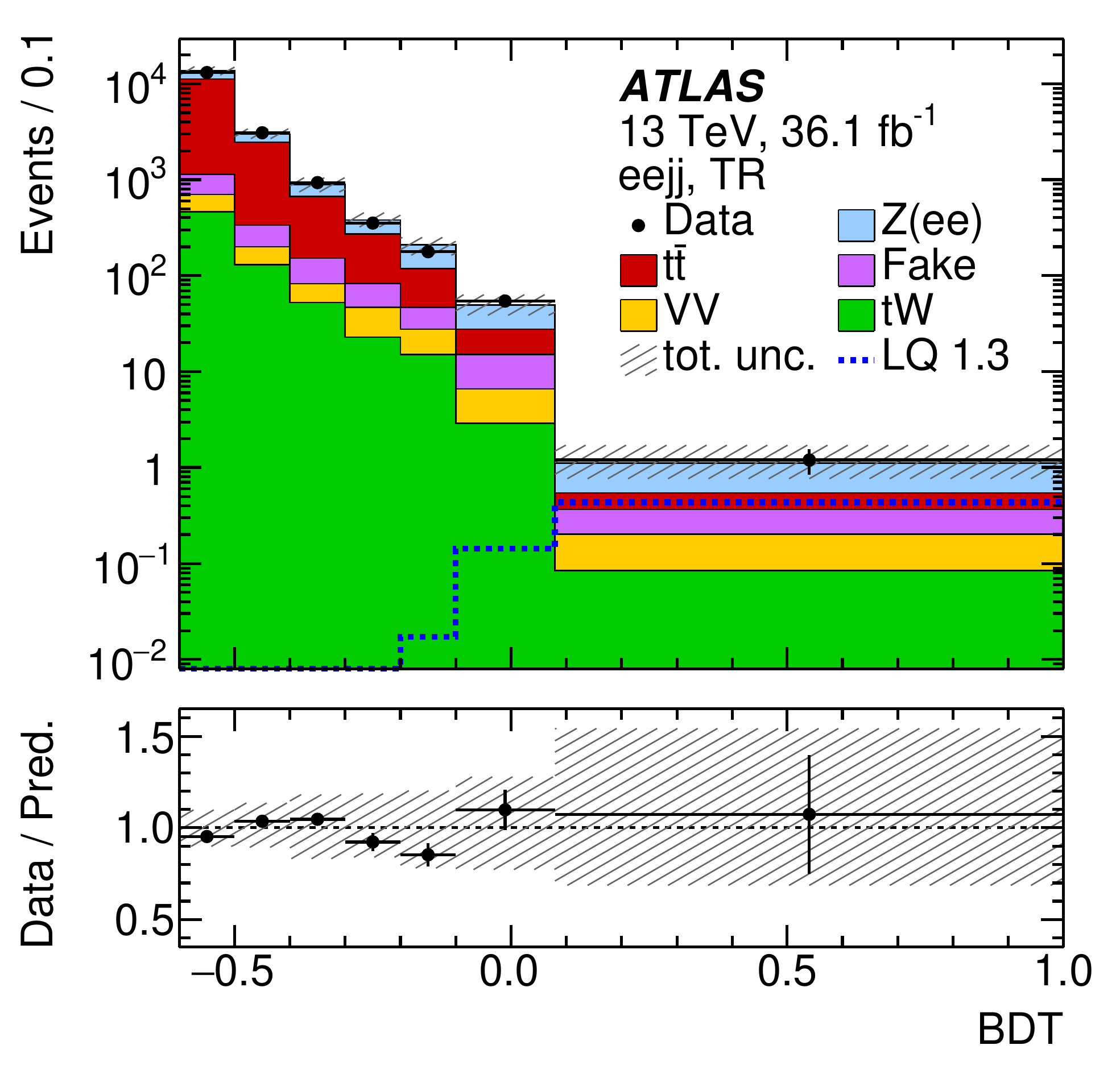} }
\subfloat[]{
    \includegraphics[width=0.42\textwidth]{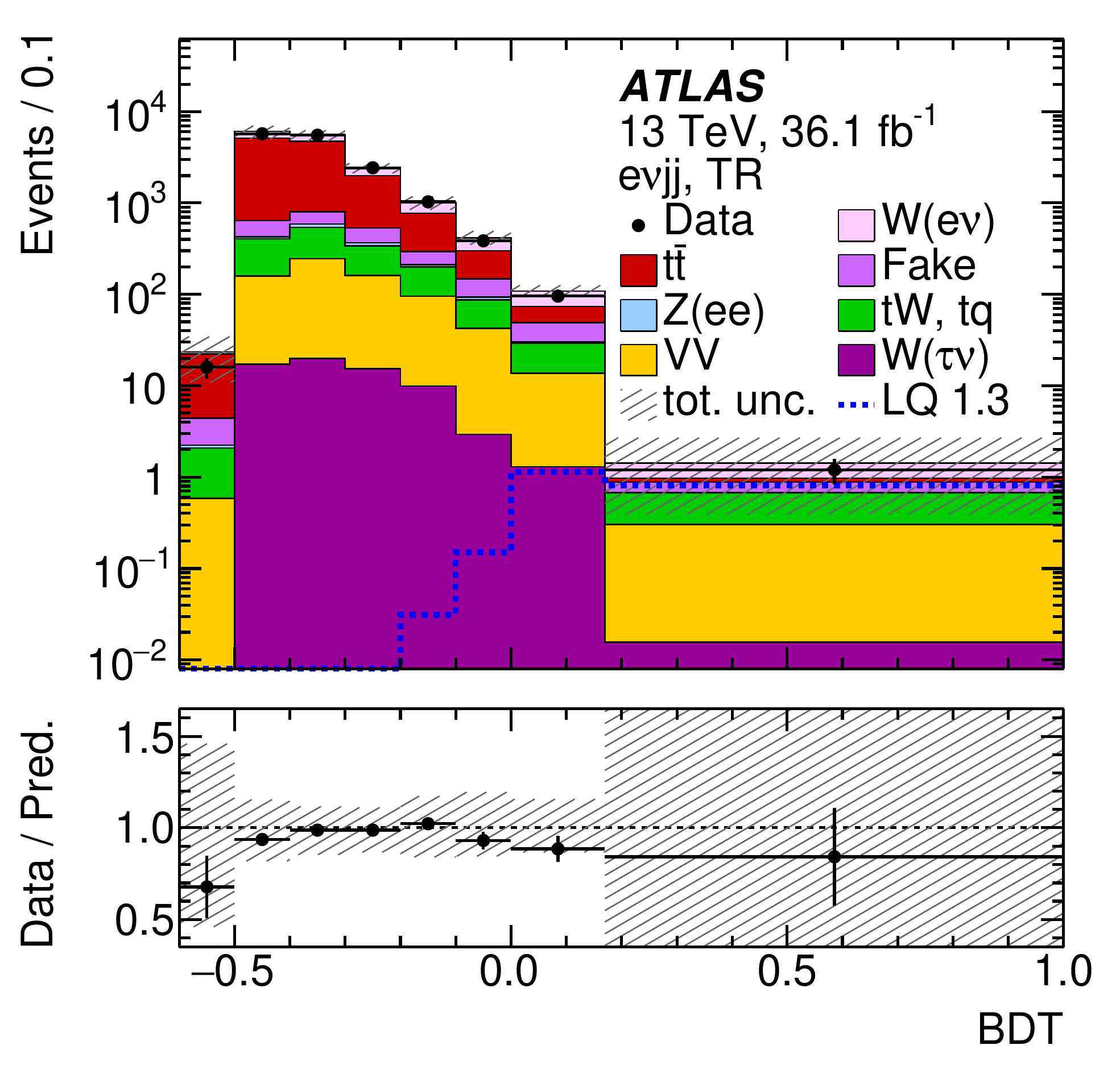}}
\\
\subfloat[]{
    \includegraphics[width=0.42\textwidth]{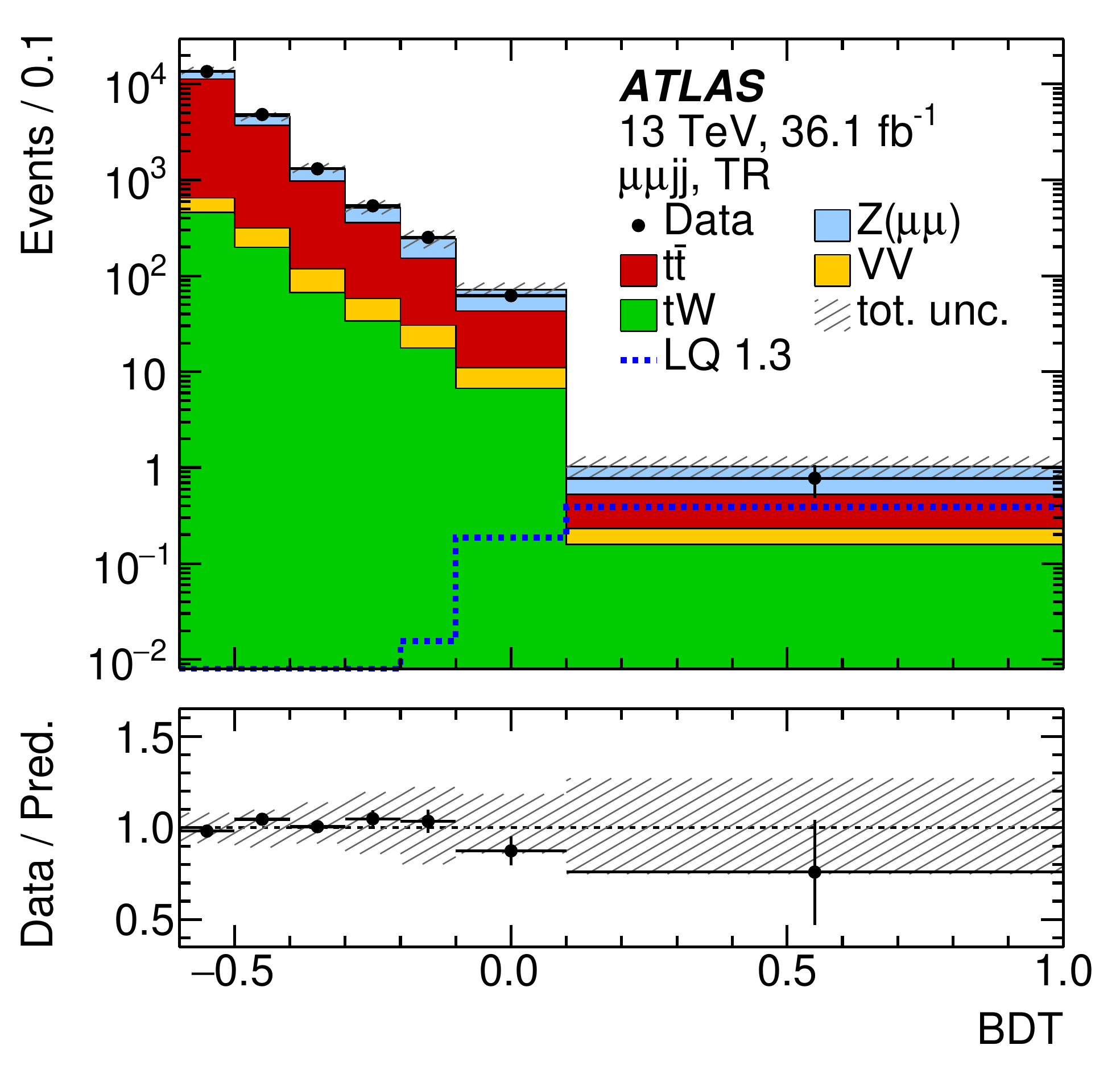}}
\subfloat[]{
    \includegraphics[width=0.42\textwidth]{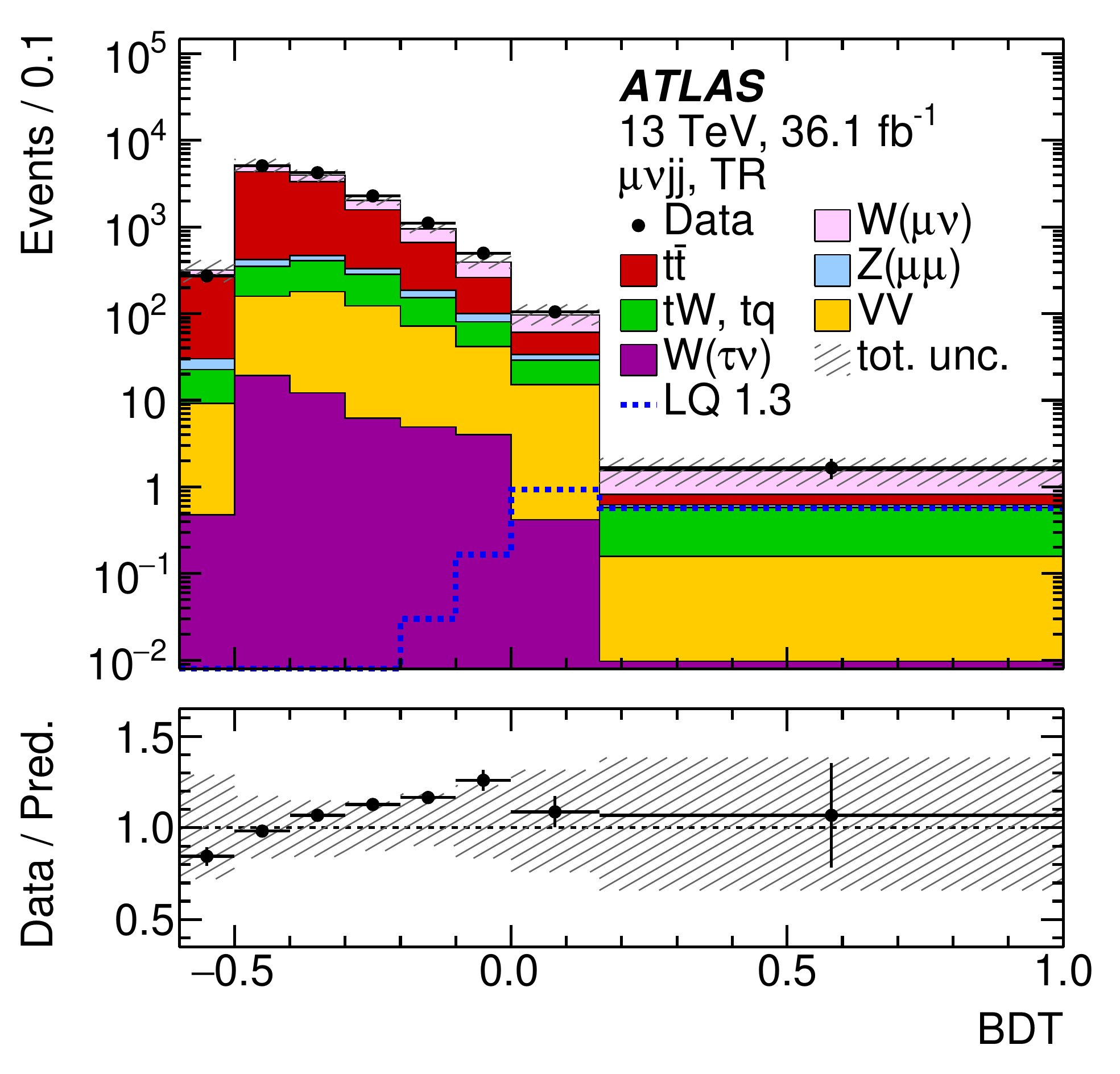}}
  \caption{Output distributions for the BDT in the training regions (TRs) for the four search channels: (a) \eejj (b) \enujj 
(c) \mumujj and (d) \munujj.
Data are shown together with pre-fit background contributions corresponding to \ttbar{}, diboson ($VV$), $Z\rightarrow ee$,  $Z\rightarrow \mu\mu$, $W\rightarrow
\mu \nu$,  $W\rightarrow e\nu$, $W\rightarrow \tau\nu$, single-top production ($tW$, $tq$) processes, and fake electrons. The prediction for a LQ with mass 1.3~TeV is also shown. The 
predicted background distributions are cumulatively stacked.
The bottom panels show the 
ratio of data to expected background.  The grey hatched band represents 
the total uncertainty. 
}
  \label{fig:BDT-out}
\end{figure}

Output distributions for the BDT discriminant in the TRs are shown in Figure~\ref{fig:BDT-out} for a signal sample of 
LQ mass 1.3~TeV. The background distributions are shown before the fit (see Sec.~\ref{sec:method}). The contribution of a LQ of mass 1.3~TeV, normalised to the production cross-section, is also shown for the four channels. The 
background includes the reweighted predictions of $Z/\gamma^*$+jets and $W$+jets. An 
increased separation between signal and background at higher values of the 
BDT output is seen.  

The SR phase-space is a subset of that considered in the TRs. 
The same object and event selections as in the TRs are made in the SRs, but in addition a requirement is placed on the BDT score. 
As stated above, each BDT is trained on half of the events in the TR and evaluated on the other half, and the obtained BDT score distribution is used to determine the SR. 
In this way, the SR determined based on a given BDT does not overlap with the TR used for that BDT. 
For the dilepton and lepton--neutrino channels separately, the SRs are defined as a single bin in the output score distribution of the respective BDT. 
The bin range is determined for each mass point separately by maximising the quantity~\cite{Cowan:2010js} 
${Z=\sqrt{(s+b)\log(1+s/b)-s}}$, where $s$ and $b$ are the signal 
and background expectations, respectively. 
In this optimisation, a number of further requirements are placed on the 
background expectations: there have to be at least two background events in the chosen region; the statistical uncertainty of the number 
of background events estimated by the simulation has to be less than 20\%; if less than ten background events are expected, the 
statistical uncertainty is required to be less than 10\%. This is done to avoid selecting bins with very low background expectation and 
thus to increase the fit stability.

In total, there are 58 SRs in the electron and the muon channel each: two SRs (\lljj{} and \lnujj{}) for each of the 29 mass 
hypotheses considered. 
The fraction of signal events that is left after acceptance selections and losses due to detector inefficiencies, is estimated from the simulation to rise from less than 1\% at low masses (around 200\,GeV) to 60\textendash70\% for masses around 1500\,GeV.

\section{Systematic uncertainties}
\label{sec:Systematics}

The systematic uncertainties for the search are divided into two categories: experimental uncertainties and theoretical 
uncertainties in the background and the signal. Most of the sources of uncertainties are common to 
both the search and the measurement. In addition 
to the common set of uncertainties, the measurement incorporates 
MC modelling uncertainties from closure tests.  The 
search applies uncertainties to normalisation factors of the major 
backgrounds and their extrapolation to the SRs; these are evaluated in the CR-only profile 
likelihood fit in which the normalisation factors are determined. 
In Sections~\ref{sec:Syst_Exp} and~\ref{sec:Syst_Th}, the sources of experimental and theoretical uncertainies are outlined, respectively. 
The impact of the major uncertainties on the search and the measurements are described in Sections~\ref{sec:un_search} and~\ref{sec:Syst_Closure}, respectively.

\subsection{Experimental uncertainties}
\label{sec:Syst_Exp}
The uncertainty in the combined 2015+2016 integrated luminosity is 2.1\%. It is derived, following a methodology similar to 
that detailed in 
Ref.~\cite{DAPR-2013-01}, and using the LUCID-2 detector for the baseline luminosity measurements \cite{Avoni:2018iuv}, from 
calibration of the luminosity 
scale using x\textendash y beam-separation scans.

An uncertainty due to the reweighting of the vertex multiplicity to describe pile-up is estimated by scaling 
the average number of interactions per bunch-crossing in data by different factors.  

The major sources of uncertainties affecting the electron measurement derive from the electron energy scale and resolution measurements.    
The analysis also considers uncertainties due to the modelling  of the efficiencies of the four electron 
selection criteria: trigger, reconstruction, identification and isolation.

Sources of experimental uncertainties related to muon reconstruction and calibration include uncertainties in the determination of the MS momentum scale, MS momentum resolution, ID 
momentum resolution and additional charge-dependent corrections. Furthermore, uncertainties in the determination of the four efficiency scale 
factors, introduced at the end of Section~\ref{sec:sigback},  
for trigger, 
identification, isolation and track-to-vertex matching are taken into account.

There are two main sources of uncertainty related to the jet reconstruction: jet energy scale (JES) and 
jet energy resolution (JER). Another source of jet-related uncertainty corresponds to corrections made for 
the $b$-tagging efficiency. The uncertainty due to the JES is largely derived from various {in situ} techniques.
Four components of this uncertainty are considered.

The uncertainties due to the electron, muon and jet energy scale and resolution are propagated to the 
estimation of \met{}.
In addition, there is an uncertainty from the soft term, which has three components: one corresponding to the \met{} scale and two components for the \met{} resolution uncertainty.
The resolution uncertainty is split into components parallel and perpendicular to an 
axis in the transverse plane which is defined along the direction of a vectorial sum of all the hard objects 
in the event (electrons, muons and jets). 

An uncertainty in the fake rate of electrons is determined by varying the MC prediction that is used to remove the contributions of prompt 
electrons 
from the fake-enriched sample by 30\%. The effect is propagated to the fake background estimate. 
The uncertainty due to neglecting the muon fake background is taken to be half the size of the electron fake background, see Section~\ref{sec::CtrlRegion}.

\subsection{Theoretical uncertainties}
\label{sec:Syst_Th}

The \textsc{Sherpa 2.2.1} $V$+jets samples include event weights 
reflecting variations of the nominal PDF set (\textsc{NNPDF3.0}) and the use of two 
different PDF sets: \textsc{MMHT2014NNLO68CL}~\cite{Harland-Lang:2014zoa} and \textsc{CT14NNLO}~\cite{Dulat:2015mca}. 
The NNPDF intra-PDF uncertainty is estimated as the standard deviation of the set of 101 \textsc{NNPDF3.0} eigenvectors.
This procedure takes into account the effect of varying $\alphas$ by $\pm 0.001$ around its nominal value of 0.119.
The envelope of the differences between the nominal \textsc{NNPDF} set and the other two PDF sets is used as an 
additional uncertainty. 

The samples include weights for seven sets of variation of the renormalisation ($\muR$) and factorisation 
($\muF$) scale, i.e.\ varying the scales up and down by a factor of two, either together or independently. 
The envelope of all these variations is taken as an estimate of the scale uncertainty. 

A further uncertainty from the reweighting of the $V$+jets simulations in 
$m_{jj}$ is considered. The full difference between the reweighted and 
the unweighted 
distributions is taken as an estimate of the uncertainty induced by the reweighting.

The uncertainties in the modelling of the production of \ttbar{} are assessed from a number of alternative simulation samples. 
Differences between different generators and different models for fragmentation and hadronisation are considered.  
In addition, 
parameters affecting initial- and final-state radiation are also varied.   
For the \ttbar{} sample the default PDF set is \textsc{NNPDF3.0} and the PDF uncertainty is estimated from the \textsc{PDF4LHC15} 
prescription~\cite{Butterworth:2015oua}.

\subsection{Uncertainties for the search}
\label{sec:un_search} 

As described in Section~\ref{sec:method} three CRs are fit
simultaneously to obtain normalisation factors. Sources of theoretical uncertainty
described above are taken into account in  this procedure. Furthermore, theory uncertainties in the SRs
correspond to variations in the output BDT spectra and are thus taken into account in the
final limit setting calculation.

The electron energy scale and reconstruction affect the 
total background yield in the SRs for the electron channels by 2--14\%. Similarly uncertainties due to the muon reconstruction and calibration lead to background uncertainties of between 
2 and 20\% for the muon channels. The impact of uncertainties related to the JES (JER) lead to uncertainties on the predicted background of 5--20\% (2--20\%).   
The effect of the soft-term uncertainties in the \met{} estimations leads to variations of the background estimate from less than 1\% to 15\%.  Generally, the experimental uncertainties in
the signal yields in the SRs are not larger than 2\%.

Uncertainties of approximately 100\% for the fake electron background are estimated 
at low values of lepton-jet mass in the TR and about 20\% at masses above 1~TeV.
The statistical uncertainty in the electron fake rate determination is below 10\%. As discussed earlier, the contribution from fake muons is typically half the size of the electron fake 
background, and this is used as a (conservative) estimate of the uncertainty in the total background due to neglecting the muon fake contribution. 

The final modelling uncertainties in the background yields for $Z$+jets and $W$+jets are around 2--10\%, with the scale uncertainty typically being the dominant
source. Signal predictions have an uncertainty of a similar size. The background from \ttbar{} production is affected by a theoretical unceratinty of up to $50\%$, the large value arising due 
to the statistical imprecision of the MC samples.

\subsection{Uncertainties for the cross-section measurements}
\label{sec:Syst_Closure}
The uncertainties described below are those which apply to the particle-level cross-section measurements in the MRs, in addition to the relevant uncertainties from the previous sections.
As is described in 
Section~\ref{sec:unfoldingmethod}, the cross-section calculation involves the subtraction of non-dominant processes from the observed 
data, and the application of bin-by-bin correction factors. The experimental uncertainties play a role in both of these steps,
 since they affect the detector-level yield of both the non-dominant processes (when subtracting from data) and the dominant process 
(when calculating the correction factors). The dominant experimental uncertainties are the JES and the lepton 
efficiencies, which have effects of up to 5\% and 2\% on the fiducial cross-sections respectively. Theory and modelling uncertainties 
in the dominant process affect both the particle-level and detector-level yields, and therefore largely cancel out when 
calculating the correction factors $T/R$ (see Eq.~\ref{eq:diffXsec}).

The theory uncertainties in the dominant processes in the $Z$+jets and \ttbar{} MRs therefore have only a small impact ($\leq 2\%$) on the fiducial cross-sections. 
The theory uncertainties in the estimation of non-dominant processes generally have a minimal impact since these processes usually account for only small fractions of the MR yield. The exception is the case of the \ttbar{} background to the $Z$ MRs, which also contribute a small but non-negligible theory uncertainty.

In addition, bin-by-bin correction uncertainties are derived by comparing the nominal 
correction factors with those obtained using re-weighted 
$V$+jets and alternative \ttbar\ samples. The size of these modelling uncertainties ranges between $0.5\%$ and $4\%$ of the fiducial cross-sections.

The overall uncertainty of the fiducial cross-sections, accounting for all sources of systematic uncertainty, is between 5\% and 9\% 
depending on the MR.

\section{Statistical procedure for the search}
\label{sec:method}
The results of the analysis are interpreted using a profile likelihood 
method as implemented in Histfitter~\cite{Baak:2014wma}, in particular using the asymptotic formulae from Ref~\cite{Cowan:2010js}. 
The signal and backgrounds are described by a binned probability density function (p.d.f.) built using either the 
three CRs described in Section~\ref{sec::CtrlRegion}, 
or the two SRs, one in the dilepton channel and one in the lepton--neutrino channel, as described in  
Section~\ref{sec:BDT&SR}.
All CRs and SRs consist of only a single bin.

The CRs are enriched in a specific background process and have a negligible signal 
contamination;  
they are thus used to normalise the predicted backgrounds to data, whereas the SRs drive the 
signal extraction. 
The p.d.f.\ for the fit to the SRs includes the signal strength, i.e. the scaling factor with respect to the predictions 
for the signal cross-section, as the parameter of interest.
Systematic uncertainties are incorporated in the p.d.f.\ as statistical nuisance parameters. They are introduced as 
shape uncertainties, i.e.\ they can only affect the relative size of a given background in different regions while preserving the overall 
number of expected events. 
This is done in 
order to remove strong correlations with the normalisation factors for the main backgrounds that are extracted from the CRs
and which are used to scale MC yields to match data. 
Experimental uncertainties are treated as being fully correlated between different physics processes and fit regions. 
Theoretical or modelling uncertainties are treated as correlated in different fit regions, but uncorrelated between processes.  
The statistical uncertainties of the MC samples are included in the p.d.f.\ as nuisance parameters by constraint terms described by 
a $\Gamma$ distribution. 
.
The fit is performed in two stages: 
first, only the CRs are included in the fit in order to extract the normalisation factors for the three main 
backgrounds. 
These normalisation factors are then applied to the main backgrounds in the second step, which is a fit to only the SRs. 
The uncertainty in the normalisation factors from the CR fit is introduced in the form of Gaussian nuisance parameters. 
The best-fit central values and constraints on other nuisance parameters are not transferred from the CRs to the SRs. 
The optimal value and error of the signal strength and 
nuisance parameters  as well as their correlations are determined 
simultaneously when the p.d.f.\ is fitted to data. 
This conservative two-step approach is chosen for its simplicity compared with the case of a simultaneous fit to all regions.

For some systematic uncertainties, certain adjustments are made before they are incorporated in the fit, as described in the following. 
Uncertainties larger than 50\% are capped at 50\%, because their large size is due to the statistical uncertainty of simulation samples used to estimate them. 
This applies in particular to the theoretical uncertainties in the \ttbar{} estimation. 
Uncertainties with very asymmetric up and down variations are symmetrised, using the larger of the two variations. 
In the SR fit, experimental uncertainties that before normalisation have an effect smaller than 2\% on the total background or the signal 
expectation are not considered in the fit for the background or the signal, respectively. 
Other nuisance parameters are not considered for certain background processes if their effect is less than 2\% on that process before normalising the uncertainty.


\section{Results}
\label{secl:FResults:StatAnalysis}

\subsection{Results of the cross-section measurements}

The measured fiducial particle-level cross-sections are shown in Table~\ref{tab:fiducialXS} for each of the six MRs. 
The values in all MRs are found to agree with the generator predictions within uncertainties.  
The uncertainties in the measured cross-sections are in all cases dominated by experimental uncertainties (in particular the jet energy scale). 
The theory uncertainties in the generator predictions are dominated by variations in the cross-sections, arising from QCD scale variations in the phase-space regions of the measurements.
In both the standard and extreme MRs, the measured fiducial cross-sections in the $Z$~MRs are found to agree between electron and muon decay modes, as expected.

\begin{figure}[h]
\centering
\includegraphics[width=0.4\linewidth]{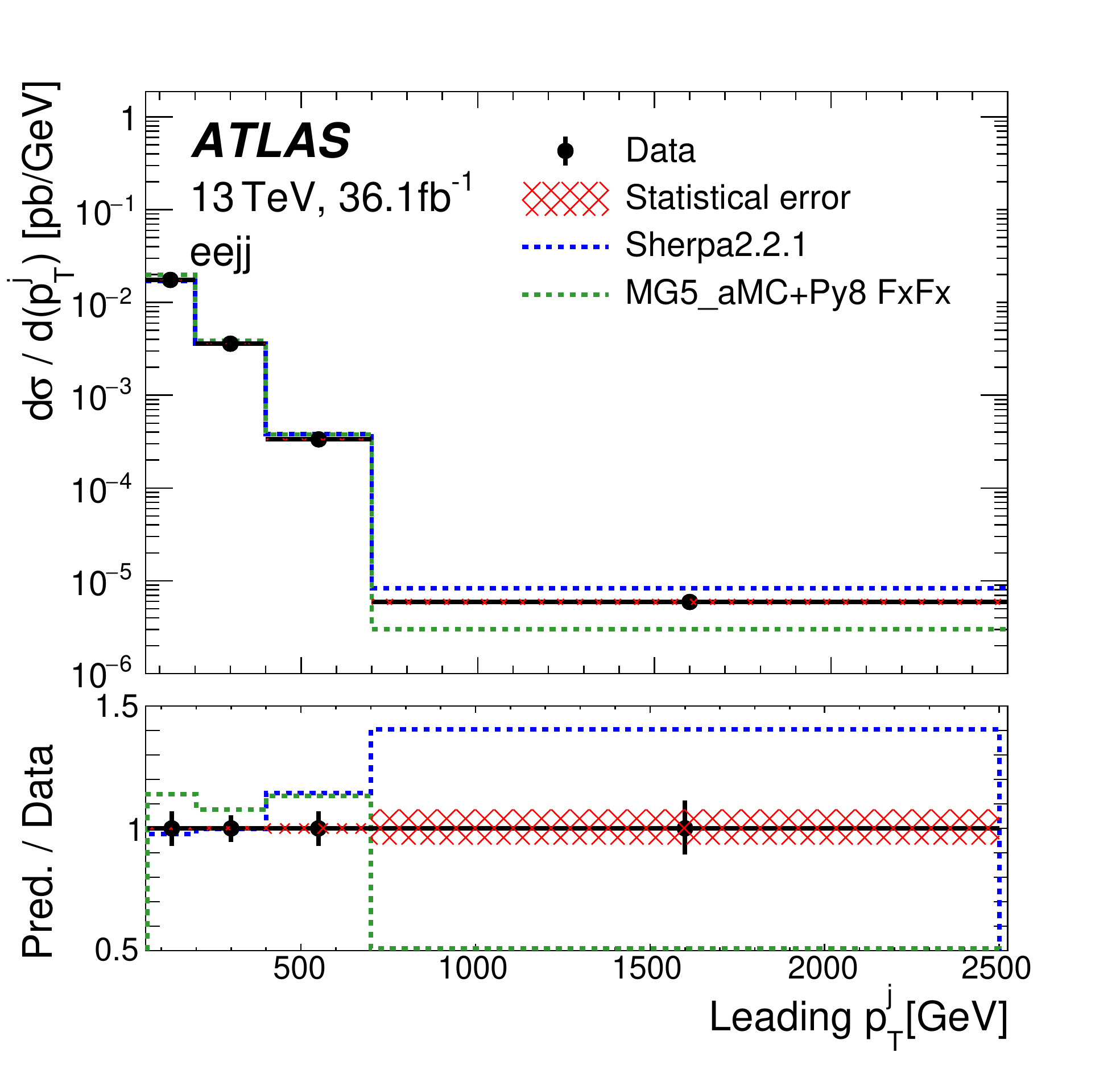}
\includegraphics[width=0.4\linewidth]{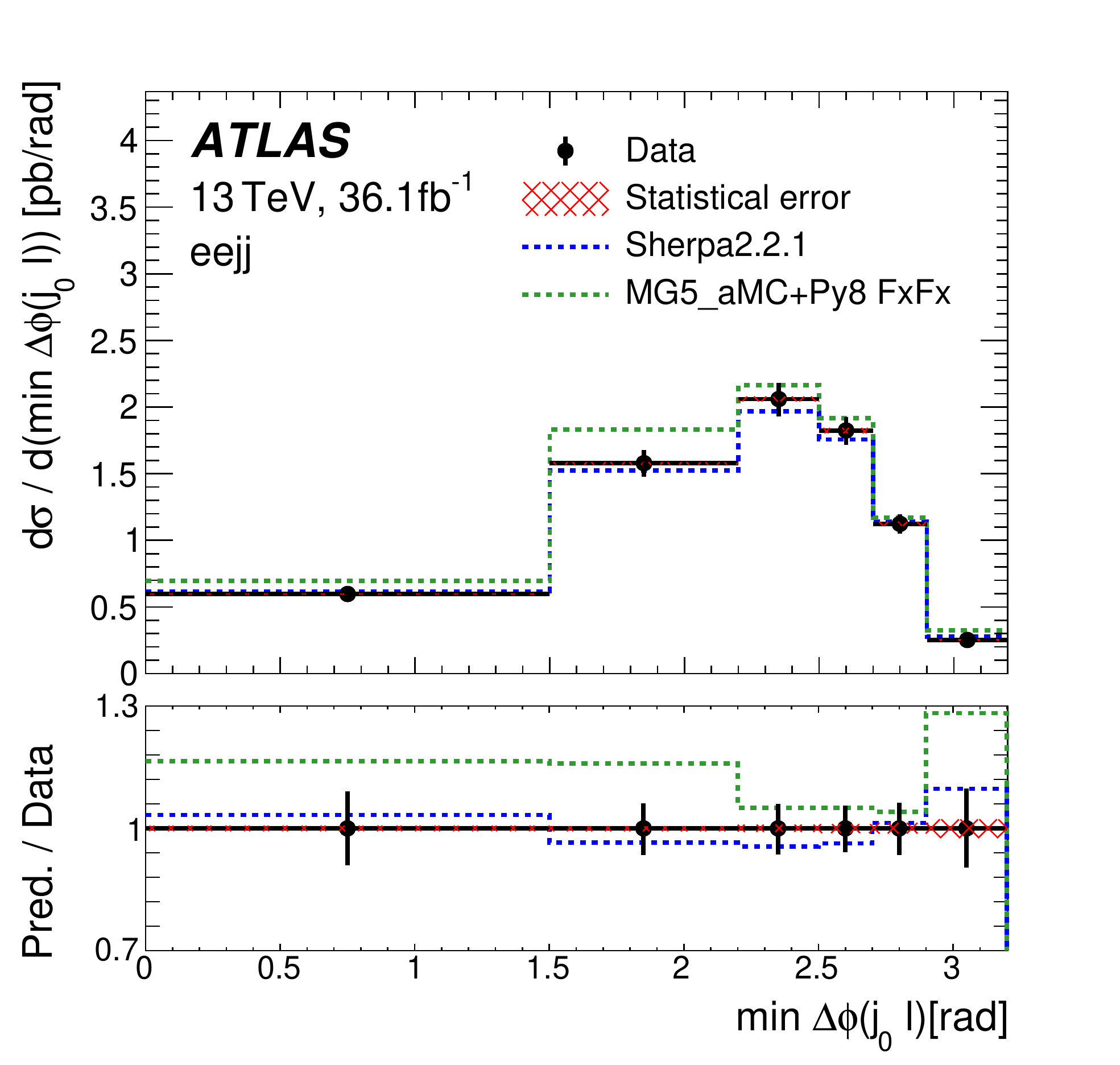}\\
\includegraphics[width=0.4\linewidth]{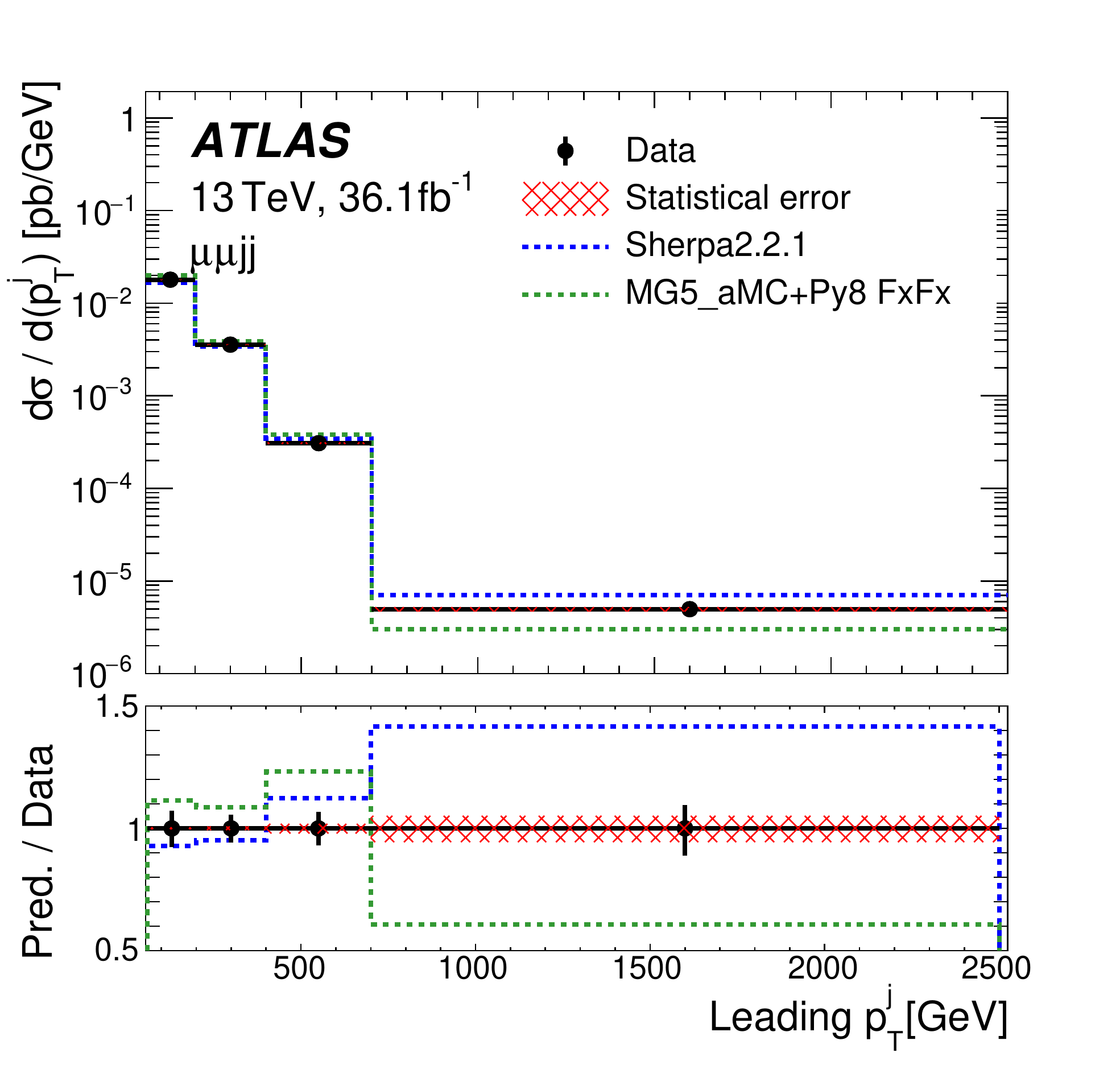}
\includegraphics[width=0.4\linewidth]{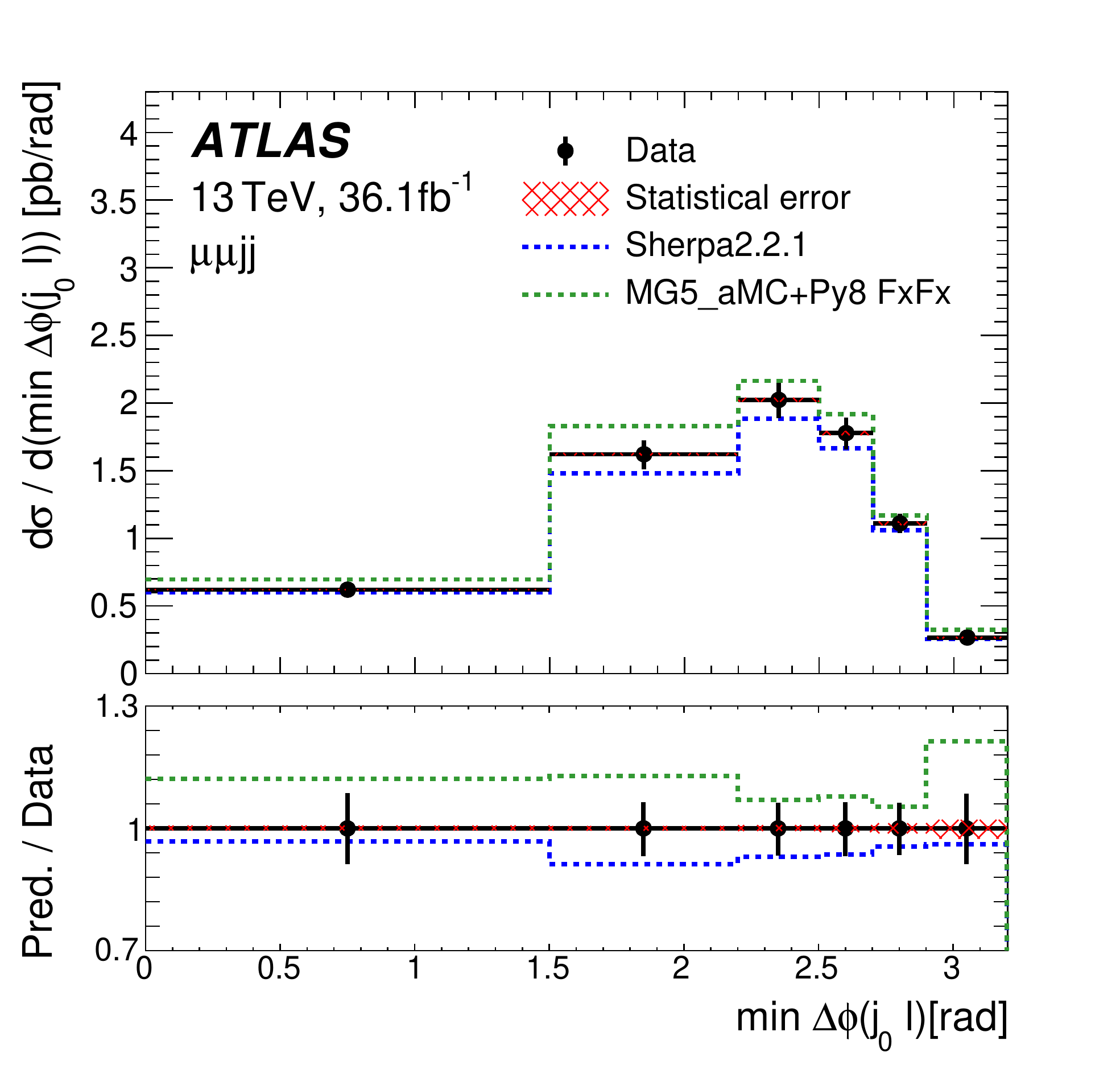}\\
\includegraphics[width=0.4\linewidth]{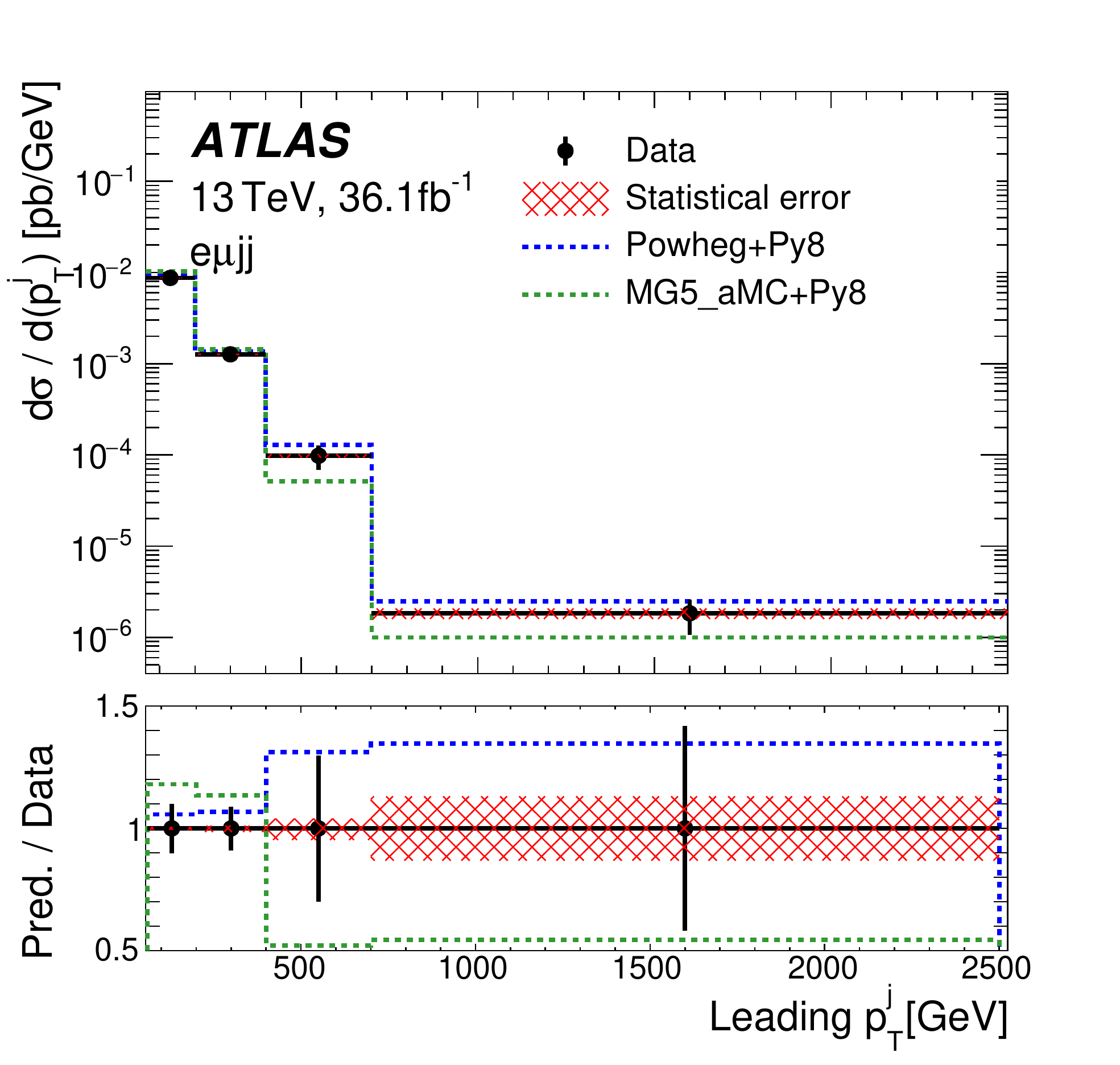}
\includegraphics[width=0.4\linewidth]{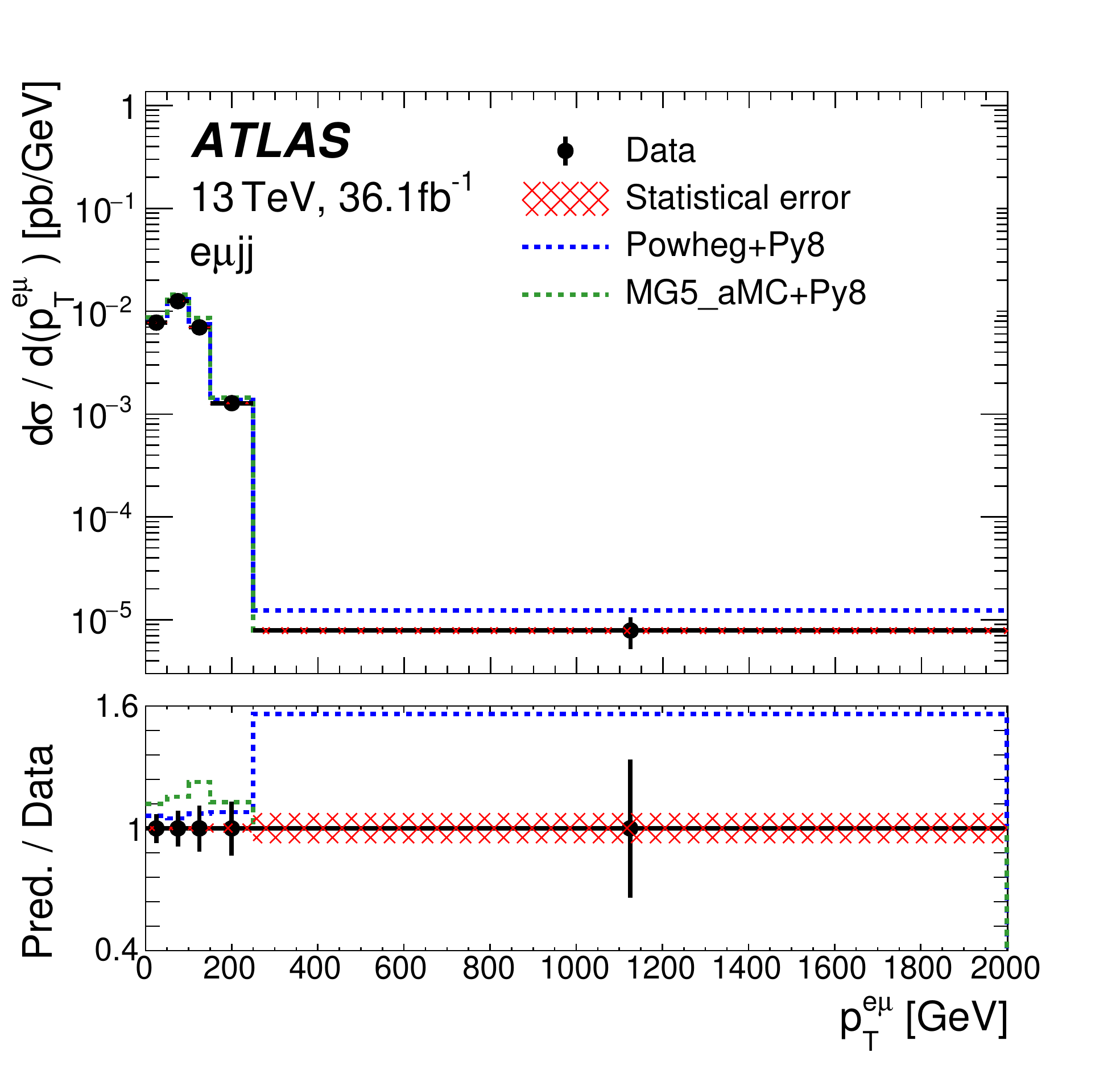}\\
\caption{Examples of the measured particle-level differential cross-sections in the \eejj, \mumujj and \emujj measurement regions, exclusively for the dominant process in each channel (\Zee, \Zmm and \ttbar{}, respectively). The MC prediction for the dominant process is also shown, with no $m_{jj}$ reweighting applied. The red band represents the statistical component of the total uncertainty on the measurement which is indicated by the error bar on each point. The variable $ min \Delta \Phi (j0,l) $ refers to the minimal difference in $\phi$ between the leading jet and a prompt lepton.}
\label{fig:unfolded_data}
\end{figure}

\begin{figure}[h]
\centering
\includegraphics[width=0.4\linewidth]{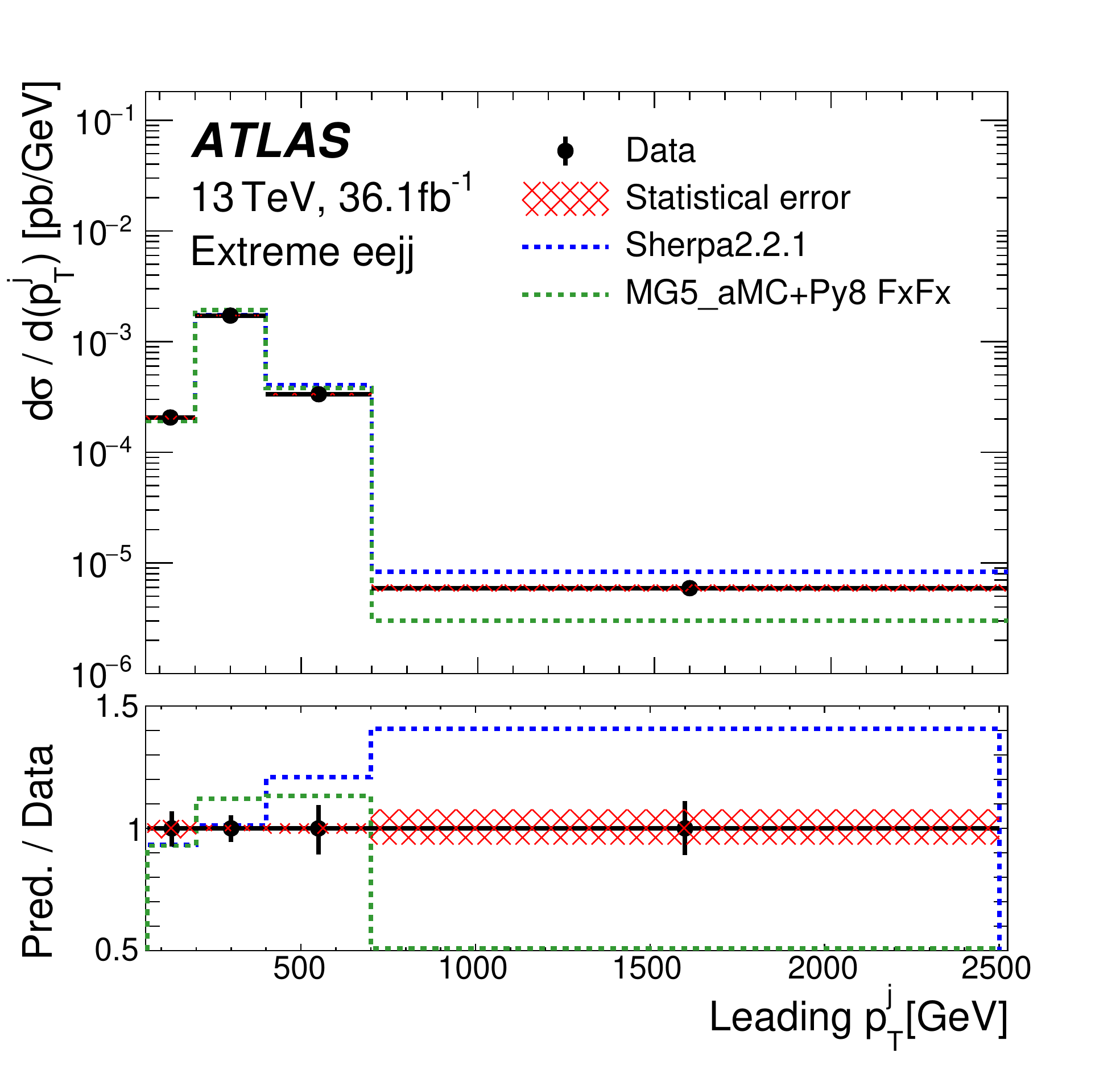}
\includegraphics[width=0.4\linewidth]{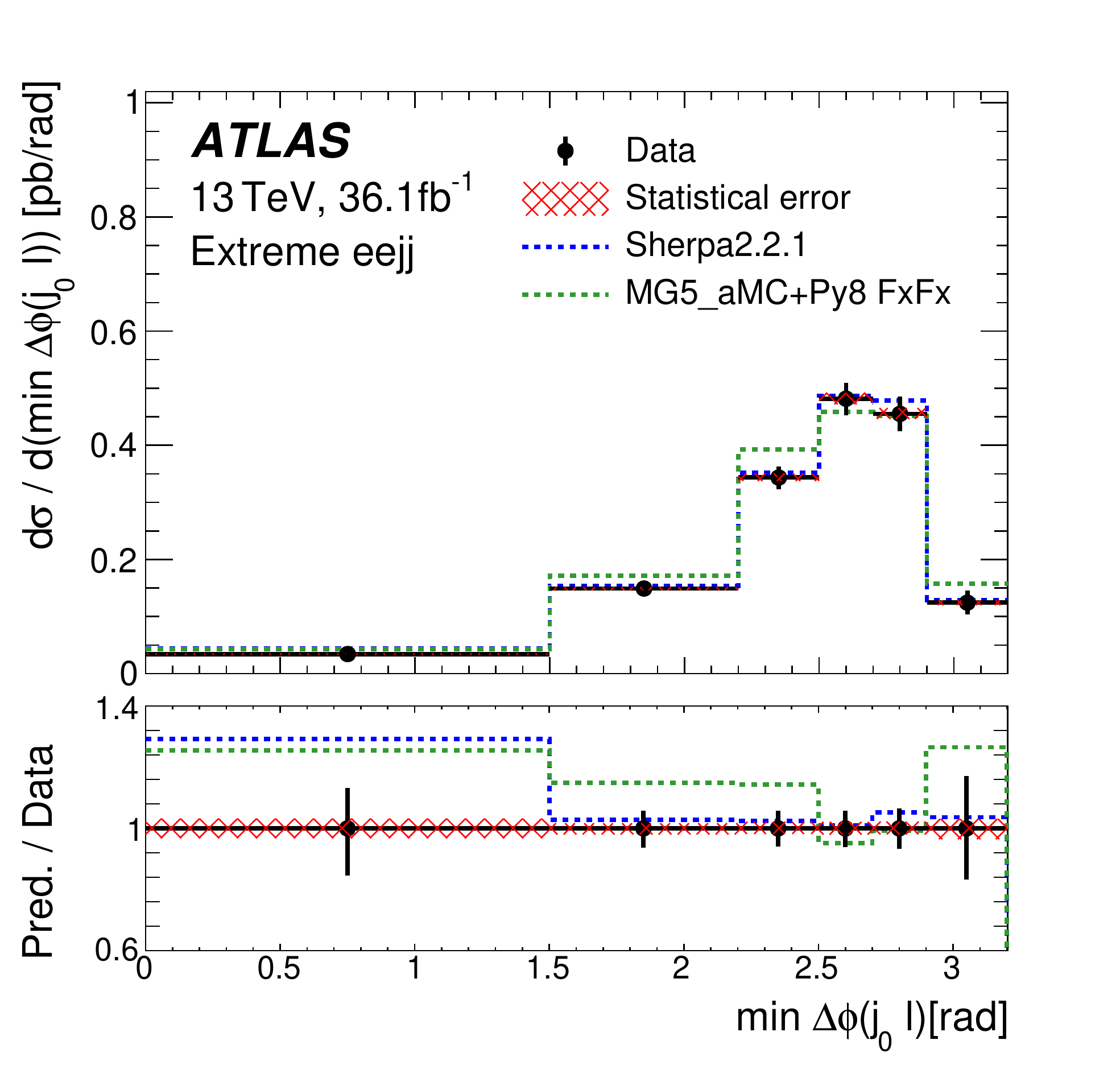}\\
\includegraphics[width=0.4\linewidth]{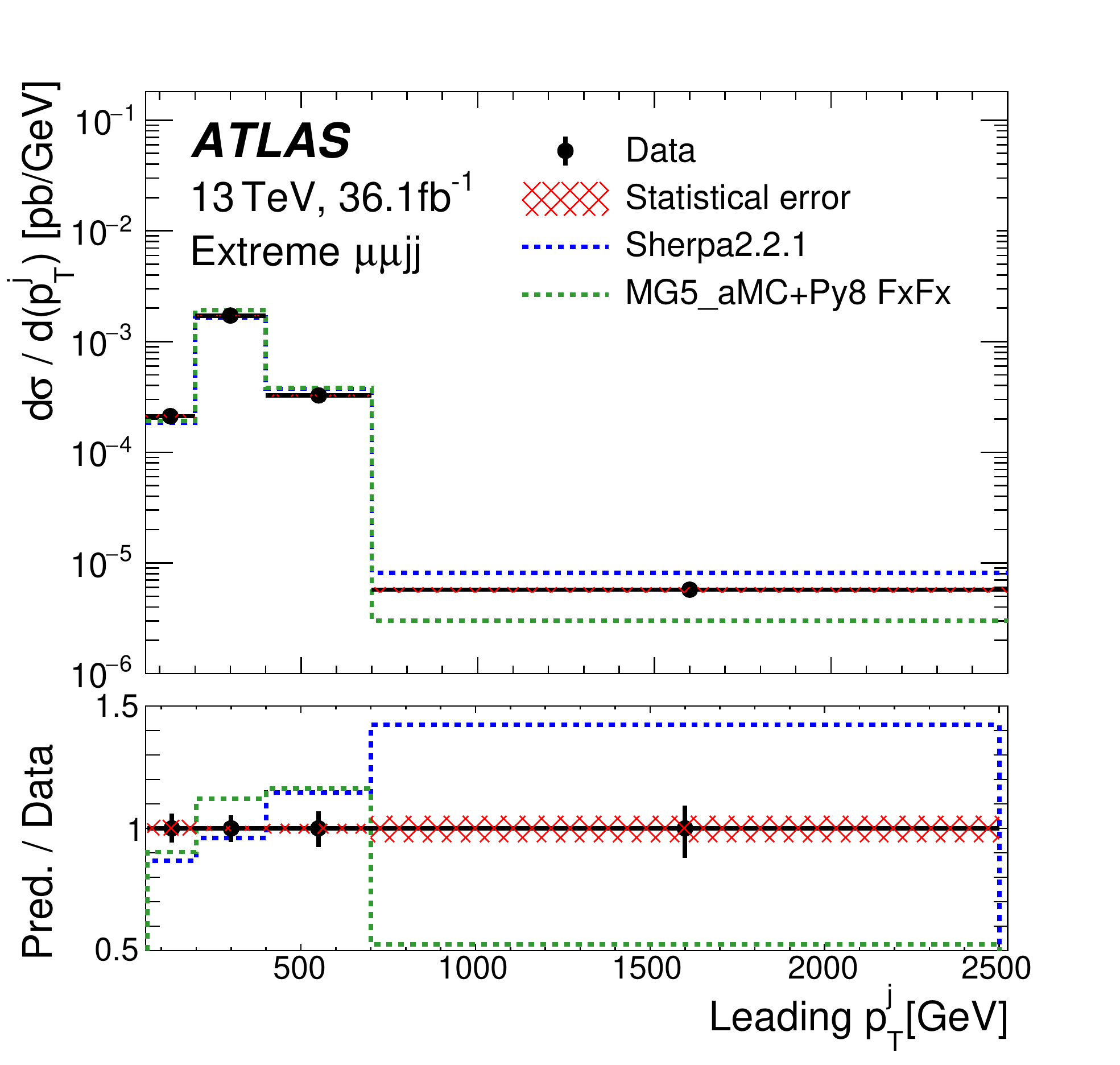}
\includegraphics[width=0.4\linewidth]{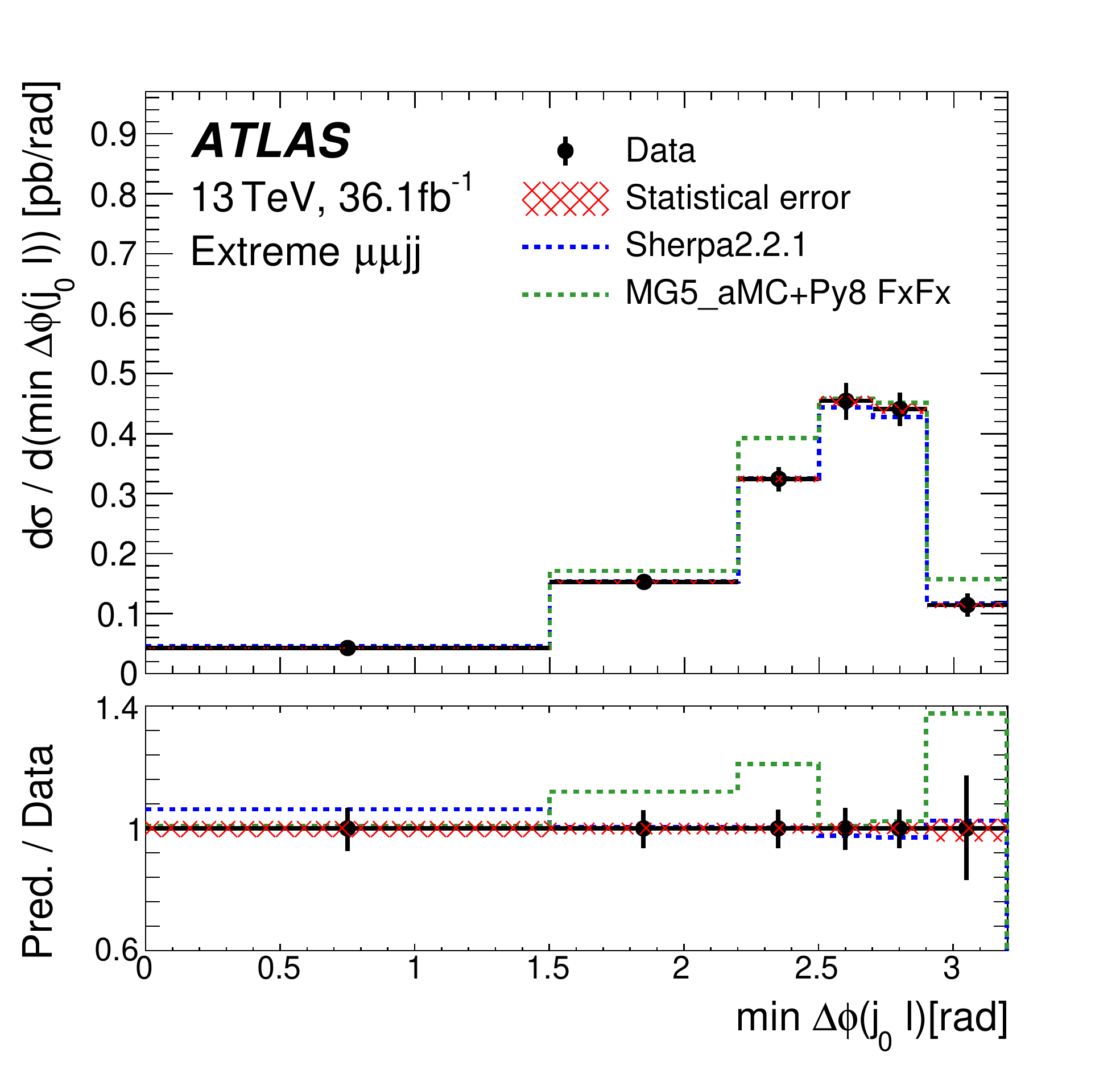}\\
\includegraphics[width=0.4\linewidth]{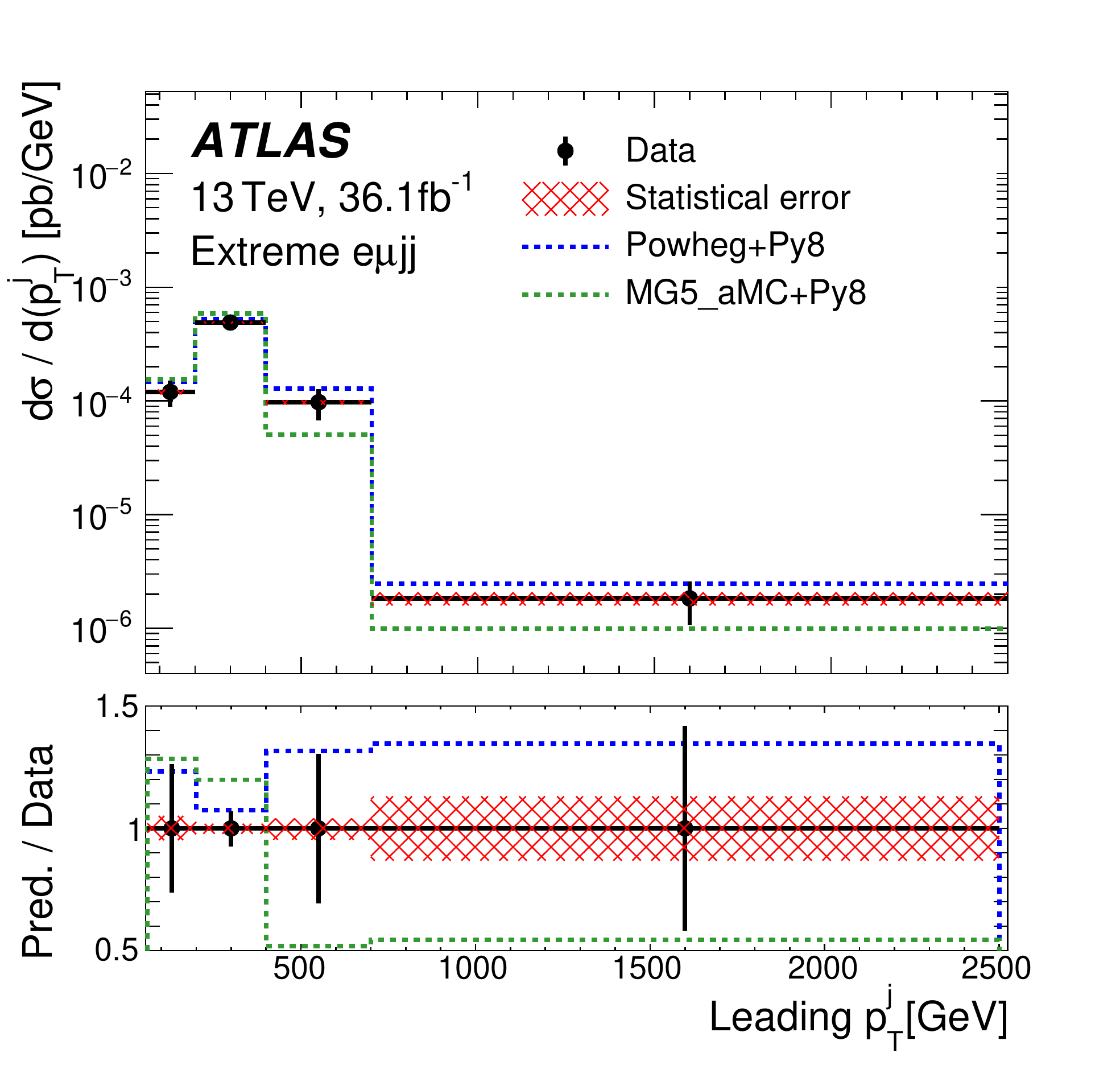}
\includegraphics[width=0.4\linewidth]{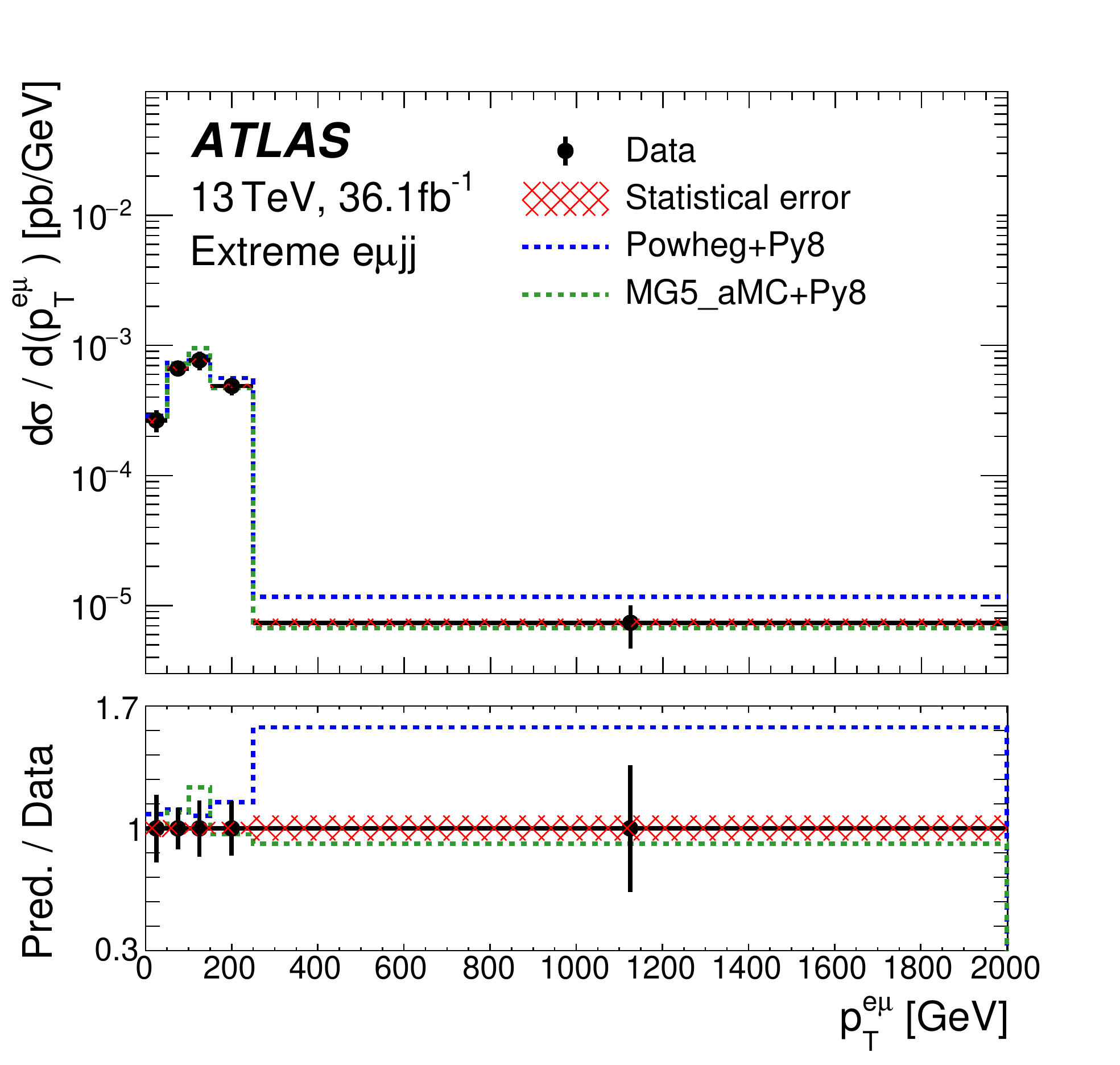}\\
\caption{Examples of the measured particle-level differential cross-sections in the extreme \eejj, \mumujj and \emujj measurement regions, exclusively for the dominant process in each channel (\Zee, \Zmm and \ttbar{}, respectively). The MC prediction for the dominant process is also shown, with no $m_{jj}$ reweighting applied. The red band represents the statistical component of the total uncertainty on the measurement which is indicated by the error bar on each point. For the leading jet \pt\ measurement, the uncertainty in the second bin is smaller because the bin-by-bin correction factor from the alternative \ttbar{} samples agrees more closely with the nominal ones than in neighbouring bins. The variable $ min \Delta \Phi (j0,l) $ refers to the minimal difference in $\phi$ between the leading jet and a prompt lepton.}
\label{fig:unfolded_dataST}
\end{figure}

Examples of the particle-level differential cross-sections are shown in Figures~\ref{fig:unfolded_data} 
and~\ref{fig:unfolded_dataST}, where they are compared with the cross-sections predicted from the 
nominal simulation without the $m_{jj}$ reweighting. Predictions from alternative generators are also shown in each case. 
Differential cross-sections for a larger 
list of variables are included in Appendix~\ref{app:unfoldedxsplots} and in the 
HEPData~\cite{Maguire:2017ypu,hepdata} record for this paper~\cite{hepdatarecord}. 

In the \eejj and \mumujj MRs (and their extreme counterparts) the predicted differential cross-sections for 
variables involving jet energies exhibit a degree of mis-modelling by the \textsc{Sherpa 2.2.1} simulation. This can be seen in 
Figures~\ref{fig:unfolded_data} and~\ref{fig:unfolded_dataST}, for the leading jet $\pt$ as an example. Similar mis-modelling is seen for $m_{jj}$, $H_\mathrm{T}$, subleading jet \pt, and $S_\mathrm{T}$ differential measurements. All other quantities are shown to be well modelled by the nominal \textsc{Sherpa 2.2.1} simulation. 
In all cases, the prediction of an alternative generator, \textsc{MadGraph5\_aMC@NLO} (with events generated at NLO accuracy for $Z+0, 1, 2 $ jets, showered using \textsc{Pythia8} with the A14 tune, and for which different jet multiplicities were merged using the FxFx prescription~\cite{Frederix:2012ps}), is also shown. Typically, the \textsc{MadGraph5\_aMC@NLO} prediction is also discrepant with the data for the same observables, but often in a direction opposite to that of the \textsc{Sherpa 2.2.1} prediction. This suggests that the discrepancies are caused by choices of the generator parameters, which would have been tuned in a different phase-space region given by previous $Z+\textrm{jets}$ measurements. The measurements provided by this paper will therefore help to validate new tunes of the generators to ensure better agreement in this region for future searches. These disagreements in modelling are covered by the reweighting uncertainty when evaluating limits for the search.

For the \emujj MRs, the predictions from the nominal \textsc{Powheg} simulation are found to agree with the data for all observables aside from the dilepton \pt, where a slight mis-modelling is observed in the high tail, as can be seen in the examples shown in Figures~\ref{fig:unfolded_data} and~\ref{fig:unfolded_dataST}. The predictions of an alternative generator (\textsc{MadGraph5\_aMC@NLO}) are also shown, and these do agree with the data for this observable. This points to the generator tuning choice as the source of the disagreement.

\begin{table}[tb]
 \caption{Measured fiducial cross-section for the dominant process in each of the MRs, with uncertainties broken down into the statistical, experimental, and theoretical components. The generator predictions with their uncertainties are obtained from \textsc{Sherpa 2.2.1} for the \Zee and \Zmm samples, and \textsc{Powheg+Pythia8} for the \ttbar{} sample. Alternative generator predictions are also provided, obtained using \textsc{MadGraph5 FxFx} and \textsc{MadGraph5+Pythia8} for the $Z$~MRs and $\ttbar$~MRs, respectively.}
  \resizebox{\textwidth}{!}{
  \centering
  \begin{tabular}{|l|l|l|l|l|}
    \hline
    \multicolumn{1}{|c}{Measurement}                      & \multicolumn{1}{|c}{Dominant}                       & \multicolumn{1}{|c}{Measured cross-section [pb]}                                                    & \multicolumn{1}{|c}{Generator prediction [pb]}                               & \multicolumn{1}{|c|}{Alternative generator}  \\ 
          \multicolumn{1}{|c}{region}                    & \multicolumn{1}{|c|}{process}                        &                                                                                           &   \multicolumn{1}{c}{(\textsc{Sherpa 2.2.1} }    &   \multicolumn{1}{|c|}{prediction [pb]}          \\ 
                                     &                                &                                                                                           &    \multicolumn{1}{c}{or \textsc{Powheg+Pythia8})}    &   \multicolumn{1}{|c|}{(\textsc{MadGraph5})}          \\ \hline
		\eejj                            & \Zee                           &  $ 3.28\,\textrm{ ~} \pm 0.02\, \textrm{(stat.)~~}\pm 0.19\,\textrm{(exp.)~~} \pm 0.09\,\textrm{(th.)} $       & $3.24\textrm{ ~} \pm 1.04 $                          & $3.59\textrm{ ~}  \pm 0.66 $ \\
		\mumujj                          & \Zmm                           &  $ 3.32\textrm{ ~} \pm 0.02\, \textrm{(stat.)~~}\pm 0.21\,\textrm{(exp.)~~} \pm 0.09\,\textrm{(th.)} $       & $3.12\textrm{ ~} \pm 1.01$                           & $3.63\textrm{ ~}  \pm 0.42  $ \\
	  \emujj                           & $\ttbar \rightarrow e\mu$      &  $ 1.503 \pm 0.010 \,\textrm{(stat.)~~}\pm 0.130 \,\textrm{(exp.)} \pm 0.002\, \textrm{(th.)} $    & $1.578 \pm 0.085$                           & $1.702   \pm 0.041  $ \\
		Extreme \eejj                    & \Zee                           &  $ 0.483 \pm 0.005 \,\textrm{(stat.)}\pm 0.024 \,\textrm{(exp.)} \pm 0.014 \,\textrm{(th.)} $   & $0.511 \pm 0.170$                           & $0.660  \pm 0.219   $\\
		Extreme \mumujj                  & \Zmm                           &  $ 0.481 \pm 0.004 \,\textrm{(stat.)}\pm 0.022 \,\textrm{(exp.)} \pm 0.014 \,\textrm{(th.)} $   & $0.483 \pm 0.161$                           & $0.484  \pm 0.069  $ \\
	  Extreme \emujj                   & $\ttbar\rightarrow e\mu$       &  $ 0.147 \pm 0.003 \,\textrm{(stat.)}\pm 0.011 \,\textrm{(exp.)} \pm 0.001 \,\textrm{(th.)} $   & $0.161 \pm 0.032$                         & $0.146  \pm 0.013  $ \\
    \hline
  \end{tabular}
  } 
 \label{tab:fiducialXS}
\end{table}

\FloatBarrier

\subsection{Results of the leptoquark search}

\begin{figure}
  \centering
\subfloat[]{
    \includegraphics[width=0.42\textwidth]{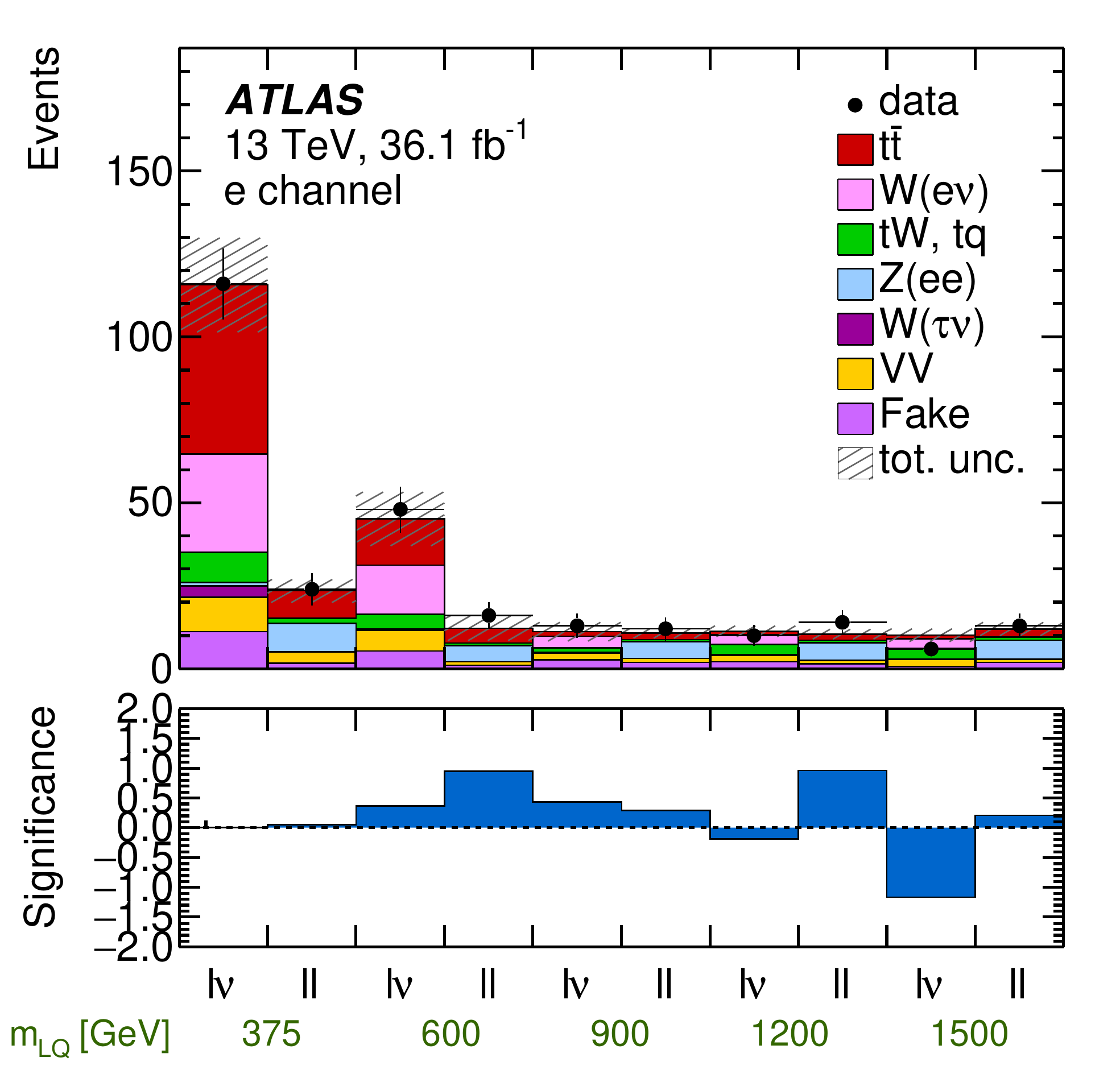} }
\subfloat[]{
    \includegraphics[width=0.42\textwidth]{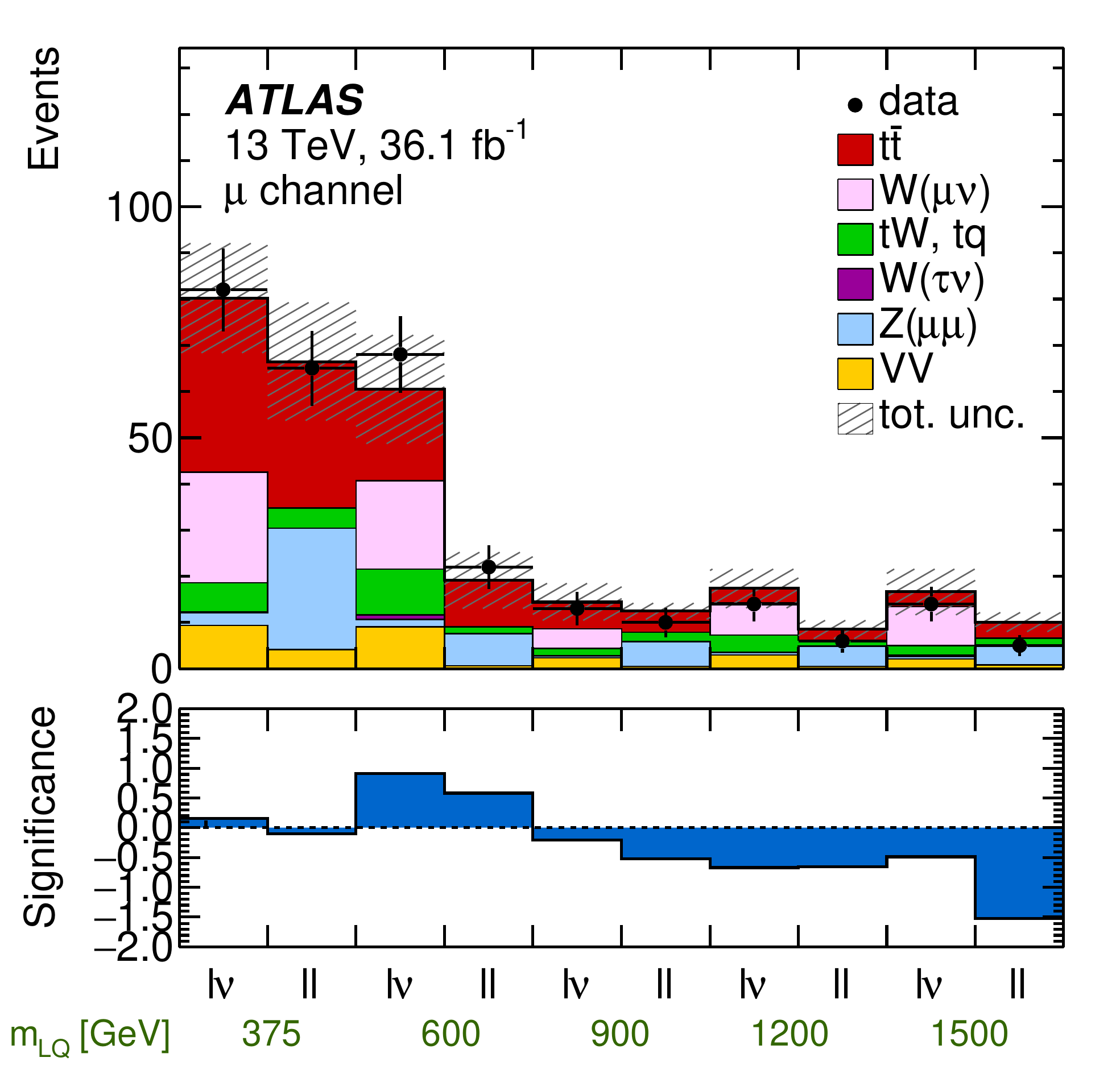}}
  \caption{Data and background yields in the SR for different values of LQ mass for (a)  first- and (b) second-generation LQs. Each bin in 
these plots corresponds to one SR, the bin label indicating the channel, i.e.\ 
dilepton or lepton--neutrino. Two consecutive bins show the two SRs for a given mass hypothesis, that is indicated 
at the bottom of the plots. Mass points from 375\,\GeV\ to 1.5\,\TeV\ are shown.
The 
background contributions correspond to ~\ttbar{}, $W$+jets, diboson 
($VV$), $Z$+jets, single-top ($Wt, Wq$) processes, and
fake electrons. The background distributions are cumulatively stacked.  
The grey band indicates the total uncertainty in the background estimate after the fit.
The bottom panel shows the significance of the deviations, taking only statistical uncertainties into account.}
  \label{fig:SR-lq}
\end{figure}

The data event yields (black points) in a representative subset of the SRs are shown in Figure~\ref{fig:SR-lq} 
together with the predicted background yields as evaluated with the fit. 
The bottom panel shows the significance~\cite{Cowan:2010js} of the 
deviations, assuming 
Poisson statistics and taking only statistical uncertainties into account. 
Even without including systematic uncertainties, no significant excess above the SM expectation is observed in
 any of the SRs. The modified frequentist
${\mathrm{CL}}_{\textrm{s}}$ method \cite{0954-3899-28-10-313} is used to set limits on the strength of the LQ signal.

Each BDT uses as inputs variables that reconstruct the LQ mass. Since these have a high discriminating power, the 
sensitivity of the SRs exhibits a strong dependence on the LQ mass below values of $\mlq$ around 600~\GeV, where the intervals between simulated mass points are greater than the typical mass resolution for adjacent masses (the converse is true for masses above around 600~\GeV). 
Therefore, in this regime a simple interpolation of acceptance between simulated mass points does not give a reliable 
estimate of the limit in the mass interval. To account for this limitation, results are presented separately for low and high 
mass regions, defined as being less than and greater than 600~\GeV, respectively. 

A coarse scan of the mass is performed in the low mass region in intervals of 50 GeV. Exclusion limits that are obtained at these scan points are used as conservative limits for 
the intermediate masses between these points. While for masses above $\sim  600$~GeV the  mass-scan intervals are  smaller than the mass resolution at the test points, the 
intervals used below $\sim 600$~GeV are  larger than the mass resolution. Therefore, above  $\sim  600$~GeV  a simple interpolation is used to obtain the limits in the mass 
intervals. In the range below $\sim 600$~GeV this is not appropriate and a different approach is used. 
Since the acceptance of a given SR is highest for the mass point which the SR is designed for and lower at the neighbouring mass points, the limit is estimated for a given mass interval, from $m_1$ to $m_2$, in the following way; two limits are evaluated  
for $m_1$ and $m_2$ using the SR defined by $m_1$ (i.e. the respective BDT trained for the mass hypothesis $m_1$). Similarly, two estimations are calculated for the SR defined by $m_2$. 
For each SR, the weaker of the two limits is retained, leaving two limits to consider, one from each SR. 
Of these two remaining results from the two SRs, the stronger limit is used.

\begin{table}[tb]
\caption{Expected and observed 95\%\,CL lower limits on first- and second-generation LQ masses in 
units of GeV for different values of the 
branching ratio  $\beta$  into a charged lepton and quark.}
\begin{center}
\begin{tabular}{|c|rr|rr|}
\hline 
                          & Expected                                    & Observed & Expected & Observed             \\  
                          &  ($m_{\text{LQ1}}$)                   & ($m_{\text{LQ1}}$)  & ($m_{\text{LQ2}}$) &  ($m_{\text{LQ2}}$)   
\\
\hline
1.0                     & 1400                                       & 1400     & 1400     & 1560                 \\ \hline
0.5                     & 1280                                        & 1290      & 1200      & 1230                  \\  \hline
0.1                     & 1020                                         & 1010      & 960      & 960                  \\ 
\bottomrule
\end{tabular}
\end{center}
\label{tab:massLimits}
\end{table}

\FloatBarrier

\begin{figure}[h]
  \centering
\subfloat[]{
    \includegraphics[width=0.42\textwidth]{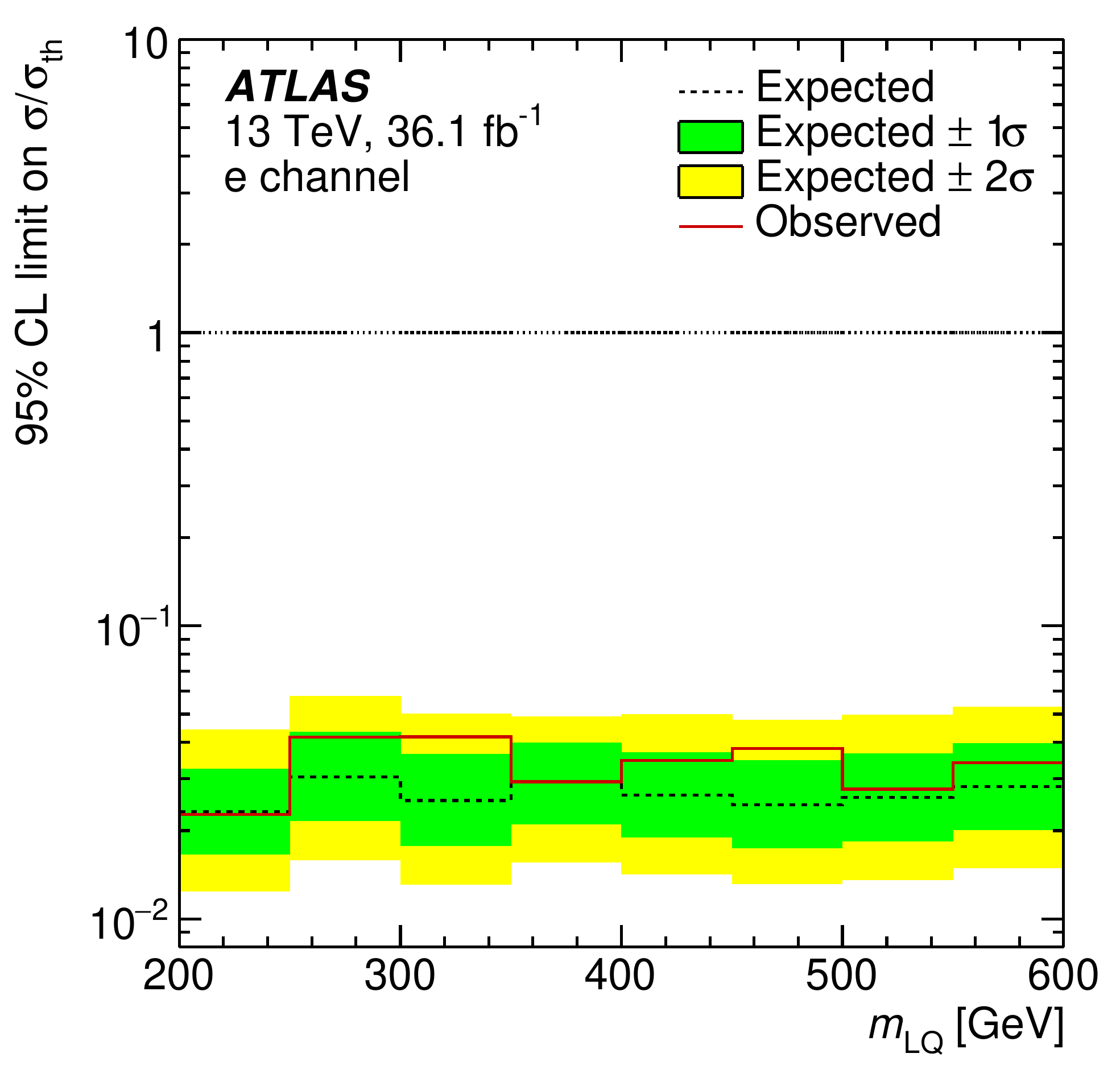}}
\subfloat[]{
    \includegraphics[width=0.42\textwidth]{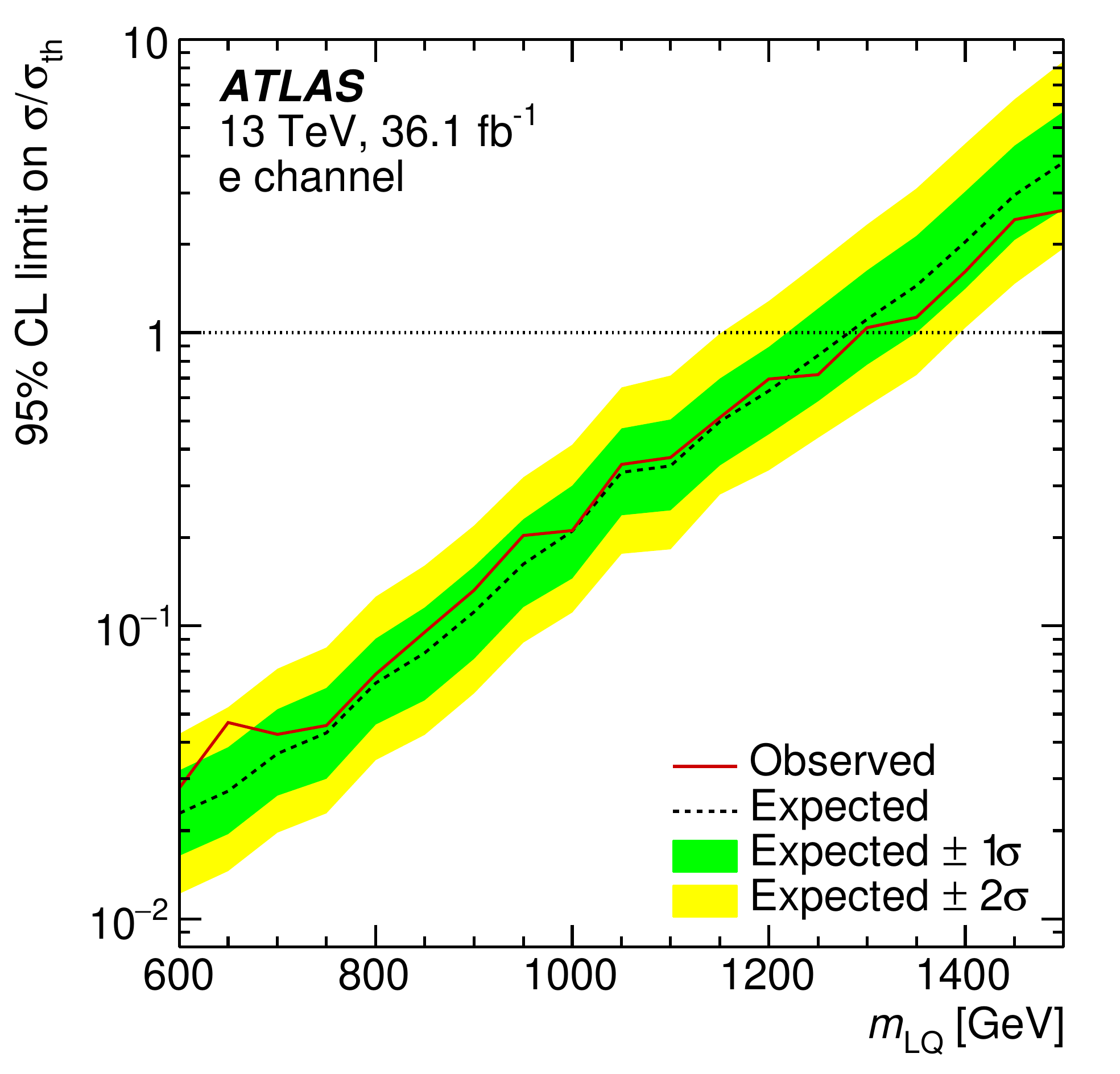}}
\\
\subfloat[]{
    \includegraphics[width=0.42\textwidth]{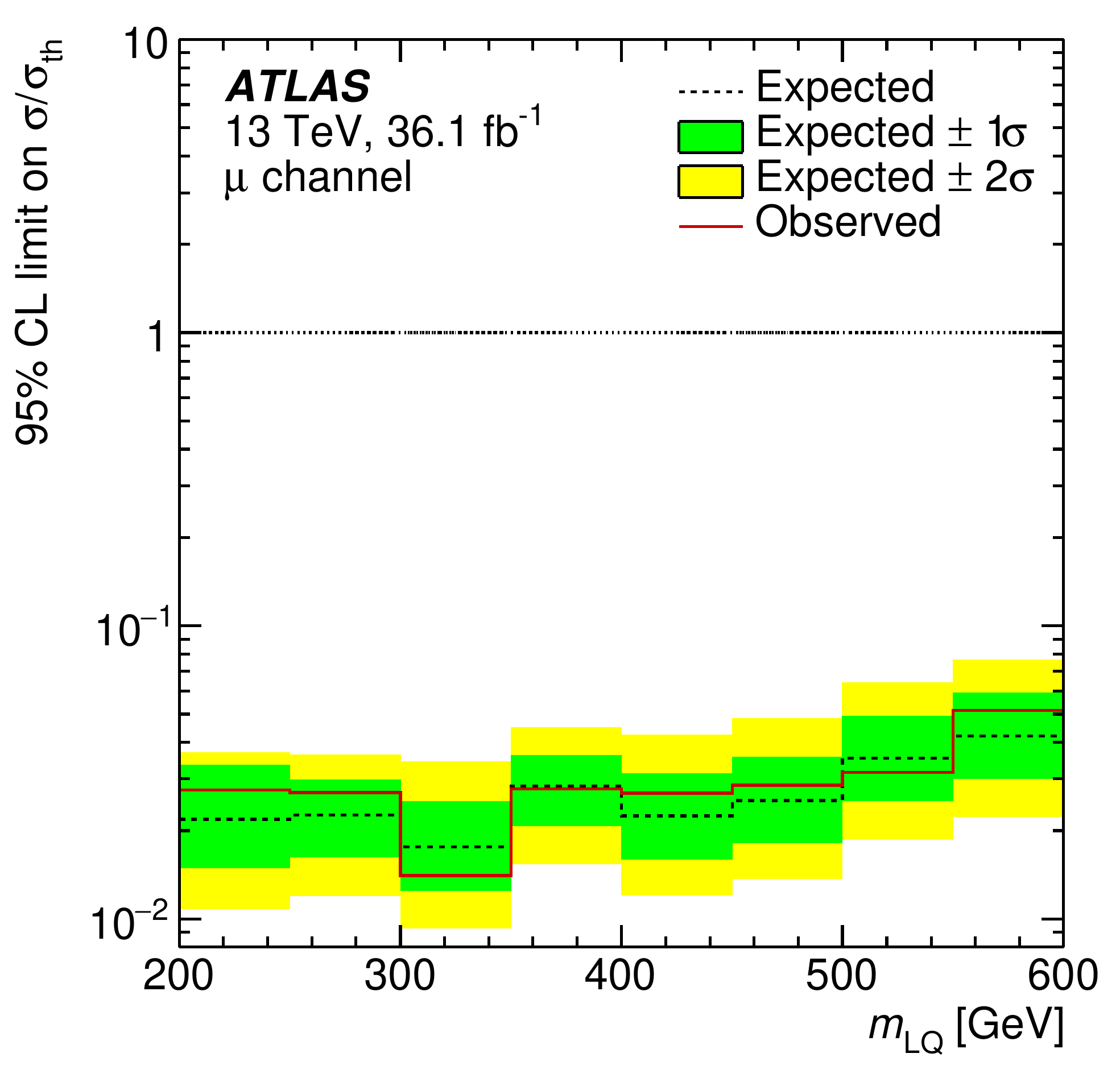}}
\subfloat[]{
    \includegraphics[width=0.42\textwidth]{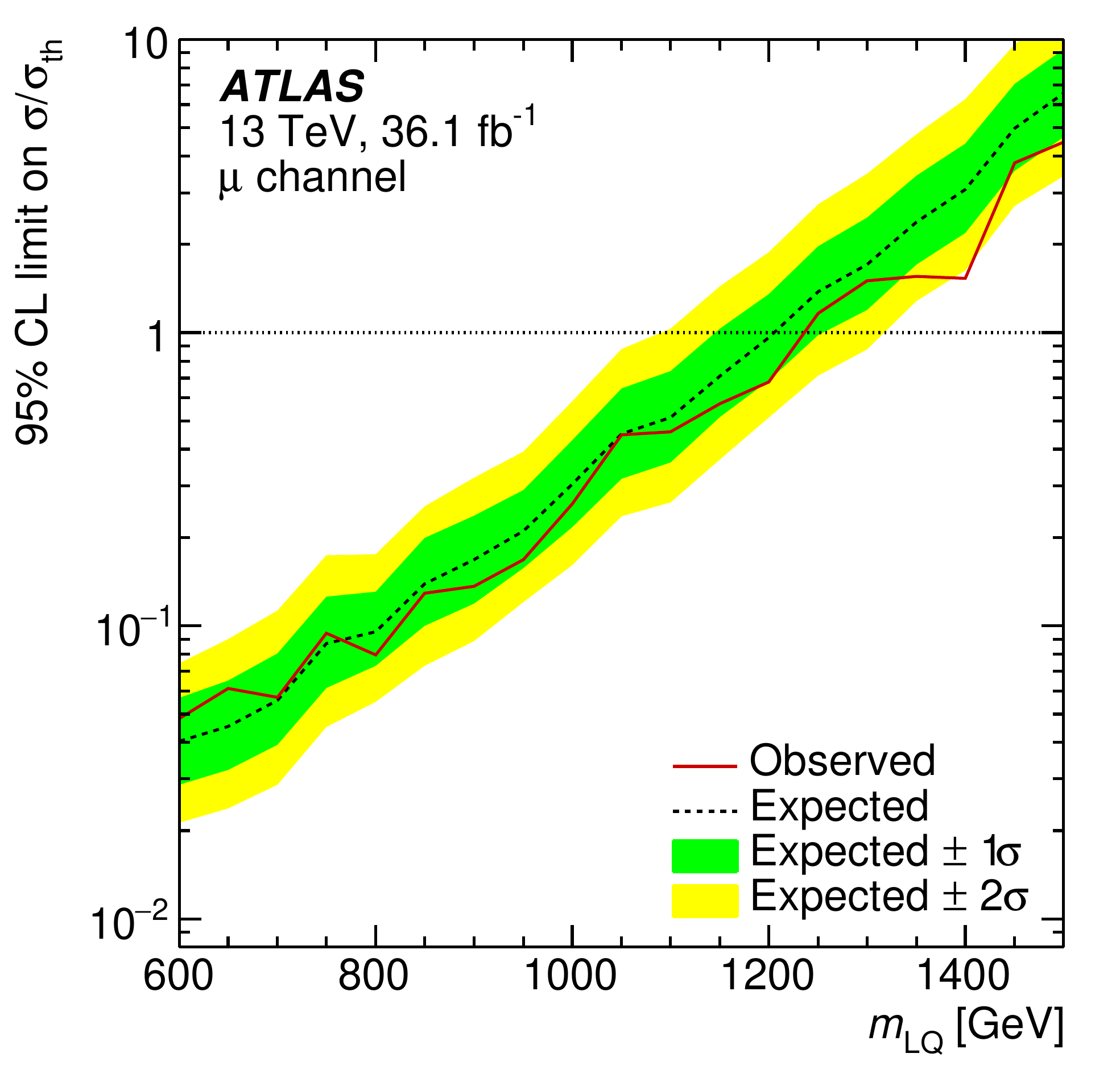}}
 \caption{Upper limits (observed and expected) on the pair-production cross-section for (top) first- and (bottom) second-generation 
LQs normalised to the predicted cross-section ($\sigma/\sigma_\textrm{th}$),
as a function of LQ
mass for an assumed value of $\beta=0.5$.  Limits are presented for low (left) and high (right) mass regions, which correspond to 
masses greater
than and less than $600$~\GeV, respectively. The $\pm~1\sigma$ ($\pm 2\sigma$) uncertainty bands on the expected limit represent all
sources of systematic and statistical uncertainty.}
  \label{fig:limits-m}
\end{figure}

\begin{figure}[h]
  \centering
\subfloat[]{
    \includegraphics[width=0.42\textwidth]{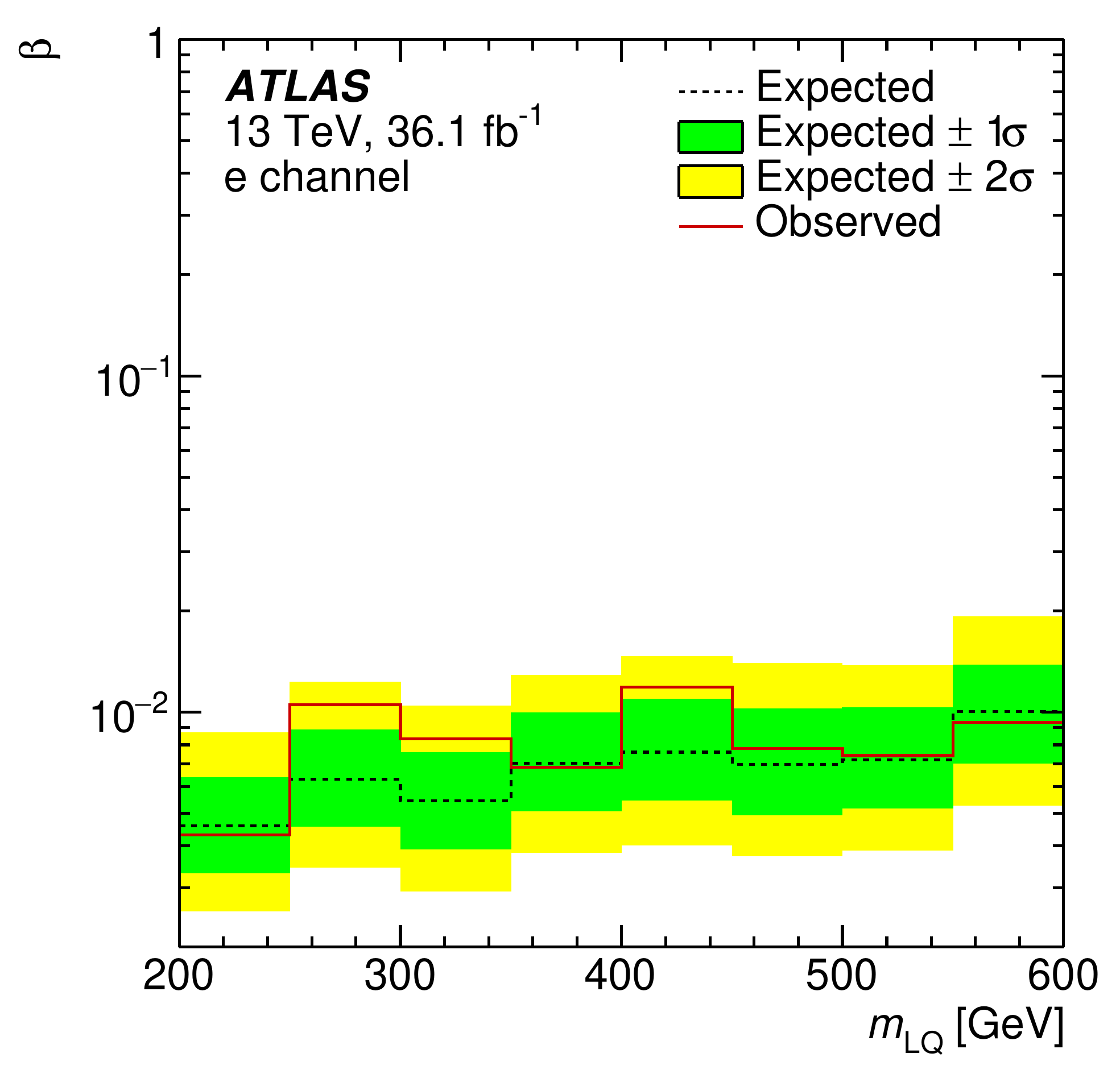}}
\subfloat[]{
    \includegraphics[width=0.42\textwidth]{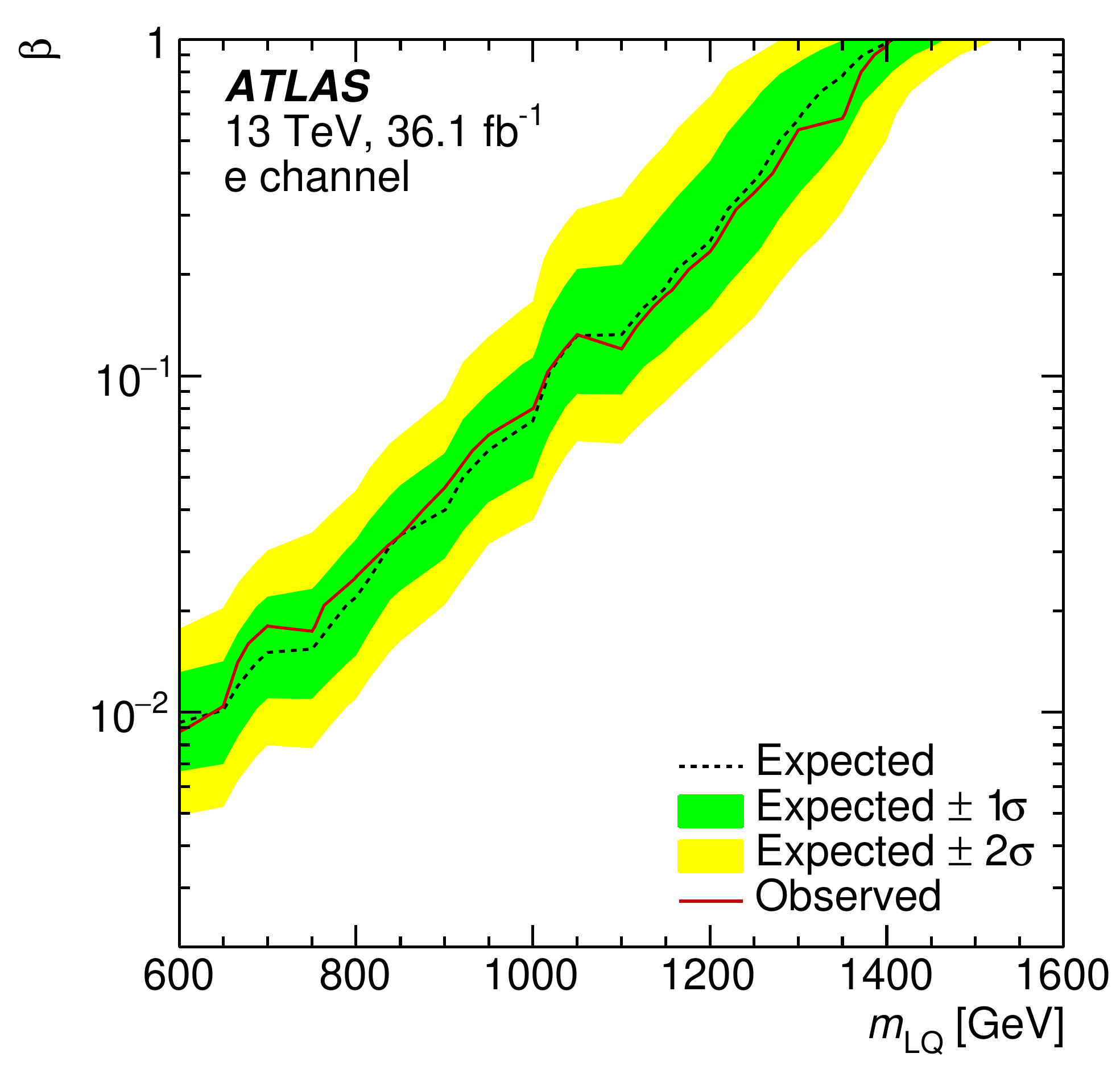}}
\\
\subfloat[]{
    \includegraphics[width=0.42\textwidth]{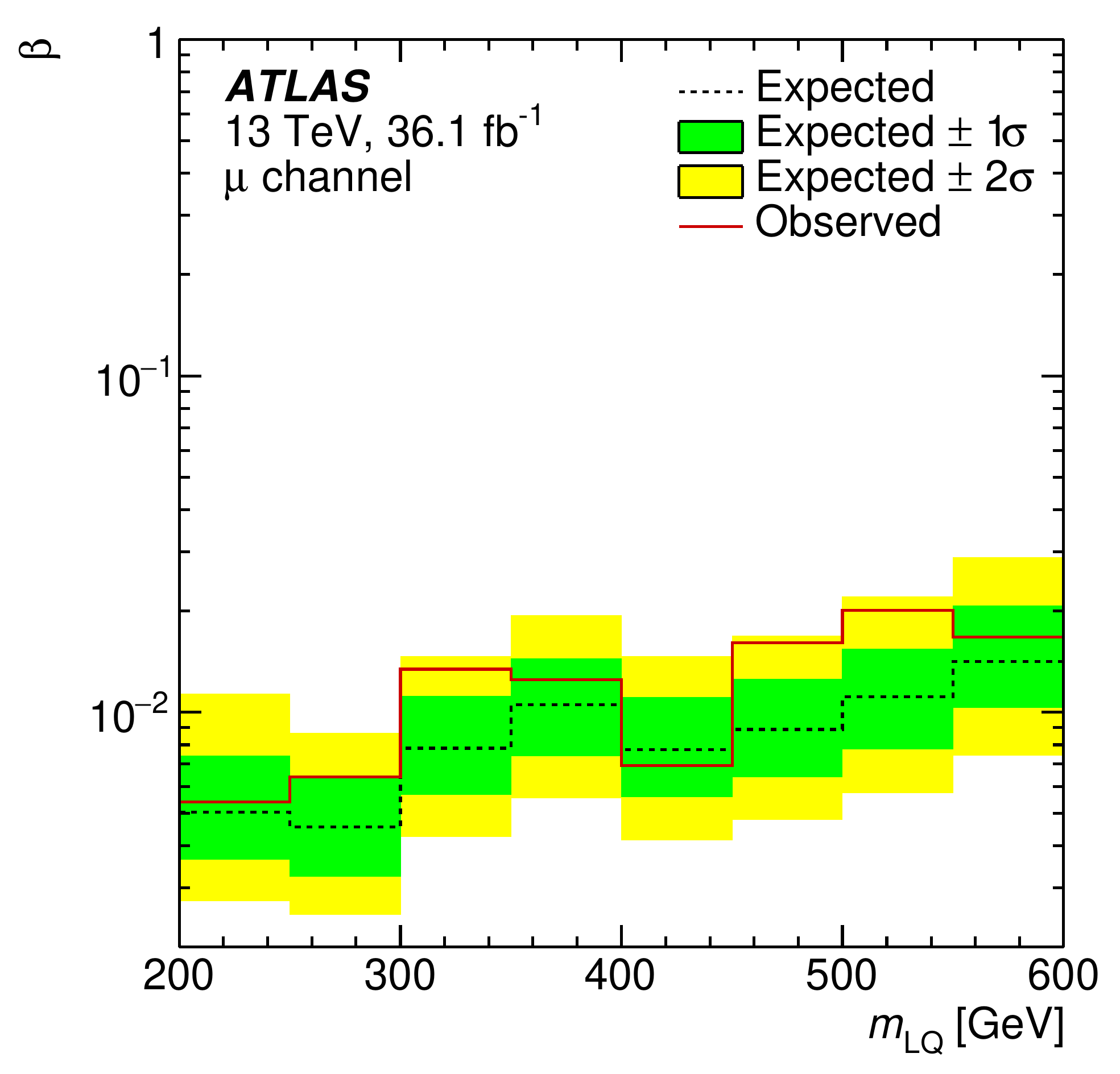}}
\subfloat[]{
    \includegraphics[width=0.42\textwidth]{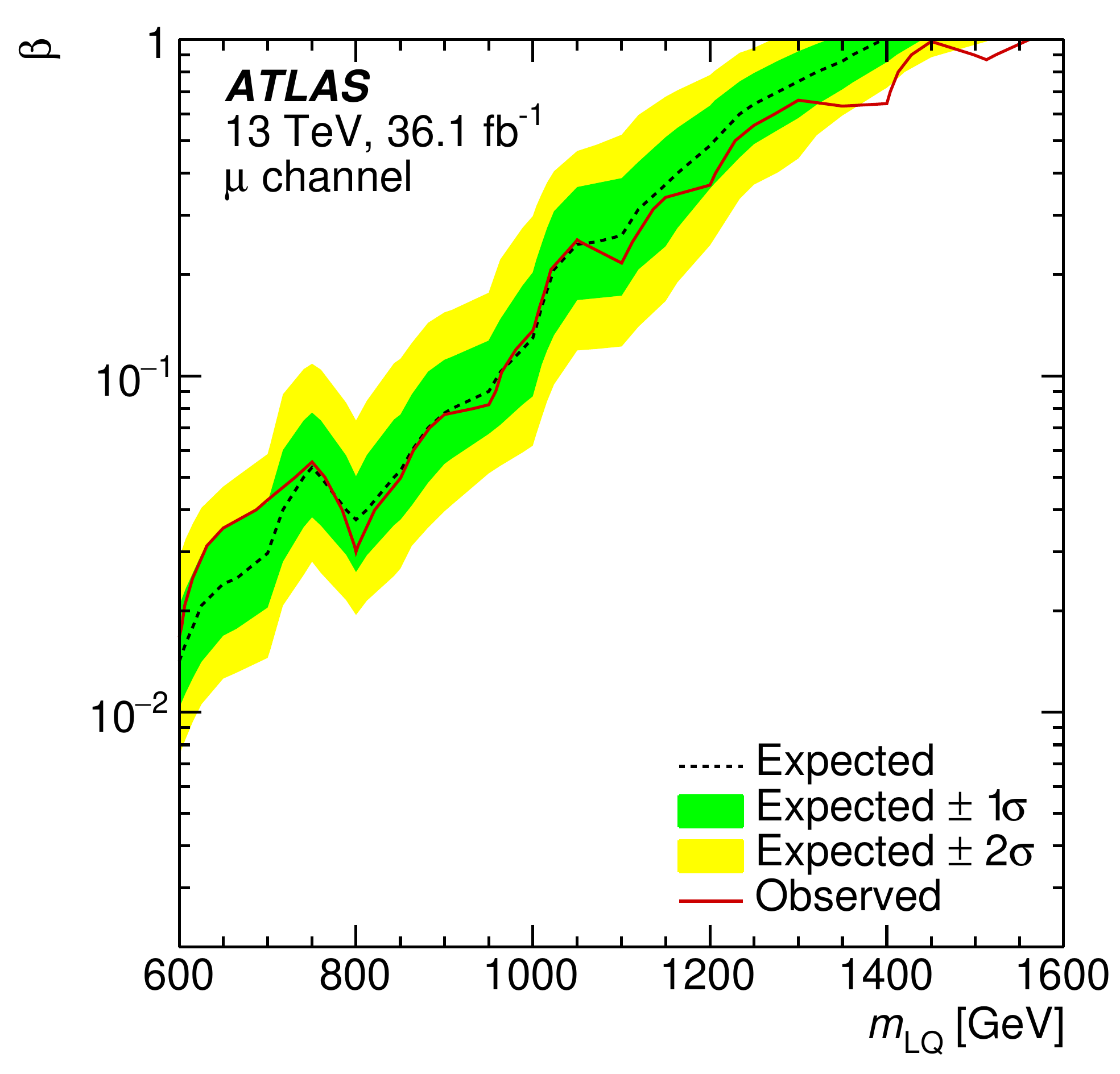}}
 \caption{Upper limits (observed and expected) on the branching ratio for (top) first- and (bottom) second-generation LQs into a lepton 
and quark as a 
function of LQ mass. Limits are presented for low (left) and high (right) mass regions, which correspond to masses greater 
than and less than $600$~\GeV, respectively. The $\pm~1\sigma$ ($\pm 2\sigma$) uncertainty bands on the expected limit represent all 
sources of systematic and statistical uncertainty.}
  \label{fig:limits-br}
\end{figure}

The 95\% CL upper limits on the cross-section for scalar-LQ pair production, normalised to
the predicted cross-section, are presented as a function
of $\mlq$ in Figure~\ref{fig:limits-m} for first- and second-generation LQs for an assumed value of  $\beta =0.5$.  Expected
limits with their one and two standard-deviation bands are also shown.  The observed limits are
consistent with the expected limits for all mass points for both channels. Lower mass limits of about 1.25~\TeV\ are obtained,
representing an increase of around 400~\GeV\ compared with earlier work~\cite{EXOT-2015-19}.

Exclusion contours in the  $\beta$--$\mlq$ plane are shown in Figure~\ref{fig:limits-br} for different values of  $\beta$  for the 
first
two LQ generations. The sensitivity in  $\beta$  is greatest for the lowest mass region
considered ($\sim$ 200~\GeV) for which the search is sensitive to  $\beta$  values around $10^{-2}$.
For  $\beta=1$, a mass of 1400~\GeV\ (1560~\GeV) can be excluded for first-generation (second-generation) LQs.
Limit values are also given for first- and second-generation LQs in Table~\ref{tab:massLimits}. The considerably stronger observed limit 
for the second generation is due to a downward fluctuation in the data in 
the relevant SRs.

\FloatBarrier

%% file: DataBG.tex
\section{Data and Monte Carlo samples}
\label{sec:DataMC}

\subsection{Data sample}
\label{subsec:trigger}

This analysis is based on proton--proton collision data at a 
centre-of-mass energy of 
\sqrts~=~13\,\TeV, collected at the 
LHC during 2015 and 2016. After imposing requirements based on beam and detector
conditions and data quality, the data sample corresponds to an integrated luminosity of 
36.1\,\ifb.
A set of triggers is used to select events~\cite{TRIG-2016-01,ATL-DAQ-PUB-2016-001,ATL-DAQ-PUB-2017-001}.
For the dielectron channel, a two-electron trigger with a transverse 
energy
(\et{}) threshold of 17\,\GeV\ for each electron was used
for the entire data set. 
In the electron--neutrino channel, 
events were selected by either of two single-electron triggers, as described below.
For the 2015 data set, one trigger required \et{} above 60~GeV.  A second trigger with a higher threshold of
120\,\GeV\ but looser identification requirements otherwise was used.
For the 2016 data set, the main difference is that the second trigger had a threshold of 140\,\GeV. The trigger efficiency 
is at least 95\% for electrons above threshold for the kinematic region used in this work.  

In the muon channel, the same triggers were used for the \mumujj{} and the \munujj{} channels.
Events were selected by either of two single-muon triggers.
For the 2015 data, one trigger required a transverse momentum of at least 26\,\GeV\ and applied an isolation criterion. To 
maintain
a high efficiency at high \pt{}, a second trigger with a threshold of 50\,\GeV\ and no further requirements was used.
The main difference for the 2016 data set is that slightly different isolation requirements were used for the lower-\pt{} trigger.
The thresholds were the same as in 2015. 
For the offline \pt{} values considered in this analysis, the trigger efficiencies have reached their plateau values, 
which  vary between 50\% and 80\% in the more central detector region and 
are around 90\% for $|\eta|>1.05$. The variation in trigger efficiency is 
due to local inefficiencies and incomplete detector coverage.

Multiple $pp$ interactions in the same or neighbouring bunch-crossings can lead to many 
reconstructed vertices in the beam collision of interest. The primary vertex of the event, from
which the leptons are required to originate, is defined as that with the largest sum of squared transverse momenta
of its associated tracks.
Selected events must contain a primary vertex with at least two associated tracks with 
$p_{\mathrm{T}} > 0.4$~\GeV.

\subsection{Signal and background simulations}\label{sec:sigback}

Samples of simulated events with pair-produced scalar LQs with masses between 
200~\GeV\ and 1700~\GeV, in steps of 50~\GeV\ up to 1500~\GeV\ and thereafter 100~\GeV,  were generated 
at next-to-leading order (NLO) in QCD~\cite{Mandal:2015lca,Kramer:2004df,Kramer:1997hh} with the \textsc{MadGraph 2.4.3}~\cite{Maltoni:2002qb,Mandal:2015lca} 
program using \textsc{MadSpin}~\cite{Artoisenet:2012st} for the decay of LQs. 
In the simulation, leptoquarks with a charge of $\frac{1}{3}e$ are used. 
Accordingly, the first generation LQ decays to either a $u$-quark and an electron, or a $d$-quark and an electron neutrino. 
The second generation LQ decays to either a $c$-quark and a muon or an $s$-quark and a muon neutrino. 
The anti-leptoquarks decay into the corresponding anti-particles. 
The generator output was interfaced with \textsc{Pythia 8.212} \cite{Sjostrand:2007gs} for the event simulation beyond the hard scattering process, i.e.~the  parton shower, hadronisation and underlying event, collectively referred to as UEPS. 
The \textsc{A14} set of 
tuned parameters (tune)~\cite{ATL-PHYS-PUB-2014-021} was used for the UEPS modelling.
The \textsc{NNPDF}3.0 NLO \cite{Ball:2014uwa} parton 
distribution function (PDF) set was used. The coupling 
$\lambda_{\mathrm{LQ} \rightarrow 
\ell q}$, which determines the LQ lifetime and width \cite{LQlq_coupling} 
was set  to
 $\sqrt{4\pi\alpha}$, where $\alpha$ is the fine-structure constant. This
value corresponds to a LQ full width of about 0.2\% of its mass; LQs can thus be considered to decay promptly.  
Samples were generated for  $\beta$  with a value of $0.5$.

Events containing $W$ or $Z$ bosons and associated jets~\cite{ATL-PHYS-PUB-2016-003} were simulated with the 
\textsc{Sherpa 
2.2.1} 
generator~\cite{Gleisberg:2008ta}. 
These event samples include off-shell production of the bosons.  
Matrix elements were calculated in perturbative QCD for up to 
two partons at NLO and four 
partons at leading order (LO) using
 the \textsc{Comix}~\cite{Gleisberg:2008fv}  and \textsc{OpenLoops}~\cite{Cascioli:2011va} matrix element generators. Merging with the \textsc{Sherpa} 
UEPS model was performed with the ME+PS@NLO method~\cite{Hoeche:2012yf}. The \textsc{NNPDF3.0} PDF set~\cite{Lai:2010vv} at 
next-to-next-to-leading order (NNLO) was used. Simulated processes in which a $Z$ boson decays into leptons are hereafter termed Drell--Yan processes.

For the simulation of 
$t\bar{t}$ 
and single-top-quark production in the $Wt$ final state and $s$-channel the \textsc{Powheg-Box 
v2}~\cite{Nason:2004rx,Frixione:2007vw,Alioli:2010xd,Alioli:2011as}  
generator was used.  
For the UEPS modelling for the $t\bar{t}$ samples, \textsc{Pythia} 8.210 was used, with the \textsc{A14} tune and the 
\textsc{NNPDF2.3
LO}~\cite{Ball:2012cx} PDF set. 
The $t\bar{t}$ samples used the \textsc{NNPDF3.0} set in the matrix element calculations, the single-top samples used the \textsc{CT10} PDF set. Electroweak $t$-channel single-top-quark events were 
produced with the
 \textsc{Powheg-Box v1} generator. For the generation of single-top samples, the UEPS modelling was done
 using \textsc{Pythia 6.428}~\cite{Sjostrand:2006za} with the \textsc{CTEQ6L1}~\cite{Pumplin:2002vw} PDF sets and \textsc{Perugia 
2012}~\cite{Skands:2010ak} 
tune. 
 The value of the top mass ($m_t$) was set to 172.5~\GeV. 
The \textsc{EvtGen v1.2.0}~\cite{Lange:2001uf} program was used for properties of
 the bottom and charm hadron decays. 

 Diboson processes with at least one charged lepton in the final state were simulated using the \textsc{Sherpa 2.1.1} 
generator. All diagrams with
 four electroweak vertices were considered. They were calculated for up to one parton at NLO and
 up to three partons at LO using the \textsc{Comix} and \textsc{OpenLoops} matrix element generators and merged
 with the \textsc{Sherpa} UEPS model using the \textsc{ME+PS@NLO} prescription. The \textsc{CT10} PDF set was used
 in conjunction with dedicated parton shower tuning developed by the \textsc{Sherpa} authors. The generator cross
 sections, calculated up to NLO, were used.

To model the effect of multiple proton--proton interactions in the same or neighbouring bunches 
(pile-up), simulated inclusive proton--proton  events were overlaid on each 
generated signal and background 
event.  The pile-up was simulated with \textsc{Pythia 8.186} using tune A2~\cite{ATL-PHYS-PUB-2012-003} and the 
MSTW2008LO PDF set~\cite{Martin:2009iq}. 
Simulated events were corrected using
per-event weights to describe the distribution of the average number of interactions per proton bunch-crossing as observed in data.

The detector response to events in the SM background samples was evaluated with the 
\textsc{Geant4}-based 
detector simulation~\cite{SOFT-2010-01, GEANT4}. A fast simulation employing a 
parameterisation of 
calorimeter response~\cite{ATL-PHYS-PUB-2010-013} and 
\textsc{Geant4} for the other detector components was used for the signal samples. 
The standard ATLAS reconstruction software was
used for both simulated and $pp$ data. Furthermore, correction factors, termed scale factors, which were derived from data were applied as event 
weights to the 
simulation  of the lepton
 trigger, reconstruction, identification, isolation, and impact parameter selection, and of  $b$-tagging 
efficiencies.

%% file: Conclusions.tex
\section{Summary and conclusions}
\label{sec:conclusion}
Searches for pair production of first- and second-generation 
scalar leptoquarks in $pp$ collisions at {$\sqrt{s} = 13$~\TeV{}} have been 
made using the ATLAS detector at the LHC. The 
searches exploit a data set corresponding to 36.1~fb$^{-1}$ of integrated luminosity and probe the lepton--quark and  
lepton--neutrino LQ decay channels.  
No significant excess above the  SM background expectation is observed in 
any channel and exclusion limits have been evaluated. 
The results presented here significantly extend the sensitivity in mass compared to previous ATLAS results, and yield an exclusion very similar to that found by the CMS experiment using a dataset of a similar size~\cite{CMS-EXO-17-009,CMS-EXO-17-003}. 
Within the minimal Buchm\"uller--R\"uckl--Wyler model and assuming a branching ratio for the decay into a charged lepton and a quark of 50\%, leptoquarks with 
masses up to 
1.29\,TeV are excluded at  95\%\,CL for first generation leptoquarks, and up to 1.23\,TeV for second generation leptoquarks. 
  
In addition to the search, measurements have been made of particle-level fiducial and differential cross-sections for six 
measurement regions related to the control regions used for the search. Two measurement regions are identical to the $Z$ 
control regions  
used in the search, and their fiducial cross-sections are measured to be $3.28\pm 0.22$~pb and $3.32\pm 0.23$~pb for the \Zee and \Zmm channels respectively. The measurements agree with the cross-sections predicted by \textsc{Sherpa 2.2.1}, which are $3.24 \pm 1.02$~pb and $3.12 \pm 1.01$~pb respectively. 
A third measurement region is dominated by $t\bar{t}\rightarrow e\mu$, where the fiducial cross-section is measured to be $1.50\pm 0.13$~pb, compared to $1.60 \pm 0.38$~pb predicted by \textsc{Powheg+Pythia8}.
Measurements are also made in three other regions where the scalar sum of the \pt{} of the jets and leptons is above $600\gev$. These so-called ``extreme'' regions may be useful for generator tuning since they represent regions which are utilised in searches but where measurements are rarely made.

In addition to the inclusive cross-section measurements, differential measurements are made for eleven variables: the transverse momenta of the leading and subleading jets, the minimum angles between the leading and subleading jets with a lepton, the dilepton transverse momentum, the opening angle between the leptons, the scalar sum of the jet transverse momenta, the scalar sum of the jet and lepton transverse momenta, and the opening angles in $\eta$ and $\phi$ between the two leading jets. In the $Z$ measurement regions, the differential cross-sections for variables involving leptons are typically found to be well-modelled by the nominal generators and the alternative generators shown typically give reasonable but weaker agreement.
The measurements involving jet energies and angles are found to exhibit a degree of mis-modelling by both the nominal and alternative simulations. The nominal and alternative generator predictions are often discrepant with respect to the data in different directions, indicating that these differences are symptoms of different choices of the generator parameters, which would have been tuned in less extreme regions of phase-space given by previous $Z$ + jets measurements. Measurements in the regions of phase space used as control regions by this and related searches have not been made directly in the past. Therefore, the measurements provided by this paper can be used to help validate new generator versions and parameter choices in this region of phase space, and thus help improve the modelling of the background for future searches.

%% file: Appendix.tex
\section{MR differential cross-section measurements}
\label{app:unfoldedxsplots}

\begin{figure}[h]
\centering
\includegraphics[width=0.30\linewidth]{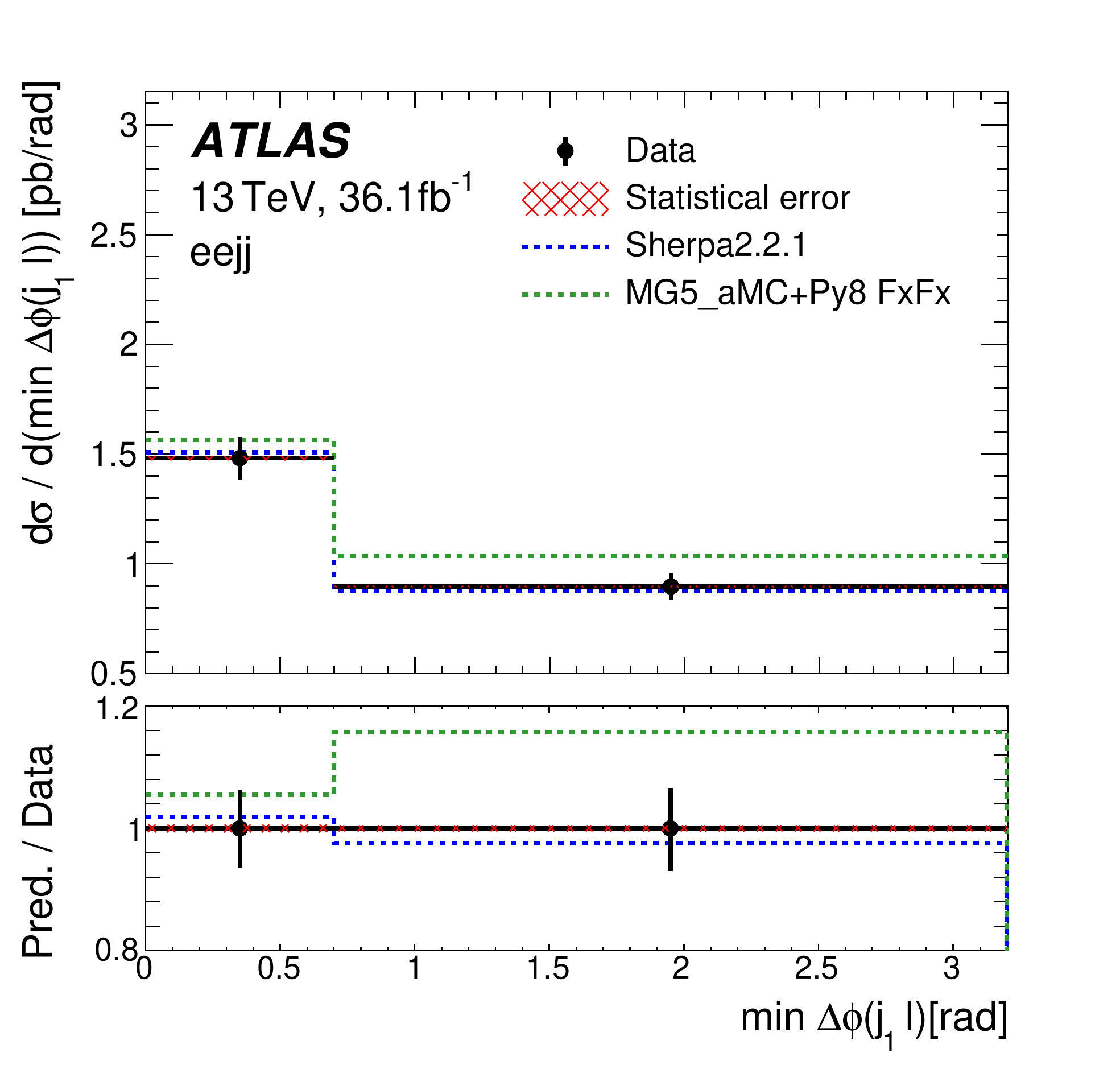}
\includegraphics[width=0.30\linewidth]{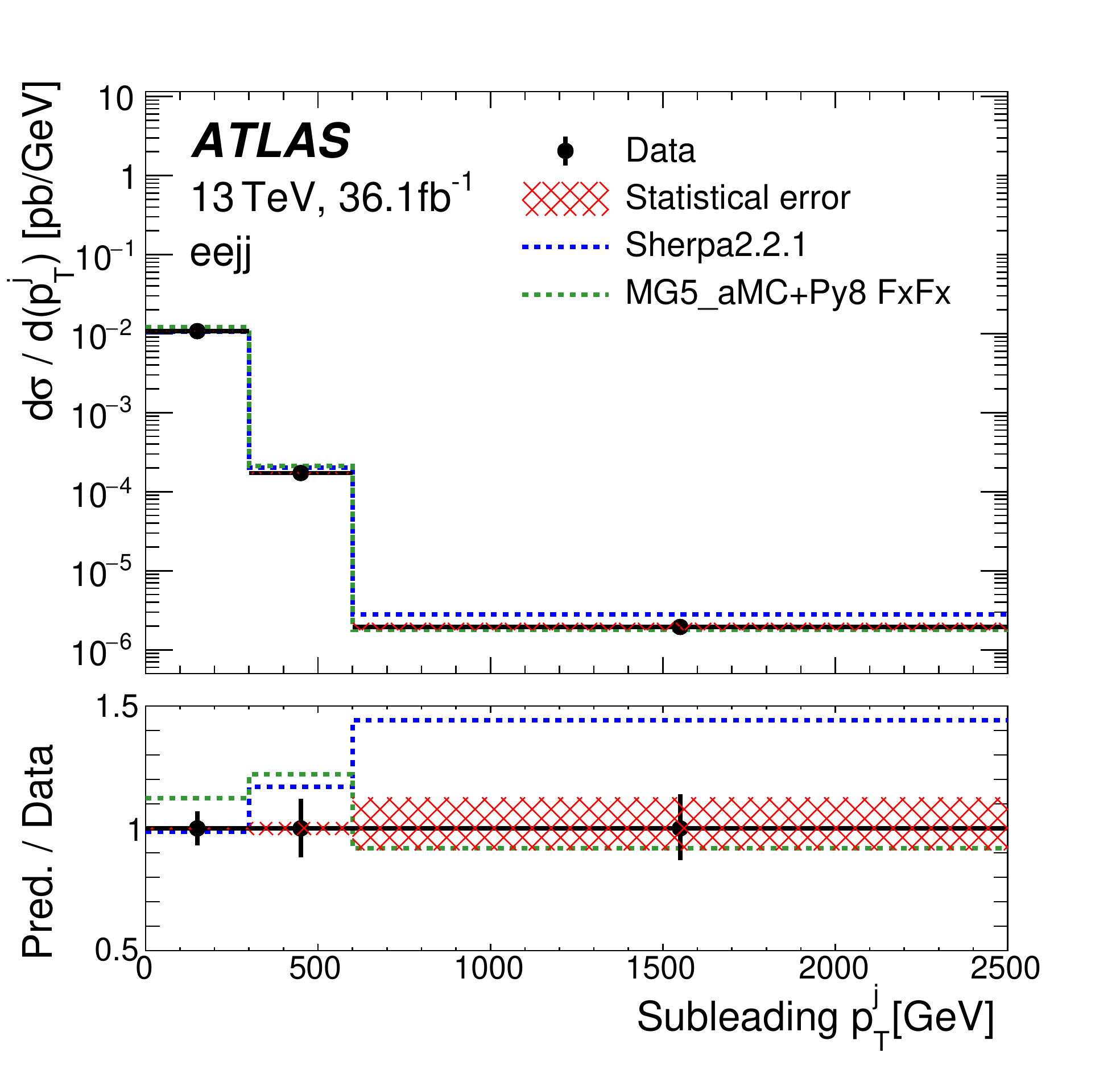}
\includegraphics[width=0.30\linewidth]{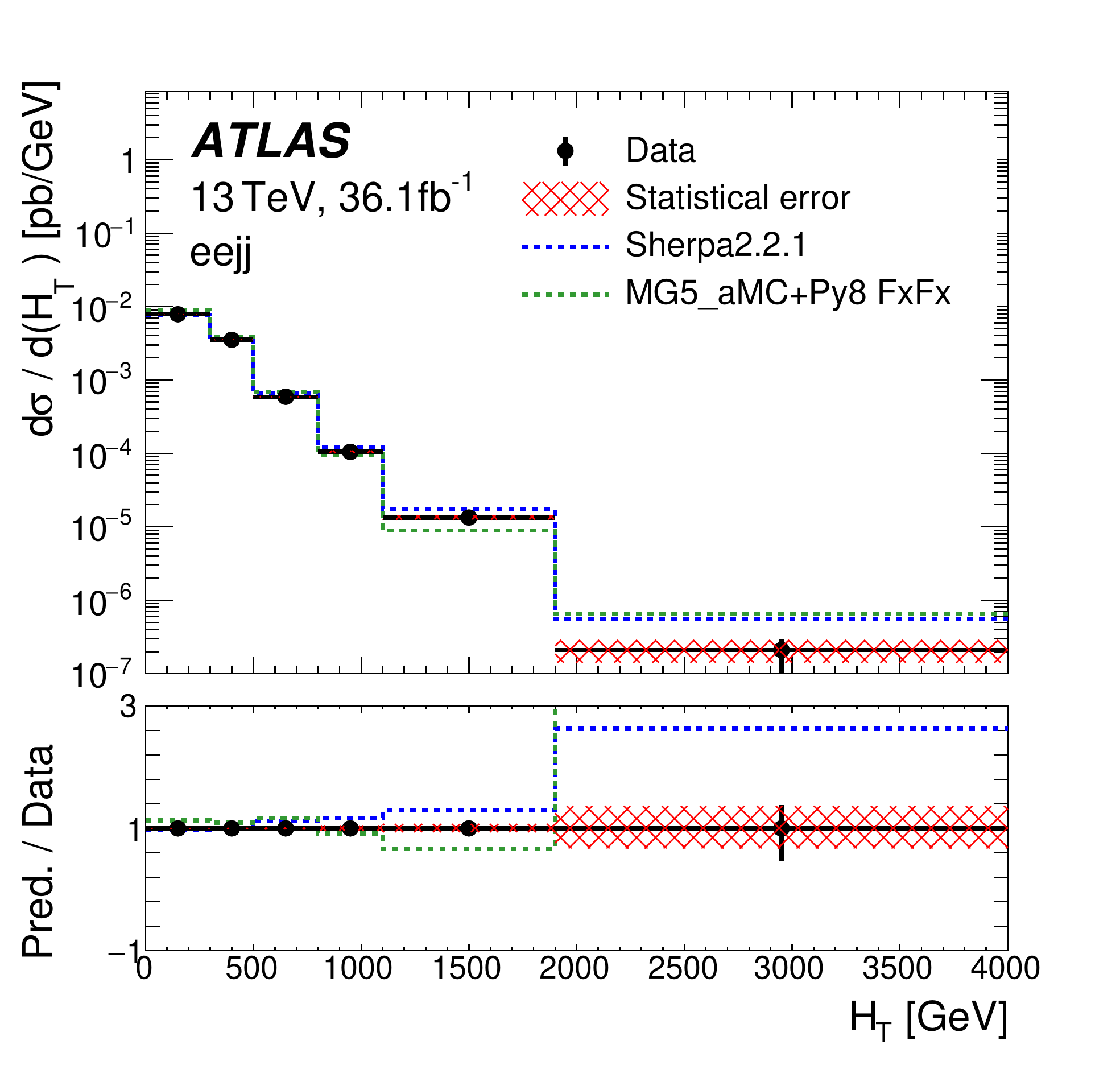}\\
\includegraphics[width=0.30\linewidth]{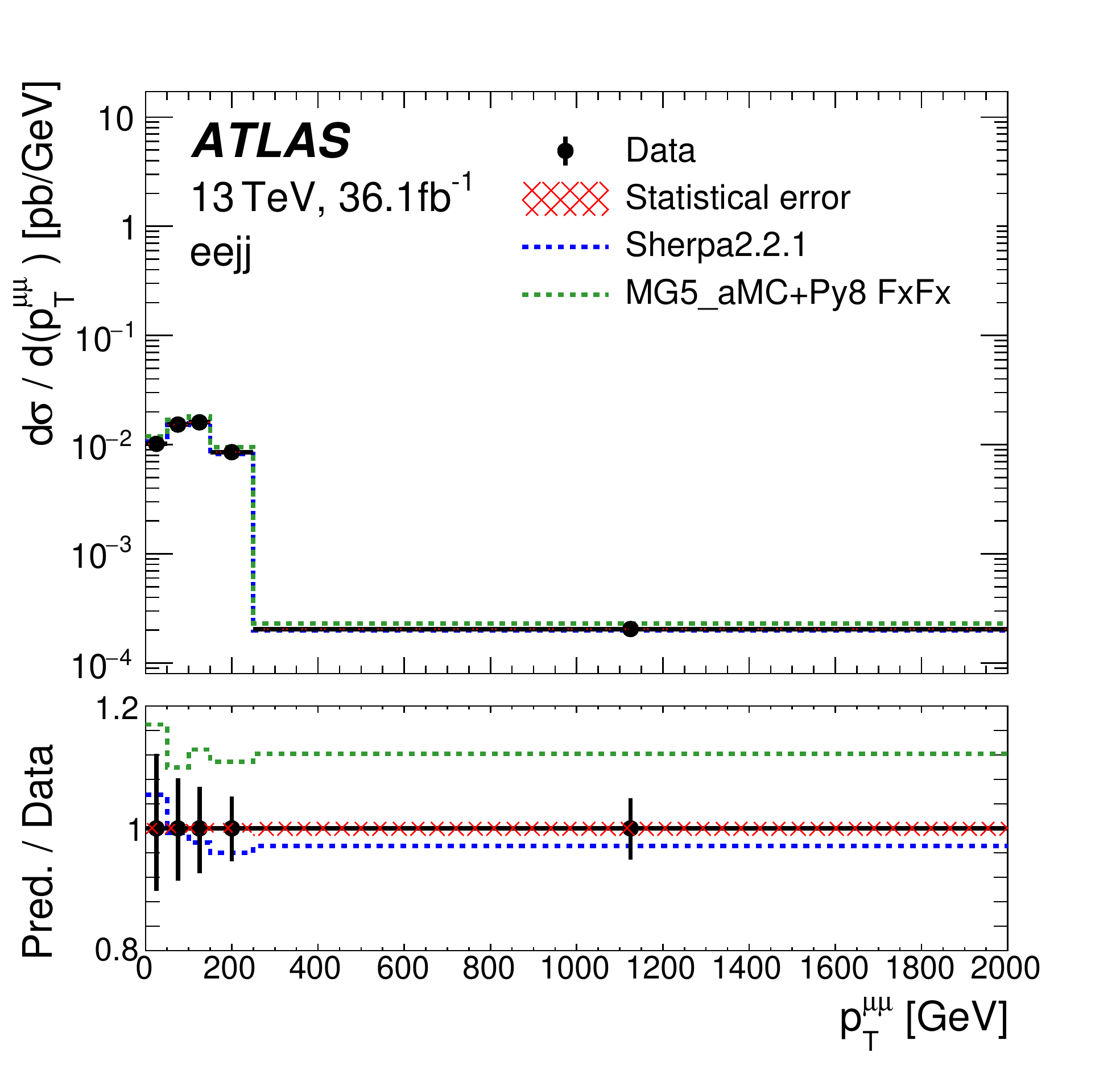}
\includegraphics[width=0.30\linewidth]{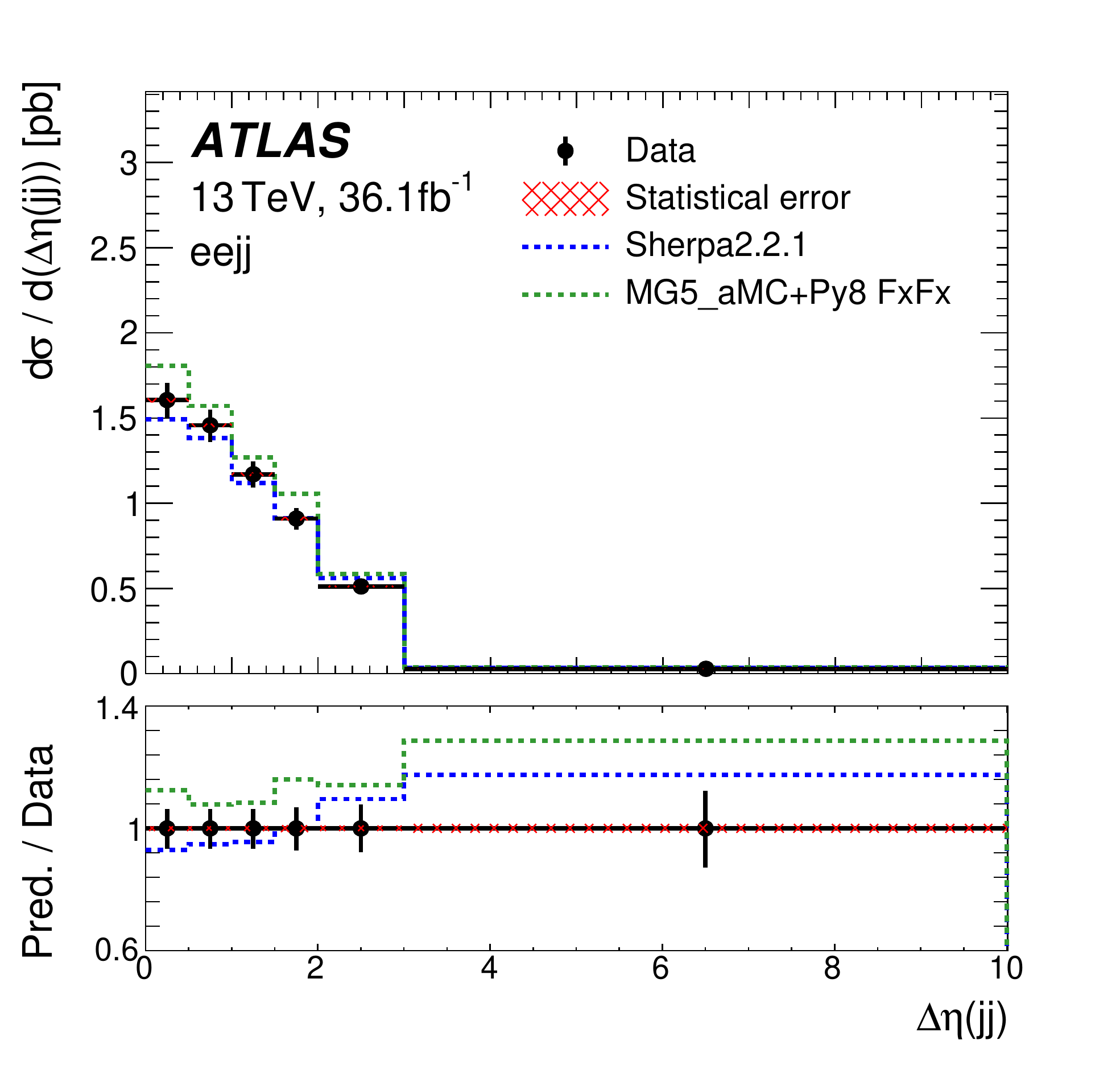}
\includegraphics[width=0.30\linewidth]{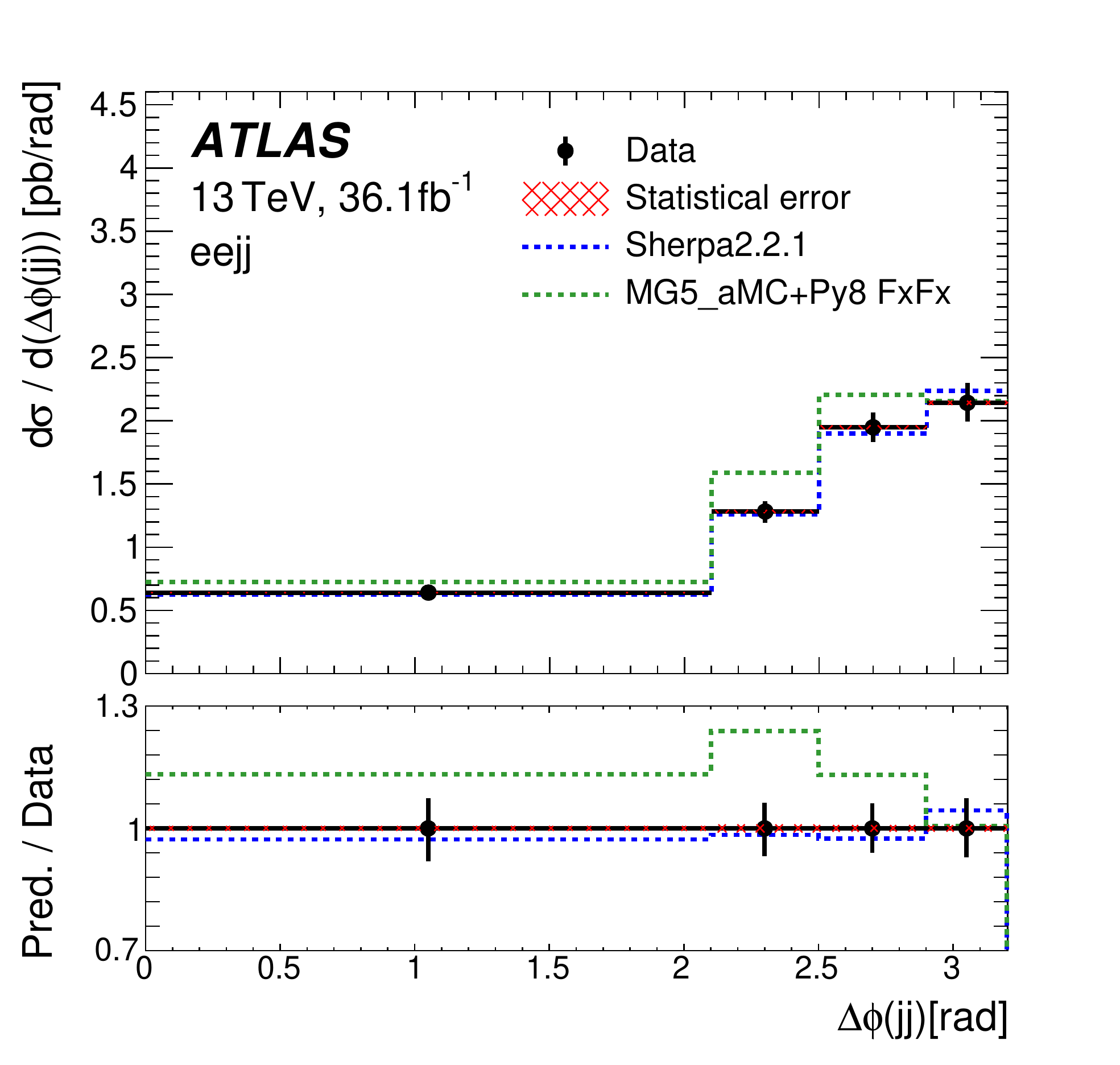}\\
\includegraphics[width=0.30\linewidth]{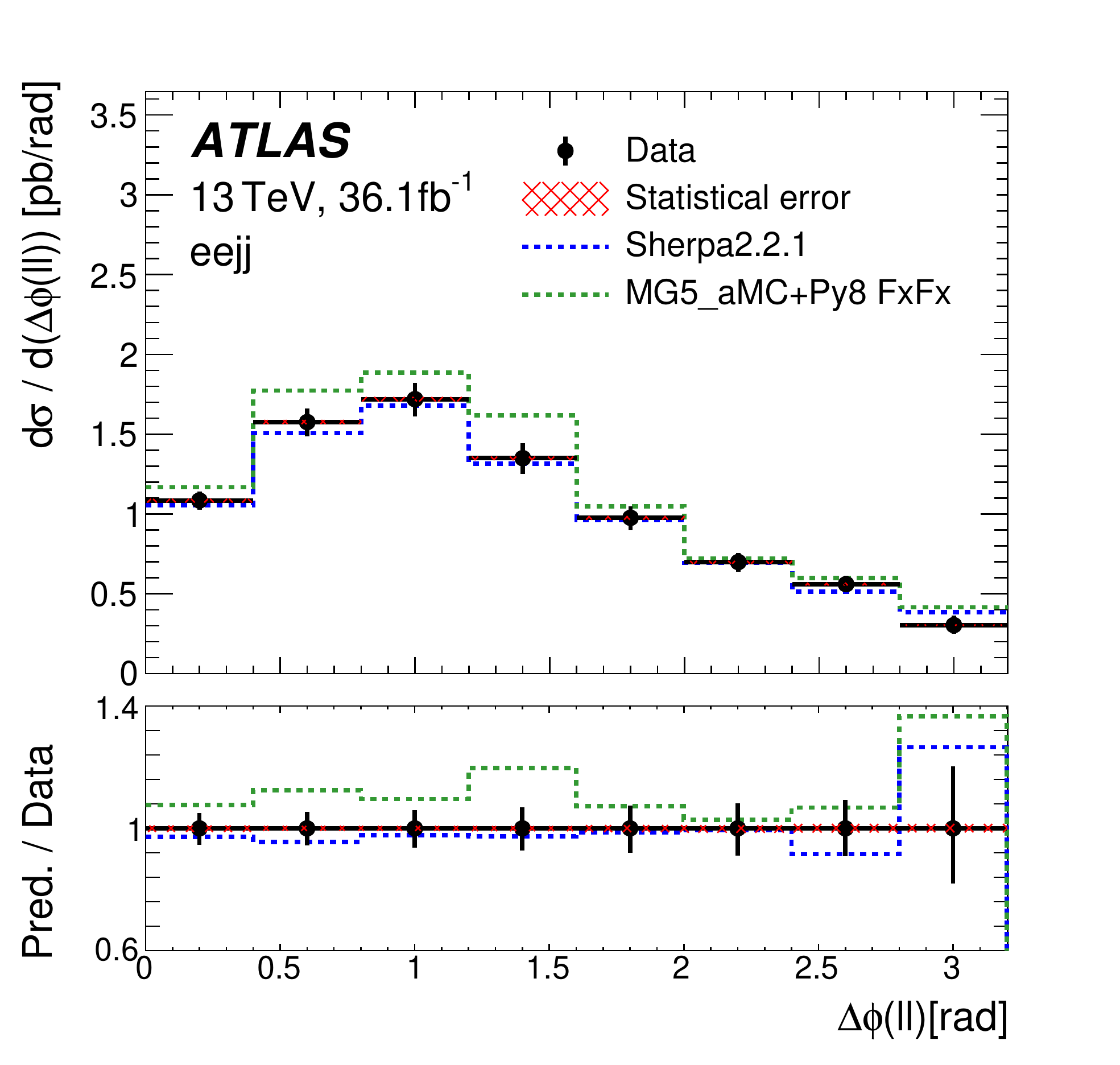}
\includegraphics[width=0.30\linewidth]{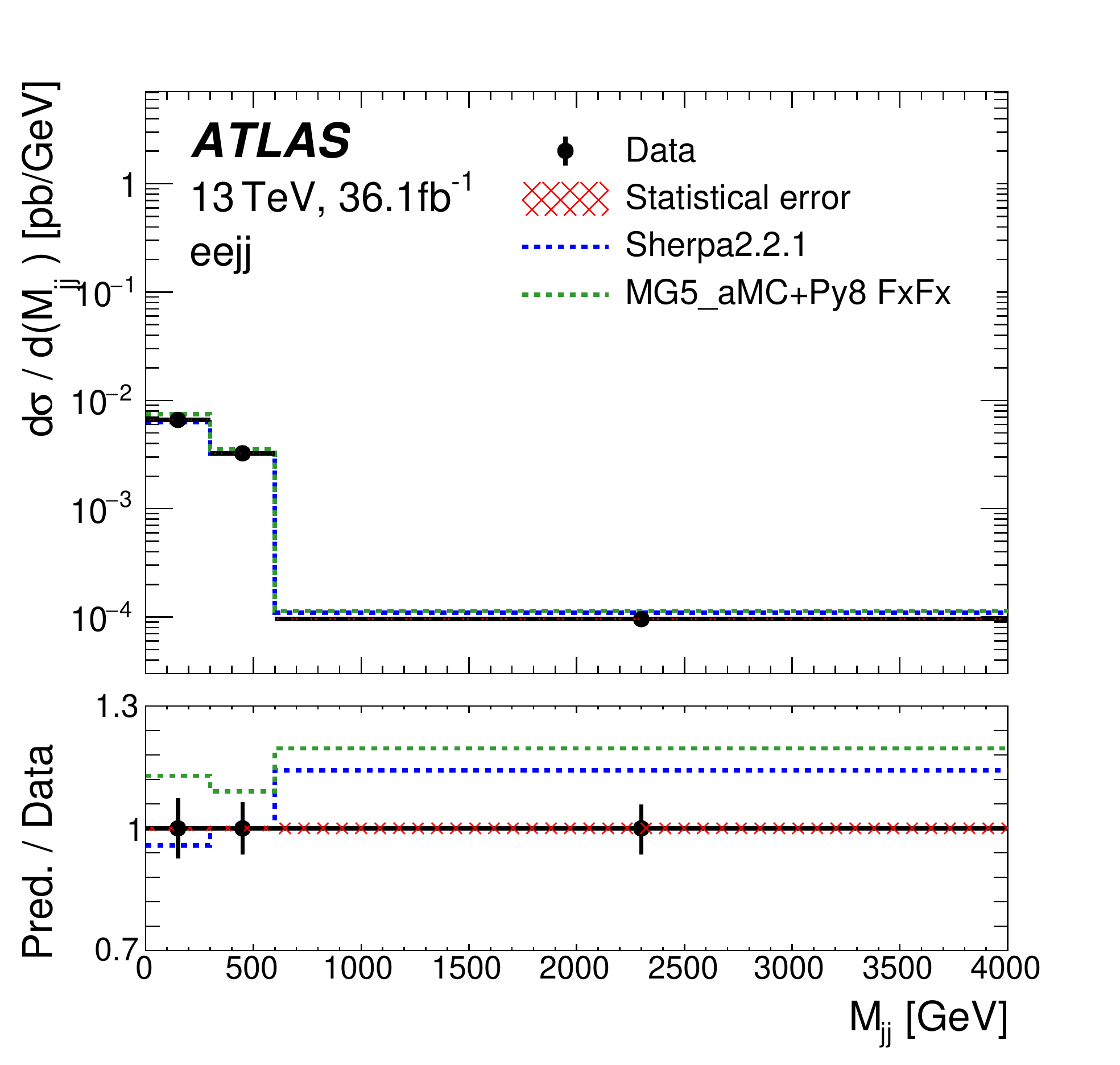}
\includegraphics[width=0.30\linewidth]{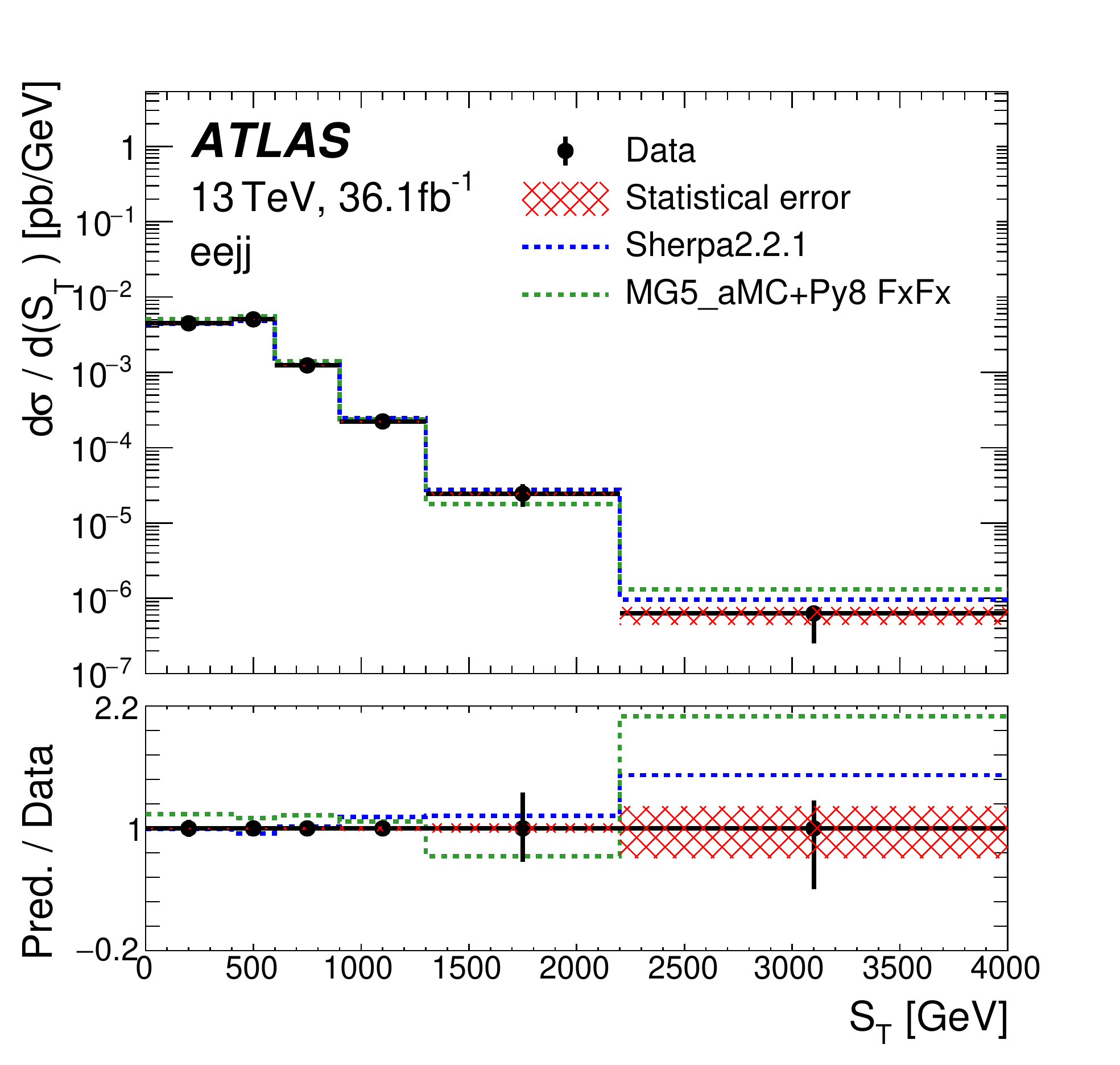}
\caption{The measured generator-level differential cross-sections in the \eejj channel, exclusively for the \Zee 
process. The MC prediction for the dominant process is also shown, with no $m_{jj}$ reweighting applied. The red band represents the statistical component of the total uncertainty which is indicated by the error bar on each point. The variable $ min \Delta \Phi (j1,l) $ refers to the minimal difference in $\phi$ between the subleading jet and a prompt lepton. The variable $H_{T}$ refers to the scalar sum of the jet \pt values. The variables $\Delta \eta(jj)$, $\Delta \phi(jj)$ and $\Delta \phi(ll)$ refer to the difference in $\eta$ and $\phi$ between the two leading jets or leptons.}
\label{fig:unfolded_data-1-isZCReejj}
\end{figure}

\begin{figure}[h]
\centering
\includegraphics[width=0.30\linewidth]{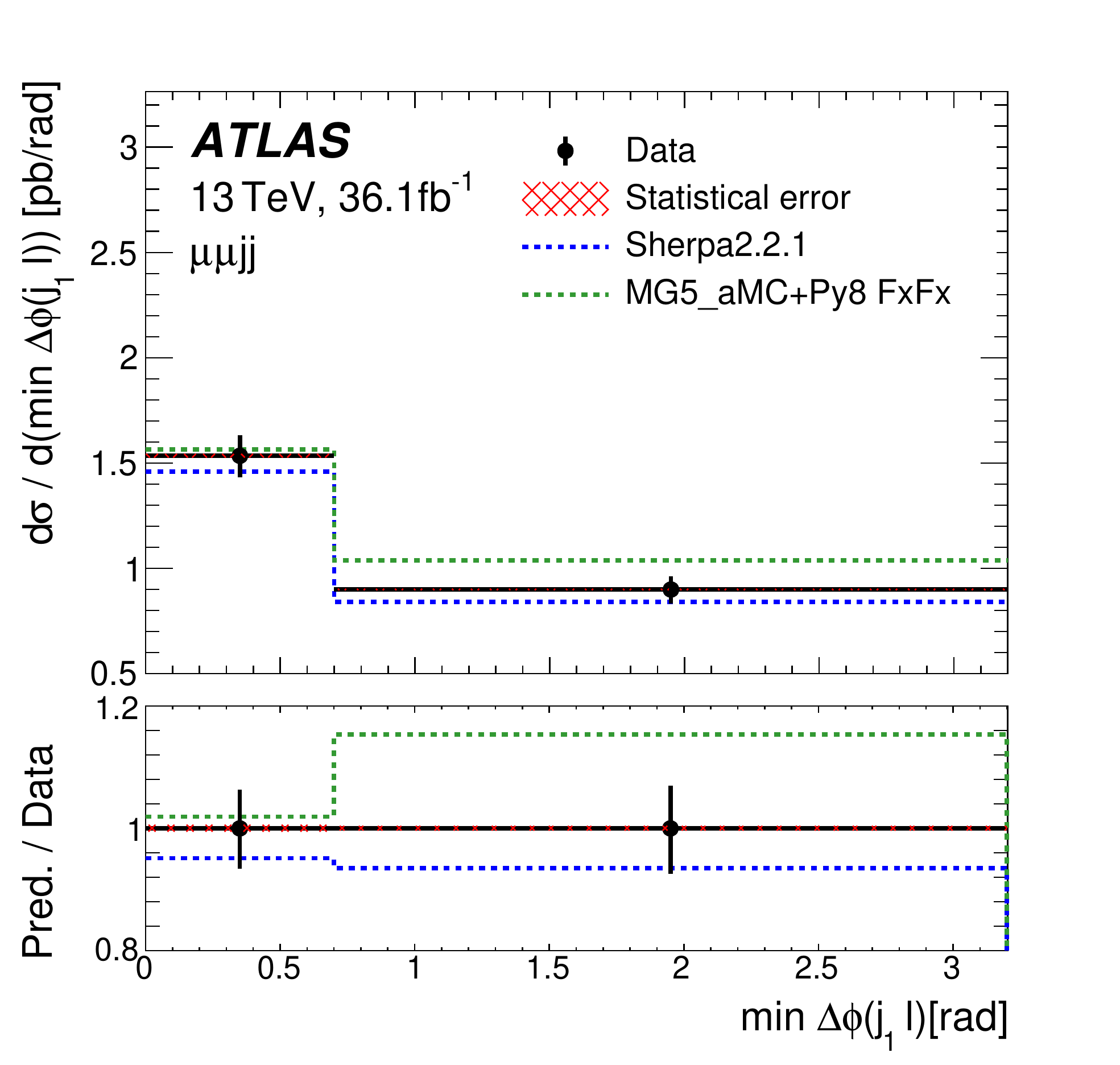}
\includegraphics[width=0.30\linewidth]{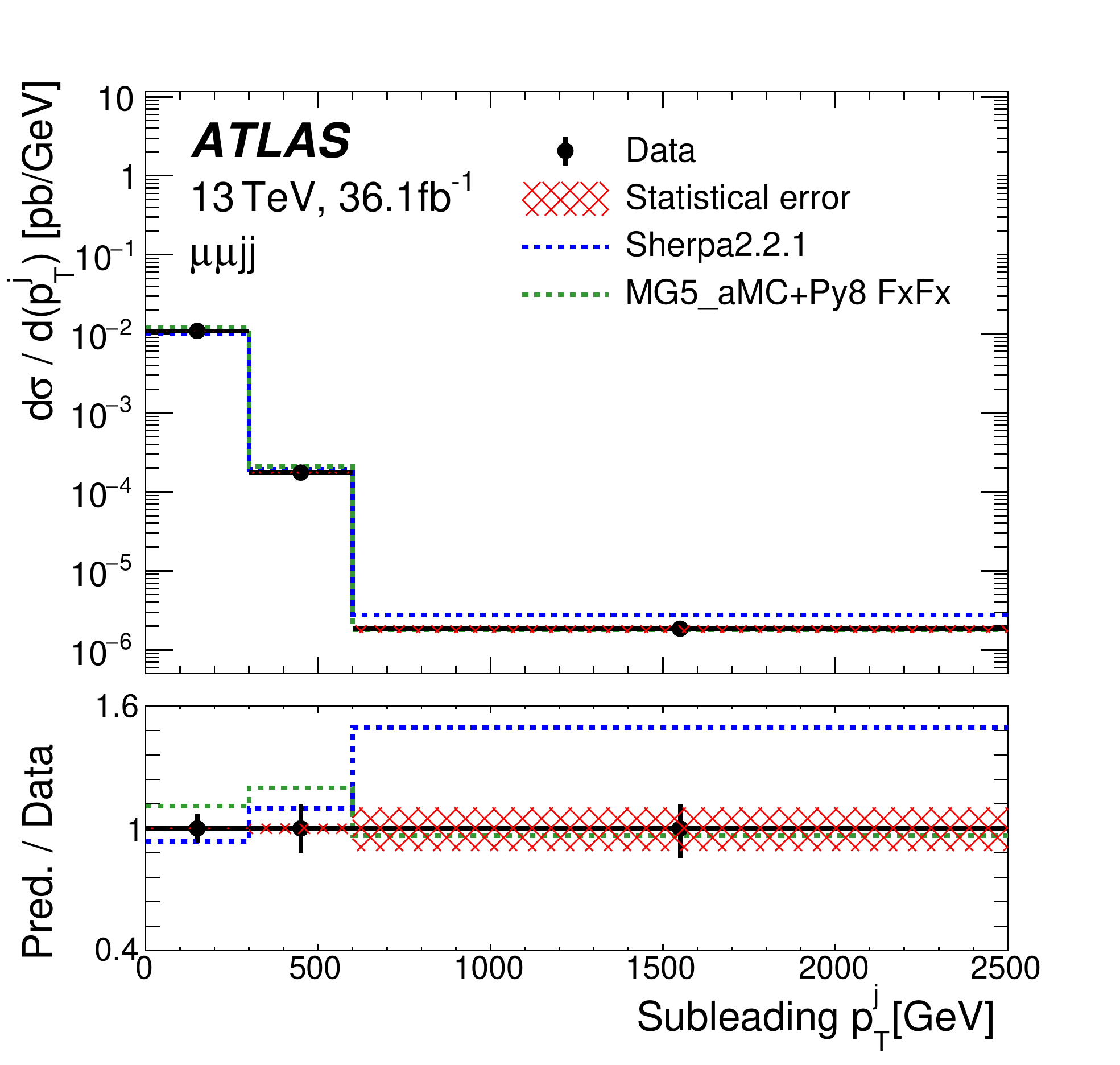}
\includegraphics[width=0.30\linewidth]{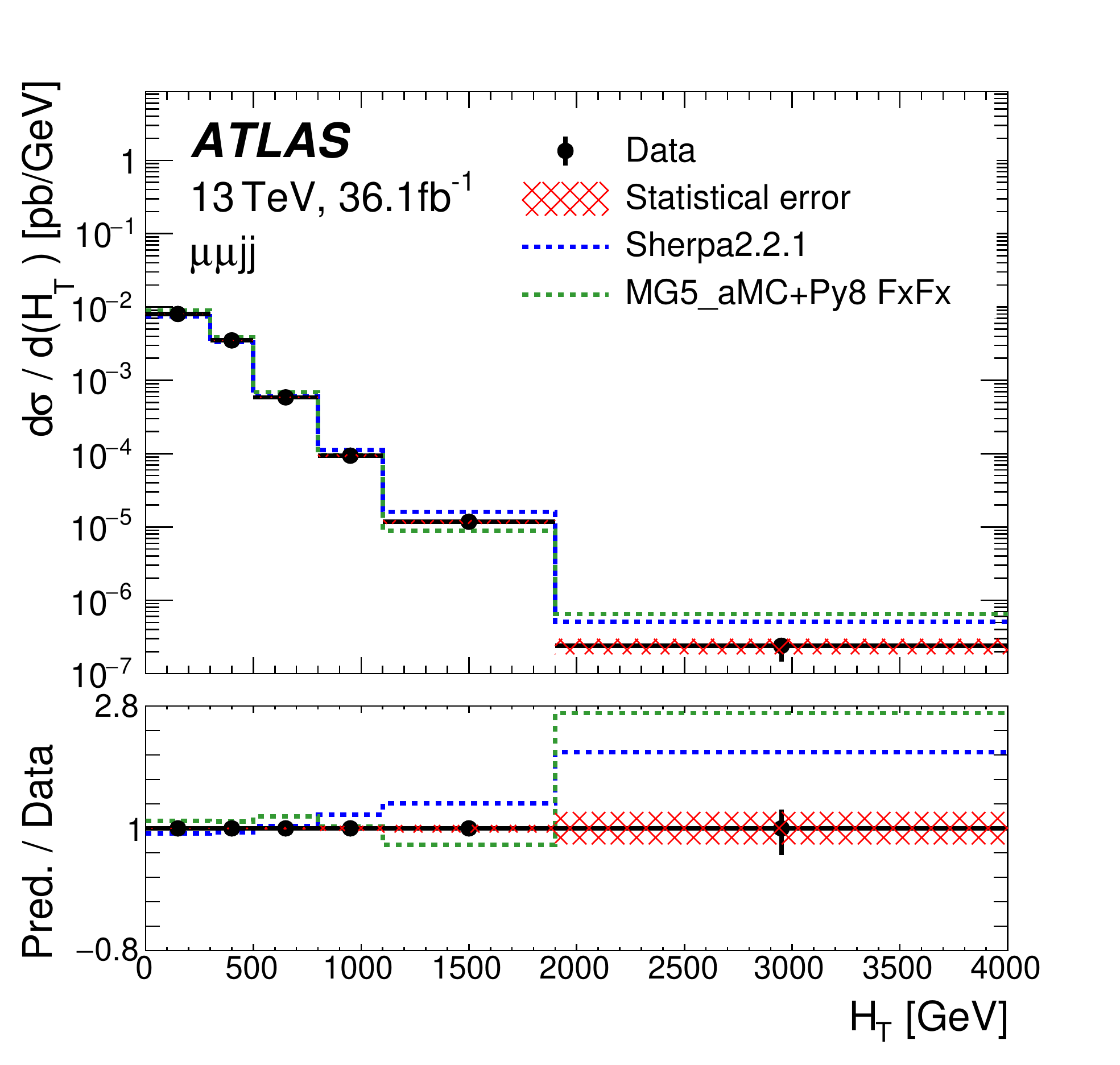}\\
\includegraphics[width=0.30\linewidth]{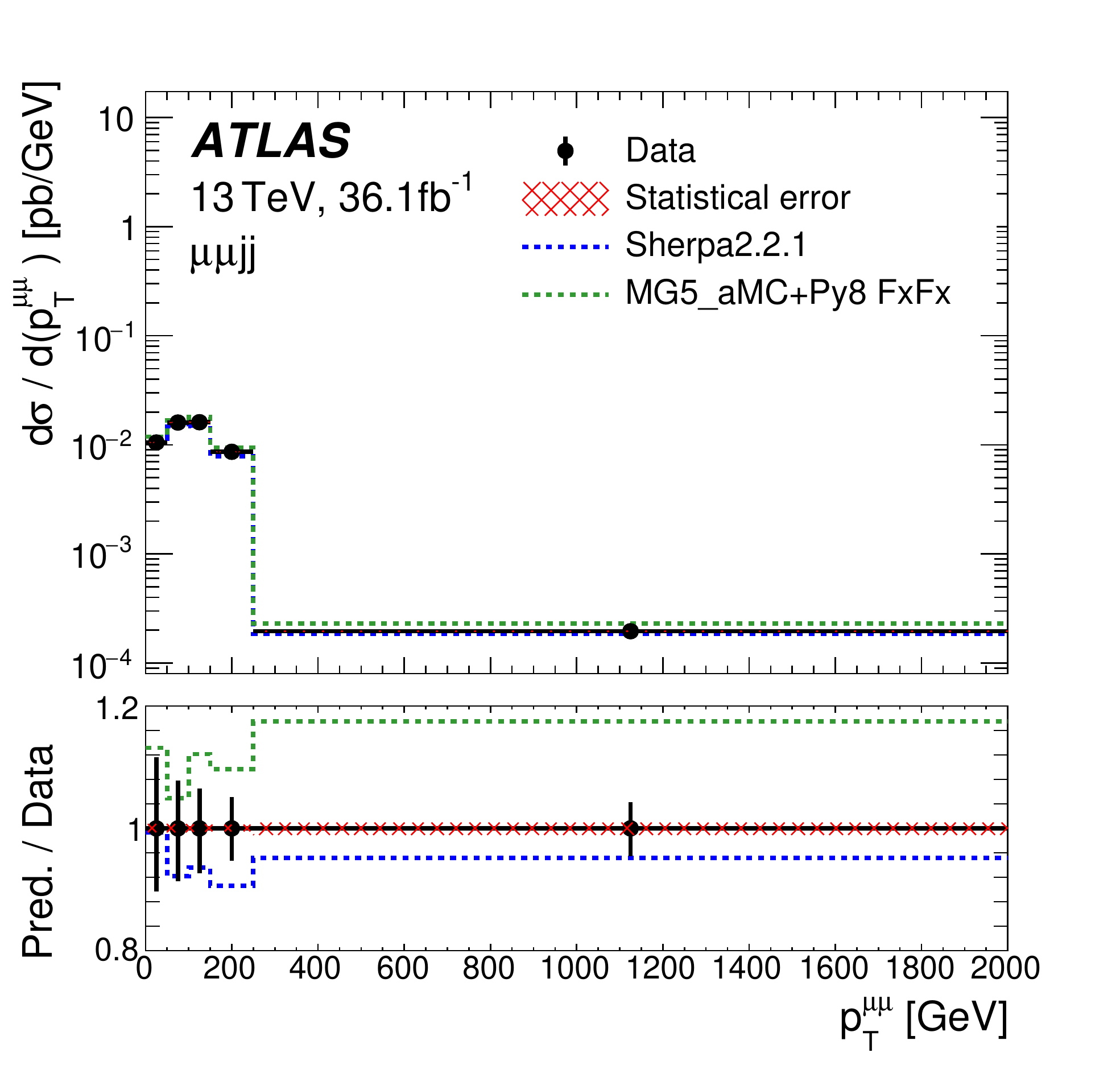}
\includegraphics[width=0.30\linewidth]{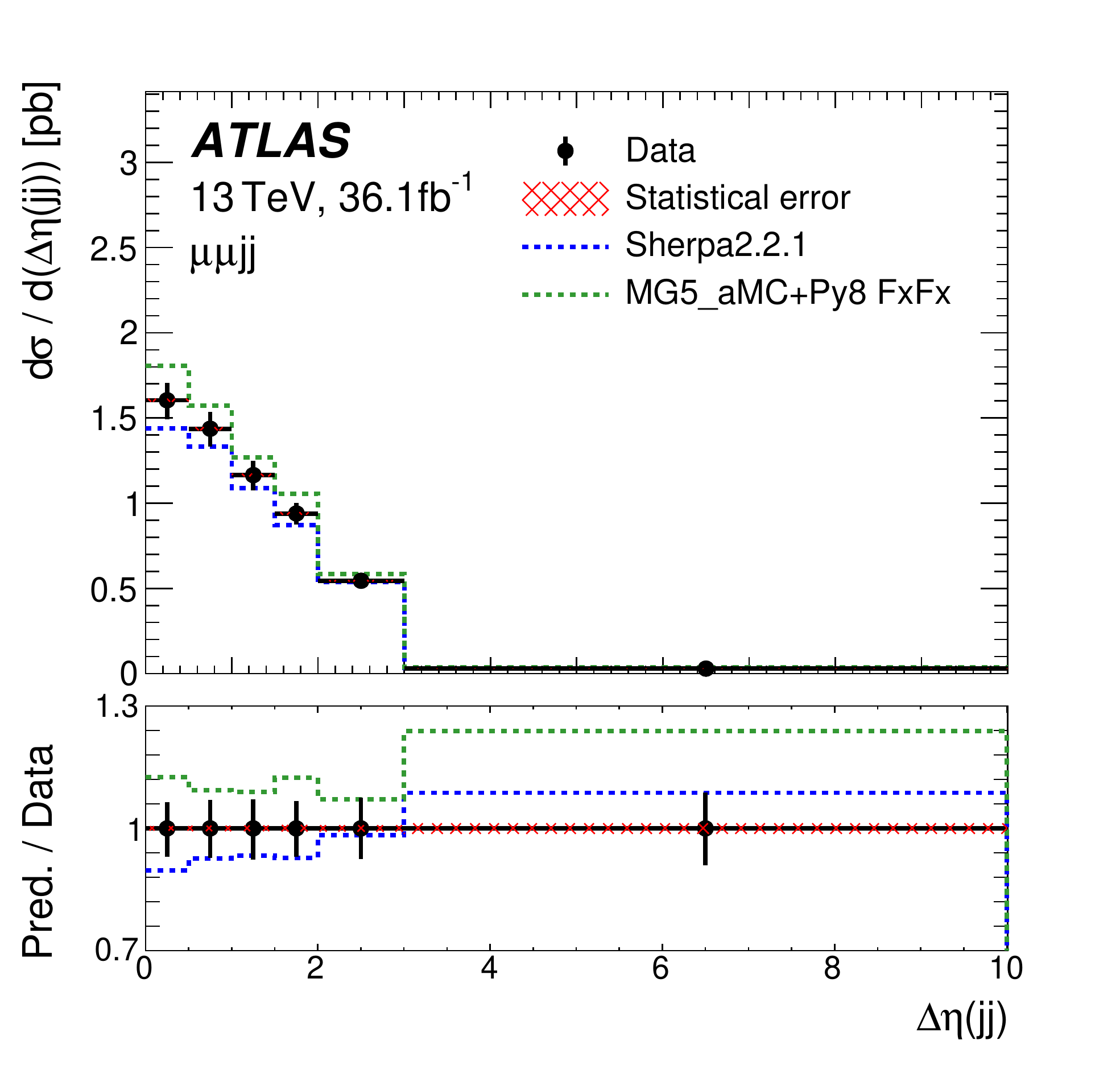}
\includegraphics[width=0.30\linewidth]{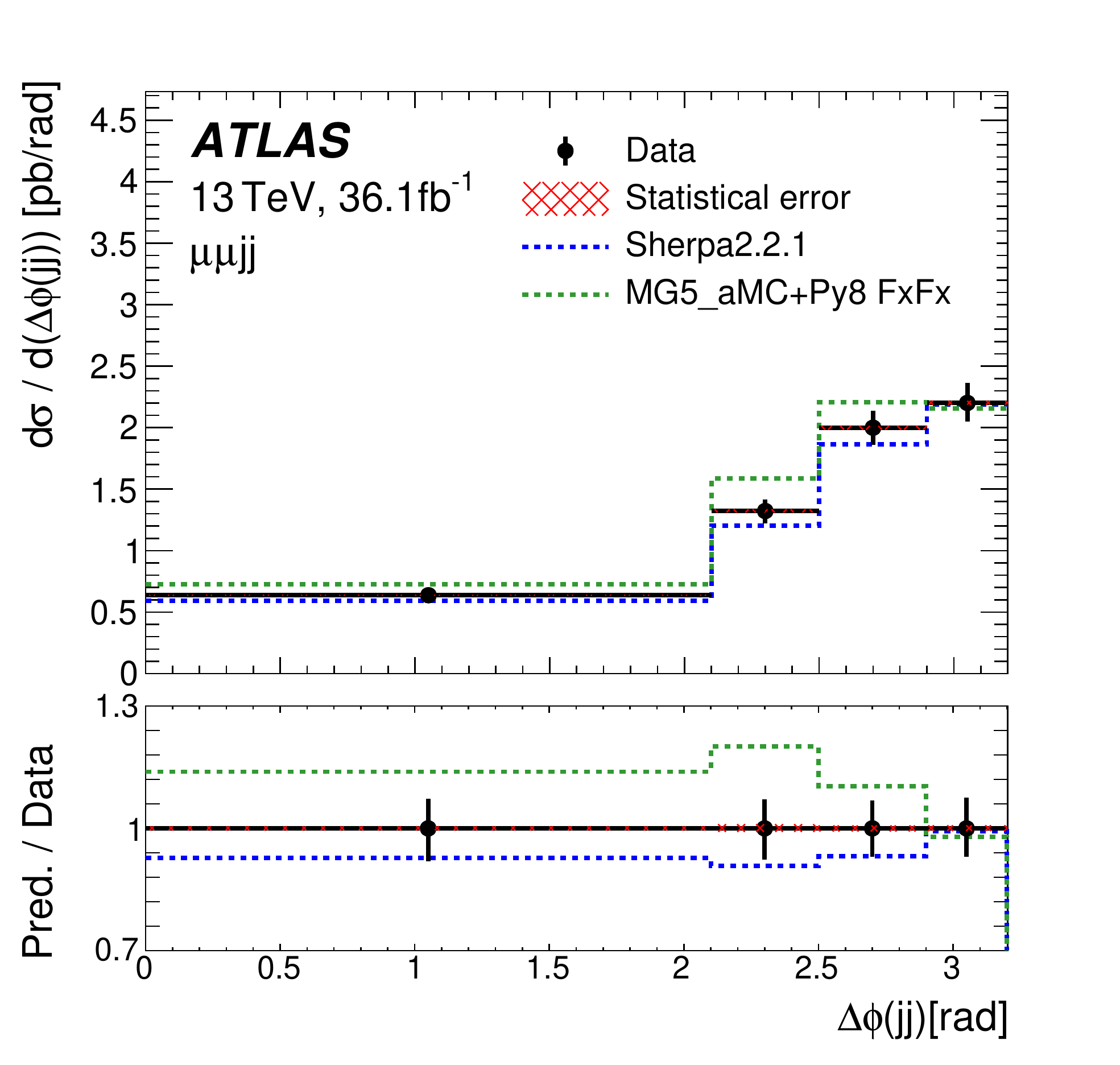}\\
\includegraphics[width=0.30\linewidth]{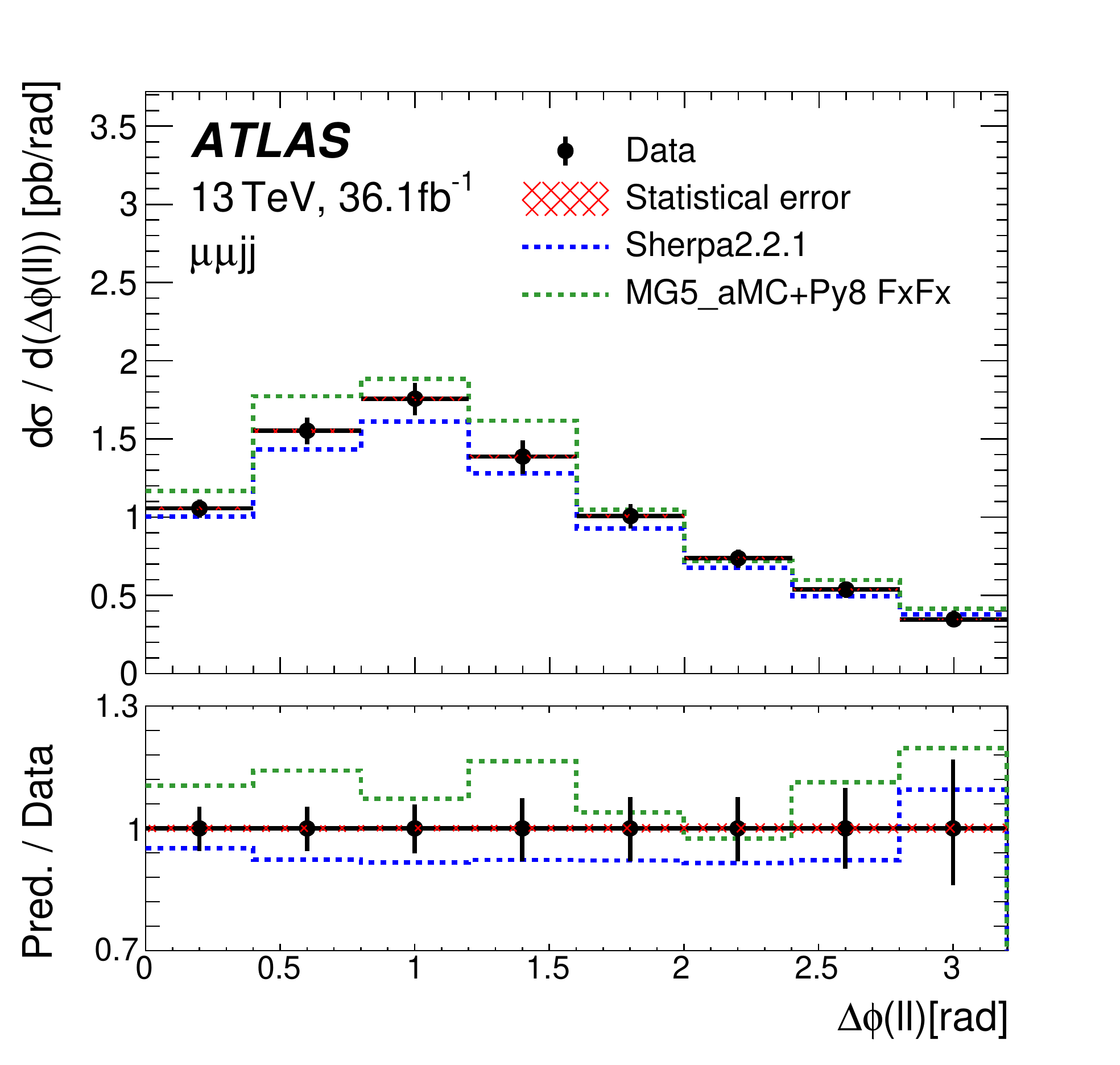}
\includegraphics[width=0.30\linewidth]{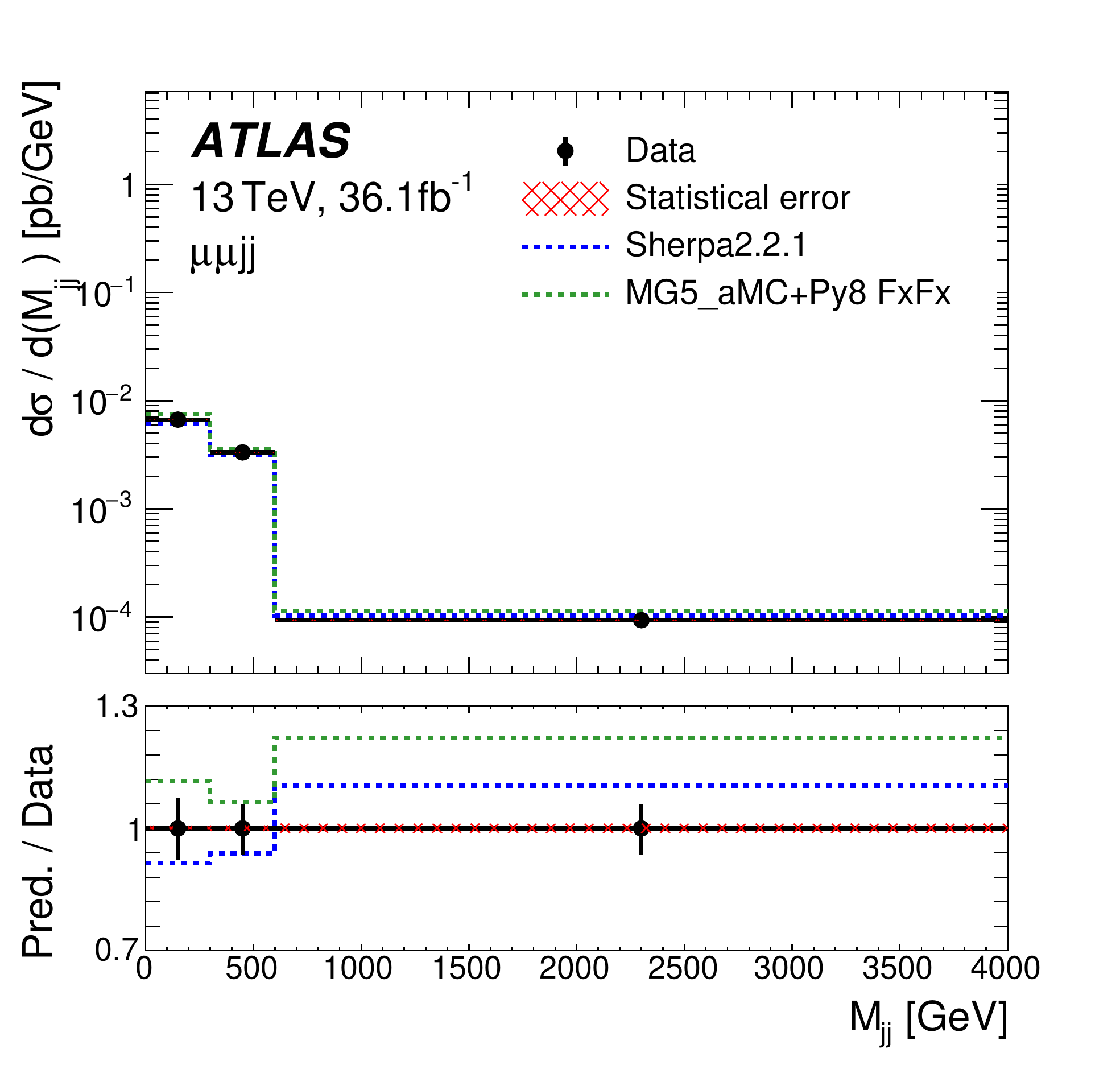}
\includegraphics[width=0.30\linewidth]{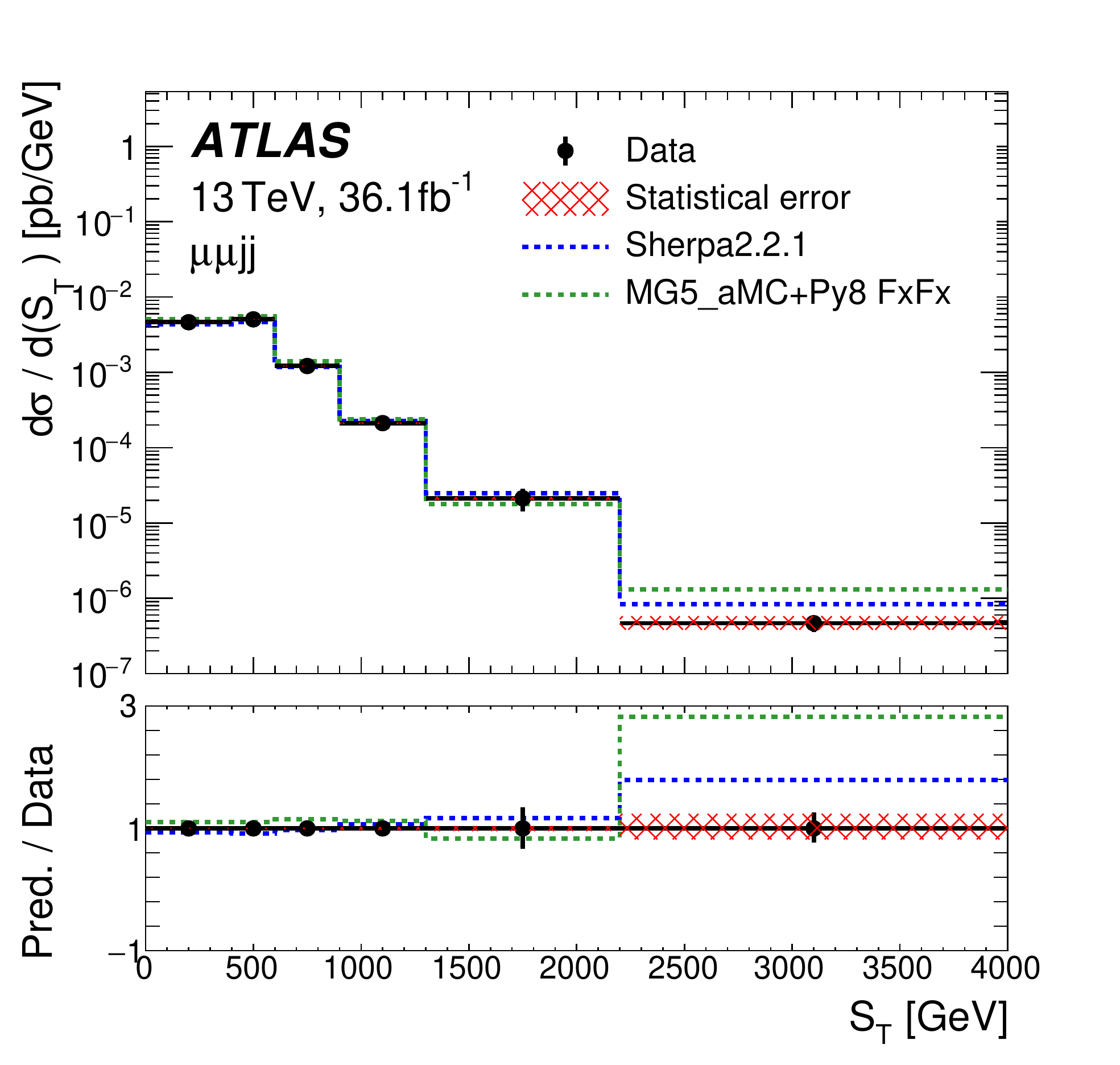}
\caption{The measured generator-level differential cross-sections in the \mumujj channel, exclusively for the \Zmm 
process. The MC prediction for the dominant process is also shown, with no $m_{jj}$ reweighting applied. The red band represents the statistical component of the total uncertainty which is indicated by the error bar on each point. The variable $ min \Delta \Phi (j1,l) $ refers to the minimal difference in $\phi$ between the subleading jet and a prompt lepton. The variable $H_{T}$ refers to the scalar sum of the jet \pt values. The variables $\Delta \eta(jj)$, $\Delta \phi(jj)$ and $\Delta \phi(ll)$ refer to the difference in $\eta$ and $\phi$ between the two leading jets or leptons. The variable $S_{T}$ refers to the scalar sum of the \pt of the jets and leptons.}
\label{fig:unfolded_data-1-isZCRmmjj}
\end{figure}

\begin{figure}[h]
\centering
\includegraphics[width=0.30\linewidth]{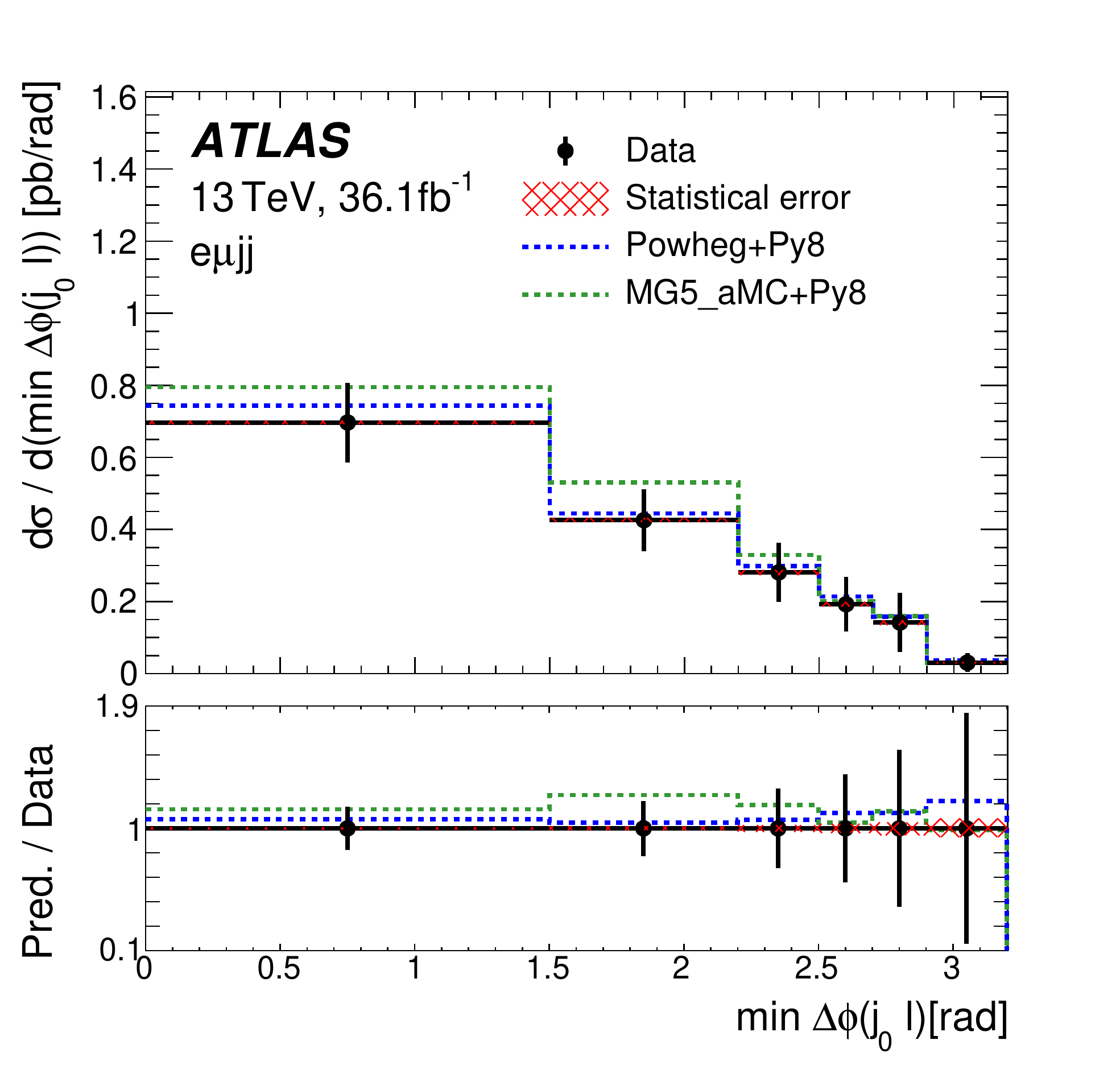}
\includegraphics[width=0.30\linewidth]{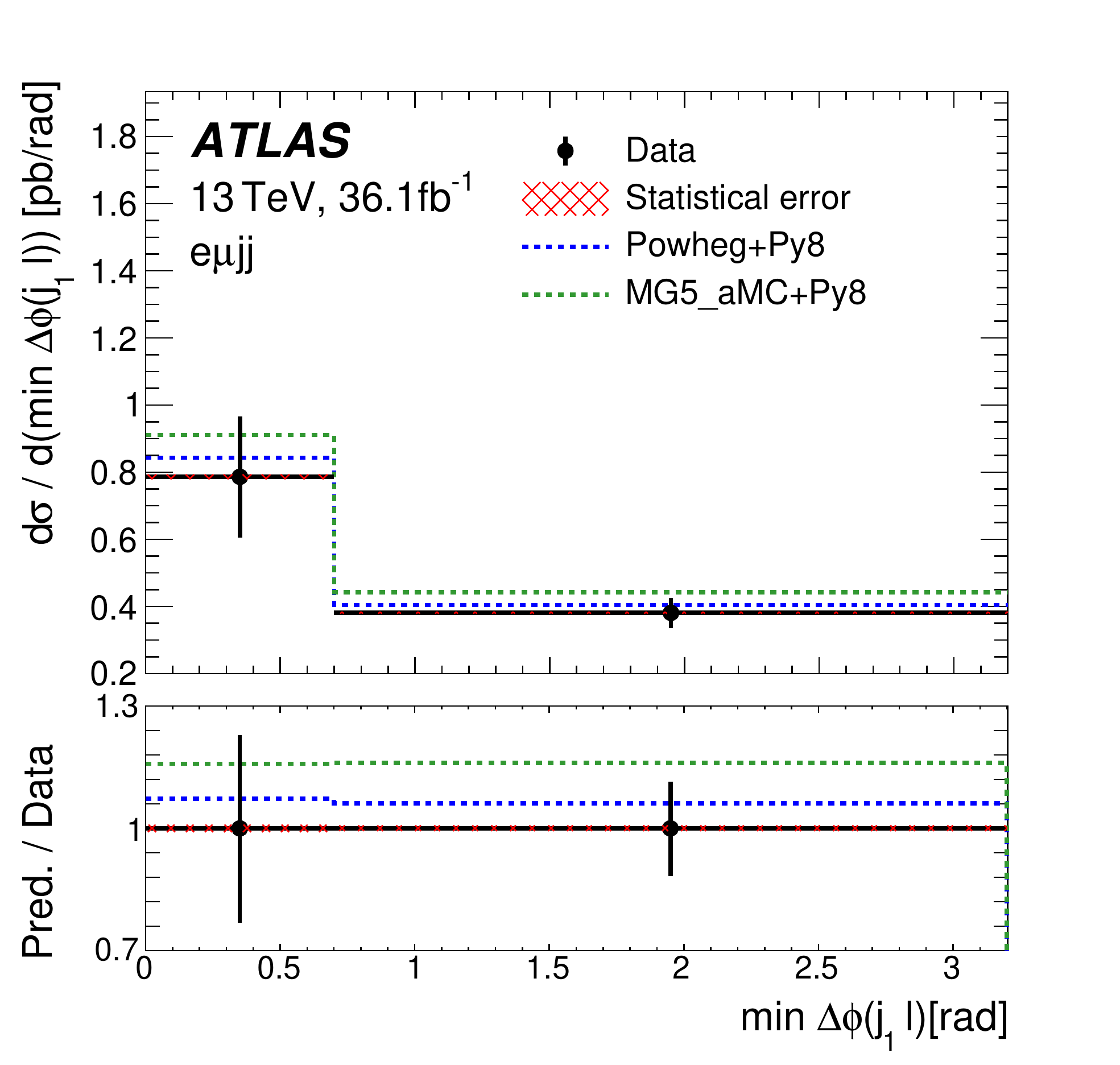}
\includegraphics[width=0.30\linewidth]{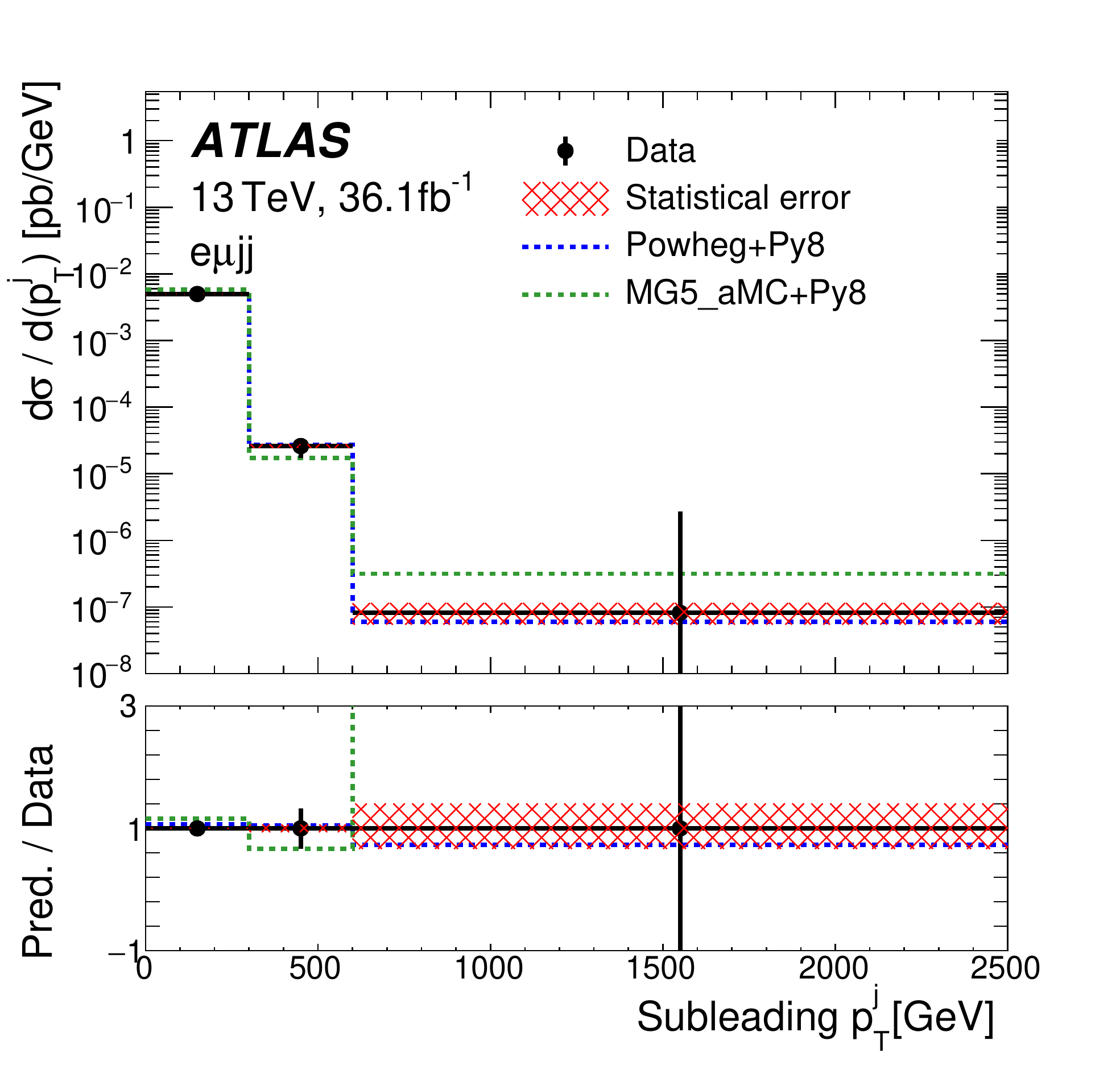}\\
\includegraphics[width=0.30\linewidth]{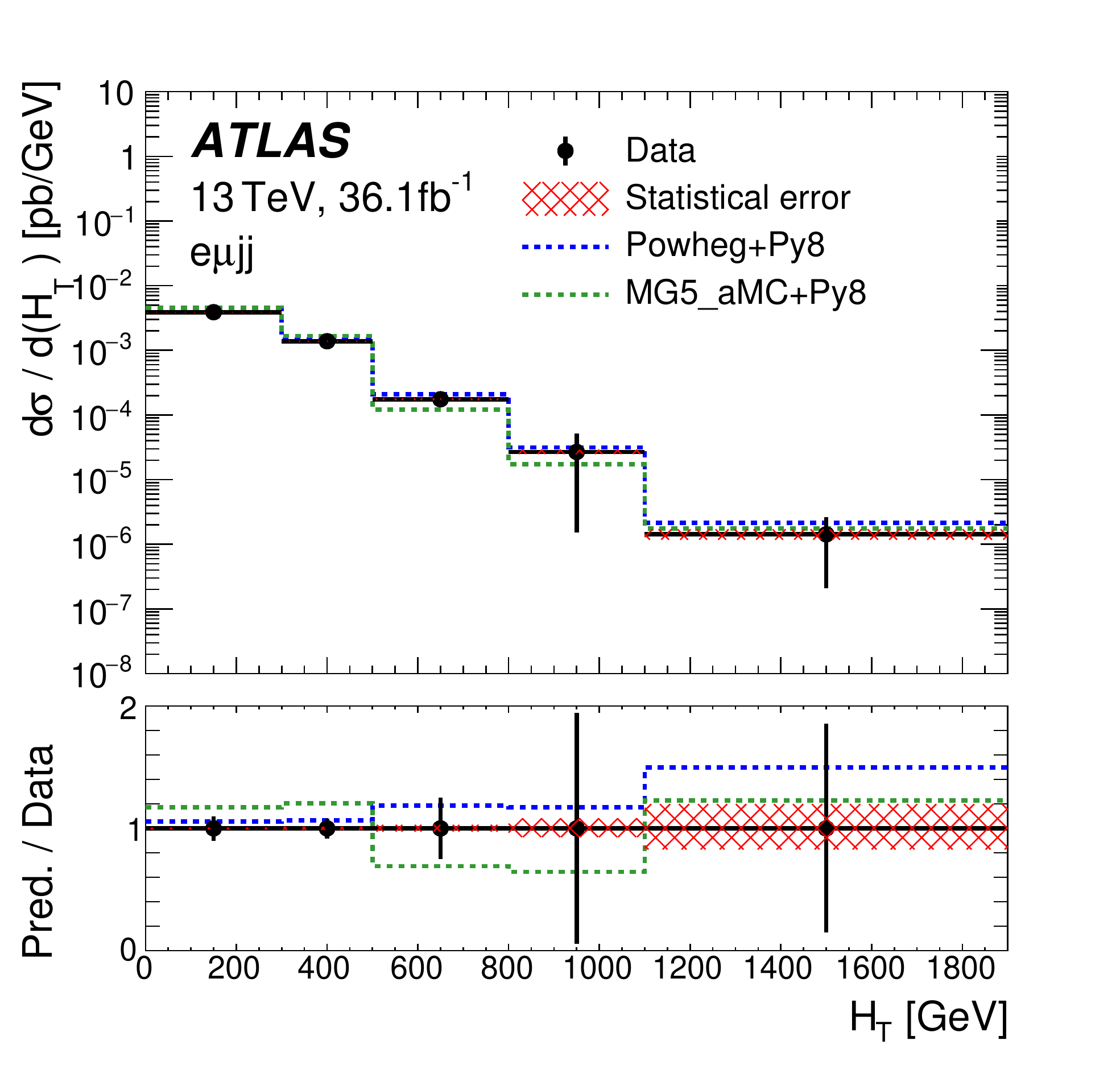}
\includegraphics[width=0.30\linewidth]{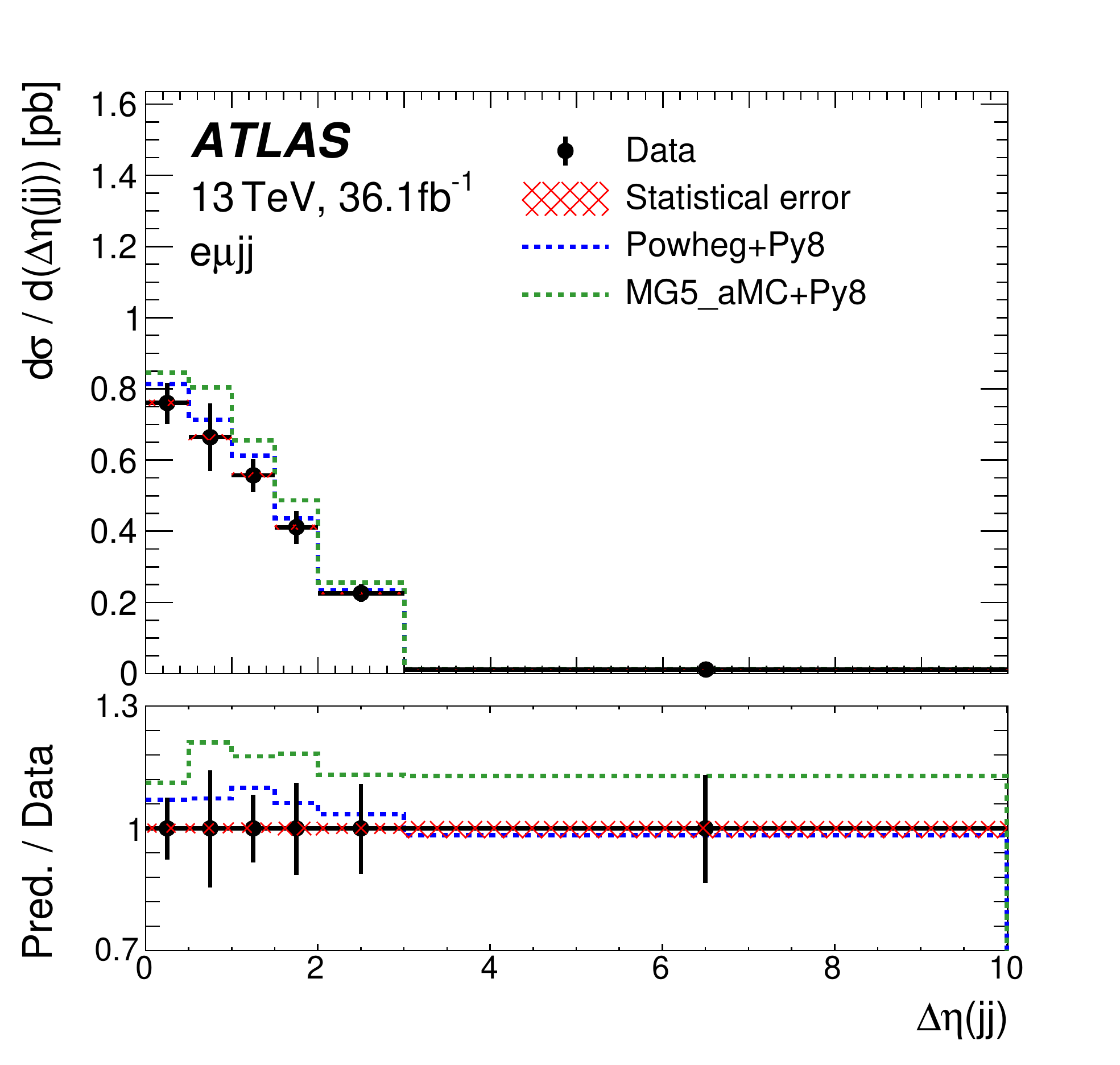}
\includegraphics[width=0.30\linewidth]{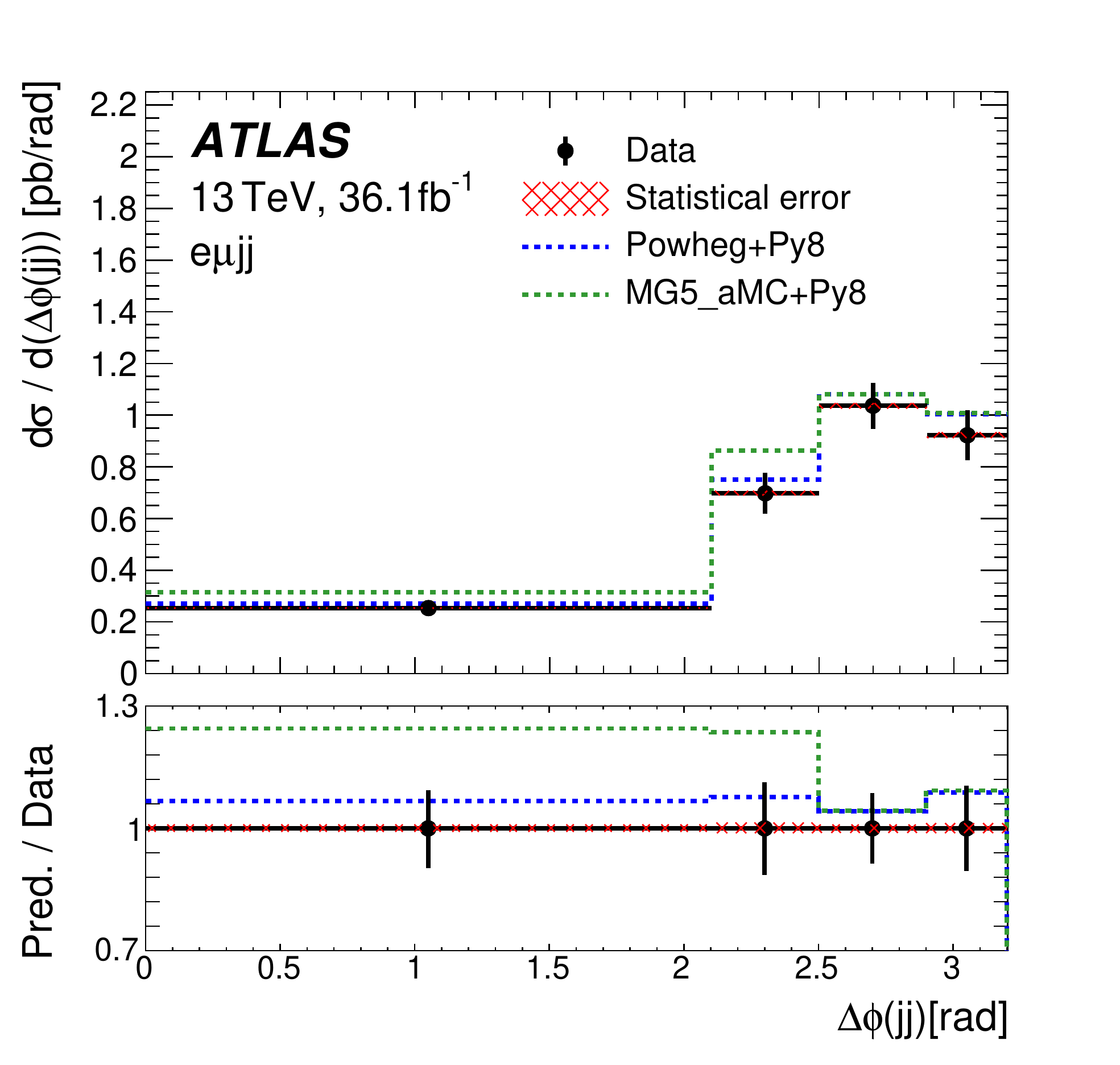}\\
\includegraphics[width=0.30\linewidth]{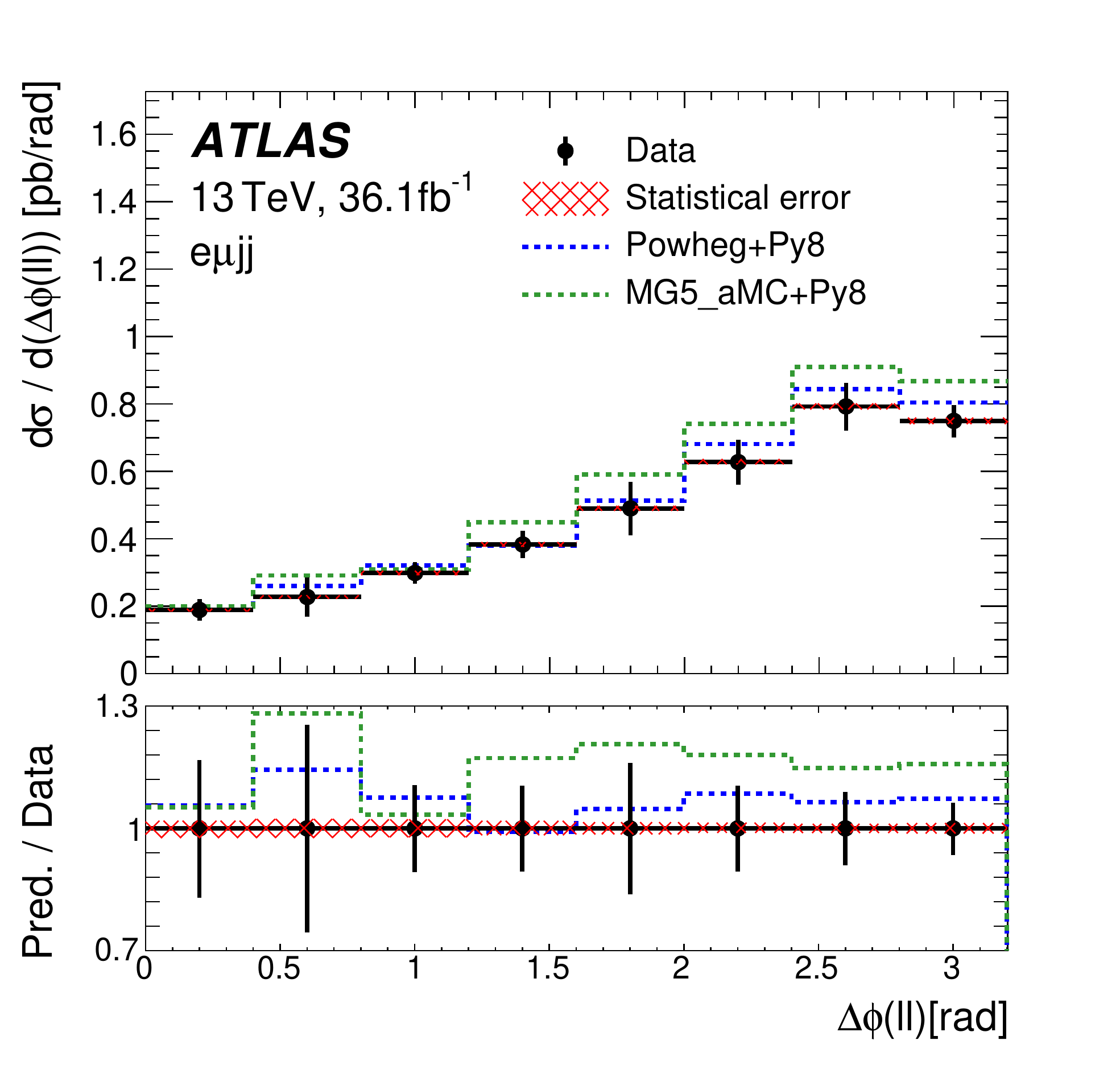}
\includegraphics[width=0.30\linewidth]{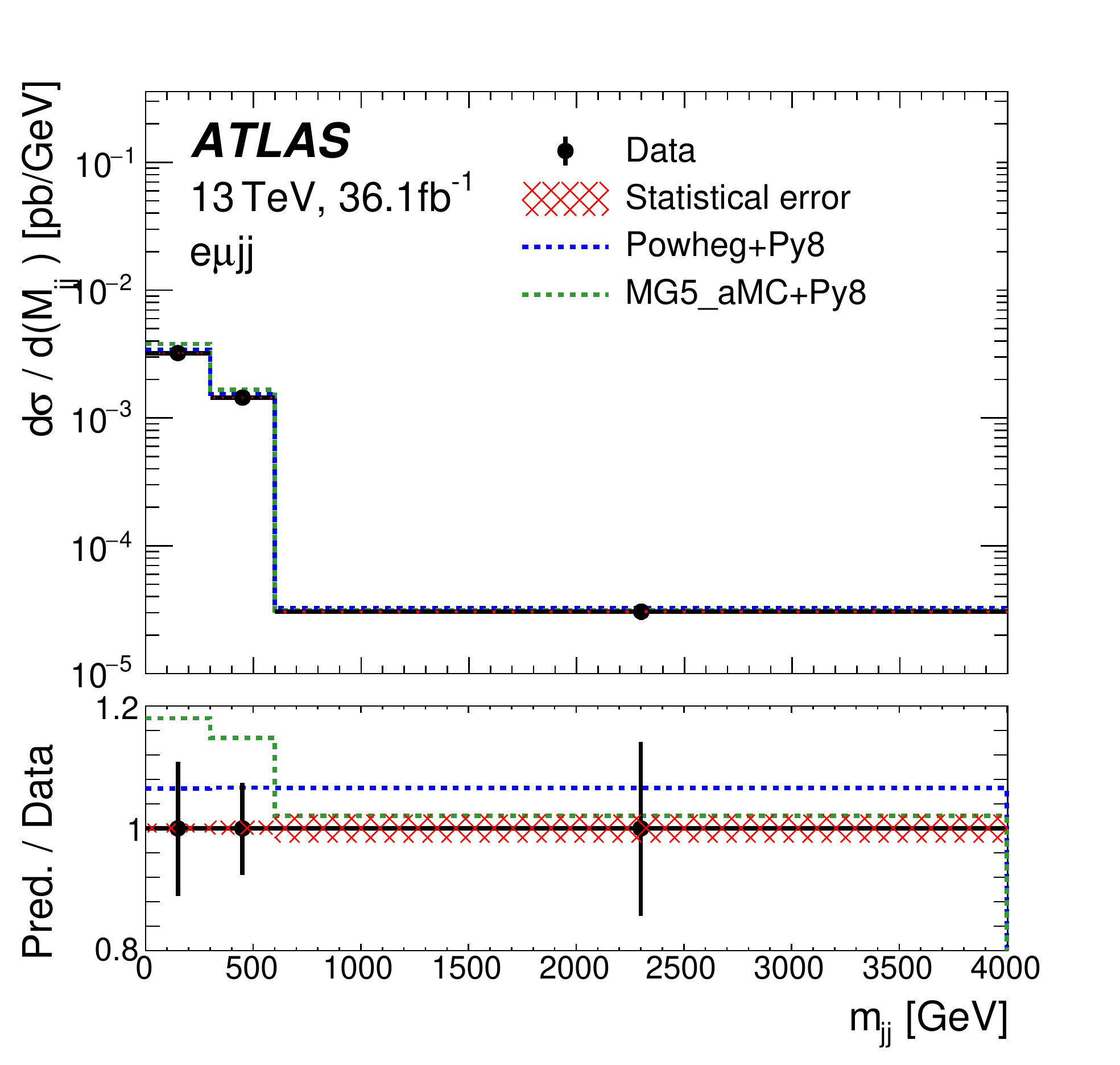}
\includegraphics[width=0.30\linewidth]{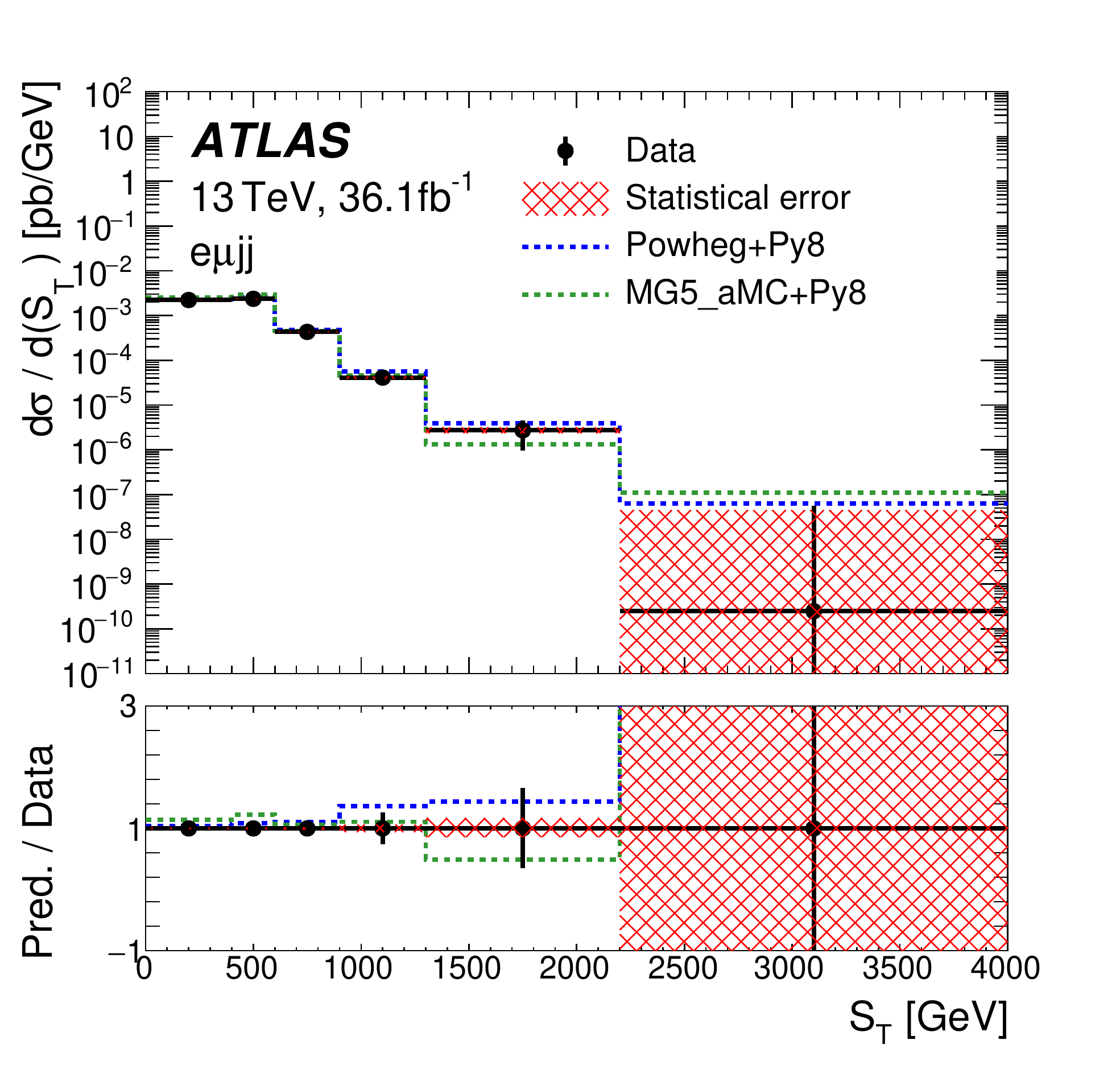}
\caption{The measured generator-level differential cross-sections in the \emujj channel, defined as for the \eejj and \mumujj MRs, but 
requiring the two leptons to have 
different flavours, and with no requirement on the dilepton mass. The measurements are made exclusively for the \ttbar~process. The MC prediction for the dominant 
process is also shown. The red band represents the statistical component of the total uncertainty which is indicated by the error bar on each point. The variables $ min \Delta \Phi (j0,l) $ and $ min \Delta \Phi (j1,l) $ refers to the minimal difference in $\phi$ between a prompt lepton and the leading or subleading jet respectively. The variable $H_{T}$ refers to the scalar sum of the jet \pt values. The variables $\Delta \eta(jj)$, $\Delta \phi(jj)$ and $\Delta \phi(ll)$ refer to the difference in $\eta$ and $\phi$ between the two leading jets or leptons. The variable $S_{T}$ refers to the scalar sum of the \pt of the jets and leptons.} 
\label{fig:unfolded_data-1-isCRemjj}
\end{figure}

The measured differential cross-sections in the MRs are shown in Figure~\ref{fig:unfolded_data-1-isZCReejj} for the \eejj channel, in 
Figure~\ref{fig:unfolded_data-1-isZCRmmjj} for the \mumujj channel and in Figure~\ref{fig:unfolded_data-1-isCRemjj} for the \emujj channel.

\begin{figure}[h]
\centering
\includegraphics[width=0.30\linewidth]{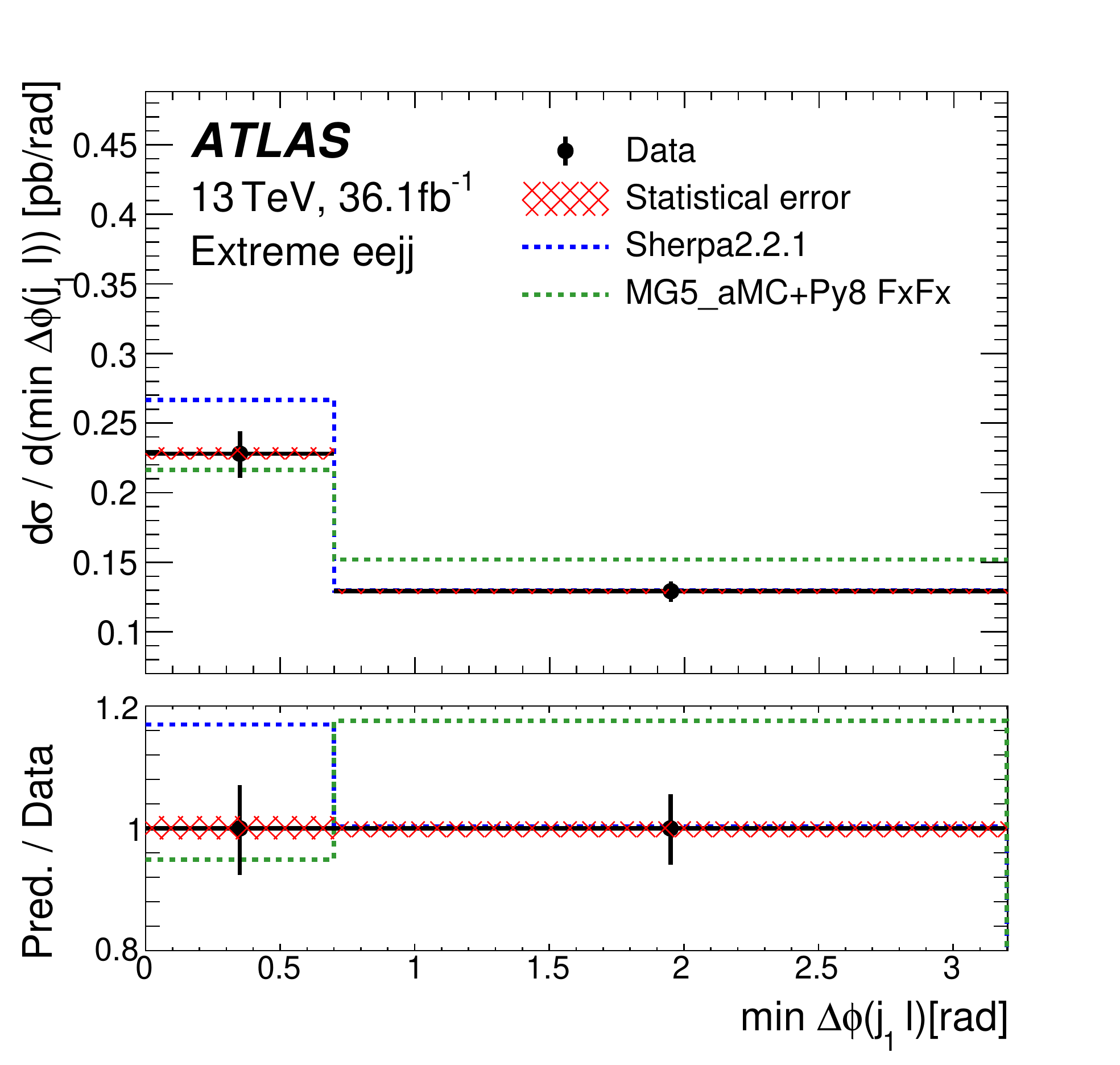}
\includegraphics[width=0.30\linewidth]{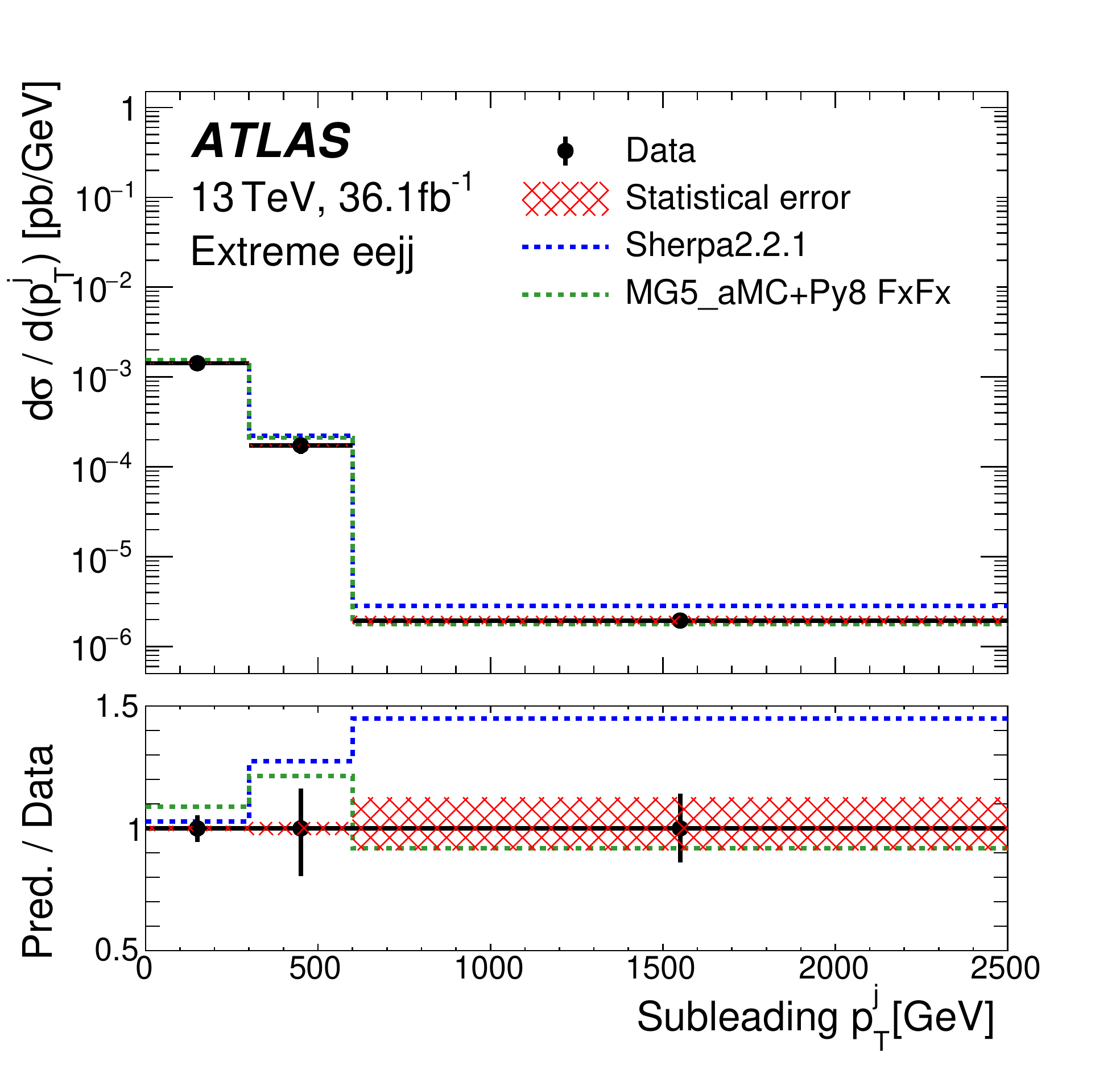}
\includegraphics[width=0.30\linewidth]{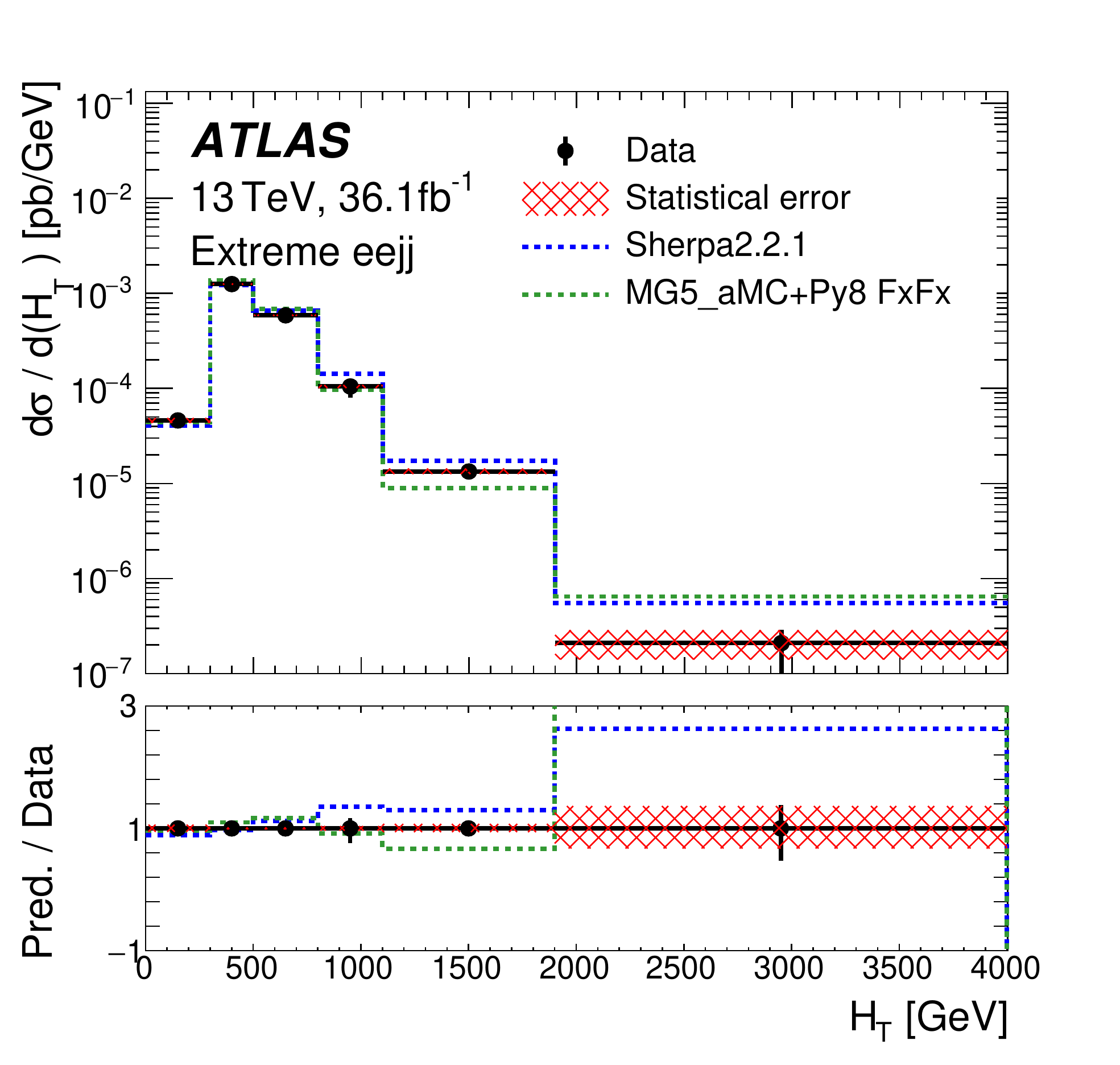}\\
\includegraphics[width=0.30\linewidth]{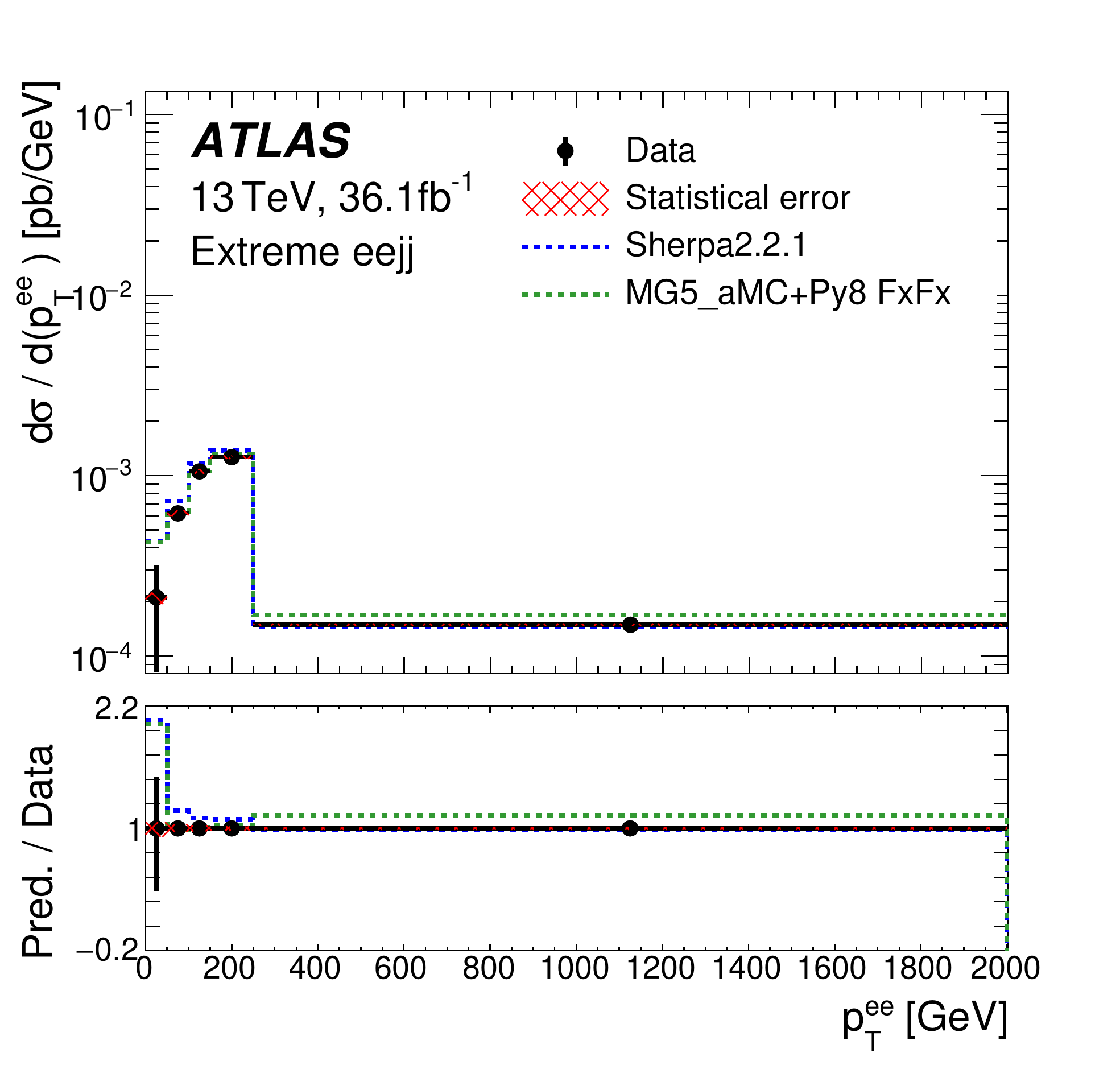}
\includegraphics[width=0.30\linewidth]{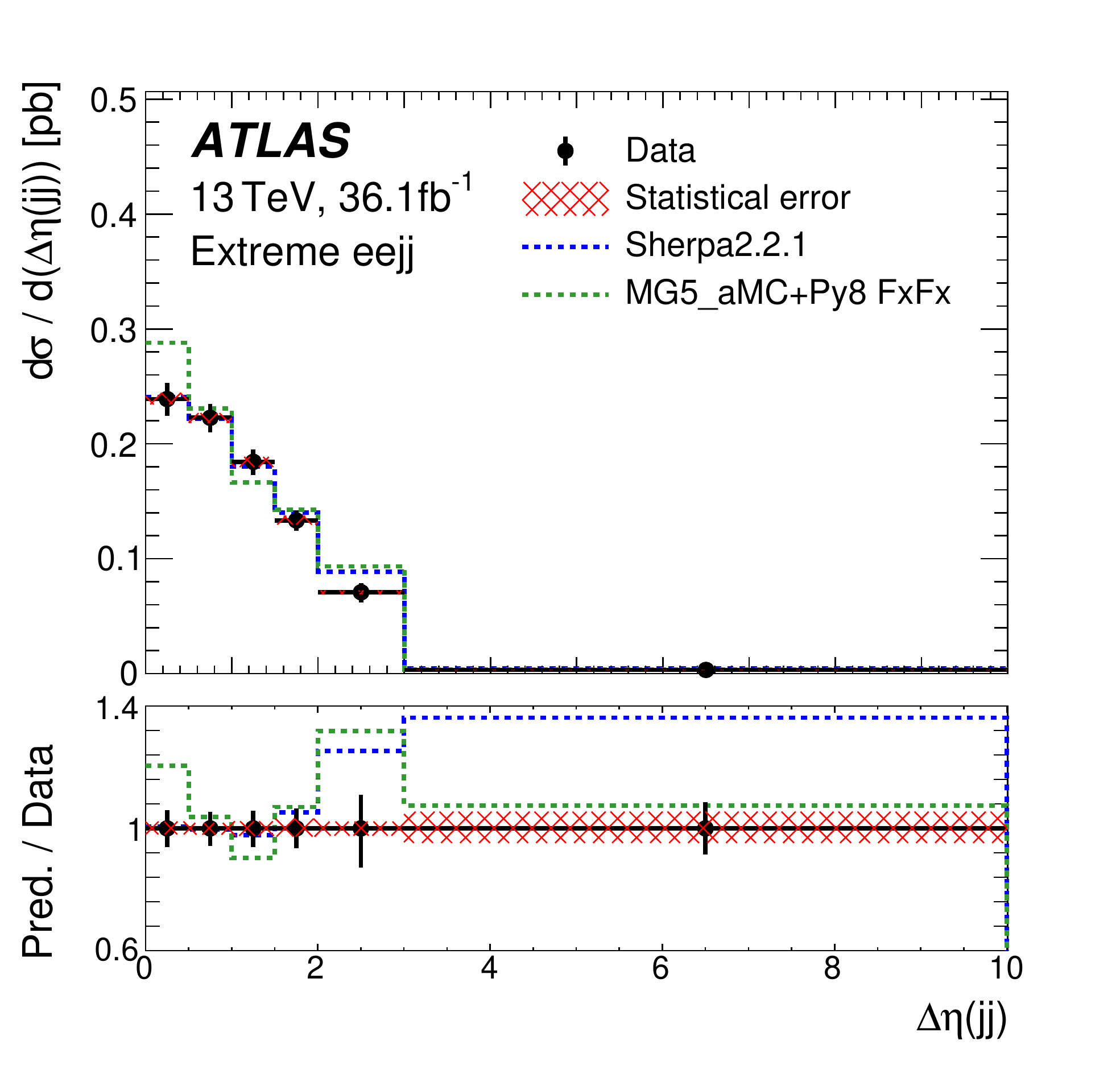}
\includegraphics[width=0.30\linewidth]{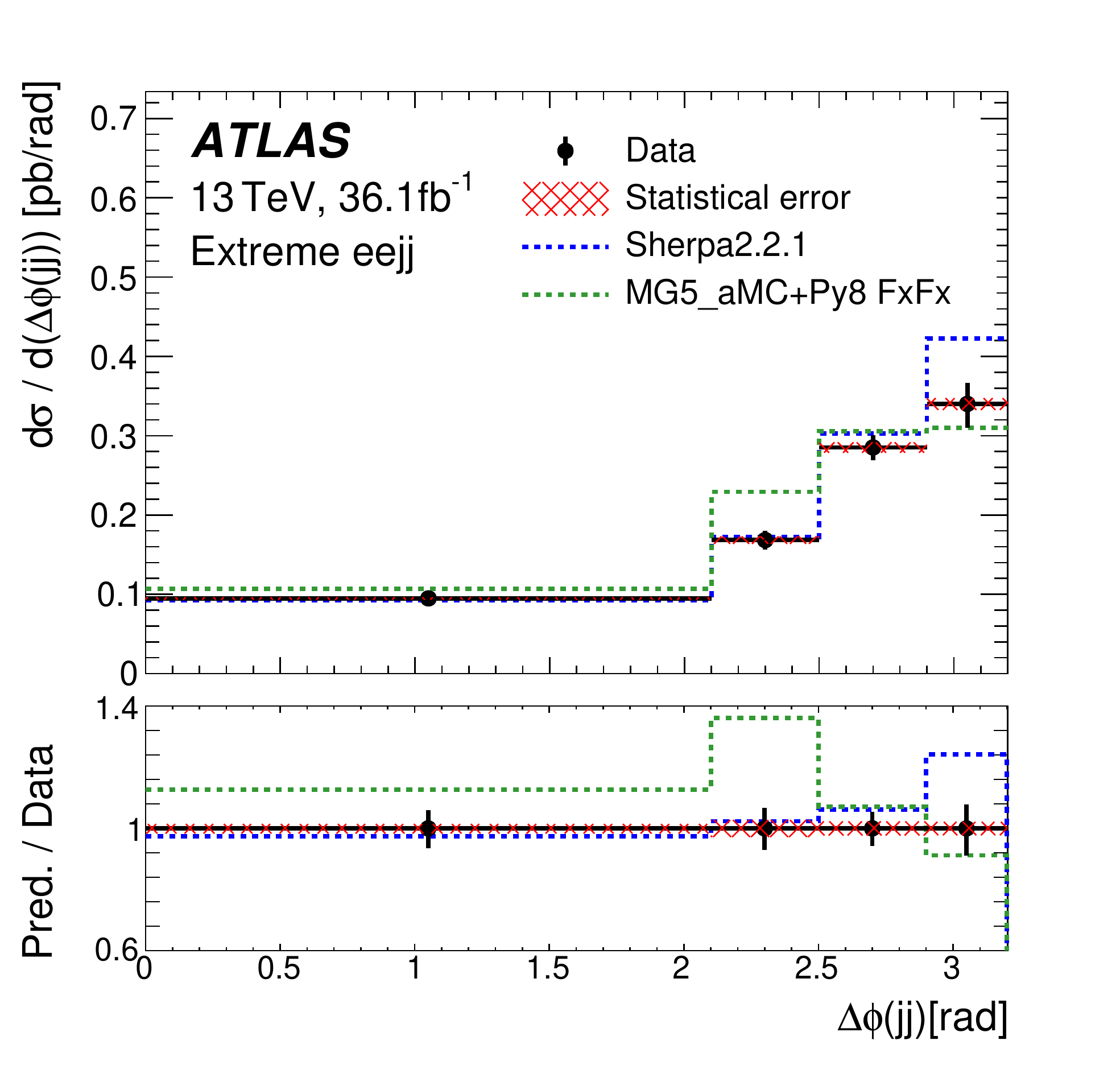}\\
\includegraphics[width=0.30\linewidth]{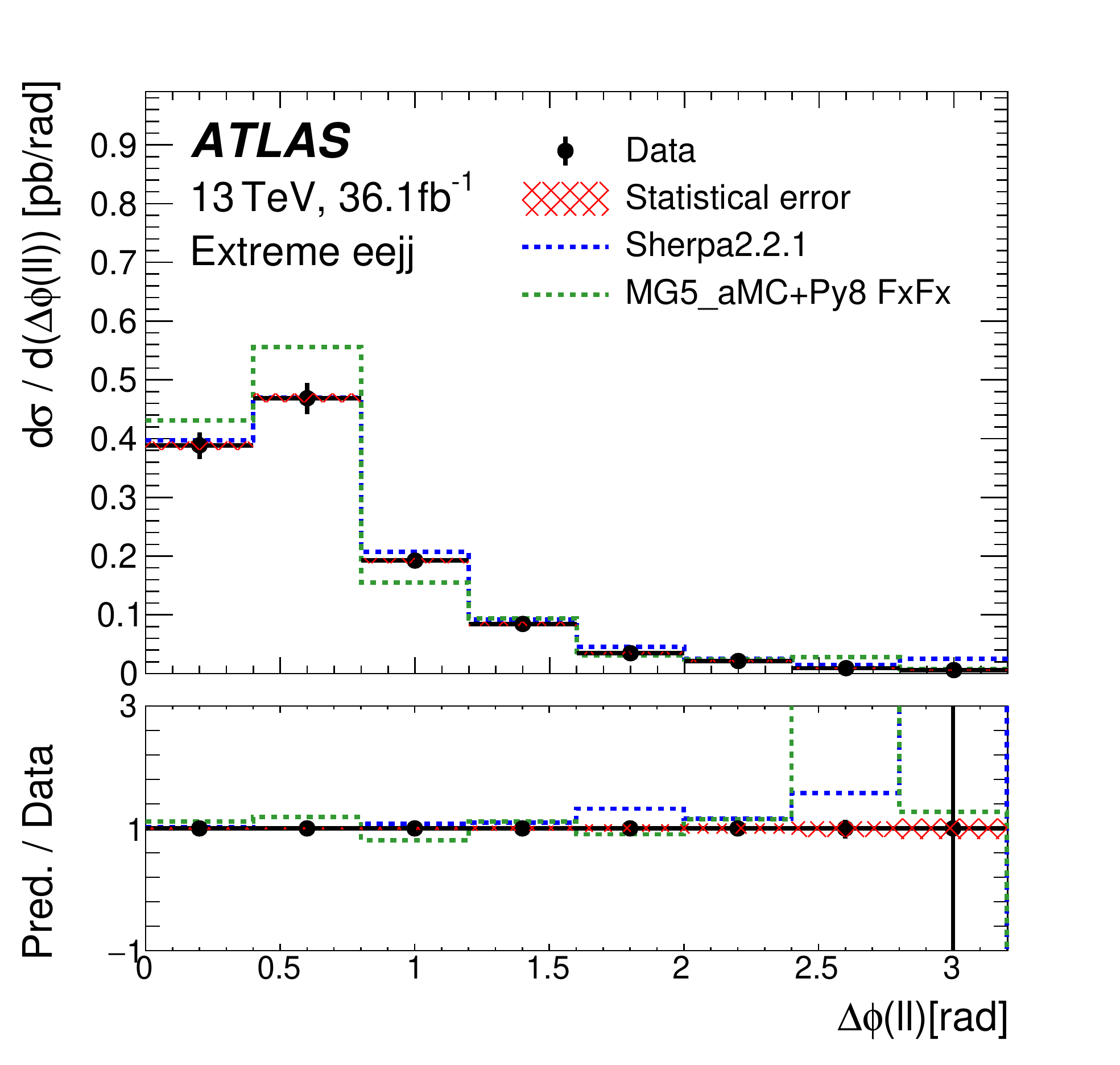}
\includegraphics[width=0.30\linewidth]{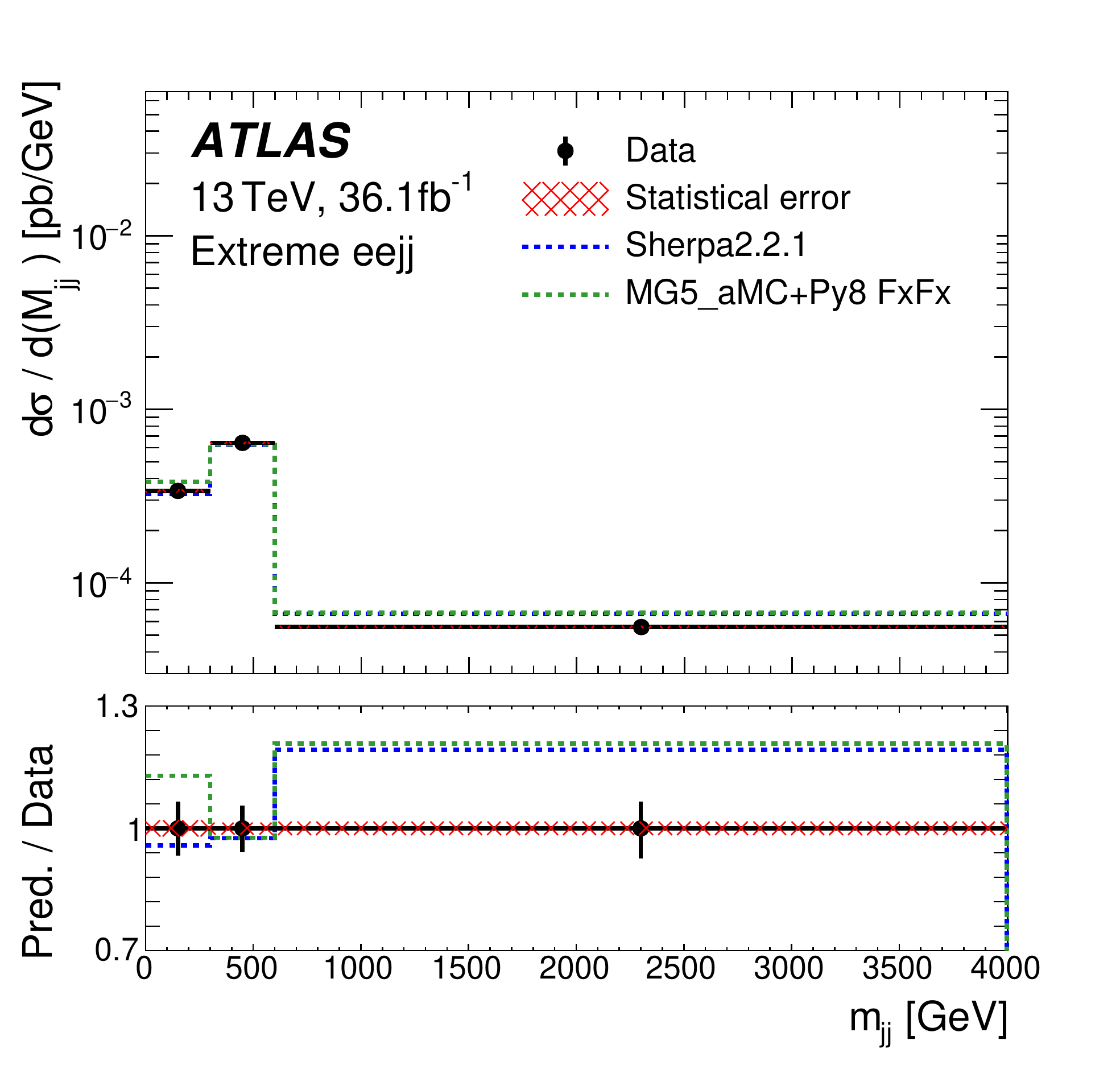}
\includegraphics[width=0.30\linewidth]{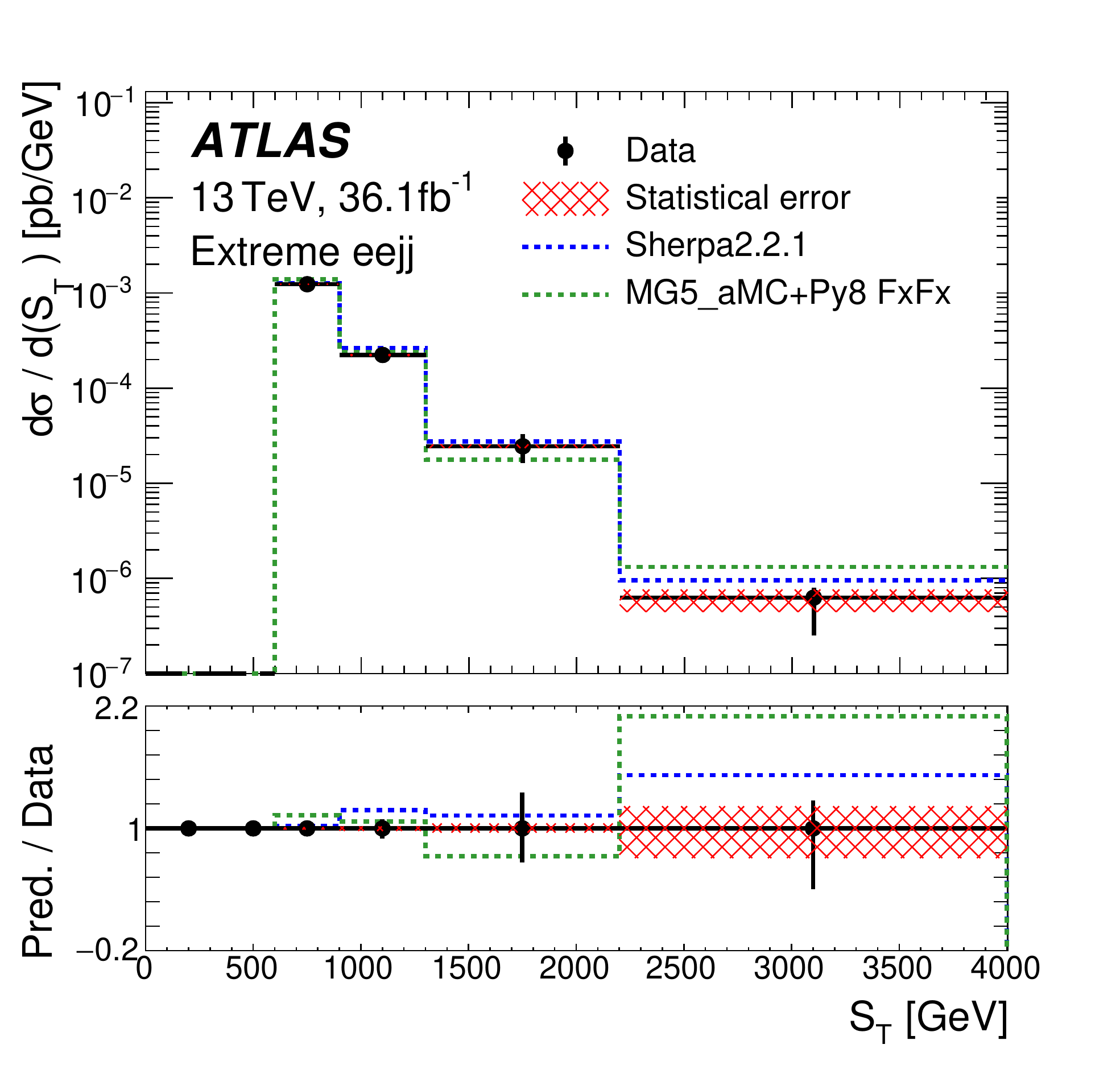}
\caption{The measured generator-level differential cross-sections in the \eejj 
channel, where an additional selection of $\st > 600~\GeV$ has been applied, exclusively for the \Zee process. The MC prediction for the dominant process is also shown, with no $m_{jj}$ reweighting applied. The red band represents the statistical component of the total uncertainty which is indicated by the error bar on each point. The variable $ min \Delta \Phi (j1,l) $ refers to the minimal difference in $\phi$ between the subleading jet and a prompt lepton. The variable $H_{T}$ refers to the scalar sum of the jet \pt values. The variables $\Delta \eta(jj)$, $\Delta \phi(jj)$ and $\Delta \phi(ll)$ refer to the difference in $\eta$ and $\phi$ between the two leading jets or leptons. The variable $S_{T}$ refers to the scalar sum of the \pt of the jets and leptons.}
\label{fig:unfolded_data-1-isZCReejjST}
\end{figure}

\begin{figure}[h]
\centering
\includegraphics[width=0.30\linewidth]{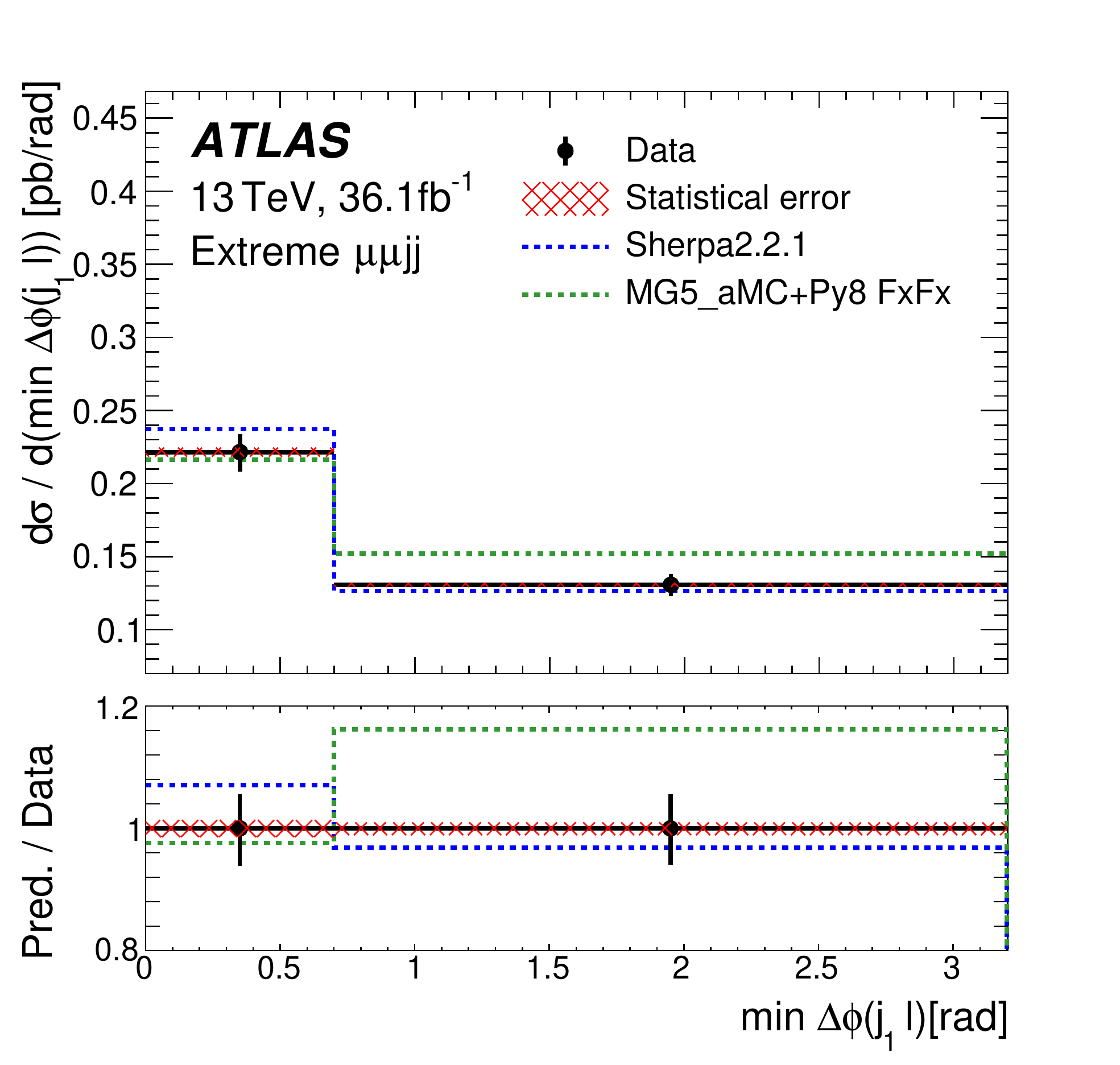}
\includegraphics[width=0.30\linewidth]{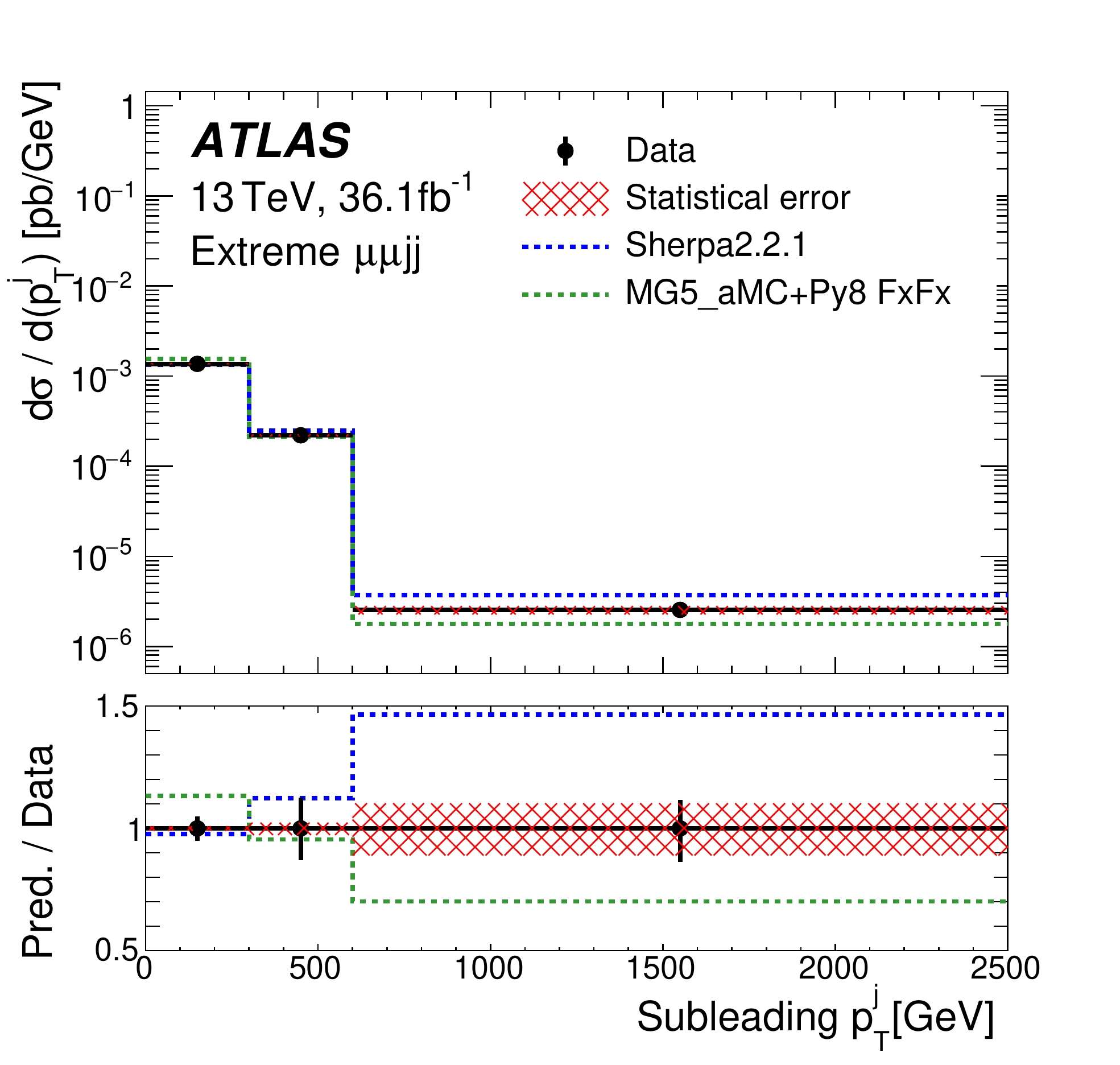}
\includegraphics[width=0.30\linewidth]{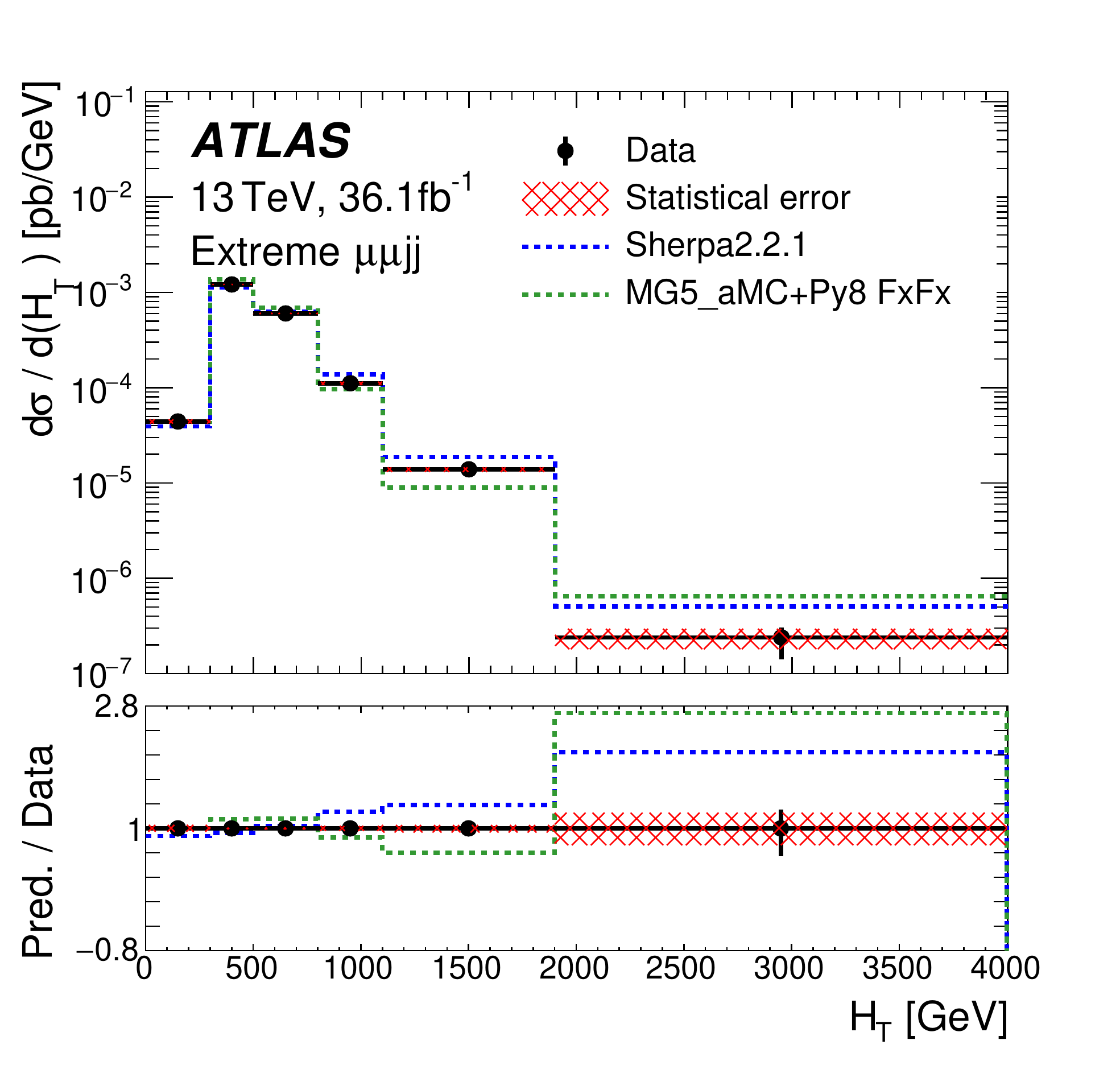}\\
\includegraphics[width=0.30\linewidth]{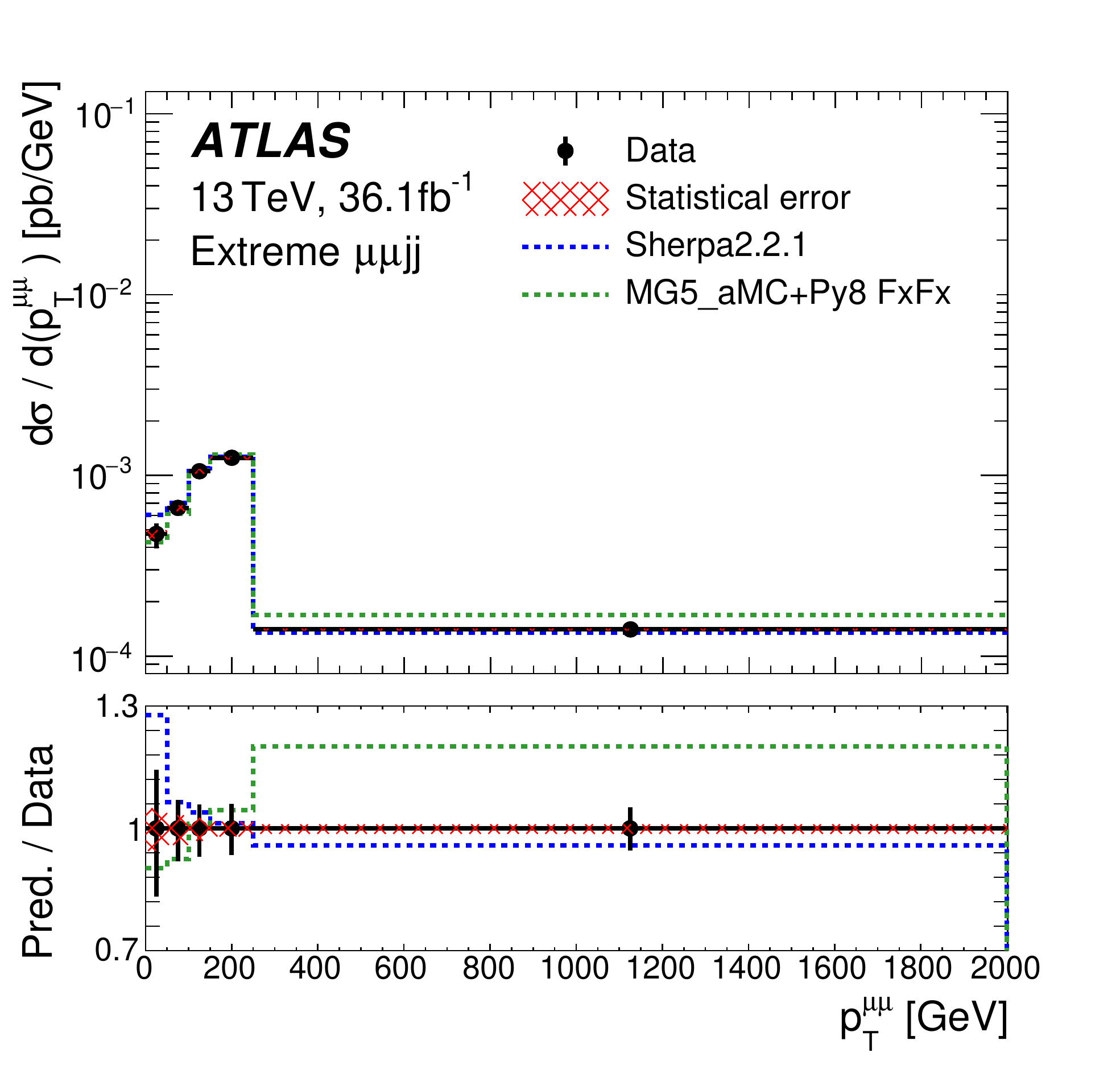}
\includegraphics[width=0.30\linewidth]{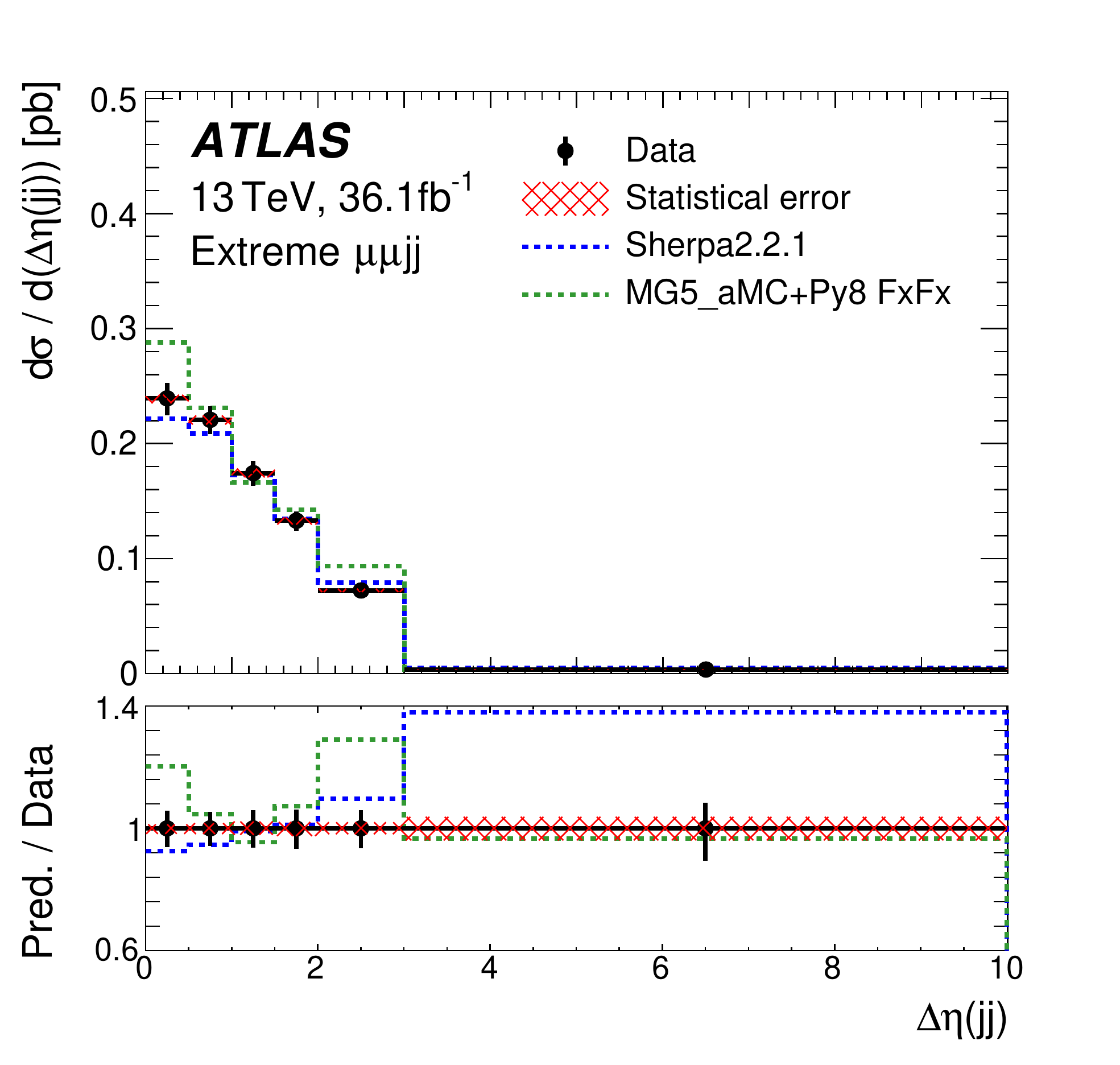}
\includegraphics[width=0.30\linewidth]{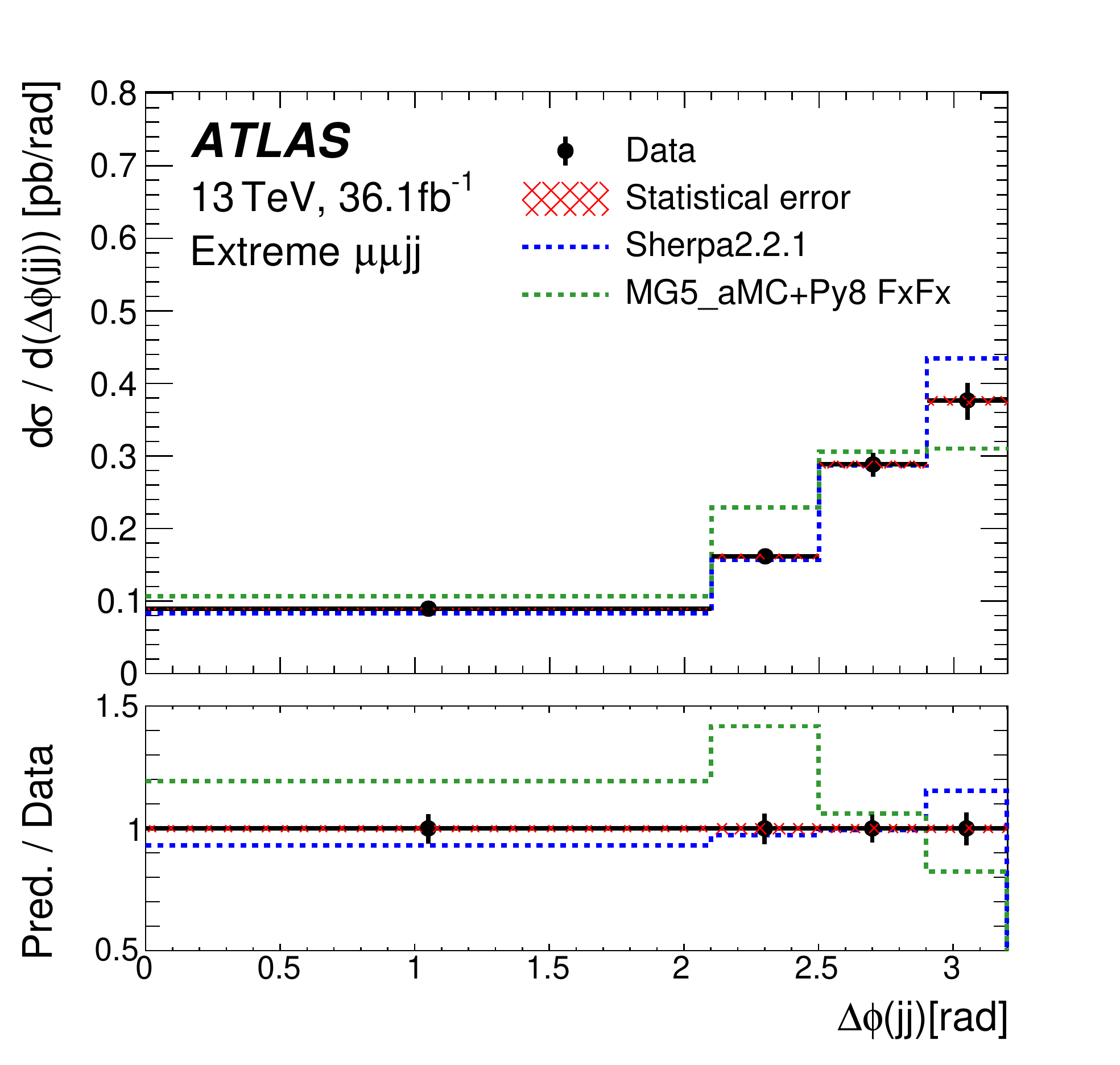}\\
\includegraphics[width=0.30\linewidth]{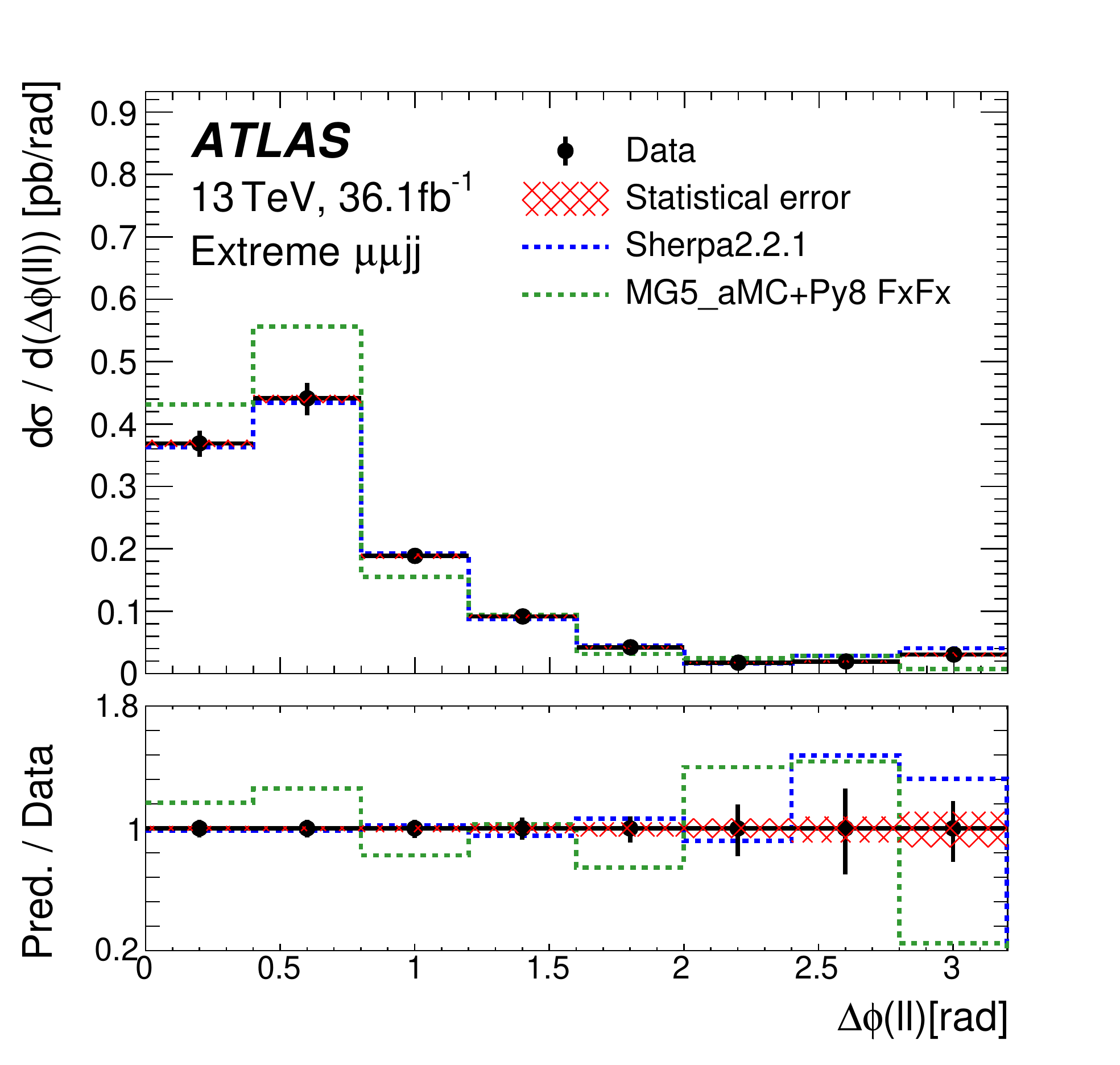}
\includegraphics[width=0.30\linewidth]{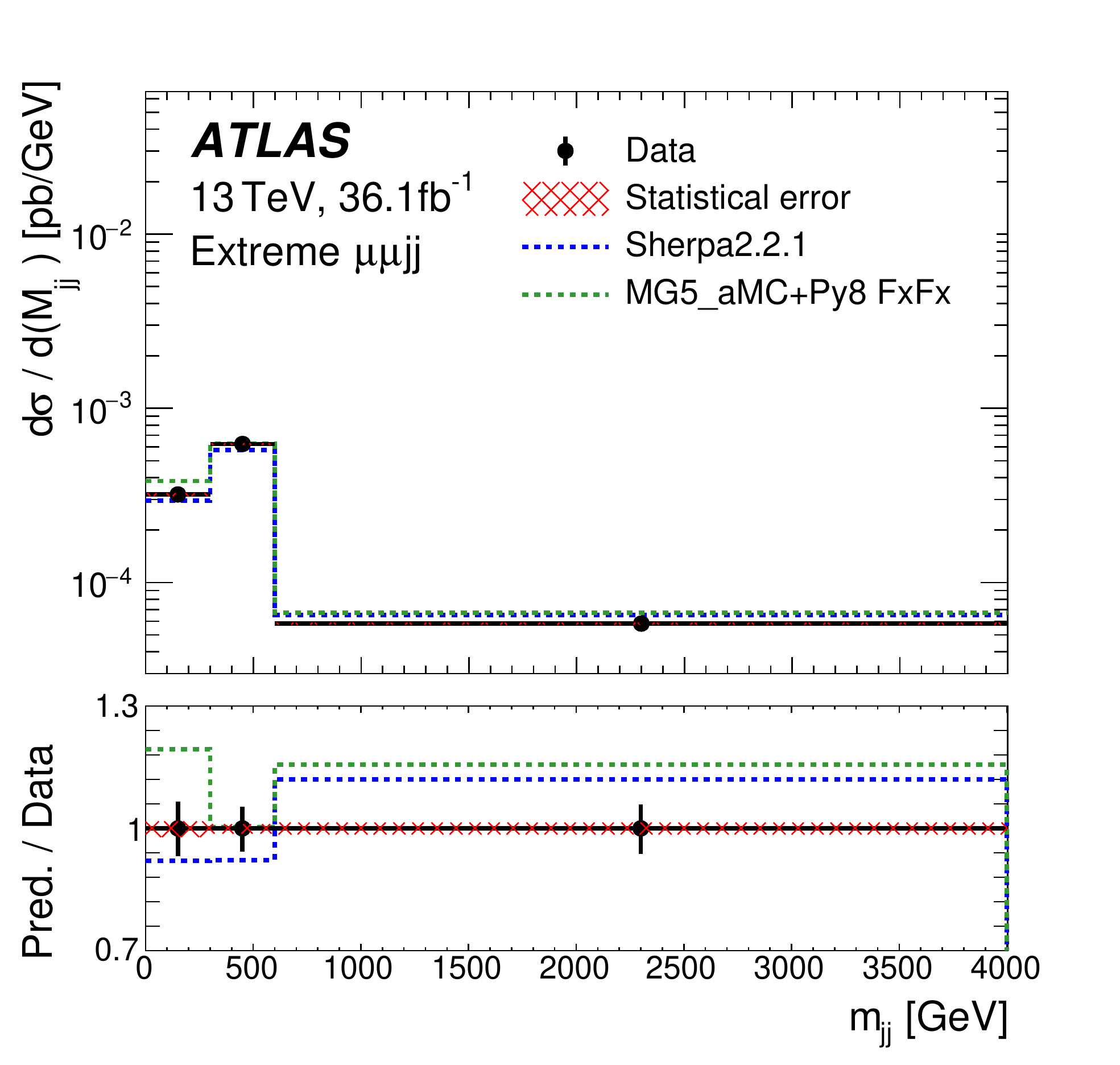}
\includegraphics[width=0.30\linewidth]{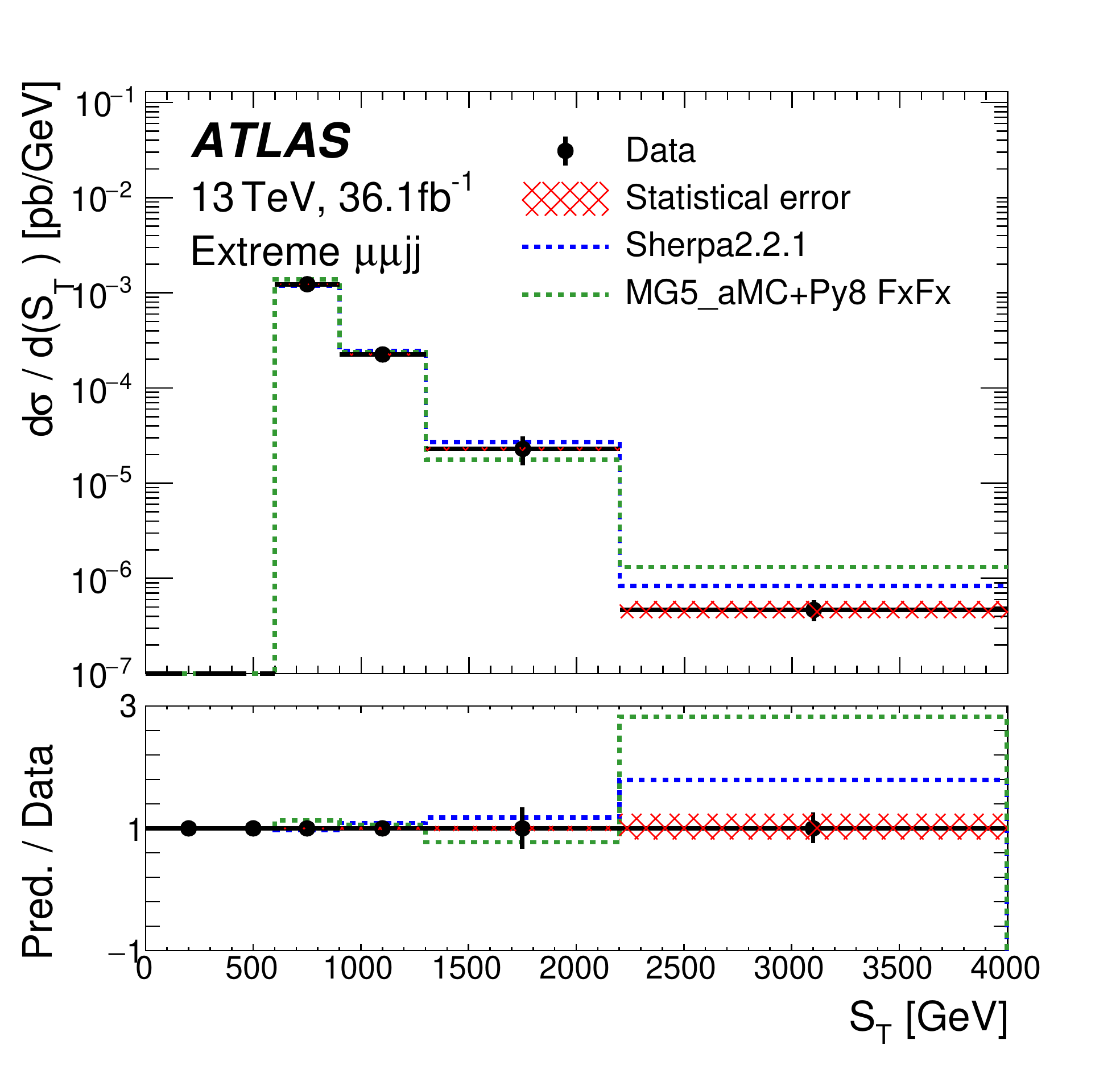}
\caption{The measured generator-level differential cross-sections in the \mumujj channel, where an additional selection of $\st > 
600~\GeV$ has been applied. The measurements are made exclusively for the \Zmm process. The MC prediction for the dominant process is also shown, with no $m_{jj}$ reweighting applied. The red band represents the statistical component of the total uncertainty which is indicated by the error bar on each point. The variable $ min \Delta \Phi (j1,l) $ refers to the minimal difference in $\phi$ between the subleading jet and a prompt lepton. The variable $H_{T}$ refers to the scalar sum of the jet \pt values. The variables $\Delta \eta(jj)$, $\Delta \phi(jj)$ and $\Delta \phi(ll)$ refer to the difference in $\eta$ and $\phi$ between the two leading jets or leptons. The variable $S_{T}$ refers to the scalar sum of the \pt of the jets and leptons.}
\label{fig:unfolded_data-1-isZCRmmjjST}
\end{figure}

\begin{figure}[h]
\centering
\includegraphics[width=0.30\linewidth]{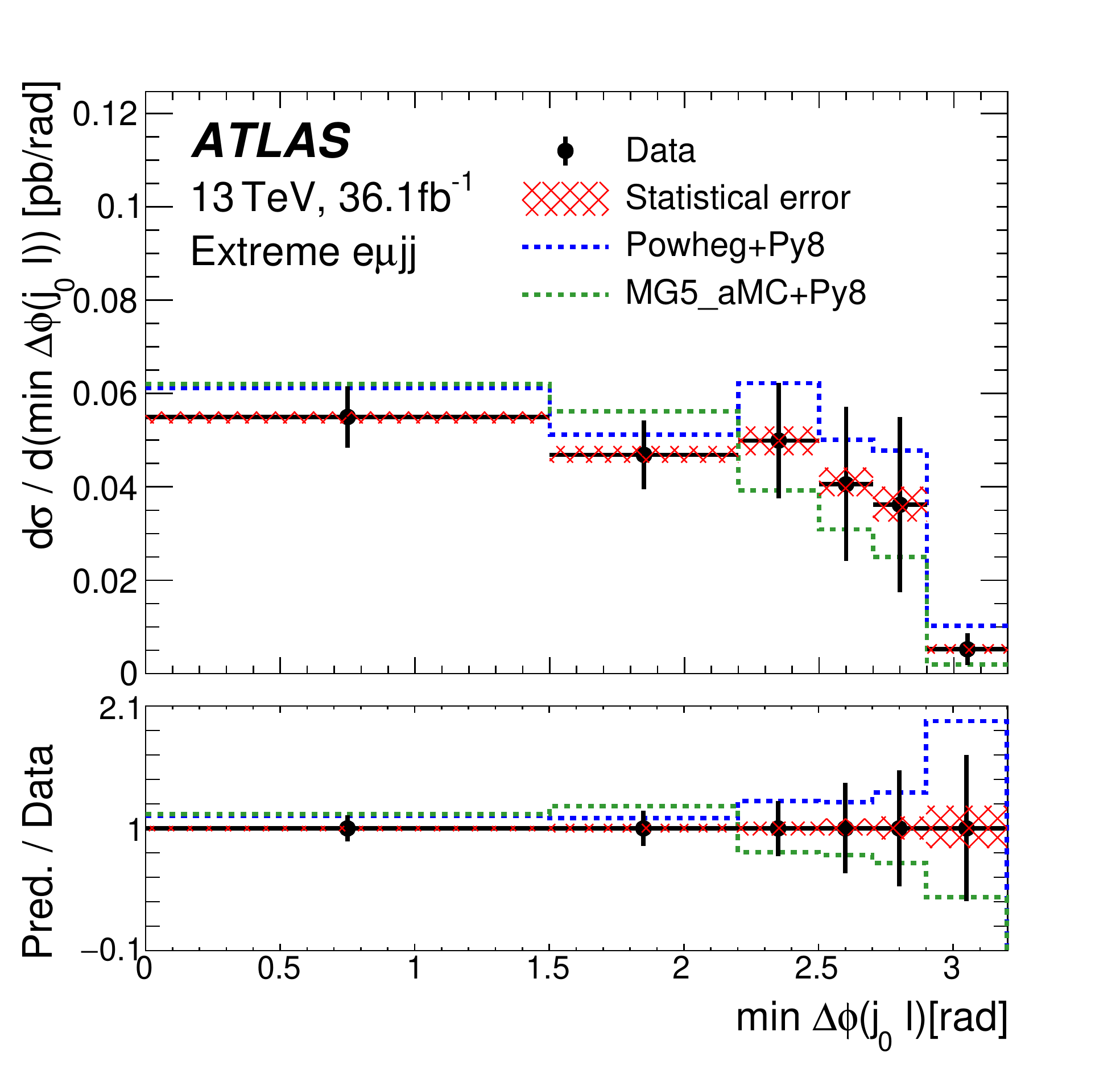}
\includegraphics[width=0.30\linewidth]{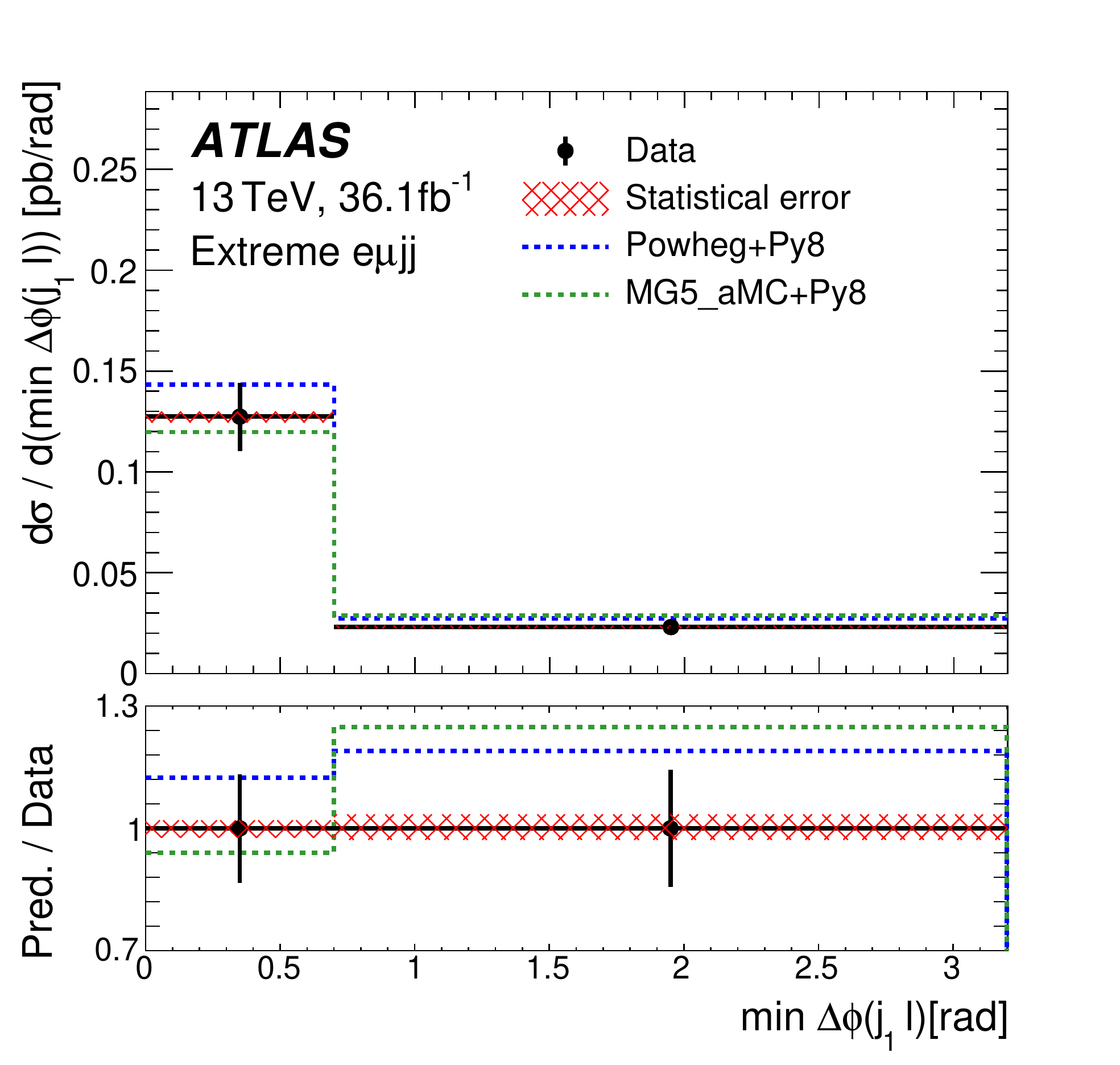}
\includegraphics[width=0.30\linewidth]{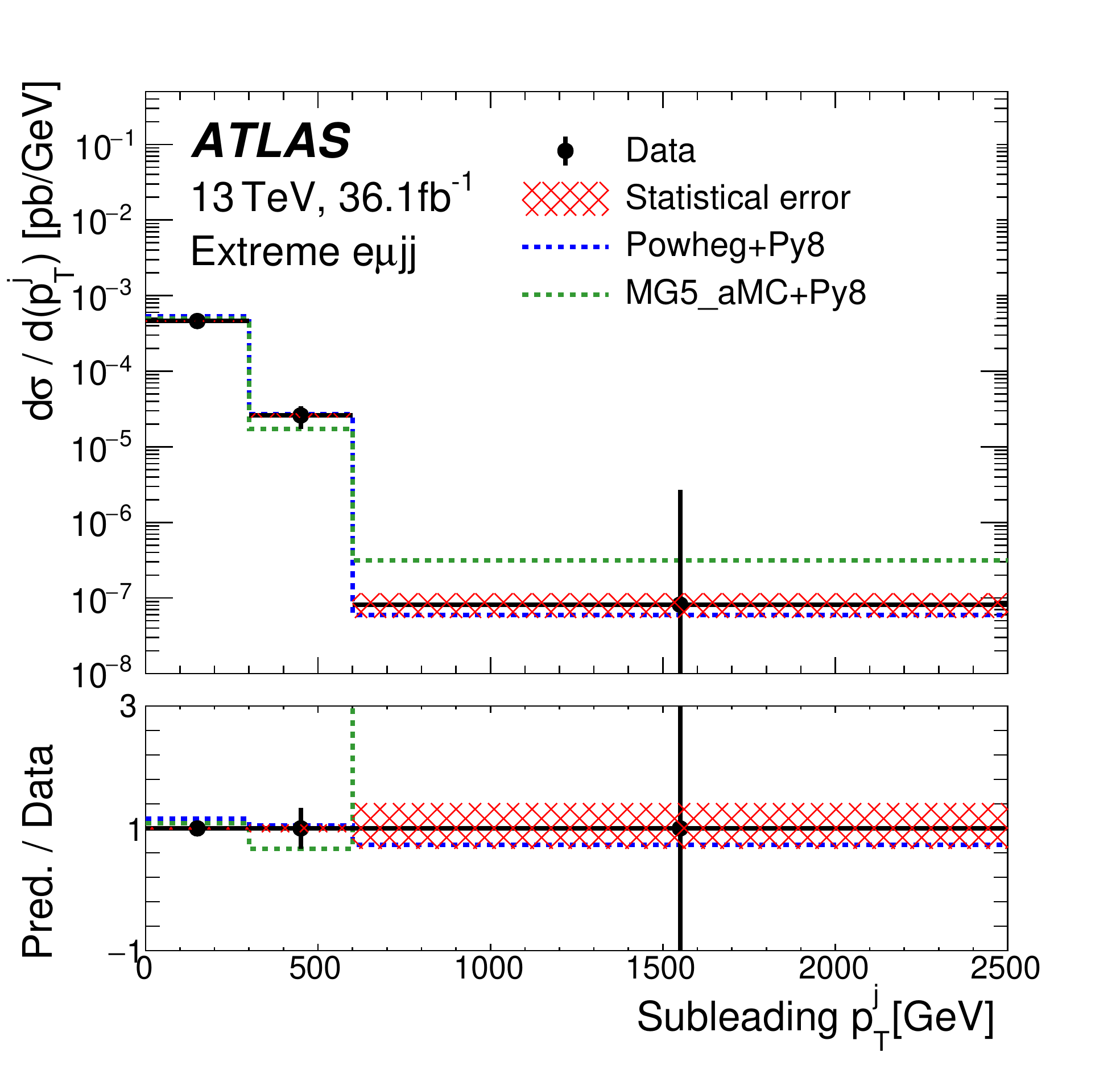}\\
\includegraphics[width=0.30\linewidth]{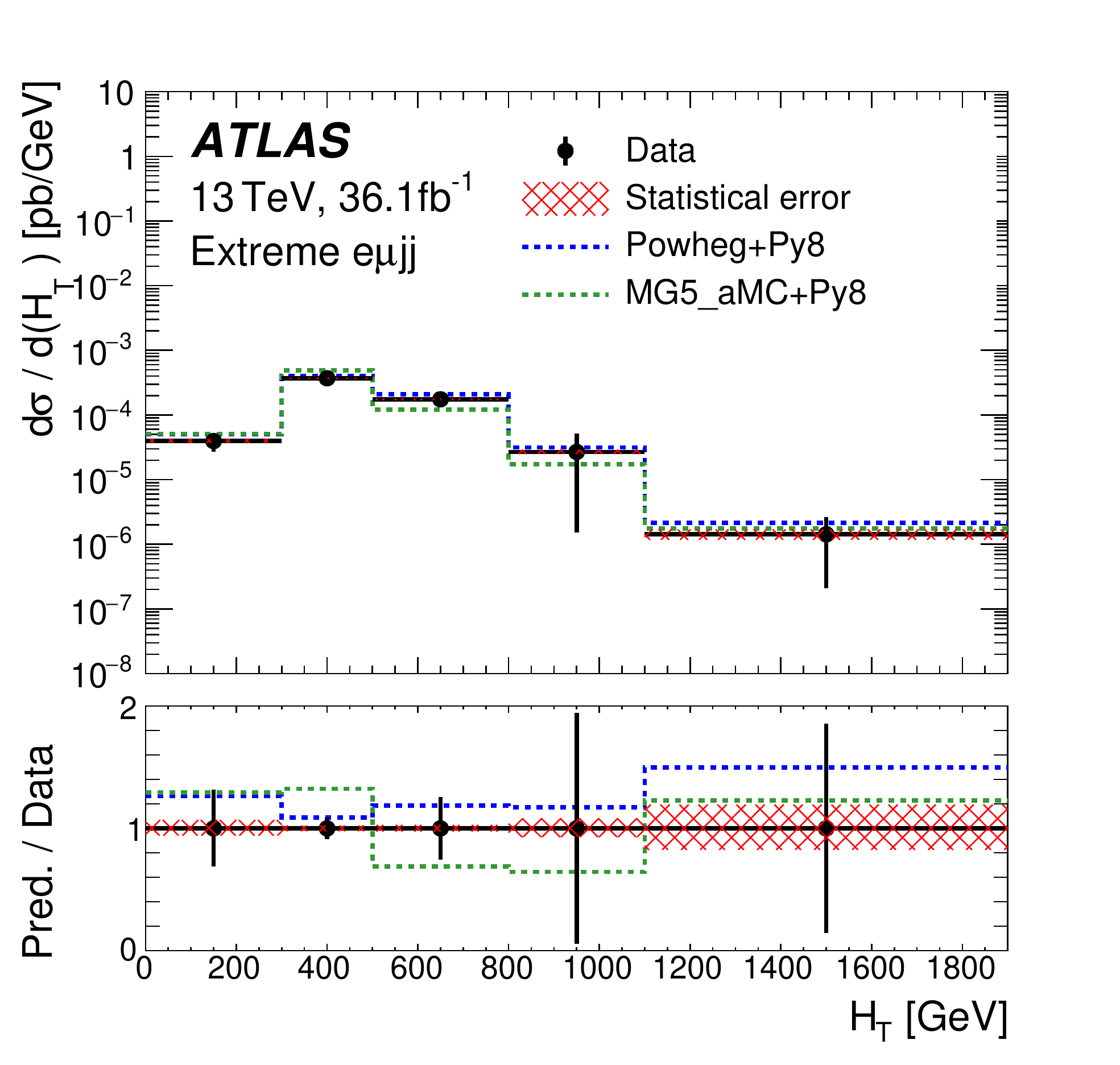}
\includegraphics[width=0.30\linewidth]{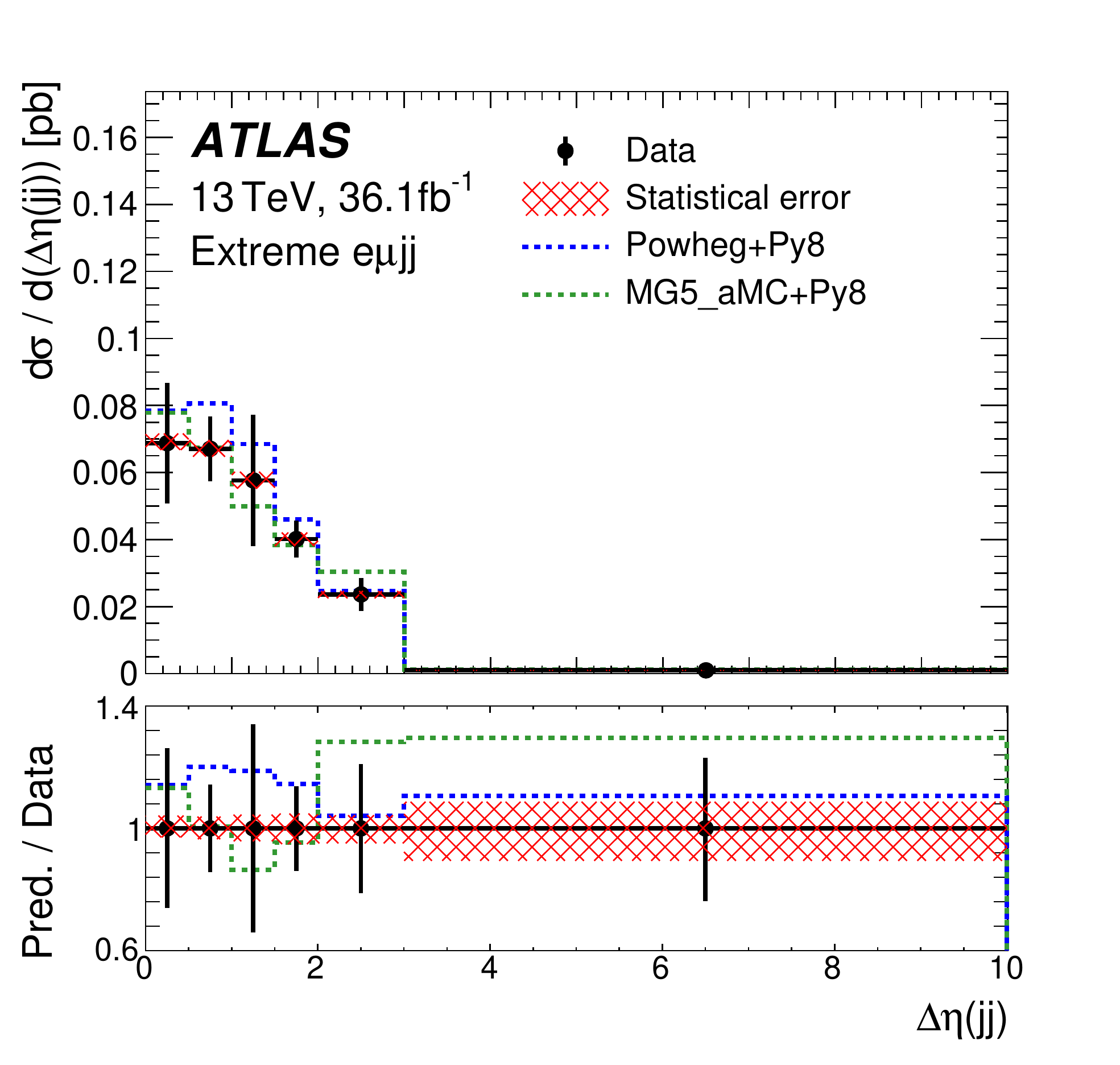}
\includegraphics[width=0.30\linewidth]{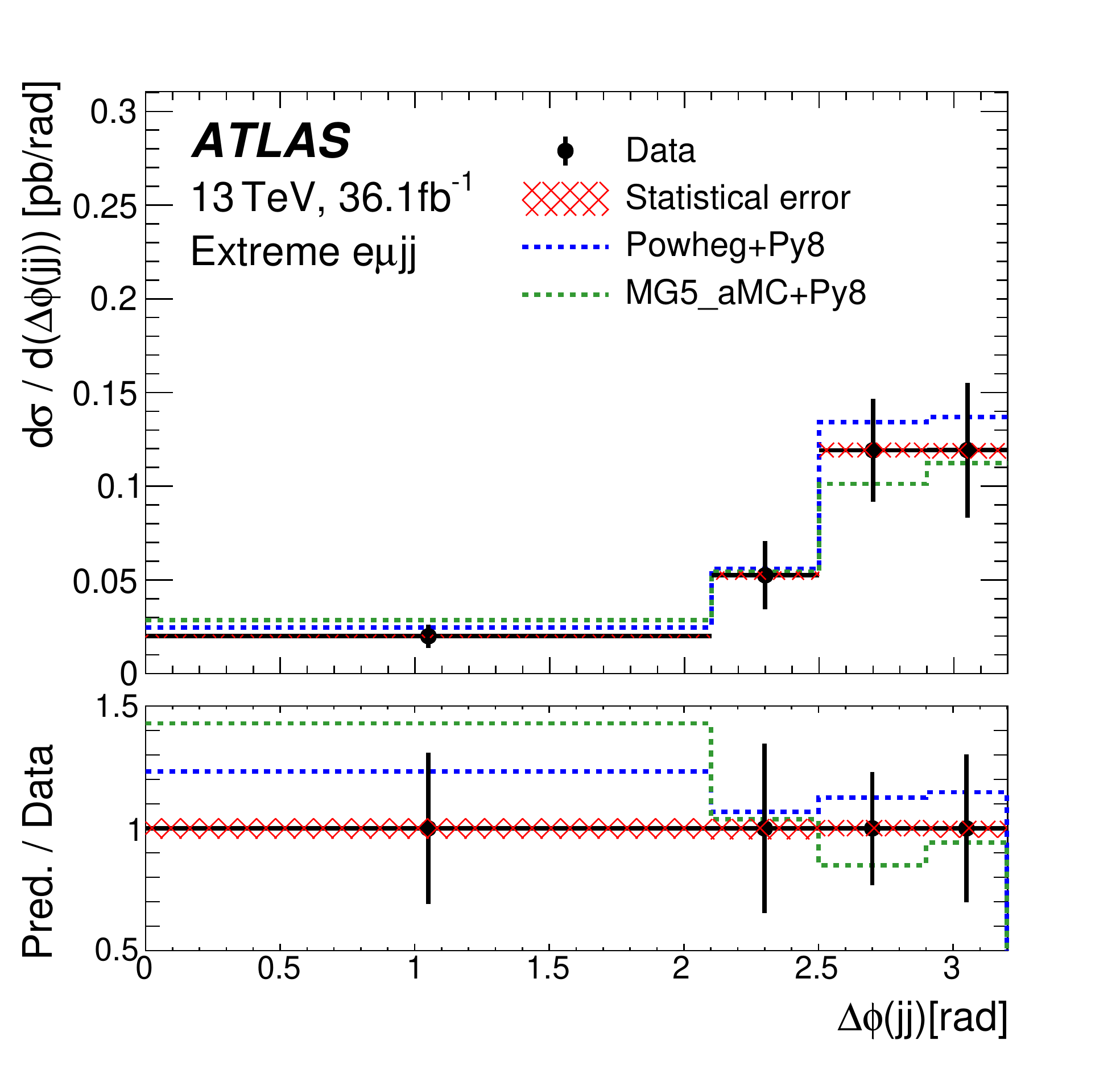}\\
\includegraphics[width=0.30\linewidth]{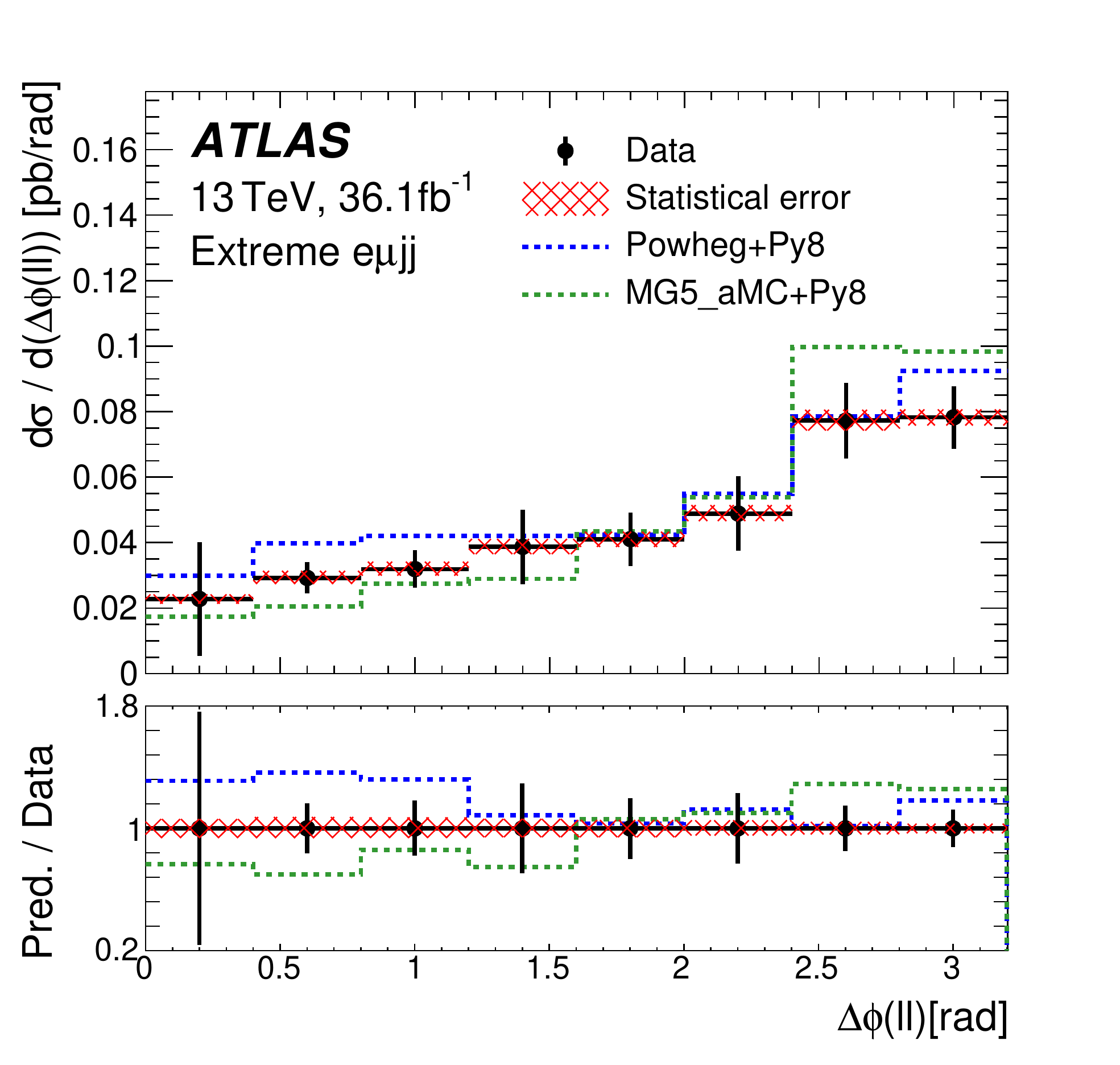}
\includegraphics[width=0.30\linewidth]{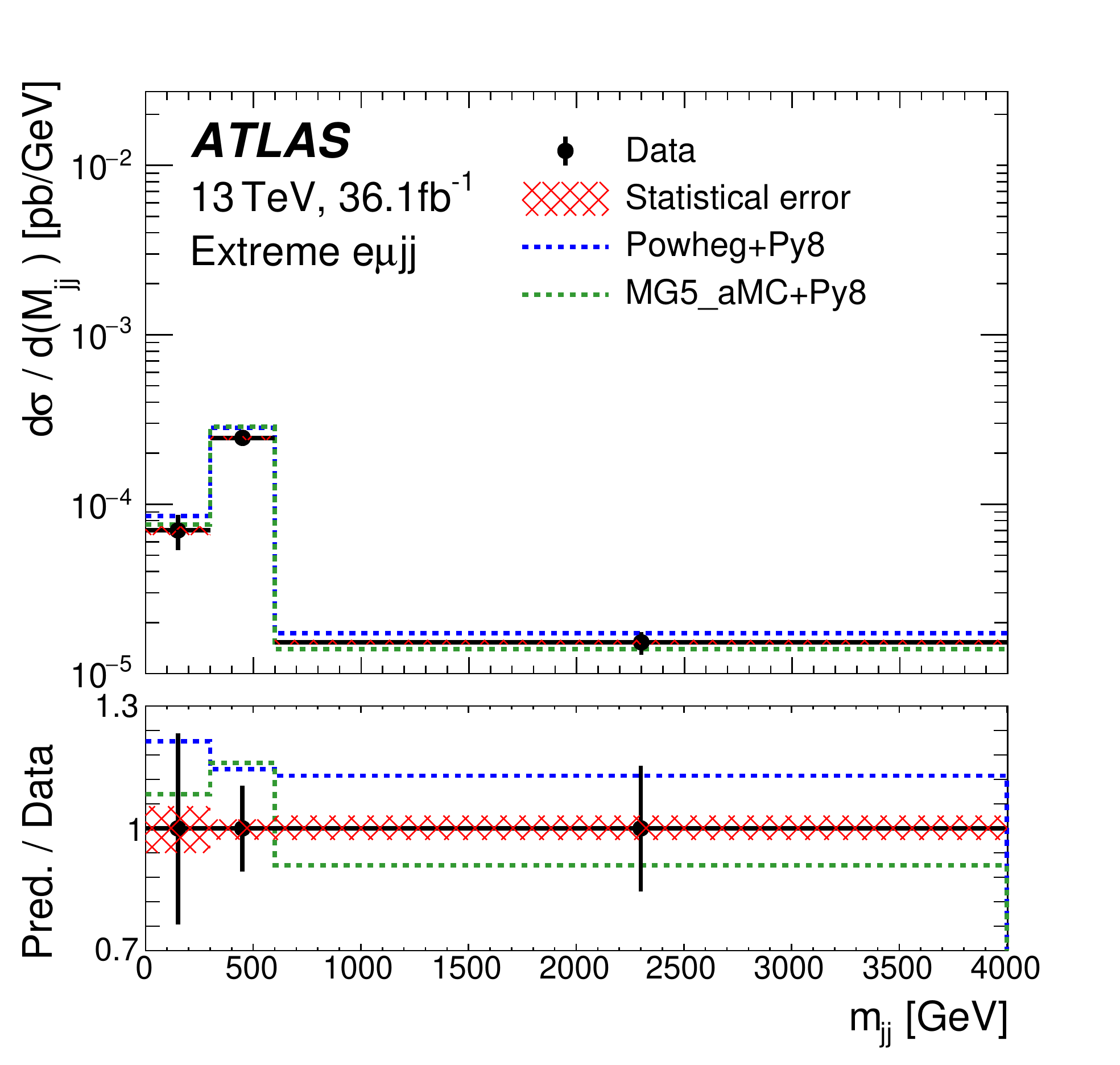}
\includegraphics[width=0.30\linewidth]{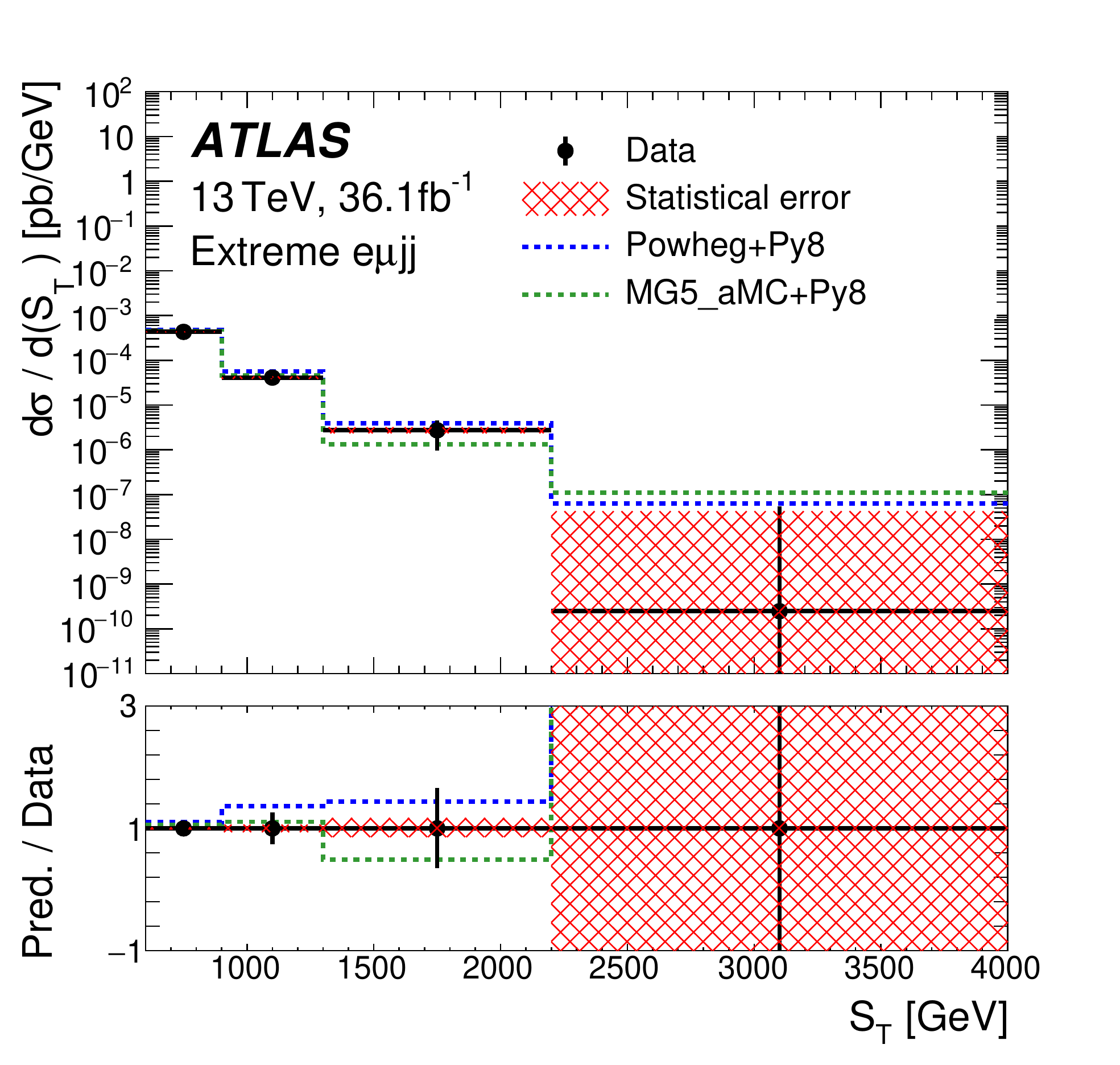}
\caption{The measured generator-level differential cross-sections in the \emujj channel, defined as for the \eejj and \mumujj MRs, but 
requiring the two leptons to have 
different flavours, and with no requirement on the dilepton mass, where an additional selection of $\st > 600~\GeV$ has been applied. The measurements are made exclusively for the \ttbar~process. The MC prediction for the dominant 
process is also shown. The red band represents the statistical component of the total uncertainty which is indicated by the error bar on each point. The variables $ min \Delta \Phi (j0,l) $ and $ min \Delta \Phi (j1,l) $ refers to the minimal difference in $\phi$ between a prompt lepton and the leading or subleading jet respectively. The variable $H_{T}$ refers to the scalar sum of the jet \pt values. The variables $\Delta \eta(jj)$, $\Delta \phi(jj)$ and $\Delta \phi(ll)$ refer to the difference in $\eta$ and $\phi$ between the two leading jets or leptons. The variable $S_{T}$ refers to the scalar sum of the \pt of the jets and leptons.} \label{fig:unfolded_data-1-isCRemjjST}
\end{figure}

The measured differential cross-sections in the MRs where an additional selection of $\st > 600~\GeV$ has been applied are shown in 
Figure~\ref{fig:unfolded_data-1-isZCReejjST} for the \eejj channel, in Figure~\ref{fig:unfolded_data-1-isZCRmmjjST} for the \mumujj channel, and in Figure~\ref{fig:unfolded_data-1-isCRemjjST} for the \emujj channel. 

\FloatBarrier

\section{Training region distributions}
\label{app:trplots}
\begin{figure}
 \centering
\subfloat[]{
  \includegraphics[width=0.42\textwidth]{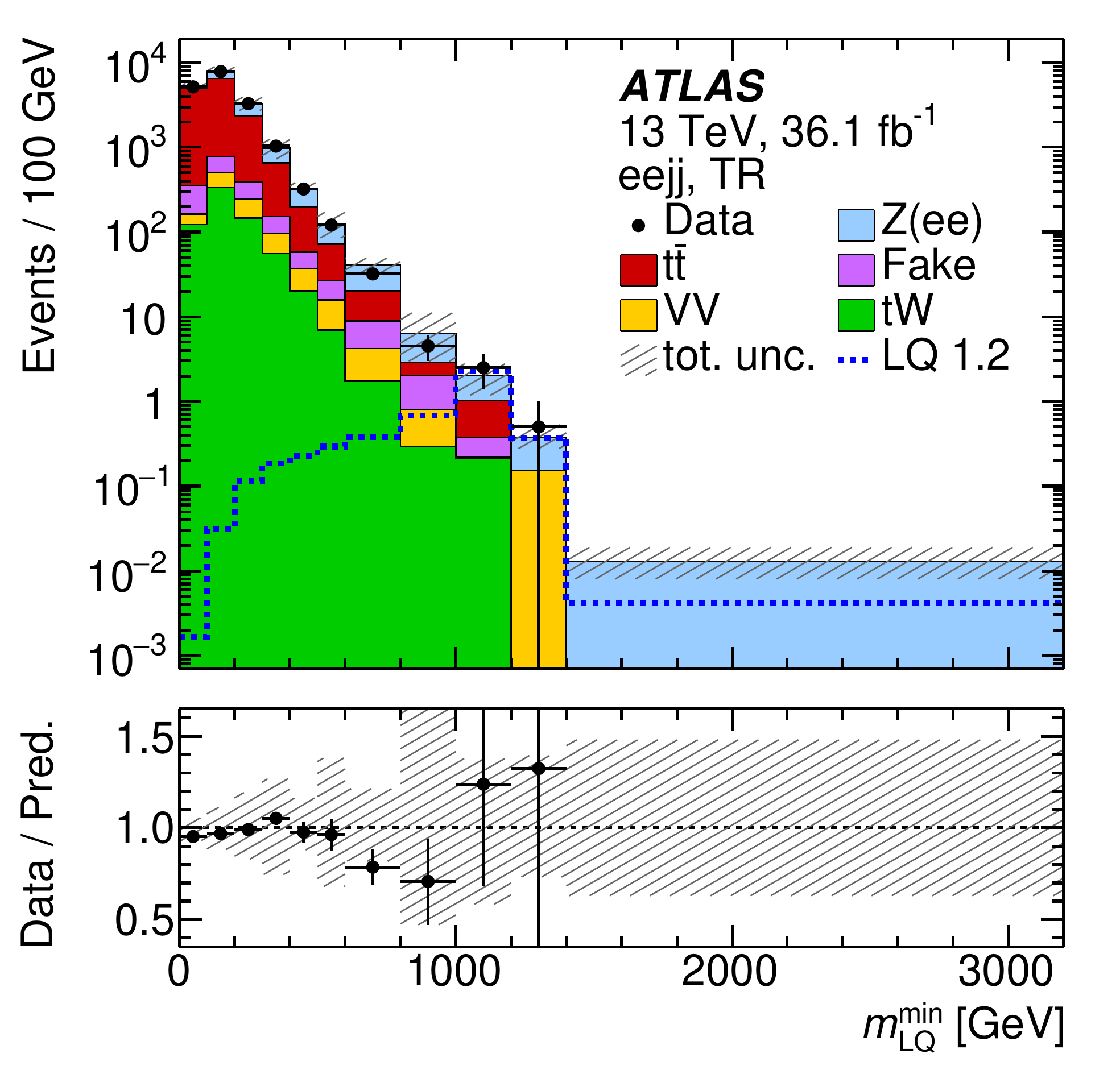}}
\subfloat[]{
  \includegraphics[width=0.42\textwidth]{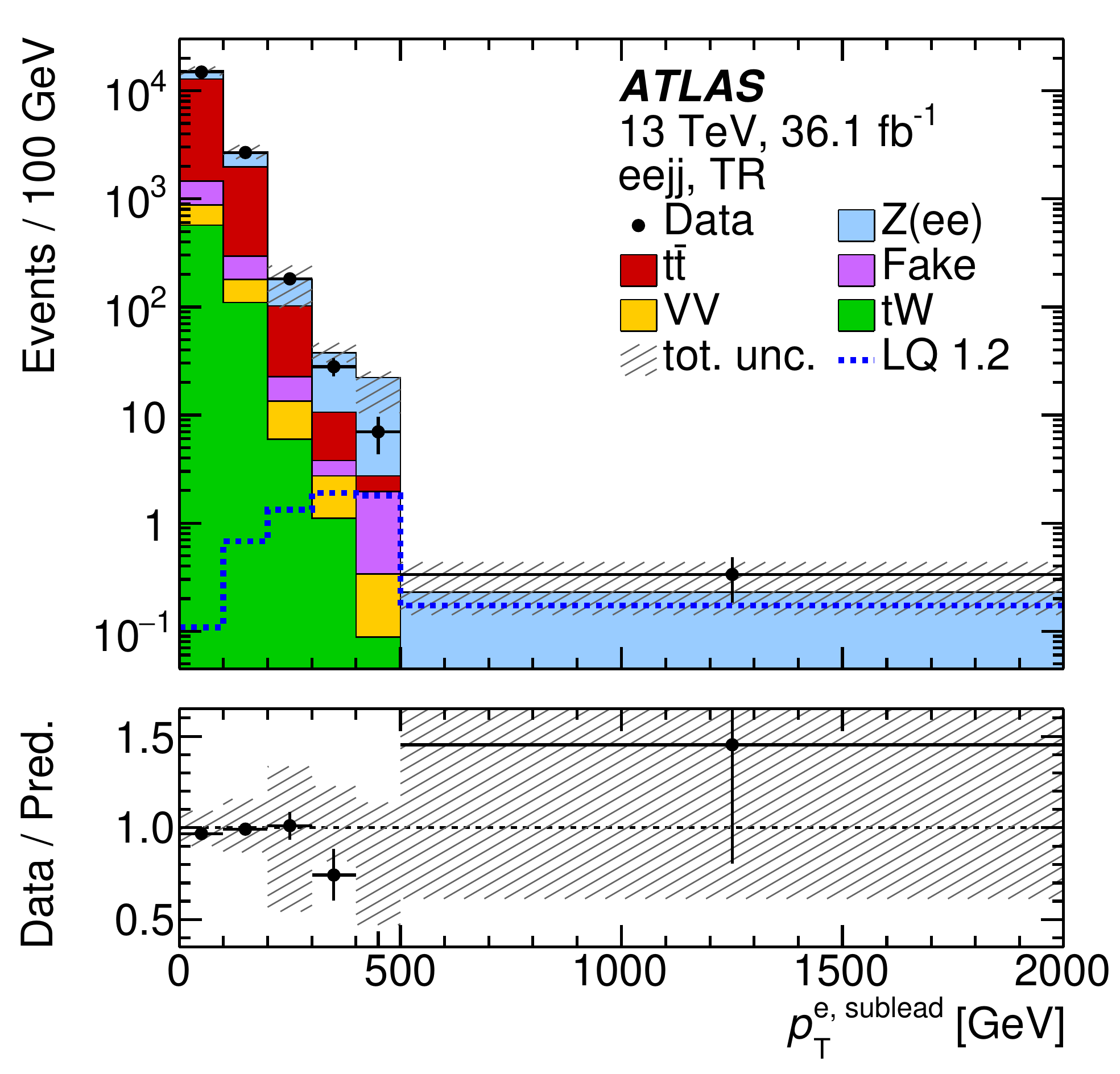}}
\\
\subfloat[]{
  \includegraphics[width=0.42\textwidth]{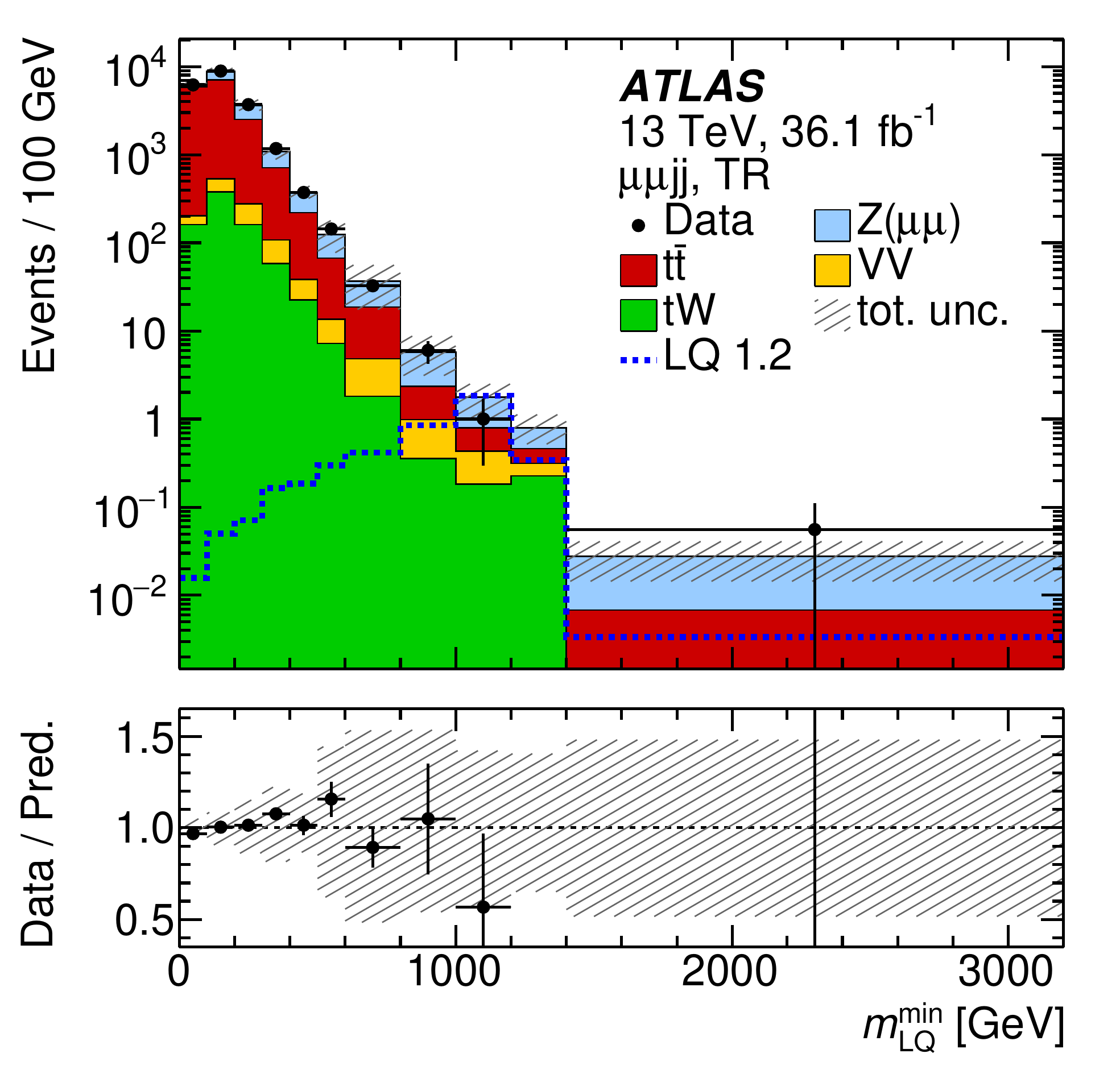} }
\subfloat[]{
  \includegraphics[width=0.42\textwidth]{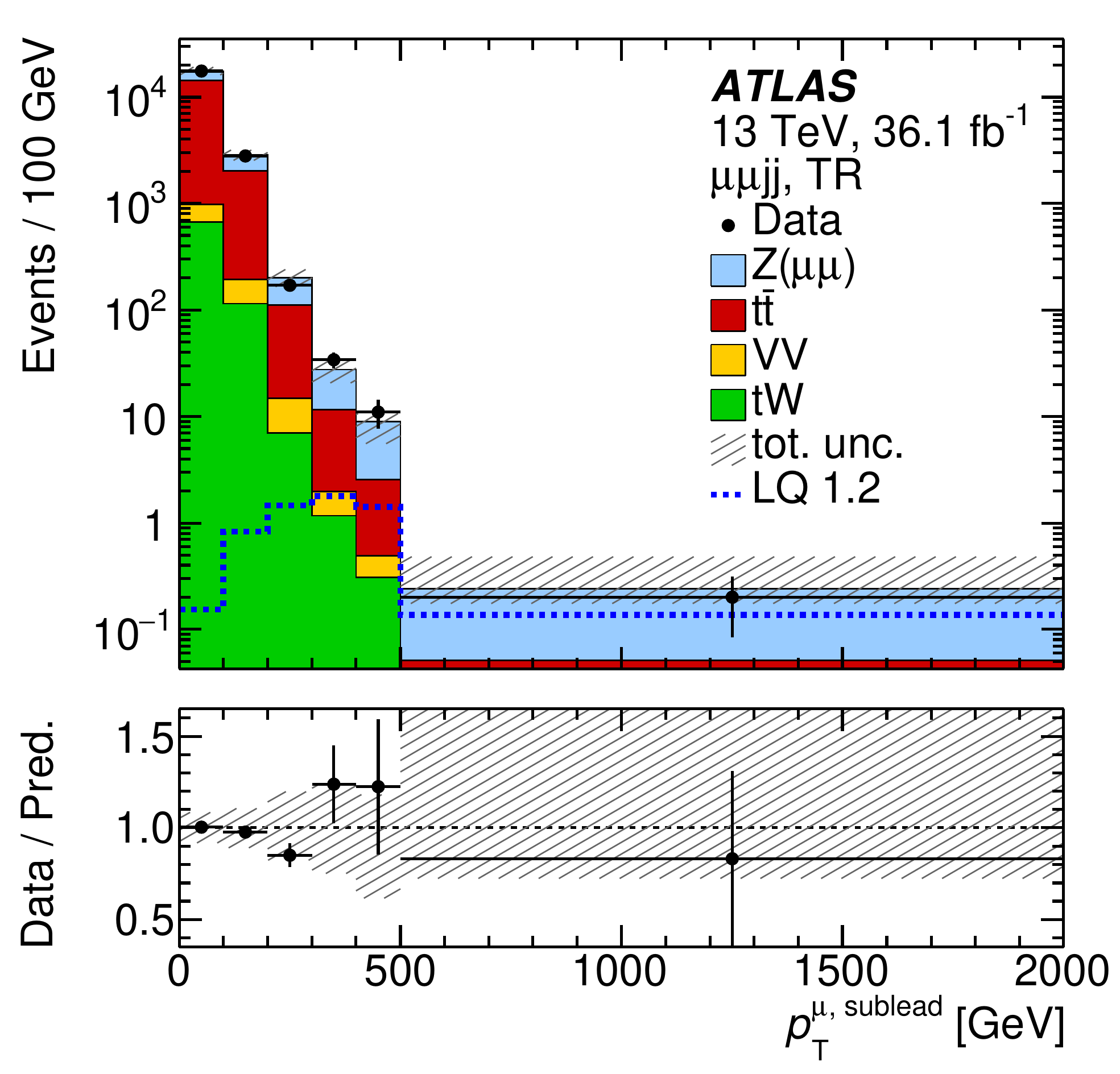}}

\caption{Distributions of \mlqmin{} (left) and \eltwopt{} (right)
 in the training regions (TRs) for the BDT for the \eejj (top) and \mumujj (bottom) channels.
Data are shown together with background contributions corresponding to \ttbar{}, diboson ($VV$), $Z\rightarrow\tau\tau$, $Z \rightarrow ee$, $Z \rightarrow \mu\mu$, single-top
($tW$) processes, and fake electrons. The background distributions are cumulatively stacked. The bottom panels show the ratio of data to expected background. The grey hatched band 
represents
the total uncertainty. Predictions for a LQ with mass of 1.2~\TeV\ are also shown.}
 \label{fig:TR-plots-lljj}
\end{figure}

\begin{figure}
 \centering
\subfloat[]{
  \includegraphics[width=0.42\textwidth]{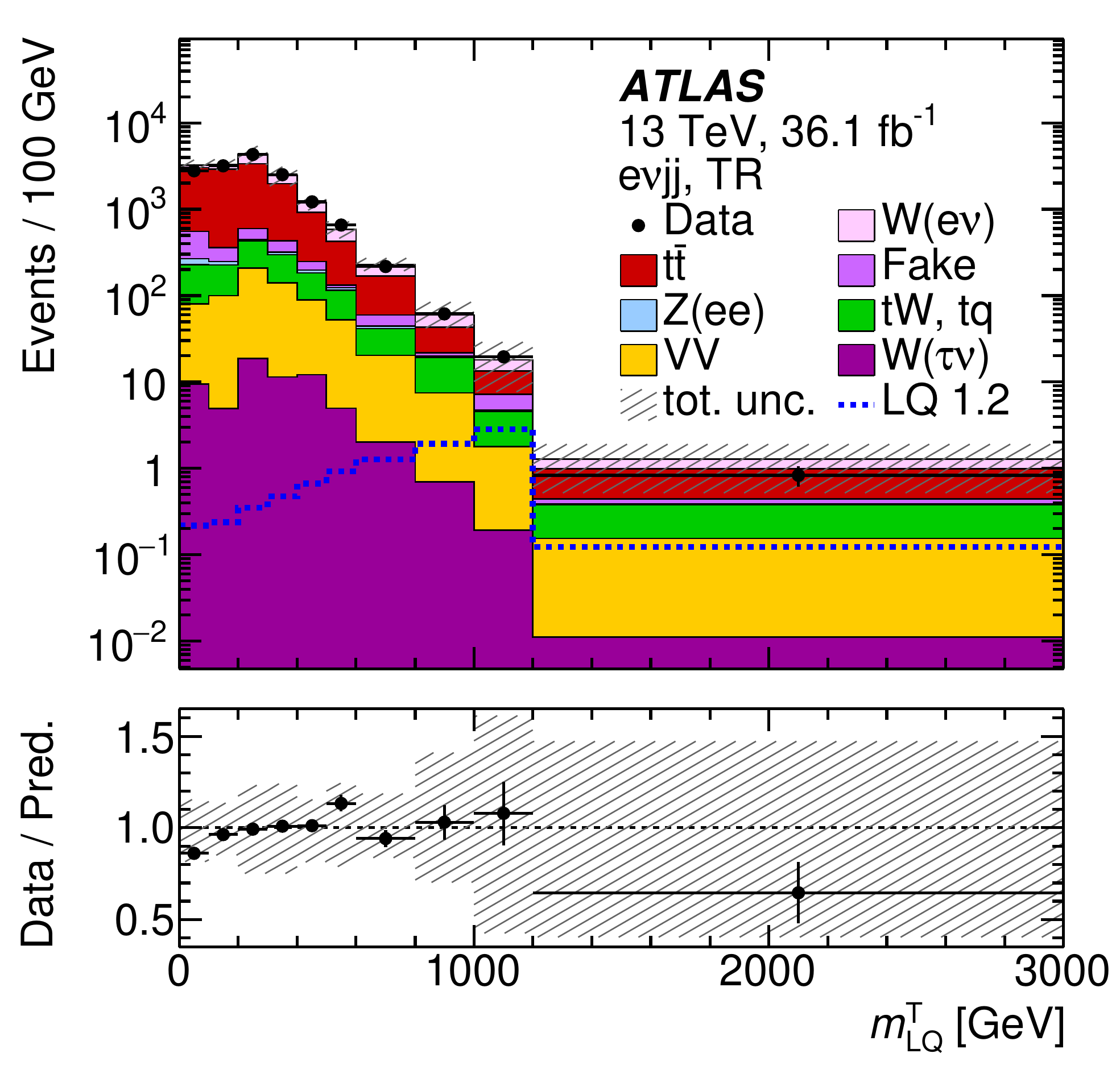}}
\subfloat[]{
  \includegraphics[width=0.42\textwidth]{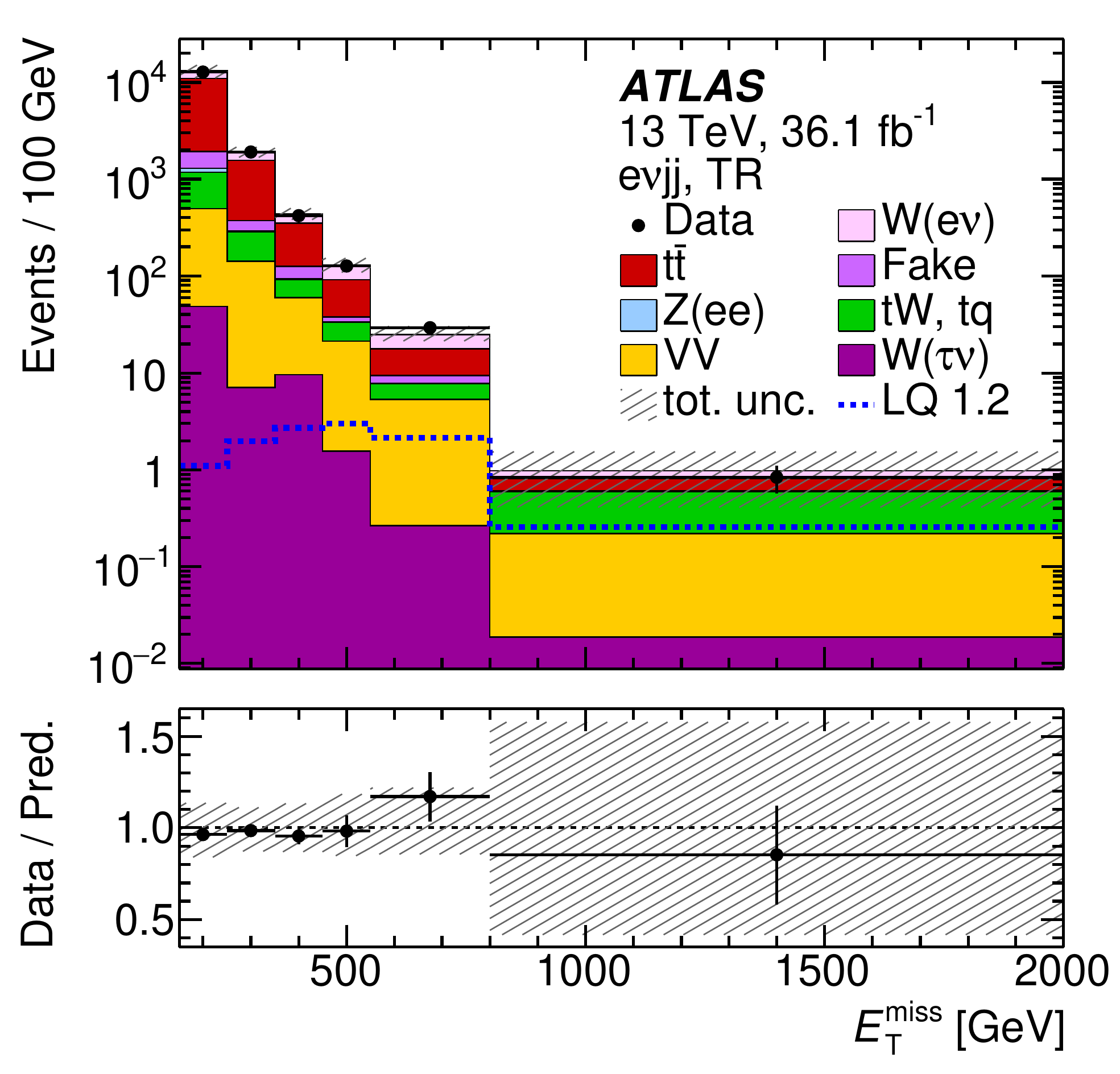}}
\\
\subfloat[]{
  \includegraphics[width=0.42\textwidth]{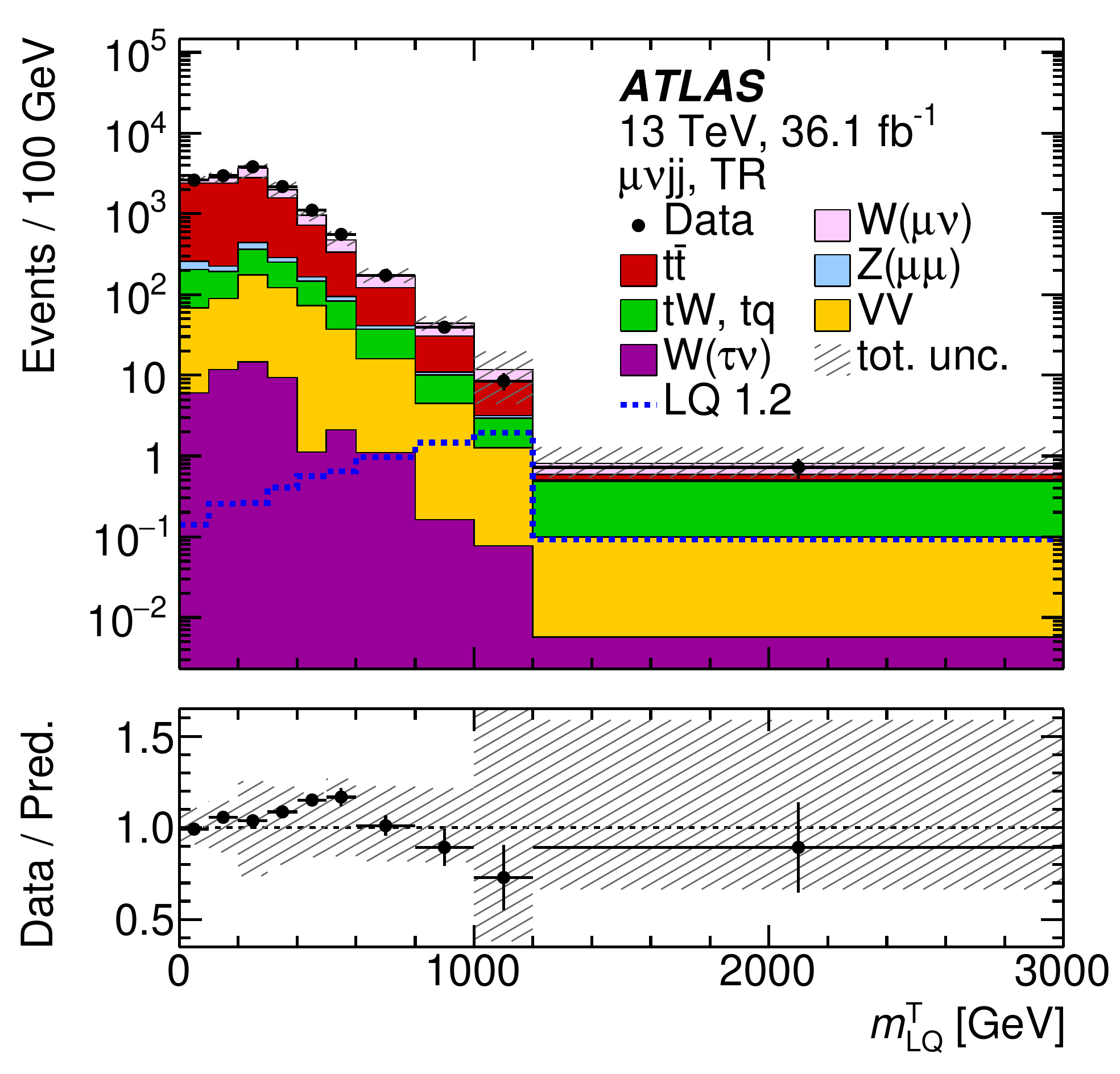} }
\subfloat[]{
  \includegraphics[width=0.42\textwidth]{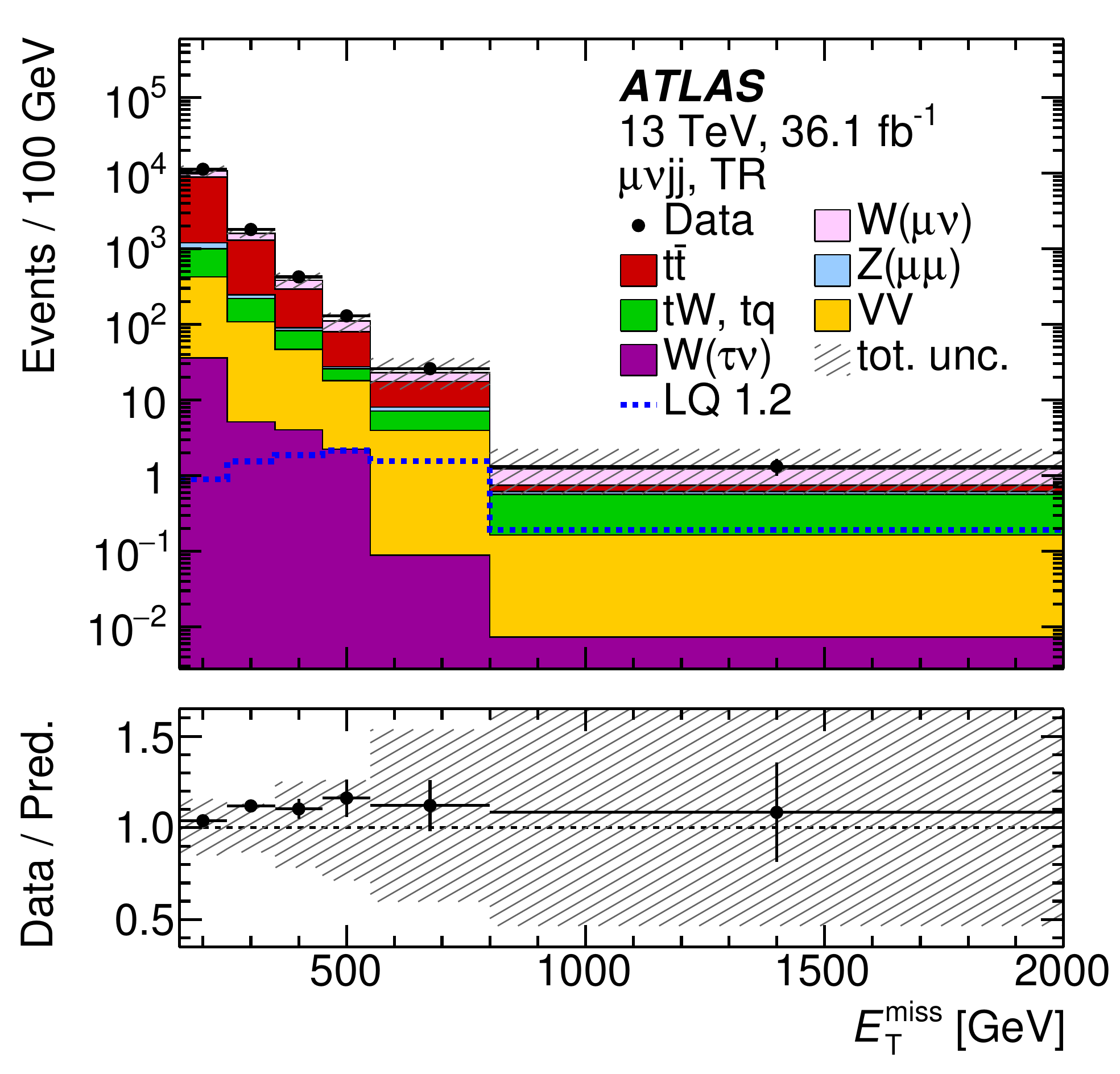}}

\caption{Distributions of \mlqt{} (left) and \met{} (right)
 in the training regions (TRs) for the BDT for the \enujj (top) and \munujj (bottom) channels.
Data are shown together with background contributions corresponding to \ttbar{}, diboson ($VV$), $Z\rightarrow\tau\tau$, $Z \rightarrow ee$, $Z \rightarrow \mu\mu$, single-top
($tW$) processes, and fake electrons. The background distributions are cumulatively stacked. The bottom panels show the ratio of data to expected background. The grey hatched band 
represents
the total uncertainty. Predictions of a LQ with mass of 1.2~\TeV\ are also shown. }
 \label{fig:TR-plots-lnujj}
\end{figure}

Figures~\ref{fig:TR-plots-lljj} and~\ref{fig:TR-plots-lnujj} show kinematic distributions for the TRs for the dilepton and lepton--neutrino channels, respectively. 
These regions corresponds to the SRs albeit with the BDT selection removed.
For each channel, one of the most discriminating variables is shown (\mlqmin{} for the dilepton, \mlqt{} for the lepton-neutrino channel). 
In addition, the \met{} distribution is shown for the lepton-neutrino channel, as this is a characteristic variable for this channel newly included in this search. 
For symmetry, the subleading lepton \pt{} is shown in the dilepton channel.

\FloatBarrier

%% file: acknowledgements/Acknowledgements.tex

We thank CERN for the very successful operation of the LHC, as well as the
support staff from our institutions without whom ATLAS could not be
operated efficiently.

We acknowledge the support of ANPCyT, Argentina; YerPhI, Armenia; ARC, Australia; BMWFW and FWF, Austria; ANAS, Azerbaijan; SSTC, Belarus; CNPq and FAPESP, Brazil; NSERC, NRC and CFI, Canada; CERN; CONICYT, Chile; CAS, MOST and NSFC, China; COLCIENCIAS, Colombia; MSMT CR, MPO CR and VSC CR, Czech Republic; DNRF and DNSRC, Denmark; IN2P3-CNRS, CEA-DRF/IRFU, France; SRNSFG, Georgia; BMBF, HGF, and MPG, Germany; GSRT, Greece; RGC, Hong Kong SAR, China; ISF and Benoziyo Center, Israel; INFN, Italy; MEXT and JSPS, Japan; CNRST, Morocco; NWO, Netherlands; RCN, Norway; MNiSW and NCN, Poland; FCT, Portugal; MNE/IFA, Romania; MES of Russia and NRC KI, Russian Federation; JINR; MESTD, Serbia; MSSR, Slovakia; ARRS and MIZ\v{S}, Slovenia; DST/NRF, South Africa; MINECO, Spain; SRC and Wallenberg Foundation, Sweden; SERI, SNSF and Cantons of Bern and Geneva, Switzerland; MOST, Taiwan; TAEK, Turkey; STFC, United Kingdom; DOE and NSF, United States of America. In addition, individual groups and members have received support from BCKDF, CANARIE, CRC and Compute Canada, Canada; COST, ERC, ERDF, Horizon 2020, and Marie Sk{\l}odowska-Curie Actions, European Union; Investissements d' Avenir Labex and Idex, ANR, France; DFG and AvH Foundation, Germany; Herakleitos, Thales and Aristeia programmes co-financed by EU-ESF and the Greek NSRF, Greece; BSF-NSF and GIF, Israel; CERCA Programme Generalitat de Catalunya, Spain; The Royal Society and Leverhulme Trust, United Kingdom. 

The crucial computing support from all WLCG partners is acknowledged gratefully, in particular from CERN, the ATLAS Tier-1 facilities at TRIUMF (Canada), NDGF (Denmark, Norway, Sweden), CC-IN2P3 (France), KIT/GridKA (Germany), INFN-CNAF (Italy), NL-T1 (Netherlands), PIC (Spain), ASGC (Taiwan), RAL (UK) and BNL (USA), the Tier-2 facilities worldwide and large non-WLCG resource providers. Major contributors of computing resources are listed in Ref.~\cite{ATL-GEN-PUB-2016-002}.

%% file: atlas_authlist.tex
 
\begin{flushleft}
{\Large The ATLAS Collaboration}

\bigskip

M.~Aaboud$^\textrm{\scriptsize 34d}$,    
G.~Aad$^\textrm{\scriptsize 99}$,    
B.~Abbott$^\textrm{\scriptsize 125}$,    
O.~Abdinov$^\textrm{\scriptsize 13,*}$,    
B.~Abeloos$^\textrm{\scriptsize 129}$,    
D.K.~Abhayasinghe$^\textrm{\scriptsize 91}$,    
S.H.~Abidi$^\textrm{\scriptsize 164}$,    
O.S.~AbouZeid$^\textrm{\scriptsize 39}$,    
N.L.~Abraham$^\textrm{\scriptsize 153}$,    
H.~Abramowicz$^\textrm{\scriptsize 158}$
% ...
\bigskip
\\

% ...
$^{13}$Institute of Physics, Azerbaijan Academy of Sciences, Baku; Azerbaijan.\\
% ...
$^{34}$$^{(a)}$Facult\'e des Sciences Ain Chock, R\'eseau Universitaire de Physique des Hautes Energies - Universit\'e Hassan II, Casablanca;$^{(b)}$Centre National de l'Energie des Sciences Techniques Nucleaires (CNESTEN), Rabat;$^{(c)}$Facult\'e des Sciences Semlalia, Universit\'e Cadi Ayyad, LPHEA-Marrakech;$^{(d)}$Facult\'e des Sciences, Universit\'e Mohamed Premier and LPTPM, Oujda;$^{(e)}$Facult\'e des sciences, Universit\'e Mohammed V, Rabat; Morocco.\\
% ...
$^{39}$Niels Bohr Institute, University of Copenhagen, Copenhagen; Denmark.\\
% ...
$^{91}$Department of Physics, Royal Holloway University of London, Egham; United Kingdom.\\
% ...
$^{99}$CPPM, Aix-Marseille Universit\'e, CNRS/IN2P3, Marseille; France.\\
% ...
$^{125}$Homer L. Dodge Department of Physics and Astronomy, University of Oklahoma, Norman OK; United States of America.\\
% ...
$^{129}$LAL, Universit\'e Paris-Sud, CNRS/IN2P3, Universit\'e Paris-Saclay, Orsay; France.\\
% ...
$^{153}$Department of Physics and Astronomy, University of Sussex, Brighton; United Kingdom.\\
% ...
$^{158}$Raymond and Beverly Sackler School of Physics and Astronomy, Tel Aviv University, Tel Aviv; Israel.\\
% ...
$^{164}$Department of Physics, University of Toronto, Toronto ON; Canada.\\
% ...
$^{a}$ Also at Borough of Manhattan Community College, City University of New York, NY; United States of America.\\
$^{b}$ Also at California State University, East Bay; United States of America.\\
$^{c}$ Also at Centre for High Performance Computing, CSIR Campus, Rosebank, Cape Town; South Africa.\\
$^{d}$ Also at CERN, Geneva; Switzerland.\\
$^{e}$ Also at CPPM, Aix-Marseille Universit\'e, CNRS/IN2P3, Marseille; France.\\
% ...
$^{*}$ Deceased

\end{flushleft}
